# C-arm inverse geometry CT for 3D cardiac chamber mapping

by

## Jordan M. Slagowski

A dissertation submitted in partial fulfillment of
the requirements for the degree of

Doctor of Philosophy

(Medical Physics)

at the

UNIVERSITY OF WISCONSIN - MADISON

2017

Date of final oral examination: August 17, 2017

Members of the Final Oral Examination Committee:

Michael A. Speidel, Ph.D., Assistant Professor, Medical Physics

Guang-Hong Chen, Ph.D., Professor, Medical Physics

Sean B. Fain, Ph.D., Professor, Medical Physics

Timothy P. Szczykutowicz, Ph.D., Assistant Professor, Radiology

Amish N. Raval, M.D., Associate Professor, Medicine

*Dedicated to my family.*

# Acknowledgements

I would like to acknowledge the many individuals, without whom I could not have completed this work. First and foremost, I would like to express my sincere gratitude to my advisor, Dr. Michael Speidel. Thank you for your guidance, encouragement, and ideas that helped me past the many challenges encountered throughout this work. It was an absolute joy learning from you and working with you on this project. I learned many lessons including how to select problems that are worthwhile of research, to apply the scientific process and make evidence-based conclusions, and to effectively communicate ideas to others.

Next, I would like to thank Dr. Guang-Hong Chen for teaching me the physics of CT imaging. Your enthusiasm for science and CT is contagious and inspired me throughout this work. I'd like to express my sincere appreciation to Dr. Amish Raval for always making me feel welcome to interrupt and ask clinical questions related to this work. I am grateful for the support and guidance you provided throughout my graduate education. I would also like to thank Dr. Tim Szczykutowicz and Dr. Ke Li for many useful suggestions and conversations that enhanced the quality of this work.

This work would not have been possible without the excellent technical support provided by Triple Ring Technologies, Inc. Thank you to Dr. Tobias Funk, Dr. Joe Heanue, Peter Howard, Dr. Jamie Ku, Willem-Jan Ouberg and other scientists and engineers for their support. In particular, thank you to Paul Kahn for deepening my understanding of the SBDX hardware and software.

The time spent working on this project would not have been nearly as enjoyable without the support of many of my friends and colleagues at UW-Madison. I am grateful to Dr. Jie Tang and Dr. Yinghua Tao who taught me to code on graphics processing units. A huge thank you to my friend and fellow labmate Dr. Dave Dunkerley for talking SBDX the past five years. Thank you to former labmates Dr. Mike Tomkowiak and Dr. Charles Hatt for showing me where to find the good coffee and providing many useful suggestions related to this work. I must acknowledge all of my friends in the CT research group who have made my time in graduate school more enjoyable. A special thanks to Dr. John Garrett and Yinsheng Li for your shared enthusiasm discussing new CT reconstruction algorithms and implementation tricks, as well as Dr. Yongshuai Ge. I look forward to reading your future work. Thank you, Daniel Gomez-Cardona, for assisting with many CT experiments. There are countless other individuals whom I must acknowledge including Dr. Mike Van Lysel, Dr. Chuck Mistretta, Dr. Charlie Strother, Dr. Frank Ranallo, Dr. Sean Fain, Dr. Sebastian Schafer, Dr. Tim Hacker, Gary Frank, Dr. Martin Wagner, Dr. Marc Buehler, and Lindsay Bodart.

Lastly, I would like to thank my family. I am grateful to my parents who have always supported and encouraged my education. I hope this work makes you proud. To my brothers, thanks for being such great friends. And most of all, I would like to thank my best friend and wife, Lauren. Your endless support, encouragement, and patience in letting me pursue this means everything. I love you, always.

-Jordan Slagowski (July 18, 2017)

# Abstract

C-arm inverse geometry CT for 3D cardiac chamber mapping

Jordan M. Slagowski

Under the supervision of Dr. Michael A. Speidel
At the University of Wisconsin-Madison

Image-guided navigation of catheter devices to anatomic targets within large 3D cardiac chambers and vessels is challenging in the interventional setting due to the limitations of a conventional 2D x-ray projection format. Scanning-beam digital x-ray (SBDX) is a low-dose inverse geometry x-ray fluoroscopy technology capable of real-time 3D catheter tracking. SBDX performs rapid tomosynthesis using an electronically scanned multisource x-ray tube and photon-counting detector mounted to a C-arm gantry. While this technology could facilitate 3D image-guided navigational tasks, SBDX currently lacks the ability to perform volumetric computed tomography from a rotational C-arm scan. C-arm CT is an expected feature of interventional x-ray systems that could provide, for example, the necessary 3D cardiac chamber roadmap during catheter ablation of left atrial fibrillation. This work develops a novel volumetric CT capability for the SBDX platform, termed C-arm inverse geometry CT, suitable for rotational scans of the beating heart. The work is divided into three tasks: development of image reconstruction algorithms, implementation on the SBDX hardware, and performance assessment for the example task of 3D cardiac chamber mapping. SBDX-CT data acquisition is performed by simultaneous x-ray source scanning at 15 scan/s and C-arm rotation over a 190º short-scan arc in 13.4 seconds. An iterative reconstruction method based on prior image constrained compressed sensing was developed to accommodate fully truncated projections and data inconsistency resulting from cardiac motion during rotation. Hardware implementation included development of a C-arm angle measurement method, development of a geometric calibration method to account for non-ideal C-arm rotations, and detector response nonlinearity correction. The geometric calibration procedure mitigated artifacts from C-arm deflection during rotation. SBDX-CT image quality was evaluated in terms of artifacts, uniformity, and spatial resolution in a series of static phantom studies. Dynamic phantom studies evaluated chamber segmentation accuracy in the presence of chamber motion and field-of-view truncation. Segmentation error was quantified as the 99th percentile of a histogram of the surface deviations from the reference. For a chest phantom containing an atrium undergoing 60-88.2 cycle/minute motion and imaged at 50% full power, segmentation errors were 3.0-4.2 mm. Feasibility of *in-vivo* SBDX-CT was demonstrated in a porcine model.

# Table of Contents

# List of figures

# List of tables

# 1  Introduction

## 1.1  Overview

Percutaneous transcatheter interventions in the cardiac catheterization lab have become a widely accepted treatment option for a range of cardiovascular diseases. These procedures are less invasive than surgical alternatives and offer several potential advantages, including treatment for inoperable patients,[1] reduced recovery times, less blood transfusion,[2] and potential for cost reduction depending on the disease and procedure.[3] Advances in catheter device technology have also enabled new types of procedures, beyond traditional coronary artery interventions. These include treatment of structural heart conditions such as atrial and ventricular septal defects or fistulas,[4–6] valvular diseases such as aortic and mitral valve stenosis,[7–9] and electrophysiological conditions such as atrial fibrillation.[10,11,12,13] However, as catheter-based interventions become increasingly complex, the demands on the interventional image guidance system have grown.

X-ray fluoroscopy is the dominant imaging modality used in the cardiac catheterization laboratory (i.e. cath lab) as it provides excellent spatial and temporal resolution that is useful for visualizing high-contrast catheter-based devices.[14] However, the 2D projection format and poor soft tissue contrast are limiting, particularly in a growing class of non-coronary based electrophysiology procedures and structural heart interventions where devices are navigated within relatively large 3D anatomic chambers and/or vessels.[15–17] In these procedures, knowledge of the catheter or device position relative to the target anatomy is critical to avoid rare but potentially fatal complications. For example, errant catheter motion during radiofrequency catheter ablation of atrial fibrillation (RFCA) can result in esophageal puncture, a potentially fatal condition referred to as atrioesophageal fistula. Similarly, during transcatheter aortic valve replacement (TAVR), the prosthetic valve must be deployed at a precise location defined relative to the aortic annulus to prevent embolization into the left ventricle or aorta.[18,19] Soft-tissue contrast provided by fluoroscopic imaging is insufficient for the necessary visualization of the atrium wall or aortic annulus in these example procedures. The integration of CT or MR derived 3D anatomic roadmaps with real-time 2D fluoroscopic imaging has the potential to improve the visualization of soft tissue targets, reduce long procedure times, and reduce the risk of complications attributed to navigation errors.

Computed tomography or MR imaging performed prior to an interventional procedure may be used to produce the 3D anatomic roadmap of a cardiac chamber.[20,21] However, *intraprocedural* 3D imaging (i.e. imaging performed in the cath lab immediately before, during, or after a procedure while the patient is on the table) is desirable as it may yield a more accurate representation of the patient's anatomy at the time of the intervention, improve registration accuracy with live fluoroscopic imaging,[22] and be used to validate device positioning post-deployment (e.g. stents placed within intracranial arteries).[23] Three dimensional rotational angiography performed by C-arm rotation of a fluoroscopic system's x-ray tube and image intensifier was demonstrated in 1997 for neuroradiologic applications by *Fahrig et al.*[24] Soon after, *Jaffray et al.* constructed a bench-top cone-beam CT (CBCT) system that utilized an amorphous silicon flat-panel detector for applications in image-guided radiotherapy.[25] Two years later, in 2002, *Jaffray et al.* mounted an x-ray tube and flat-panel detector to a slowly rotating (2 °/s) medical linear accelerator.[26] In 2005, *Siewerdsen et al.* replaced the image intensifier on a mobile C-arm fluoroscopic system with a CsI:Tl flat-panel detector and demonstrated C-arm CT for image-guided spinal surgery in animal studies.[27] Following the introduction of flat panel detectors, C-arm CT has become an expected and critical feature of commercially available fluoroscopy systems. C-arm CT products offered by major vendors now include XperCT (Philips Healthcare, Andover, MA), DynaCT (Siemens Healthcare, Forchheim, Germany), Innova CT (GE Healthcare, Waukesha, WI), and Infinix (Toshiba Medical Systems Corporation, Tochigi, Japan).

C-arm CT has become a reliable imaging modality widely used for applications in image guided surgery and neuroradiology focused on static anatomy. Commercial solutions for imaging dynamic organs such as the heart, however, are not yet readily available. Projection data inconsistency resulting from long C-arm data acquisition times (e.g. 6-14 sec.) that span several heart beats manifest as image artifacts. To mitigate artifacts observed in cardiac C-arm CT, *Lauritsch et al.* proposed retrospective ECG-gating of multiple rotational acquisitions.[28] In 2012, *Chen et al.* presented an iterative reconstruction method that can be used to generate images at arbitrary cardiac phases from a single C-arm rotational acquisition.[29] The development of cardiac C-arm CT could be used to provide anatomic roadmaps during interventional procedures. As the complexity and number of cardiac interventions performed annually continues to rise,[30] the need for improved 3D image guidance and reduction in fluoroscopy associated radiation dose is likely to increase. The conventional 2D fluoroscopic projection format limits the ways in which live imaging can be visualized with a 3D roadmap. Long fluoroscopic imaging times may result in peak skin doses exceeding 2 Gy that produce immediate and/or delayed deterministic skin injuries including erythema, epilation, and telangiectasia.[31] Of equal concern, radiation dose to the interventional staff, resulting primarily from x-ray scatter, is associated with an increased probability of cancer and cataract formation.[32–34]

Scanning Beam Digital X-ray (SBDX) is a novel *inverse geometry* x-ray fluoroscopy imaging system designed for cardiac interventions that addresses both the radiation dose concerns and limitations of the 2D projection format associated with conventional fluoroscopy.[35,36] Shown in Figure 1.1, SBDX performs fluoroscopic imaging using a rapidly scanned narrow x-ray beam, high speed photon counting detector, and real-time image reconstructor.

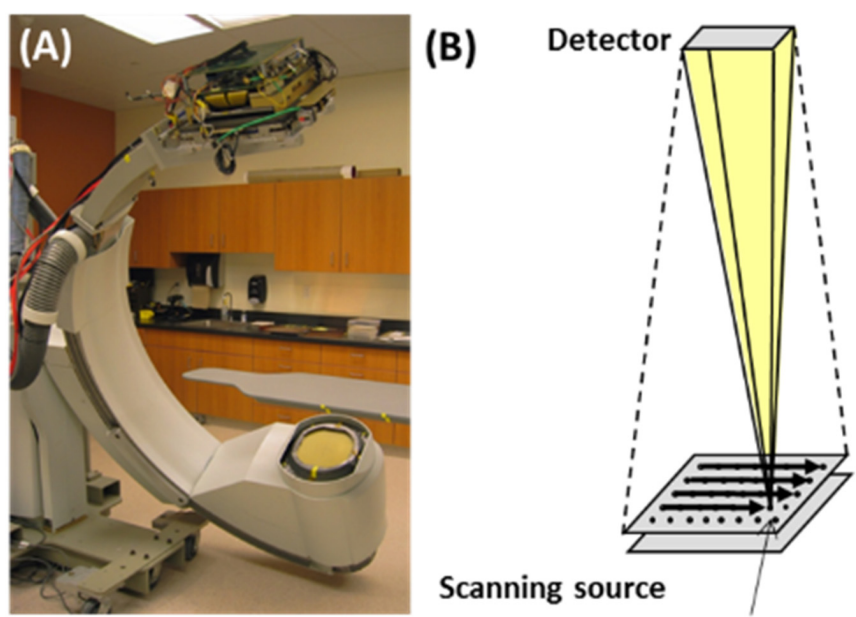

Figure 1.1: (A) Scanning Beam Digital X-ray (SBDX) is an inverse geometry x-ray fluoroscopy system. (B) SBDX utilizes a multi-focal spot scanning source to produce x-rays that are focused by a multi-hole collimator onto a photon counting detector array.

Improved dose efficiency relative to a traditional cone-beam design is achieved through i) the use of narrow x-ray beams and a large airgap, that minimize image-degrading scatter, ii) a 2 mm thick CdTe detector which maintains high detection efficiency at high kV, and iii) an "inverse geometry" design, which spreads x-rays out over a large entrance area and reduces dose to points on the skin. SBDX has been shown capable of achieving a 3-7 times reduction in entrance exposure at equivalent SNR versus a conventional fluoroscopy system.[37] SBDX is inherently a real-time tomosynthesis system due to the use of an electronically scanned source array and stationary detector.[36] Real-time tomosynthesis enables SBDX to perform frame-by-frame 3D tracking of high-contrast objects, such as cardiac catheters.[38,39] Notably, *Dunkerley* recently developed and validated a method for real-time 3D tracking and visualization of catheters to support electrophysiology ablation and structural heart interventions (Figure 1.2).[40] In addition to catheter tracking, SBDX tomosynthesis has been exploited for a number of applications in interventional cardiology including calibration-free vessel measurements for device sizing,[41] stereoscopic fluoroscopy,[42] and registration of transesophageal echocardiography to x-ray fluoroscopy.[43]

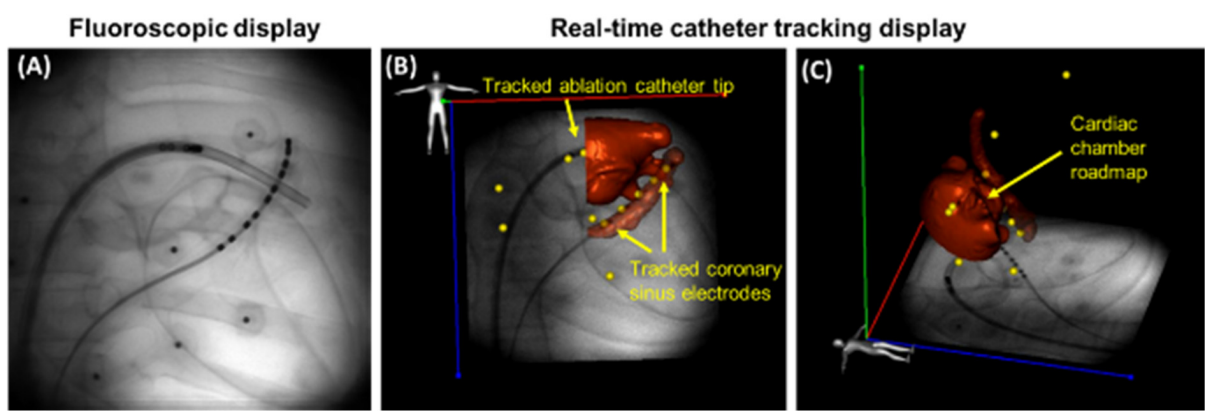

Figure 1.2: (A) An example SBDX fluoroscopic image is shown. High-contrast catheter devices are displayed with excellent spatial and temporal resolution. (B) Cather devices are tracked using SBDX tomosynthesis imaging and displayed in real-time relative to an imported cardiac chamber model. The integration of three-dimensional models may supplement poor soft tissue contrast present in fluoroscopy and improve visualization of anatomic targets relative to catheter tip position. (C) Anatomic models may be rotated and translated within the SBDX catheter tracking display. Figure provided with permission of the author of Ref.[40].

While SBDX is a promising technology for both dose reduction and 3D device guidance in the cath lab, SBDX currently lacks the ability to perform 3D C-arm CT in the interventional setting. Cone beam CT is an expected and critical feature of commercially available x-ray fluoroscopy systems. Development of a 3D CT capability for SBDX could be used, for example, to produce cardiac chamber roadmaps for the SBDX catheter tracking display (Figure 1.2) *during* interventional procedures. Inverse geometry CT (IGCT) imaging using a rotating multi-focal spot x-ray source and smaller-area detector array has been investigated previously for *diagnostic* applications.[44–48] The concept of IGCT was originally proposed by *Schmidt at al.* to increase volumetric coverage with minimal cone-beam artifacts and reduced x-ray scatter.[44,45,49] Comparable to the SBDX system, IGCT systems utilize either a large-area scanned x-ray source array or multiple discrete x-ray sources collimated to a smaller-area detector (Figure 1.3A). This contrasts with conventional C-arm CT systems (Figure 1.3B) that utilize a single source focal spot opposite a larger-area detector. Since the introduction of IGCT, *Baek et al.* (8 focal spots) and *De Man et al.* (32 focal spots) have demonstrated inverse geometry CT on a rotating gantry similar to that used in diagnostic CT.[46,47]

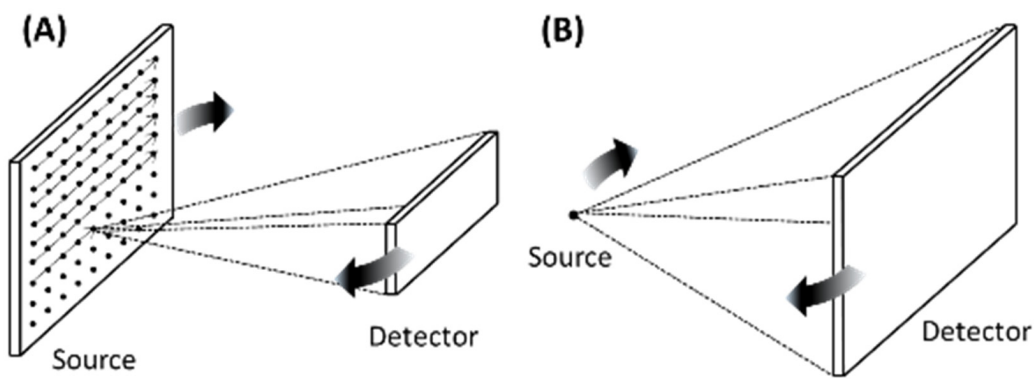

Figure 1.3: Inverse geometry CT systems (A) utilize a large area scanned source or multi-source array opposite a smaller area detector. Each of the individual focal spots produce a narrow-width cone-beam focused on the detector array. In contrast, third generation cone beam CT systems utilize a single point like source focused on a larger area detector.

Implementation of inverse geometry CT on the C-arm based SBDX system presents unique challenges not encountered with previous gantry-based scanners. First, the SBDX fluoroscopic field-of-view was designed to cover a smaller area containing the heart, as opposed to the full thorax coverage available with diagnostic CT scanners. As a result, IGCT image reconstruction methods must be developed to reduce truncation artifacts.[50] Second, C-arm rotation times for CT data acquisition may be more than an order of a magnitude longer than gantry based systems which introduces data inconsistency from cardiac motion.[29] The resulting artifacts can compromise the accuracy of a 3D cardiac chamber roadmap derived from an SBDX rotational CT scan. Third, similar to conventional C-arm CBCT, the geometric uncertainty introduced by a flexion prone C-arm during rotation may decrease spatial resolution and create unacceptable image artifacts that are avoided by more stable gantry-based systems.[51] A C-arm geometric calibration method specific to the SBDX system must be developed. And finally, previous IGCT prototypes have focused on small phantoms (e.g. cadaver ears) and small animals (e.g. rats, rabbits).[45–47] Larger phantoms or animals comparable in size to potential human cath lab patients must be considered. Each of these challenges for C-arm IGCT is addressed in this dissertation work.

## 1.2  Dissertation organization

*The purpose of this dissertation was to develop a computed tomography capability for the SBDX system. The target application for this SBDX CT method is 3D cardiac chamber mapping of the endocardial surface to support cardiac ablation and structural heart interventional procedures.* The development of a CT imaging capability for SBDX was divided into three parts.

i)    Development of an image reconstruction method that handles the unique multi-source geometry, fully truncated projection data, and data inconsistency introduced by cardiac motion during slow C-arm rotation.

ii)   Implementation of hardware calibration techniques to deal with geometric uncertainty, detector non-linearity, and the potential effects of beam hardening and x-ray scatter.

iii)  Validation of the SBDX CT imaging method for the task of 3D cardiac chamber imaging through phantom and animal studies.

A brief overview of the completed work is described here on a chapter by chapter basis. *Chapter 2* provides the clinical motivation for this work, reviews the current methods available to generate 3D cardiac chamber maps, and describes the operating principles of the Scanning Beam Digital X-ray system. *Chapter 3* investigates the feasibility of performing CT imaging with SBDX. SBDX-CT data acquisition and sampling are discussed, and several previously proposed IGCT reconstruction techniques are implemented and evaluated for the cardiac chamber imaging task. *Chapter 4* presents an iterative reconstruction method for SBDX-CT that accounts for the reconstruction challenges identified in chapter 3, including field-of-view truncation and cardiac motion induced data inconsistency. *Chapter 5* presents techniques for implementation of SBDX-CT on the imaging hardware. This includes a novel single-view geometric calibration method for the rotating C-arm, and an investigation of the effects of nonlinear detector response, beam hardening, and x-ray scatter on cardiac chamber map accuracy. *Chapter 6* presents initial SBDX-CT results obtained with a rotating C-arm for a static phantom. SBDX-CT performance is characterized in terms of spatial resolution, image uniformity and artifacts. *Chapter 7* reports phantom studies performed to evaluate the segmentation accuracy in imaging scenarios with simulated cardiac motion. The results of an *in vivo* animal study are also detailed. *Chapter 8* summarizes the development of inverse geometry CT with a rotating C-arm and areas for future work.

# 2  Background

## 2.1  Introduction

This chapter details three example cardiac interventional procedures which may benefit from 3D imaging via intraprocedural C-arm CT: radiofrequency catheter ablation (RFCA), transcatheter aortic valve replacement (TAVR), and targeted transendocardial injection of therapeutic agents. The limitations of conventional 2D fluoroscopic imaging and advantages of inverse geometry fluoroscopy are discussed. Finally, the construction and operating principles of the SBDX system are described.

## 2.2  Cardiac interventional procedures

### 2.2.1  Radiofrequency catheter ablation of atrial fibrillation

Radiofrequency catheter ablation of atrial fibrillation is an example procedure that may benefit from C-arm CT and is considered the model application for this work. Atrial fibrillation (AF) is the most common cardiac arrhythmia affecting 2.7 million to 6.1 million individuals in the United States with the expectation that AF prevalence will double by 2050.[52] In a healthy individual, an electric signal propagates throughout the heart and initiates contractions that pump blood throughout the body (Figure 2.1A). Patients suffering from AF experience irregular or abnormal heartbeats due to disorganized electrical impulses that occur in the atria (Figure 2.1B). Ineffective contractions resulting from the arrhythmia may lead to blood pooling in the left atrium or left atrial appendage thereby increasing the potential for clot formation which has been shown to increase the risk of stroke.[53] Ultimately, AF is responsible for an estimated 15-25% of all stroke cases in the United States making AF a major contributor to one of the Centers for Disease Control and Prevention's top ten public health concerns.[54–57]

Atrial fibrillation is attributed to a number of factors including genetics, obesity, excessive alcohol consumption, abnormalities due to pre-existing structural heart disorders, and inflammation which results in atria dilation.[58–60] The exact mechanisms responsible for arrhythmia formation remain under investigation,[61] however, evidence suggests at least two major triggers exist. First, changes in the cardiac substrate may lead to abnormal electrical signal propagation referred to as reentry. During reentry, an abnormal signal circuit may form that interrupts the normal signal propagation from the SA node to the AV node triggering irregular heartbeats.[57,61] For the second mechanism, ectopic activity that originates at a number of different locations in the cardiac anatomy, including the pulmonary veins (PV), atrial wall, interatrial septum, coronary sinus and crista terminalis, may trigger the arrythmia.[62] The most common sites of ectopic activity are within or at the base of the pulmonary veins.[63]

Treatment is prescribed on a patient specific basis dependent on the disease progression and individual risk factors and symptoms. For certain individuals, treatment may consist solely of antiarrhythmic drugs and/or lifestyle changes. Radiofrequency catheter ablation of (for) atrial fibrillation is an alternative and effective treatment option to restore sinus rhythm for AF patients deemed unsuitable or for which antiarrhythmic drugs are ineffective.[64,65] Interventional treatment methods aim to eliminate the AF focal spot triggers or electrically isolate the spontaneous signals originating at the pulmonary veins from the normal signal pathway. During a RFCA procedure, a catheter is inserted into a femoral vein and advanced to the patient's right atrium using fluoroscopic imaging to assist with navigation. A trans-septal puncture is then performed to advance the catheter to the left atrium where the ablation catheter tip is steered towards the focal spot trigger (e.g. pulmonary vein ostia) or target. After reaching the target location, radiofrequency energy is delivered to the myocardium tissue, as shown in Figure 2.1C, to create a series of lesions that either isolate the abnormal electrical signals responsible for the arrhythmia from propagating and interfering with the normal signal pathway or destroy the trigger mechanism itself.[66] Lesions may be administered in a number of different patterns including linearly between the superior pulmonary veins, linearly between the left inferior pulmonary vein and mitral valve annulus, or to regions within the right atrium.[55] However, the most common strategy is to create a circumferential series of lesions around the pulmonary vein openings.[66,67]

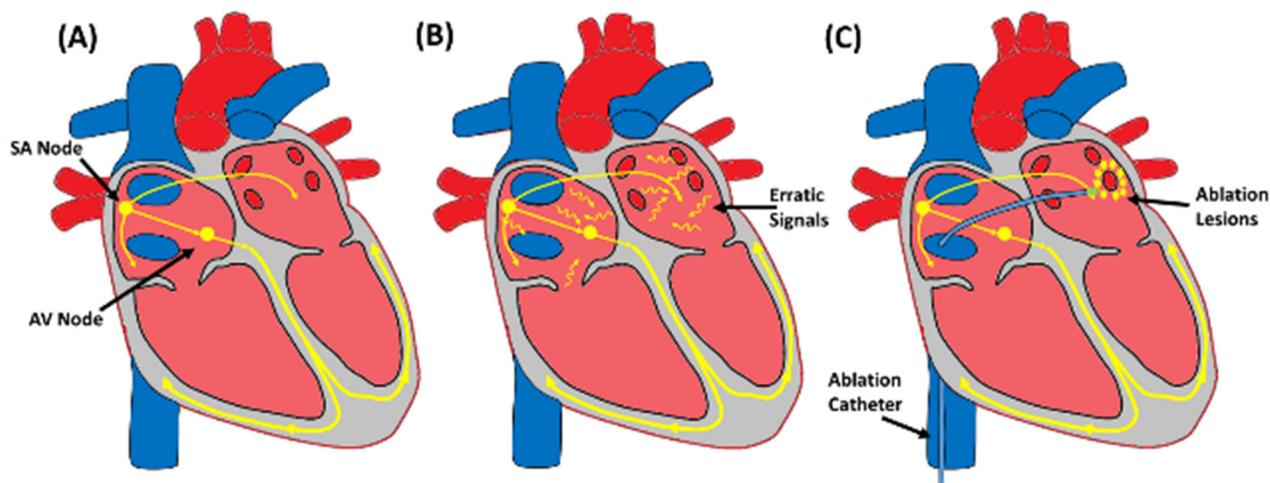

Figure 2.1: (A) Electrical impulses originating at the sinoatrial node signal contraction of the left and right atria for a normal heart. After a slight delay at the atrioventricular node, the electrical signal triggers ventricular contractions. (B) In the case of atrial fibrillation, erratic signals, often originating from the left atrium pulmonary veins, disrupt normal signal propagation resulting in a cardiac arrhythmia. (C) Radiofrequency catheter ablation of atrial fibrillation is portrayed. A circumferential series of lesions around the pulmonary vein ostia serve to isolate focal spot triggers and erratic signals from normal signal propagation.

While x-ray fluoroscopy provides useful anatomic context relative to catheter position during RFCA, it is insufficient for catheter navigation to target ablation sites on the complex interior surface of the atrium wall, due to limited soft tissue contrast and lack of 3D device visualization. Incomplete circumferential pulmonary vein isolation is associated with higher rates of AF recurrence.[68] Inadequate imaging is also a contributing factor to long RFCA procedure times that increase costs. RFCA procedures have been reported to last up to 348 +/- 69 minutes.[69] Furthermore, complications may arise from errant catheter navigation within cardiac chambers including cardiac tamponade, pulmonary vein stenosis, fistula formation, and organ perforation.[70] Cardiac tamponade (i.e. blood or fluid within the pericardium) is the leading cause of death during RFCA and may occur during catheter manipulation, administration of ablation lesions, or transeptal puncture.[71] Accidental ablation lesions administered *within* the pulmonary vein ostia may cause pulmonary vein stenosis, a potentially fatal condition.[72] Precise identification of the target anatomy surrounding the PV ostia is critical to avoid PV stenosis. Long procedure times, complications attributed to errant catheter manipulation, and inaccurate identification of cardiac anatomy that may contribute to incomplete pulmonary vein isolation motivate the development of improved intraprocedural imaging techniques for device tracking and 3D soft tissue visualization.

Non-fluoroscopic electroanatomic mapping systems that track 3D catheter position, discussed in Sec. 2.3, have been developed to guide electrophysiology procedures. The integration of static CT, MR, or CBCT derived anatomic roadmaps offer improved visualization of soft tissue anatomy such as the pulmonary veins and surrounding structures like the esophagus. Identification of the esophageal anatomy is important to avoid atrioesophageal fistula formation, a rare but potentially fatal complication of RFCA.[73] Three dimensional anatomic models may also be used to define target ablation sites and use is associated with reduced procedure times.[20] Additional benefits and the specific imaging modalities used to acquire 3D cardiac chamber models are discussed in Sec. 2.3 below.

### 2.2.2  Transcatheter aortic valve replacement

Aortic stenosis (AS) is the most common cardiac valve disease in the United States affecting more than 25% of adults over the age of 65 years.[74] AS is attributed to progressive or congenital valve calcification which leads to valve narrowing (i.e. stenosis) requiring increased left ventricle pressure to compensate for the reduced blood outflow. With time, the associated left ventricular hypertrophy may lead to heart failure.[75] Patients suffering from severe symptomatic AS deteriorate rapidly with death occurring in more than 50% of untreated cases within two years.[76]

Currently, cardiothoracic open heart surgery and valve replacement is the preferred treatment option for AS.[77] Surgery may not be an option, however, for certain high-risk patients. As an alternative to surgery, *Cribier et al.* demonstrated the feasibility of percutaneous transcatheter aortic valve replacement for an inoperable human subject.[7] Since then, the TAVR procedure has been refined and is now an accepted treatment option for high-risk or inoperable human subjects recommended by the American College of Cardiology and American Heart Association.[1,77] Clinical studies are currently underway to evaluate TAVR as a potential alternative treatment option for AS in lower risk patients. A large randomized clinical trial published in March of 2017 showed TAVR to be a viable alternative to surgery in a cohort of 1,660 intermediate-risk patients, supporting TAVR as an emerging treatment option for AS.[78]

Transcatheter aortic valve replacement procedures are performed in the cath lab and utilize a combination of x-ray fluoroscopy, echocardiography, and potentially C-arm CT for image guidance.[79–81] The first step in a TAVR procedure is to obtain vasculature access. Transfemoral access is typically preferred, although subclavian and direct aortic routes are also possible. A prosthetic valve crimped to a balloon catheter is advanced through the vasculature under fluoroscopic guidance to the location of the diseased aortic valve. The replacement valve is then positioned at the aortic annulus and deployed by balloon expansion.

A successful valve implantation is dependent on precise valve positioning relative to the plane of the aortic annulus. The workflow for the TAVR procedure is limited by a lack of full 3D image guidance to visualize prosthetic valve deployment relative to the aortic valve anatomy. X-ray fluoroscopy provides only a 2D representation of the 3D prosthetic valve and aortic anatomy. Incorrect valve deployment is associated with a number of serious risks. For example, if the valve is positioned too superior or inferior to the plane of the annulus it may embolize into the aorta or left ventricle, respectively. Additional complications from incorrect positioning may include paravalvular regurgitation or obstruction of the coronary arteries.[82] Thus, intraprocedural imaging techniques that offer improved visualization of the prosthetic valve relative to the aorta and left ventricular outflow tract are needed.

Currently, prior to valve deployment, the C-arm must be rotated to provide a 2D fluoroscopic view in which the three cusps (i.e. leaflets) of the aortic valve appear co-linear. The optimal C-arm angle is often estimated from an iterative series of angiograms with iodine contrast injection which can be time consuming and increase radiation and contrast exposure.[83] The feasibility of C-arm CT to determine the ideal fluoroscopic projection angle to visualize valve deployment has also been demonstrated.[84] The intraprocedural 3D CT volume may be manipulated to virtually determine the optimal projection angle as the volume is inherently registered with the fluoroscopic system. The benefits of the C-arm CT approach include potential contrast reduction by minimizing the number of repeated aortagrams, as well as the option to overlay a 3D roadmap of the aortic anatomy with fluoroscopic imaging.

### 2.2.3   Targeted transendocardial injection of therapeutic agents

Targeted transendocardial injection of therapeutic agents is a third potential application for C-arm CT and real-time catheter tracking. Acute myocardial infarction occurs when the blood supply to the myocardium is restricted or blocked resulting in myocardial necrosis.[85] Following necrosis, ventricular remodeling may occur that is associated with the progression of heart failure.[86] Catheter-based injection of therapeutic agents such as stem cells or gene transfer agents is currently being explored to repair tissue at the infarct border following myocardial infarction. Intraprocedural imaging is essential to accurately deliver the stem cells or biologic agents to the ischemic yet viable tissue or target regions adjacent to the scar zone. Knowledge of the 3D catheter position relative to the myocardium is also important to avoid perforation of the fragile necrotic tissue. Three-dimensional anatomic roadmaps may be used to improve visualization of target injection sites or regions-of-interest to avoid during transendocardial drug delivery.[87] SBDX-CT combined with real-time catheter tracking could potentially be used to assist catheter navigation to injection targets and also to record the points corresponding to injection sites on a CT derived roadmap. In addition to generating 3D roadmaps, C-arm CT could potentially be used for cardiac perfusion imaging as a means to identify target injection sites by determining tissue regions with perfusion defects. In the scheme proposed by *Girard et al.*, iodine contrast agent was administered and four C-arm CT data acquisitions, with recorded ECG signal, were acquired simultaneously using back and forth C-arm rotation.[88] Eight multi-sweep C-arm CT acquisitions were performed in total, immediately after iodine contrast injection and at intervals 1, 5, 10, 15, 20, 25, and 30 minutes after contrast injection. Contrast enhancement in the cardiac tissue was then quantified to identify regions of perfusion defects. The late enhancement of contrast agent in infarcted scar tissue could be used to identify and denote target injection sites on a 3D roadmap to guide the injection process.

## 2.3  Three-dimensional endocardial surface mapping

Radiofrequency catheter ablation, TAVR, and catheter-based stem cell injection procedures could each benefit from improved visualization of soft tissue anatomy and 3D device tracking. As RFCA is the target application for this work, a discussion of current imaging technologies available for device tracking and 3D endocardial surface imaging is provided here. Electroanatomic mapping systems developed to assist in electrophysiology procedures play three major roles during RFCA.[89] First, EM systems are used for activation or voltage mapping to identify target ablation sites. Second, EM systems may be used to track the 3D position of catheters in real-time. And third, a mapping catheter may be maneuvered around the endocardial surface and tracked to collect a series of points that are combined to generate a 3D anatomic roadmap. Currently, the two most widely used commercial EM systems are CARTO (Biosense Webster, Baldwin Park, CA, USA) and Ensite NavX (St Jude Medical, St. Paul, MN, USA) which use either magnetic field generators in combination with catheter-embedded sensors or electric field generators and catheter electrodes for tracking.[90]

While EM systems are widely used in electrophysiology procedures, several limitations exist. EM systems i) introduce additional equipment into the cath lab, ii) may bound the operator to specific EM compatible catheter devices, iii) may require catheter device modification with EM compatible sensors that change the mechanical properties of the device, and iv) do not provide the anatomic context relative to catheter position that exists with x-ray fluoroscopy. Most relevant to this work however, 3D chamber models provided by EM mapping exhibit poor spatial resolution as the models are rendered from a discrete number of points corresponding to the tracked mapping catheter position. As a result, EM provided roadmaps may fail to provide the necessary delineation of complicated anatomical structures.[91] Additionally, the manual mapping procedure may require long acquisition times that increase the likelihood of mapping errors due to patient motion. For demonstration, Figure 2.2 presents a 3D ventricle model reconstructed from a series of 77 tracked points using the NOGA (Biosense Webster, Baldwin Park, CA,

USA) electroanatomic mapping system at University Hospital in Madison, WI. The map required 43 minutes to generate and shows limited spatial resolution.

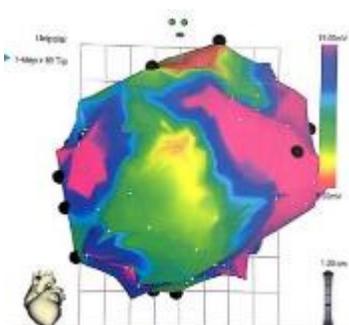

Figure 2.2: A left ventricle model generated from a series of 77 tracked points with an electroanatomic mapping system.

Recognizing the shortcomings of point-by-point 3D electroanatomic chamber mapping, *Dong et al.* integrated 3D cardiac chamber models segmented from previously acquired CT and MR images with EM catheter tracking for use in RFCA.[21,92] The integration of 3D endocardial surface roadmaps with real-time catheter tracking has since been shown to reduce procedure times, reduce radiation exposure from fluoroscopic imaging, and improve procedure outcomes.[28,93,94] For example, *Kistler et al.* observed an improvement in sinus rhythm six months after treatment in 83% of patients treated with an integrated 3D roadmap versus 60% for patients treated using EM mapping techniques.[95]

Following Dong's initial experience, others continued to investigate image fusion to guide catheter ablation procedures. Three dimensional anatomic roadmaps may now be derived from preprocedural CT images,[15,20,95] preprocedural MR images,[21] intraprocedurally acquired 3D ultrasound,[96] intraprocedural MRI, or intraprocedurally acquired C-arm CT images.[97–100] Intraprocedurally acquired roadmaps (defined as roadmaps acquired after the patient has been placed on the table in the cath lab) offer several potential advantages compared to pre-acquired roadmaps. First, changes in the patient's anatomy that may occur between the time of a preprocedure imaging study and the intervention are avoided with intraprocedural CBCT. *Starek et al.* showed a statistically significant difference in esophagus position when comparing preprocedural CT images versus intraprocedural CBCT.[101] Distortions caused by the insertion of catheter devices may also alter the anatomy. Second, C-arm CT performed in the cath lab could improve procedure workflow by eliminating the need to schedule and transport the patient to and from an imaging study in a separate location.[102] And third, the registration accuracy of the anatomic roadmap with x-ray fluoroscopy could potentially be improved by acquiring the roadmap in the same coordinate system or through 2D-3D registration techniques that utilize catheter device constraints that are not possible with preprocedural images.[22]

Given the advantages of generating 3D anatomic roadmaps immediately prior to a procedure, intraprocedural MR, echocardiography and CBCT are all appealing imaging techniques for roadmapping. MRI provides excellent soft tissue contrast but suffers from potentially long data acquisition times. The use of MRI in the cardiac cath lab has been demonstrated but requires special MR compatible equipment and may not be an option for patients with preexisting non-compatible devices.[103] Ultrasound offers excellent spatial and temporal resolution but studies have shown that it may not provide acceptable representations of the pulmonary veins.[104,105] Meanwhile, CBCT offers high spatial resolution, short data acquisition times (5-14 seconds), and is compatible with existing devices without modification. Intraprocedural CBCT imaging with iodine contrast agent has been shown to offer sufficient image quality for the task of atrium mapping, while also reducing errors due to changes in patient anatomy observed with pre-procedure imaging. [99,106] Further potential advantages of CBCT demonstrated by *Kriatselis et al*. in a human subjects study (70 patients) include reduced radiation dose, shorter RFCA procedure times, and arguably lower cost from eliminating pre-procedural imaging studies.[106]

## 2.4  Scanning Beam Digital X-ray

Scanning Beam Digital X-ray is an inverse geometry x-ray fluoroscopy system designed for dose reduction and depth-resolved imaging during cardiac interventions.[35–37] The system consists of a C-arm mounted multi-focal spot scanning source, a multi-hole collimator, photon counting detector array, and a real-time image reconstructor (Figure 2.3). The original SBDX system concept dates back to work performed in 1996 by Cardiac Mariners, Inc. in collaboration with Dr. Michael Van Lysel of the University of Wisconsin - Madison.[107] Since then, SBDX has undergone several design and development cycles, with the most recent prototype being presented at the 2015 SPIE Medical Imaging conference and summarized below.[36] The SBDX system used in this project was constructed and supported by Triple Ring Technologies, Inc. (Newark, CA).

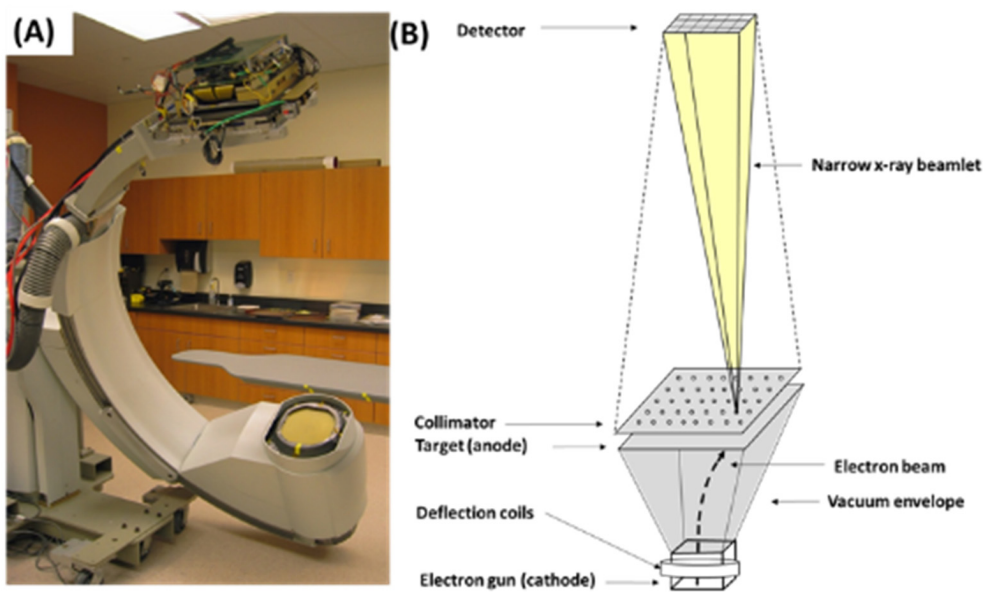

Figure 2.3: Scanning Beam Digital X-ray consists of a multi-focal spot scanned source array, multi-hole collimator, and photon-counting detector. An electron beam is swept, using deflection coils, across discrete dwell positions on a transmission-style tungsten anode to produce x-rays. The x-rays are focused as narrow beams on the detector by a multi-hole collimator.

### 2.4.1 Operating principles

SBDX performs fluoroscopic imaging using high speed x-ray beam scanning in an "inverse geometry" (Figure 2.3). X-rays are produced by electronically steering an electron beam over a 2D array of focal spot positions on a planar transmission-style tungsten target. A multi-hole collimator just beyond the target plane defines a series of narrow overlapping x-ray beams, each directed toward a small-area CdTe photon-counting detector array. In each fluoroscopic frame period (e.g. 1/15 s), the electron beam performs a complete scan of the source. Detector images are acquired for each focal spot position. The detector images are then reconstructed in real time to provide a full field-of-view image for the fluoroscopic frame. The fluoroscopy reconstruction pipeline has two stages. First, shift-and-add digital tomosynthesis (unfiltered backprojection) is performed to generate a stack of image planes parallel to the target plane and centered on the cardiac volume. Second, a pixel-by-pixel plane selection algorithm is applied to the plane stack to generate a display analogous to conventional fluoroscopy.[35,108] The resulting image, termed a multiplane composite, shows the pixel from the plane of highest sharpness and contrast for each pixel position in the 2D output display. A full plane stack and composite image are reconstructed for each scan of the x-ray tube. An individual plane or the composite image can be viewed in real time. Furthermore, as described in Sec 2.4.4, the tomosynthesis plane stack can be analyzed in real-time to perform 3D catheter tracking.

Similar to conventional x-ray fluoroscopy systems, the SBDX source and detector arrays are mounted to a C-arm gantry so that imaging may be performed at arbitrary orientations about the patient. The SBDX C-arm can rotate the source from 100º left anterior oblique (LAO) to 90º right anterior oblique resulting in a 190º range-of-motion. The C-arm is also free to rotate in the cranial-caudal dimension. Figure 2.4 shows the SBDX geometry drawn to scale. Table 2.1 summarizes the system geometry. The nominal source-to-detector distance (SDD) is 150 cm. The nominal source-to-isocenter distance is 45.0 cm. The detector array measures 10.6 cm by 5.3 cm. The source has a 100 x 100 array of focal spot positions (excluding some corner positions), and the focal spots are spaced 2.3 mm on the source target surface. As a result, the extent of the 2D source array measures 23.0 cm by 23.0 cm for the 100 by 100 scan mode. For a typical cardiac imaging scan mode, a 71 by 71 subset of the total 100 by 100 focal spot positions are scanned during each 1/15 second imaging frame period. The 71 by 71 cardiac scan mode provides an 11.4 cm by 11.4 cm fluoroscopic field-of-view at system iso-center which is sufficient for imaging a typical human heart.

The large-area scanned source array and smaller-area detector array make SBDX an inverse geometry x-ray system. The entrance area of the x-ray field increases as the distance to the source decreases. In other words, the area of the x-ray field is larger at the patient entrance surface than at the patient exit surface. This is in contrast to conventional x-ray fluoroscopy systems where the fluence decreases as $1/r^2$, where r is measured from the source. For SBDX, the x-ray fluence decreases as $1/r^2$, where r is measured from the detector. The inverse geometry design increases the system dose efficiency for reasons discussed in section 2.4.4 below.

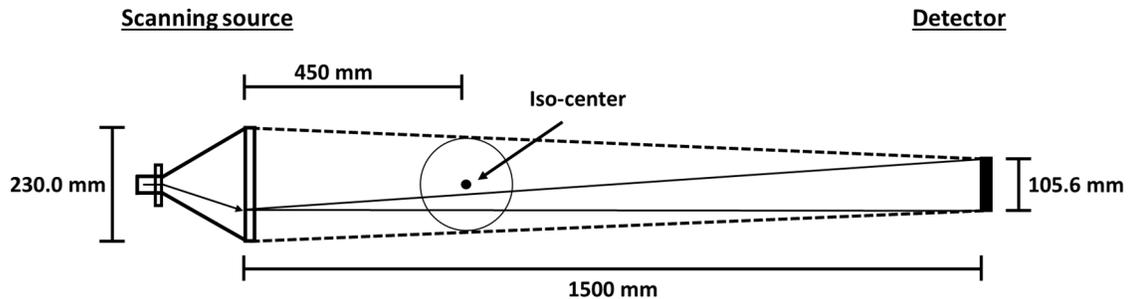

Figure 2.4: The SBDX system geometry is shown to scale. For inverse geometry x-ray fluoroscopy, the x-ray fluence decreases as $r^{-2}$ where r is the distance from the detector.

Table 2.1: SBDX nominal system geometry.

| | |
|---|---|
| *Nominal system geometry* | |
| Source-detector-distance (SDD) | 1500 mm |
| Source-axis-distance (SAD) | 450 mm |
| C-arm rotation angular range | 190º |
| *Scanning-source array* | |
| Native focal spot positions | 100 x 100 |
| Focal spots used in cardiac scan mode | 71 x 71 |
| Focal spot pitch | 2.3 x 2.3 mm |
| *Detector array* | |
| Native number of detector elements | 320 x 160 |
| Native detector element pitch | 0.33 mm |
| Detector bin mode | 2 x 2 |
| Detector converter thickness | 2 mm |

## 2.4.2  Scanning x-ray source

The SBDX source and collimator are shown in Figure 2.5. Electrons generated from a grid-controlled cathode are accelerated toward a planar target (anode).  Electromagnetic coils are used to focus and scan the electron beam across the target.  The target is an approximately 15 μm thick layer of tungsten-rhenium bonded to a 5-mm beryllium plate.  The beryllium plate is both the exit window and a structural component of the vacuum envelope. A 1.2-mm thick layer of water is actively circulated over the beryllium window to remove heat generated in the target. The water is contained with a 0.812 mm thick sheet of aluminum that provides additional inherent filtration.[36]

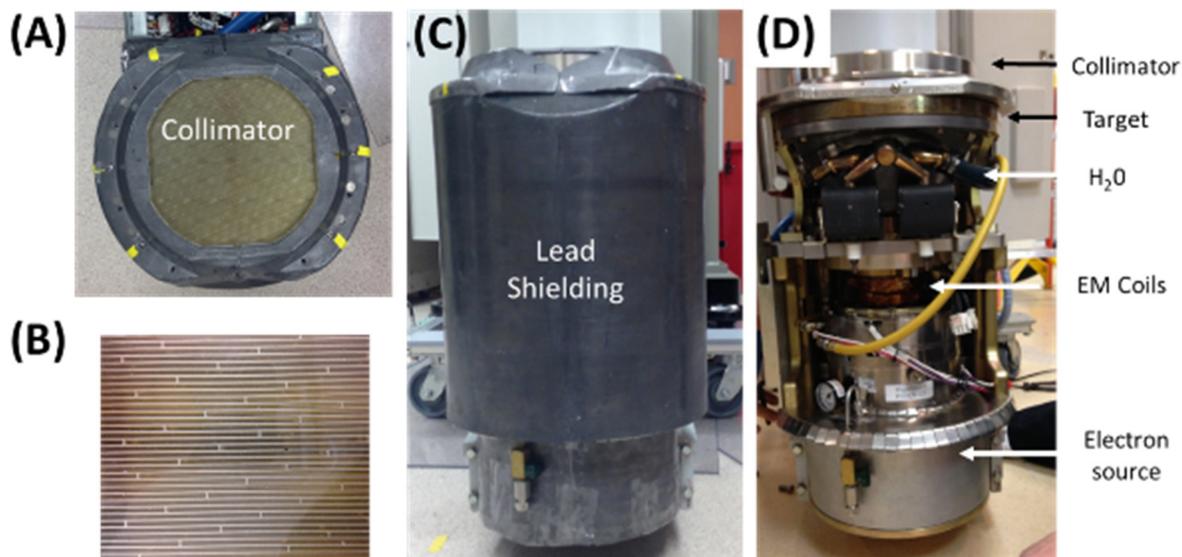

Figure 2.5: The SBDX source is shown. (A) The SBDX multi-hole collimator. (B) A zoomed view of the collimator shows the slotted structure. (C) Lead shielding is placed around the source array. (D) The SBDX source array with shielding removed shows the general locations of the electron source, deflection coils, water coolant, tungsten target and collimator.

In addition to water cooling, a multi-pass raster scan technique is utilized to reduce the peak target temperature. In the 71 by 71, 15 frame/s scanning mode, each focal spot position is visited by the electron beam 8 times per frame. The raster scanning proceeds blockwise, with all 8 passes performed on a block of 3 rows before proceeding on to the next block. The electron beam dwell time at a focal spot position is 1.04 μs and the travel time between positions is 0.24 μs. Thus, the total beam on time per frame is 41.9 ms. Additional time is allotted for row retrace which equals 21.8 μs per row. The time difference between the first and last x-ray beam illumination of a fixed point in the field-of-view is referred to as the effective pulse width. For a point at iso-center, the effective pulse width is 8.9 ms.

Source operating points are specified in terms of the tube potential (kV) and electron beam current (mA$_{peak}$) during each focal spot dwell. Tube potential may be set from 80-120 kV. The source power (kWp) is the product of the tube potential and the current during a focal spot dwell. The maximum source power is 24.4 kWp at 120 kV (203 mA$_{peak}$). Tube current is adjusted via a grid voltage. The grid voltage may be varied at each of the source dwell positions to turn individual x-ray beamlets "on" (operating at some mA$_{peak}$ value) or "off" (no current). This feature of the SBDX source has been exploited for dose-reduced region-of-interest catheter tracking using dynamic beam collimation,[39] as well as regional adaptive exposure (RAE) control for dose-reduction in fluoroscopic imaging mode.[109] Of particular interest for CT imaging, dynamic beam collimation and adaptive exposure control could be utilized to tailor the x-ray fluence field for dose reduction and/or targeted region-of-interest specific image quality (e.g. signal-noise-ratio). A number of recent studies have investigated fluence-field modulated CT using dynamic beam attenuators,[110–113] multiple aperture devices,[114] and a multi-source x-ray tube.[47,48]

A multi-hole collimator restricts the x-rays produced at each focal spot position to form a series of narrow beamlets directed at the detector. The 43.8 mm thick collimator is constructed of layers of tungsten, brass and lead plates. The inner layers are each fabricated with an array of rectangular holes such that, when stacked, they form tapered holes tilted toward the detector. The outermost layer consists of tungsten slots.[36] The geometric relationship among the narrow beam projections is constrained by the precise and rigid geometry of the collimator.

### 2.4.3 Detector and imaging pipeline

The SBDX system uses a direct-conversion photon-counting detector array with a 2 mm thick CdTe converter layer (Figure 2.6).[36,115,116] To accommodate the rapid rate of source scanning, the detector was designed to have a fast readout rate (781,250 detector images per second, corresponding to a focal spot illumination every 1.28 μsec). The detector area measures 10.6 cm by 5.3 cm. The detector is constructed from modules arranged in 4 rows of 8 in a stepped pattern. Each detector module contains 40 x 40 detector elements on a 0.33 mm pitch. As a result, the full SBDX detector array consists of 320 x 160 individual detector elements. In the 71 x 71 hole 15 frame/s scan mode, there are 8 detector images captured per collimator hole and therefore 40,328 raw detector images generated per 1/15 s frame period.

The detector uses a single discriminator threshold to convert charge signal to photon counts. With the very short capture time per detector image, the number of photons detected per detector element is typically very small (0.03-3 photons per detector element on average).[35] To facilitate high speed counting, each detector elements consists of 4 sub-elements. Veto logic linking neighboring sub-element outputs is used to mitigate crosstalk among sub-elements, caused, for example, by re-absorption of a K-fluorescence photon following an initial photoelectric absorption event.[36,115] Occasional double-counting due to fluorescence re-absorption within the same detector element represents a stochastic gain which can reduce DQE. The veto logic combats this problem through deactivation of neighboring sub-elements following a primary count event. Note this represents a potential source of detector output nonlinearity at higher fluences (explored further in Sec. 5.3).

The detector data is streamed to the real-time image reconstructor via fiberoptic links. Prior to reconstruction, front-end electronics perform summing of the 8 detector images per collimator hole. Additionally, 2 x 2 detector element binning is applied to reduce downstream bandwidth. Following this processing, the detector data for a 1/15 s scan frame is reduced from 40,328 to 5,041 detector images, and each detector image is a 160 x 80 array of 0.66 mm wide detector elements. This data is simultaneously transferred via a PCIe bus to the real-time image reconstruction pipeline and a 32 terabyte disk array (Conduant Corporation, Longmont, CO, USA). In this project, the data stored on the disk array was used for offline CT reconstruction. Note that, for IGCT, the set of projections acquired during a single frame period is referred to as a *superview*. An SBDX superview consists of the projection rays resulting from the combination of 5,041 focal spot dwell positions and 160 x 80 detector elements (i.e. 64,524,800 rays per superview).

The first stage of the real-time fluoroscopy reconstructor consists of 8 graphics processing units (GPUs) that perform tomosynthesis (shift-and-add algorithm) at 32 planes spaced by 5 mm. Plane stacks are generated at 15 frames per second. The plane stacks are then sent to two additional GPUs. One GPU is configured to generate a multi-plane composite image at 15 frames per second (alternatively, this GPU can be configured to pass-through a single tomosynthesis plane). The second GPU can be programmed for other tasks performed in parallel, such as tomosynthesis-based 3D catheter tracking at 15 Hz.[40] The live outputs from both GPUs are displayed on a monitor. A review of the SBDX tomosynthesis reconstruction method is provided in chapter 5 and detailed in Refs. [35] and [36].

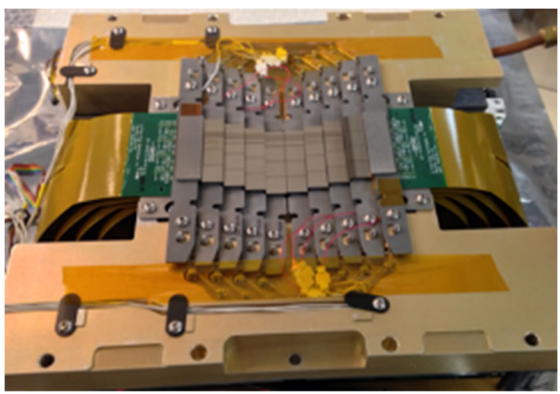

Figure 2.6: The SBDX photon-counting detector is shown. The detector consists of 32 modules (4 x 8) arranged in a stepped fashion to accommodate interconnects to a controller board. Each detector module consists of 40 x 40 elements. Detector elements are 0.33 mm wide with 2-mm thick CdTe x-ray converter material. The total detector size is 10.6 cm by 5.3 cm.

### 2.4.4 Dose efficiency and real-time tomosynthesis

SBDX achieves high dose efficiency relative to conventional cone-beam based fluoroscopy

systems through a combination of three main mechanisms. First, the use of a narrow x-ray beam and a large

air gap reduces the number of scattered photons incident on the detector. The reduction of image degrading

scatter allows a target signal-to-noise-ratio to be achieved with fewer primary x-rays. Scatter fraction

measured on the SBDX system varies from 2.8%-7.8% under a range of realistic imaging conditions.[36,117]

Second, the inverse geometry style of x-ray beam delivery spreads primary x-rays out over a larger patient

entrance area in comparison to a conventional cone beam geometry. Spreading the incident fluence over a

larger entrance area reduces dose to points on the skin entrance. And third, the 2-mm thick CdTe detector

maintains high DQE at high tube potentials. A study performed for a previous SBDX prototype

demonstrated potential for a 3-7 time reduction in skin entrance exposure versus a conventional geometry.[37]

Additional reductions to dose-area-product may be achieved using the regional adaptive exposure method discussed in section 2.4.2. Regional adaptive exposure control was shown in phantom studies to reduce the dose-area-product from 21-45% for adult sized phantoms.[109]

In addition to potential dose savings, the inverse geometry beam scanning technique provides SBDX with an inherent real-time tomosynthesis capability. SBDX tomosynthesis provides depth resolution not available with conventional fluoroscopy.  This has been exploited for a number of clinical applications. *Speidel et al.* presented a method for 3D tracking of high-contrast cardiac catheters. SBDX catheter tracking and display were recently implemented in real-time by *Dunkerley et al.*[38,118] *Tomkowiak et al.* developed a method for calibration-free vessel measurements for device sizing.[41] *Tomkowiak et al.* invented a stereoscopic imaging technique for 3D visualization during cardiac interventions.[42] *Hatt et al.* showed SBDX tomosynthesis can be used to register transesophageal echo with x-ray fluoroscopy.[43]

## 2.5  Summary

Increasingly complicated structural heart interventions have created a need for advanced image guidance techniques in the cardiac cath lab.  Conventional x-ray fluoroscopy remains the workhorse modality of the cardiac cath lab but fails to provide the 3D perspective and soft tissue contrast necessary to navigate catheter devices to precise anatomic targets located in large cardiac chambers or great vessels. Longer procedure times, resulting from increased complexity, motivate a need for dose reduction to reduce the risk of radiation induced skin injuries and to lower scatter exposure to the interventional staff.

Radiofrequency catheter ablation of atrial fibrillation is one example procedure that requires precise 3D localization of anatomic targets such as the pulmonary vein ostia that are not visible under x-ray fluoroscopy. The use of electroanatomic mapping systems that track the 3-D position of cardiac catheters in real-time, relative to imported cardiac chamber models derived from pre-procedural CT or MRI, is associated with improved procedure efficacy and RFCA outcomes. While EAM systems have demonstrated great utility in the electrophysiology lab, an EAM system requires catheter device modification and does not present the real-time anatomic overview provided by fluoroscopy. Scanning Beam Digital X-ray is a novel inverse geometry x-ray fluoroscopy system designed to overcome the shortcomings of current fluoroscopic systems. SBDX performs rapid narrow-beam scanning in an inverse geometry to improve dose efficiency and achieve a unique real-time tomosynthesis capability. SBDX tomosynthesis enables 3-D catheter tracking which can be utilized to display 3D catheter position relative to simultaneously acquired fluoroscopic images and/or 3-D imported anatomic models. Therefore, this technology is well positioned to handle RFCA, as well as other interventions requiring 3D guidance (e.g. structural heart interventions), for which EAM technology does not provide convenient solutions.

Currently, the 3-D anatomic models must be imported from CT or MR imaging studies performed prior to an interventional procedure. However, changes in organ size and position between pre-procedural imaging studies and time of intervention decrease 3-D model accuracy. Anatomic model accuracy and registration could be improved by acquiring and segmenting the 3-D models immediately prior to the start of an intervention after the patient has been placed on the table. The current SBDX prototype was designed for fluoroscopic imaging and lacks the capability to perform C-arm based inverse geometry CT. C-arm CT has become an expected and critical feature of commercial x-ray fluoroscopic systems. The next chapter investigates the feasibility of performing CT data acquisition and reconstruction with SBDX for the task of 3D cardiac chamber mapping.

# 3 Feasibility of IGCT with the SBDX system

*A subset of the results reported in this chapter were presented at SPIE-Medical Imaging 2015.*[50]

## 3.1 Introduction

For the purposes of this work, inverse geometry CT (IGCT) is defined generally as a data acquisition and associated reconstruction method in which an x-ray source and detector are oriented at different angles around the patient (e.g. a short-scan arc covering a total of 200 degrees), and which, for each orientation, the patient is imaged from multiple focal spot positions. As discussed in Chapter 1, this technique has been previously explored for several geometries similar to SBDX, although none of them were specifically designed for 3D cardiac chamber mapping with the C-arm based SBDX system. Therefore, the purpose of this chapter is to investigate the feasibility of performing IGCT data acquisition and image reconstruction with SBDX for the target task of cardiac chamber mapping. To facilitate this investigation, Section 3.2 begins with an exploration of the SBDX sampling of Radon space for a short-scan CT acquisition. Section 3.3 then reviews several reconstruction methods previously proposed for IGCT. Each method was implemented for the SBDX geometry and evaluated for its suitability in the SBDX-CT task using simple numerical simulations. The simulation work is divided into three sections. Section 3.4 evaluates reconstruction accuracy of a Shepp-Logan phantom in two-dimensional reconstruction. Section 3.5 discusses the challenges encountered when the reconstruction methods are extended to three dimensions. Section 3.6 then investigates feasibility of cardiac chamber mapping with SBDX using a simplified thorax phantom. The chapter concludes with a discussion of the existing IGCT techniques which are well-suited to the SBDX-CT cardiac imaging task, as well as the deficiencies which must be addressed in the remainder of this work.

## 3.2 SBDX-CT acquisition and sampling

IGCT sampling requirements and image reconstruction can best be described after examining the ray sampling in Radon space. For an IGCT system (see Figure 3.1), a ray connecting a source focal spot position ($s_x$, $s_y$, $s_z$) to a detector element position ($d_x$, $d_y$, $d_z$) is described by four parameters ($u$, $v$, $\theta$, $\varphi$). Following the notation of *Schmidt et al.*, $u$ and $v$ describe the location of a ray of interest relative to the ray connecting the center of the source array to the center of the detector (the iso-ray).[119] The distance $u$ is measured in the x-y plane, after projecting the ray of interest down to the x-y plane, and $v$ is measured from the ray of interest to the projected ray. The detector coordinates $u$ and $v$ can be combined into a single parameter $\rho$ describing the radial distance of a ray from the system iso-center.

$$\rho = \sqrt{u^2 + v^2} \tag{3.1}$$

The orientation of a ray is described by the tilt angle $\theta$ and the azimuthal angle $\Psi$. Note that Figure 3.1 assumes a specific orientation of the gantry. When the gantry is rotated by angle $\varphi_{gantry}$ about the z-axis, then then azimuthal angle is converted to angle $\varphi = \Psi + \varphi_{gantry}$. Each IGCT ray can be mapped to a point in 4D Radon space by its coordinates ($u$, $v$, $\theta$, $\varphi$) according to equations (3.2)-(3.6).

$$\Psi = \tan^{-1}\left(\frac{s_y - d_y}{d_x - s_x}\right) \tag{3.2}$$

$$\varphi = \Psi + \varphi_{gantry} \tag{3.3}$$

$$\theta = \frac{\pi}{2} - \tan^{-1}\left(\frac{s_z - d_z}{\sqrt{(s_x - d_x)^2 + \left(s_y - d_y\right)^2}}\right) \tag{3.4}$$

$$u = d_y \cos(\Psi) + d_x \sin(\Psi) \tag{3.5}$$

$$v = d_z \sin(\theta) + \left[d_x \cos(\Psi) - d_y \sin(\Psi)\right] \cos(\theta) \tag{3.6}$$

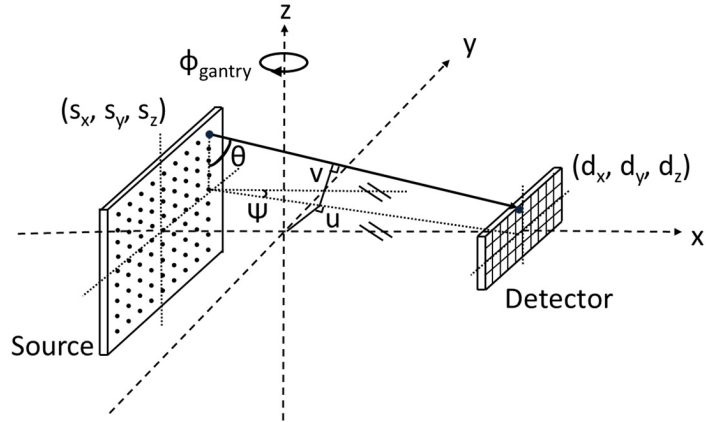

Figure 3.1: An IGCT ray maps to a 4D Radon space described by four parameters (u, v, θ, φ). Two distance parameters, u and v describe the ray's location relative to system iso-center. Two angular coordinates, θ and φ describe the rays tilt and view angle.

SBDX CT sampling in Radon space can be examined by application of equations (3.2)-(3.6). For demonstration, the 2D case (single-slice CT) will be considered here. First we note that single-slice IGCT is equivalent to a conventional fan beam geometry if the number of source points is reduced to one and the detector width is increased to cover the desired scanner field-of-view (FOV). Applying Equations (3.2)-(3.6) in this case reveals that the rays acquired at a given source and detector position map to a curve in sinogram space that resembles a slanted line; the slant occurs because the angle of a ray relative to the iso-ray increases towards the periphery of the detector (Figure 3.2A). As the source and detector rotate, the curve sweeps through sinogram space until sufficient sampling is achieved (Figure 3.2B). For the fan beam geometry, there is the well-known condition that the scanner must rotate at least 180 degrees plus the fan angle in order to acquire data spanning 180 degrees in sinogram space.

In a more general IGCT system with multiple focal spots and small detector, the collection of rays originating from a single focal spot and falling on all detector elements (for fixed gantry angle) also maps to a slanted curve in sinogram space, resembling a miniature fan-beam. However, due to the small detector width, the line spans a very narrow range of ρ values. Figure 3.3A demonstrates the sinogram space lines for the 1st and 71st focal spot positions, as well as those in between, for a single source and detector row. The SBDX geometry described in Table 2.1 for the 71x71 scan mode was assumed here.

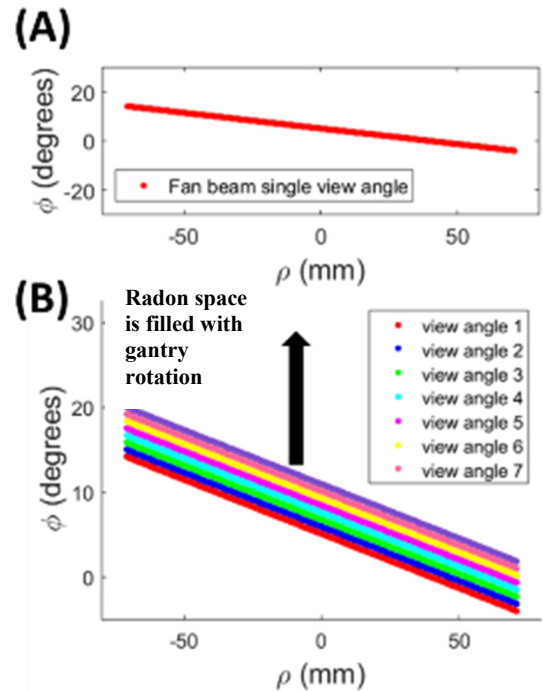

Figure 3.2: Fan beam CT data acquisition is mapped to Radon space for multiple view angle (A) and a single view angle (B). For the 2D case (single slice CT) each fan beam samples a line in Radon space.

Following IGCT convention, the term "superview" is used to refer to the collection of rays for all focal spot positions at a given gantry angle. Thus one superview maps out a parallelogram-shaped region in sinogram space. An SBDX superview spans 10.1 degrees along the $\varphi$ axis and 144.2 mm along the $\rho$ axis. As the gantry is rotated, this parallelogram shifts along the $\varphi$ axis (Figure 3.3B). The spacing between superviews is determined by the fluoroscopic scan rate and gantry rotation speed. A fluoroscopic scan rate of 15 scan/sec and a SBDX C-arm rotation speed of 14.5 degree per second were assumed for Figure 3.3. Filling of sinogram space must proceed until a region spanning at least 180 degrees has been covered. For SBDX-CT, this requires the gantry be rotated 180 degrees plus the difference in $\varphi$ for the ray corresponding to focal spot 1, detector element 1, and focal spot 71, detector element 160. This corresponds to a minimum short-scan rotation of 182.1 degrees.

After the IGCT data acquisition is completed, there is an irregular distribution of sinogram samples. As shown in Figure 3.3C, peripheral regions of the FOV are less densely sampled, the central portion of the FOV is more densely sampled, and there is a small-scale structure in the sampling pattern. Conventional analytical reconstruction algorithms cannot be directly applied to IGCT systems due to this unique sampling in sinogram space. This key difference between IGCT and conventional CT must be addressed during image reconstruction.

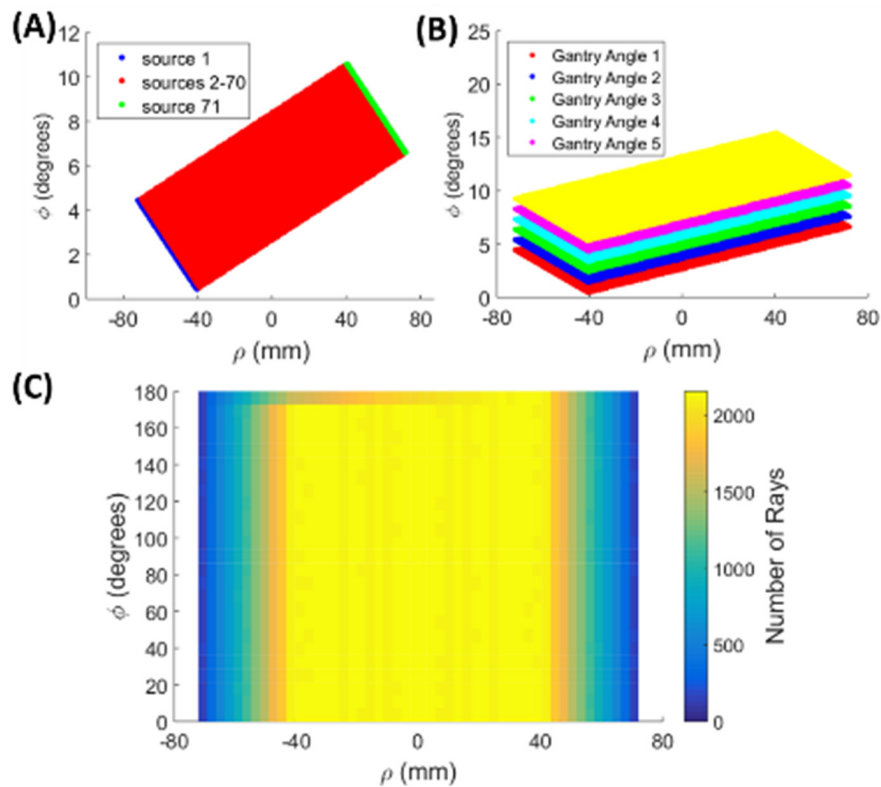

Figure 3.3: IGCT projection data is mapped to sinogram space for a single view angle (A) and multiple view angles (B). Peripheral regions in IGCT sinogram space are sampled more sparsely than central regions (C). For the 2D case (single slice CT) each IGCT superview samples a parallelogram shaped region of Radon space.

## 3.3 Inverse geometry CT reconstruction

A number of reconstruction methods have been proposed for IGCT. Analytical techniques include parallel-ray gridding based filtered backprojection (gFBP) methods,[119–121] cone-beam based gridding methods,[46,122] Fourier rebinning algorithms,[123–125] and a direct fan beam method with a non-uniform sampling correction.[126] Iterative reconstruction algorithms including penalized weighted least squares[127] (PWLS) and ordered subset expectation maximization[128] (OSEM) have also been investigated. In this chapter, the gFBP,[120] direct fan beam with non-uniform sampling correction,[126] and PWLS[127] methods were implemented for the SBDX geometry to investigate the feasibility of SBDX CT reconstruction. A review of the reconstruction methods considered and implementation details are presented below.

### 3.3.1  Gridded filtered back projection

*Schmidt et al.* proposed a 3D filtered backprojection algorithm for IGCT.[119] The proposed algorithm estimates a complete set of parallel ray projections from the IGCT projections. The parallel ray projections are then convolved with a reconstruction kernel before being backprojected through the 3D image volume. This approach has the advantages of simplicity and speed, although the parallel-ray estimation procedure ("gridding") can potentially reduce spatial resolution.

For IGCT reconstruction the gridding is performed in the 4D projection domain. The non-uniformly spaced IGCT data are rebinned to a uniformly spaced 4D Cartesian grid corresponding to the parallel ray projections. Each parallel ray projection is estimated as a weighted sum of the raw IGCT sinogram points falling within a defined bin centered on a parallel ray projection grid point. Schmidt proposed to weight the contribution of an IGCT ray sample to a parallel-ray projection sample based on a separable 4D kernel consisting of 1D Hanning functions for each of the four dimensions of Radon space (i.e. *u, v, θ, φ*). In this case, the 1D "gridding kernel" is defined as,

$$h(x) = \frac{1}{w}\left(1 + \cos\left(\frac{2\pi x}{w}\right)\right) rect\left(\frac{x}{w}\right) \tag{3.7}$$

where *x* represents the distance from the parallel ray point in Radon space to the IGCT ray and *w* represents the kernel bin-width. The 4D gridding kernel is the product of each of the one dimensional gridding kernels. The estimated parallel ray projection value is normalized by the sum of the kernel values of contributing IGCT ray samples. The selected kernel width must be large enough that each of the parallel ray projections can be estimated from the input IGCT projection values, but small enough to mitigate blurring and loss of spatial resolution.

The gFBP algorithm was implemented to investigate its application to SBDX-CT reconstruction. The gridding parameters considered here are summarized in Table 3.1. Here, the grid dimensions and kernel widths were chosen empirically to demonstrate feasibility. Later, in Chapter 4, the gridding parameters are optimized for the task of 3D cardiac chamber mapping with SBDX. A Hanning-windowed ramp filter was applied during filtered backprojection unless stated otherwise.

Table 3.1: Gridding parameters used with gFBP method

| | |
|---|---|
| Number of parallel ray view angles over 180º | 240 |
| Rays per 2D view [columns x rows] | 240 x 385 |
| 2D view sampling pitch | 0.5 mm x 0.23 mm |
| Radial kernel | 1.4 mm x 2.8 mm |
| Angular kernel | 1 degree x 1 degree |

### 3.3.2 Direct fan beam reconstruction with non-uniformity correction

A second FBP type algorithm for IGCT was proposed by *Baek et al.*[126] Figure 3.4 shows that the rays connecting a single IGCT detector element to each of the individual source elements resembles an off-axis small area fan beam. This suggests that IGCT reconstruction could be performed through direct application of the traditional fan-beam FBP algorithm to each miniature fan beam resulting from the collection of rays convergent on an individual detector element.[129] However, the combination of multiple source points results in non-uniform sampling of the image object, as described in section 3.2. Baek demonstrated the non-uniform sampling produces ringing artifacts at certain radial locations following filtered backprojection (Figure 3.5).[126]

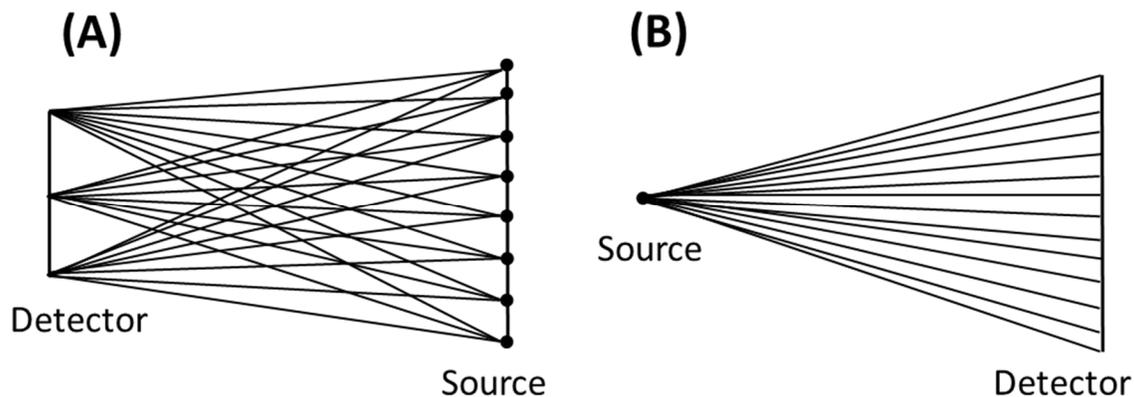

Figure 3.4: A schematic IGCT system with three detector elements and eight source elements is shown at left (A). The rays corresponding to a single detector element connected to all source elements resemble a miniature fan beam. Shown at right is the conventional fan beam geometry (B).

To eliminate ring artifacts, *Baek et al.* proposed a non-uniform sampling density correction that is applied prior to filtered backprojection of each individual detector element to source array fan beam.[126] In Baek's approach, because the source array is wider than the detector array, each IGCT detector element is viewed as a pseudo source element while the IGCT source elements are viewed as pseudo detector elements. Each of the small-width fan beams is undersampled since the (true) source spacing is much larger than the finite-sized focal spot. As a result, the source array is zero-padded between source focal spots before filtering to prevent blurring. The standard ramp filter with a Hanning window is applied to the zero-padded pseudo detector data. Next, the filtered projection data are multiplied by a gain vector to correct for the non-uniform sampling. Finally, the filtered and weighted projection data are backprojected for each detector element using the standard fan-beam FBP algorithm accounting for the IGCT geometry. The gain vector is system specific and does not depend on the imaging object. To determine the non-uniform weighting factors (gain vector) empirically, a uniform calibration cylinder may be imaged and reconstructed with a conventional CT system and then again with the IGCT system assuming unity gain values. An optimization procedure is then performed to determine the gain vector that minimizes the root-mean-square error between the two calibration images.

The key contribution of the direct reconstruction algorithm is a method to correct for the non-uniform sampling observed in IGCT. This replaces the rebinning step required by the gFBP algorithm which results in blurring and loss of spatial resolution. Baek's direct fan beam reconstruction algorithm with non-uniformity correction was implemented for the SBDX geometry.

### 3.3.3 Statistical iterative reconstruction

Iterative CT reconstruction is a flexible approach that allows for accurate modeling of an arbitrary acquisition geometry and varying noise reduction strategies. Additionally, iterative reconstruction is well suited for irregularly-organized projection data resulting from *cardiac-gated* short-scan reconstruction (i.e. a method based on the projection data collected within a specific window of each cardiac cycle in the rotational scan). These properties are desirable for SBDX-CT cardiac chamber mapping.

The PWLS algorithm, derived in Refs. 130 and 131, is an example of a statistical iterative reconstruction (SIR) algorithm that incorporates a noise model into the image reconstruction problem.[130] PWLS was first investigated for IGCT by Bequé et al.[127] A monochromatic x-ray spectrum is assumed here to simplify the introduction of the PWLS algorithm. In this case, taking the natural logarithm of the Beer-Lambert law yields the well-known relationship between the measured photon number $I$ for a ray, the input photon number $I_o$, and the line integral of attenuation coefficient $x(\vec{r})$:

$$y = \ln\frac{I_o}{I} = \int x(\vec{r})dl \tag{3.8}$$

This relationship is discretized and expressed for the entire collection of measured rays using the matrix notation $\mathbf{y} = \mathbf{Ax}$, where $\mathbf{y}$ is a column vector of line integral measurements, $\mathbf{x}$ is a column vector of voxel attenuation coefficients, and the element $a_{ij}$ of the system matrix $\mathbf{A}$ describes the path length of the $i^{th}$ ray through the $j^{th}$ image voxel with attenuation coefficient $x_j$. The detected x-ray fluence can be described by a Poisson distribution with the mean number of detected photons denoted by $\overline{I_i}$. The conditional probability for the collection of independent line integral measurements $\boldsymbol{I}$ given an image $\mathbf{x}$ is then stated by equation (3.9).

$$P(\boldsymbol{I}|\boldsymbol{x}) = \prod_i \frac{(\overline{I_i})^{I_i}}{I_i!}\mathrm{e}^{-\overline{I_i}} \tag{3.9}$$

The log-likelihood function can be derived by taking the natural logarithm of equation (3.9).

$$L(\boldsymbol{I}|\boldsymbol{x}) = \ln P(\boldsymbol{I}|\boldsymbol{x}) = \sum_i [I_i \ln \bar{I}_i - \bar{I}_i - \ln I_i!] \tag{3.10}$$

The expected number of detected photons for a ray, $\overline{I_i}$, is related to the image, $\mathbf{x}$, according to the Beer-Lambert law stated in equation (3.8) as, $\bar{I}_i = \bar{I}_{i,0}e^{-[\boldsymbol{y}]_i} = \bar{I}_{i,0}e^{-[\boldsymbol{Ax}]_i}$. Substitution into equation (3.10) and dropping irrelevant terms not containing $\overline{I_i}$ yields a relationship between the photon measurements and image, $\mathbf{x}$.

$$L(\boldsymbol{I}|\boldsymbol{x}) = \ln P(\boldsymbol{I}|\boldsymbol{x}) = \sum_i \left[ -I_i[\boldsymbol{Ax}]_i - \bar{I}_{i,0} e^{-[\boldsymbol{Ax}]_i} \right] \tag{3.11}$$

The image reconstruction problem can then be solved by finding **x** that maximizes the log-likelihood function. However, direct solution does not take into account known prior information of the imaging subject that may improve image quality by mitigating noise or data inconsistency present in the measured projection data. As an example of prior information, the distribution of linear attenuation coefficient in a human subject is expected to vary smoothly for neighboring pixels and known to be non-negative. Prior information may be incorporated by reformulating the reconstruction problem within a Bayesian framework.

$$P(\boldsymbol{x}|\boldsymbol{I}) = \frac{P(\boldsymbol{I}|\boldsymbol{x})P(\boldsymbol{x})}{P(\boldsymbol{I})} \tag{3.12}$$

Taking the natural logarithm of equation (3.12), substituting the relationship for $\ln P(\boldsymbol{I}|\boldsymbol{x})$ derived in equation (3.11), and dropping irrelevant or constant terms yields the following log-likelihood function:

$$L(\boldsymbol{x}|\boldsymbol{I}) = \ln P(\boldsymbol{x}|\boldsymbol{I}) = \sum_i \left[ -I_i[\boldsymbol{Ax}]_i - \bar{I}_{i,0} e^{-[\boldsymbol{Ax}]_i} \right] + \ln P(\boldsymbol{x}) \tag{3.13}$$

The third term of equation (3.13), $\ln P(\boldsymbol{x})$, may be replaced by a regularization function, R(**x**), selected to incorporate prior information of the image into the reconstruction procedure. Regularization functions typically penalize adjacent pixels exhibiting large differences in image values and may suppress image noise and artifacts. By performing a Taylor series expansion on equation (3.13), as detailed in Appendix A of Ref. 130, a quadratic approximation to the negative log-likelihood function can be derived and the image reconstruction problem is reduced to solving the following optimization problem.[130]

$$\hat{\boldsymbol{x}} = \arg min_x \left[ \frac{1}{2}(\boldsymbol{y} - \boldsymbol{Ax})^T \boldsymbol{D}(\boldsymbol{y} - \boldsymbol{Ax}) + \beta R(\boldsymbol{x}) \right] \tag{3.14}$$

The diagonal matrix **D** is a noise matrix typically set equal to the inverse of the variance of the measured line integral data.[131] The noise matrix **D** serves to emphasize or de-emphasize differences between the measured projection data $\boldsymbol{y}$ and the forward projection data $\boldsymbol{Ax}$, depending on the reliability of the projection measurements. The scalar β determines the relative weight of the regularization function, $R(\mathbf{x})$.

For application to SBDX-CT reconstruction, a nonlinear conjugate gradient method was used to minimize the objective function, equation (3.14). The system matrix **A** was approximated using a ray-driven forward projector and the transpose operation was modeled using pixel-driven backprojection with linear interpolation at the detector. Since the SBDX system uses a photon counting detector, the elements of **D** were set equal to the measured photon counts. The total variation (TV) function was used for the regularization function. Note that TV has been shown to introduce "patchy" artifacts that may be unacceptable for diagnostic imaging. However, that is less of a concern for anatomic surface mapping as the final display will be a volume rendered chamber model. The parameter β was varied, such that reconstructions with β = 0 are referred to as weighted least squares (WLS), while reconstructions with β > 0 are termed PWLS-TV.

### 3.3.4   Gridded statistical iterative reconstruction

IGCT projection data sets can be large due to the multiple source elements. Consequently, statistical iterative reconstruction times may be limiting for intraprocedural CT imaging with SBDX. For example, the total number of projection rays within the SBDX-CT dataset is $1.16 \times 10^{10}$ if operated in the 71 x 71 source scan mode with a 160 x 80 element detector array (2 x 2 bin mode) and 180 acquired superviews. The increased number of projection rays compared to conventional cone beam or parallel ray geometries results in additional computations for both forward and back projection. For comparison, a cone-beam CT system utilizing a 512 x 512 detector array readout and 420 view angles acquires $1.10 \times 10^{8}$ rays.[132] A voxel driven backprojector requires an iteration over the full image grid for each individual source point, which further increases backprojection times by a factor of 71 x 71 for SBDX. A potential solution is to grid the projection data to a complete set of parallel ray projections prior to iterative reconstruction.  In this initial feasibility work (Table 3.1) this approach reduced the size of the datasets by a factor of 524. To investigate this more computationally efficient approach, statistical iterative reconstruction was performed with SBDX projection data gridded to a parallel ray projection dataset followed by minimization of equation (3.14). For the gridded SIR case, the system matrix **A** was adapted to describe the parallel ray geometry. The gridded iterative reconstructions were termed gPWLS-TV (β > 0) and gWLS (β = 0).

## 3.4 Evaluation of SBDX-CT reconstruction accuracy

A computer-simulator was developed to investigate the feasibility of performing CT imaging with SBDX. The simulator performs analytical ray-tracing to compute exact path lengths through defined phantoms for the SBDX scanning geometry. The analytical simulator allows for reconstruction performance to be evaluated versus a known ground truth. Numerical simulations were performed to examine SBDX-CT reconstruction accuracy for each of the implemented algorithms (gFBP, Baek's direct FBP method, PWLS-TV, and gPWLS-TV).

### 3.4.1 Methods

The goal of this first investigation was to compare the relative accuracy of the different reconstruction methods under basic conditions. Noise-free SBDX projection data were simulated for a two-dimensional high contrast Shepp-Logan phantom. A short-scan acquisition consisting of 180 superviews uniformly distributed over 200 degrees was simulated for a single row of source and detector elements. Gating and field-of-view truncation were not simulated. The relative root mean square error (rRMSE) between the reconstructed images, **x**, and ground truth, **x$^{truth}$**, with N image pixels was calculated for each of the implemented algorithms according to equation (3.15).

$$rRMSE = \frac{1}{\max(\boldsymbol{x^{truth}}) - \min(\boldsymbol{x^{truth}})} \sqrt{\frac{1}{N}\sum_{i=1}^{N}(x_i - x_i^{truth})^2}$$

(3.15)

### 3.4.2 Results

The noise-free two-dimensional Shepp-Logan phantom was reconstructed using *Baek*'s direct reconstruction method without (Figure 3.5A) and with (Figure 3.5B) the non-uniformity correction.

Ringing artifacts were successfully removed by application of the non-uniformity correction. The rRMSE was 3.12% and 2.62% before and after gain correction, respectively.

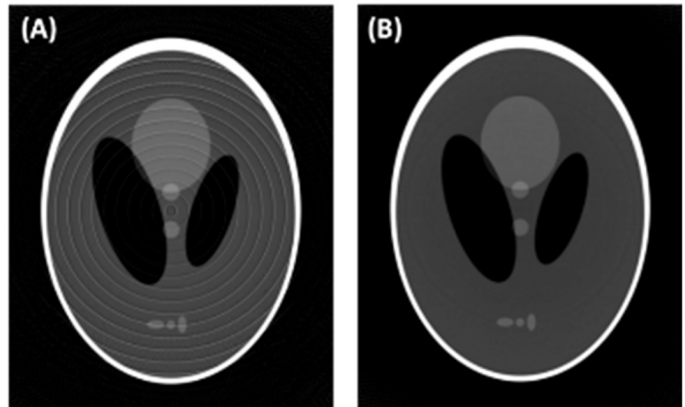

Figure 3.5: Simulated noise free IGCT high contrast Shepp-Logan phantom projection data was reconstructed using Baek's direct method without a gain vector (A) and with a gain vector (B). The gain vector is applied directly to the projection data before backprojection and corrects for the IGCT non-uniform sampling. Application of the gain vector successfully removed the ringing artifacts.

Next, the two-dimensional Shepp-Logan projection data were reconstructed using gridded filtered backprojection with a non-apodized ramp filter (Figure 3.6A). The rRMSE for the gFBP reconstruction measured 2.18%. Finally, reconstruction was performed using PWLS-TV (Figure 3.6B) and gridded PWLS-TV (Figure 3.6C). The rRMSE of the PWLS-TV reconstruction was 0.58%. The gridding procedure in the gPWLS-TV method decreased reconstruction accuracy to a level of 1.85% rRMSE.

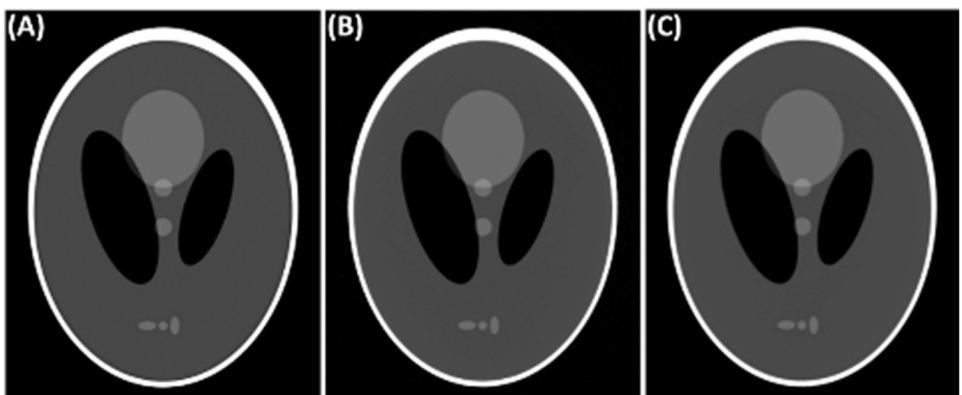

Figure 3.6: Noise-free IGCT Shepp-Logan phantom projection data was reconstructed using gridded filtered backprojection (A), penalized weighted least squares with total variation regularization (B), and gridded penalized weighted least squares with total variation regularization (C). The relative root mean square errors versus a ground truth image were 2.18%, 0.58% and 1.85% respectively.

## 3.5  Extension of SBDX-CT reconstruction to three dimensions

A major challenge with extending SBDX-CT reconstruction to three-dimensions is the increased computational load, discussed in section 3.3.4. A noise-free simulation of a high-contrast 3D Shepp-Logan phantom positioned at gantry isocenter was performed to investigate IGCT reconstruction for the full SBDX geometry. The 3D phantom size was reduced to 80 x 60 x 60 mm$^3$ to avoid data truncation as a confounding factor. Projection data were binned at the detector (2 columns x 4 rows) prior to reconstruction to reduce reconstruction times. For each of the reconstruction methods outlined below, the forward and backprojection operations were performed on a Windows PC with an Intel Xeon 3.60 GHz processor and an NVIDIA GeForce GTX 670 graphics processing unit. The inverse geometry to parallel ray gridding procedure was performed on the CPU.

The PWLS-TV algorithm was implemented for the three-dimensional SBDX geometry to evaluate iterative reconstruction times for intra-procedural IGCT. The Shepp-Logan phantom was reconstructed using equation (3.14) with β=0 (no regularizer) as shown in Figure 3.7 for a transverse (axial) slice and sagittal slice. The reconstruction was performed using a 512 x 512 x 385 image volume with 0.234 mm$^3$ isotropic pixel resolution. The rRMSE for the full 3D reconstruction versus the known ground truth was 1.4%. Figure 3.7C shows an intensity profile drawn vertically though the center of Figure 3.7A.

The PWLS-TV reconstruction took nearly 30,000 minutes due to the large amount of projection data and increased computation required for IGCT backprojection. Example reconstruction times are presented in Table 3.2 for gridded FBP, gridded PWLS-TV and PWLS-TV. Gridding the projection data to a parallel ray dataset prior to reconstruction reduced the reconstruction time to 83 minutes. Reconstruction time is an important consideration in the interventional setting for the target application of intra-procedural cardiac chamber mapping. Initial timing results indicate that the most promising IGCT reconstruction methods for intra-procedural cardiac chamber mapping will involve gridding.

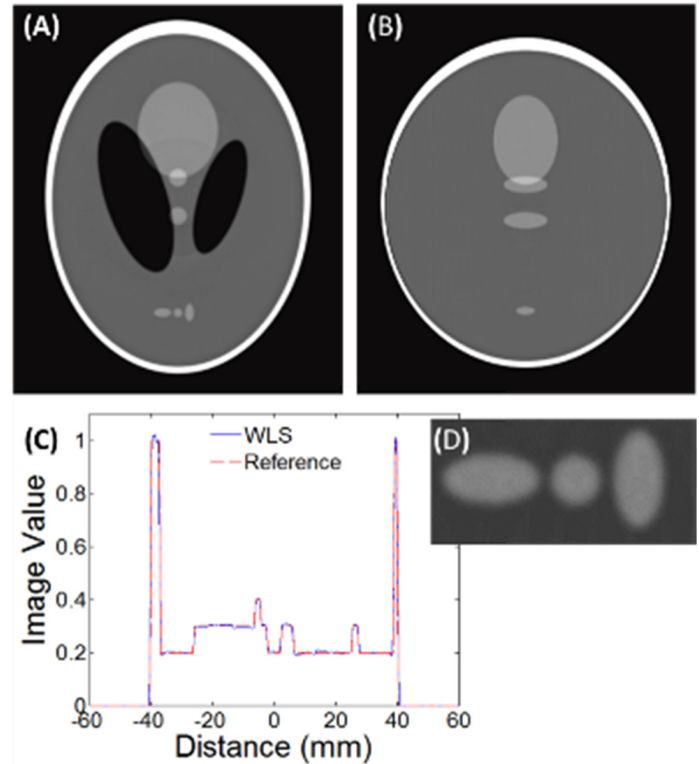

Figure 3.7: A reduced-size 3D Shepp-Logan phantom reconstructed using weighted least squares with the native SBDX geometry. A transverse slice through isocenter (A) and a sagittal slice are shown (B). A vertical profile through (A) is plotted versus the reference profile in (C). A magnified view of the low contrast objects is shown in (D).

Table 3.2: Average 3D reconstruction times

| Method | Minutes |
|---|---|
| PWLS-TV | 29,170.5 |
| gFBP | 0.1 |
| gPWLS-TV | 83.1 |
| Gridding time | 16.2 |

## 3.6 Feasibility of SBDX-CT for 3D anatomic mapping

The feasibility of 3D anatomic mapping via SBDX-CT data acquisition and reconstruction was investigated next. High contrast objects representative of contrast-enhanced chambers were segmented from reconstructed SBDX-CT images of a numerical thorax phantom. The effects of noise level, incomplete angular sampling from simulated cardiac gating, and data truncation, on the segmented surface accuracy were evaluated.

### 3.6.1 Methods

#### *3.6.1.1 Numerical simulations*

A thorax based on the FORBILD phantom was simulated to evaluate the 3D surface accuracy of a segmented contrast-enhanced chamber in the presence of noise.[133] Two phantom sizes were simulated in order to investigate performance without and with data truncation. The full sized thorax phantom contained a 3.6 cm diameter sphere positioned 2.4 cm anterior to isocenter. Sphere enhancement was set to +727 HU relative to background to simulate an arterial injection of iodinated contrast agent. A lower contrast 1.2 cm diameter cylinder (+200 HU) and resolution patterns (0.4, 0.5, 0.6, 0.7 mm) were also included. A reduced-size version of the thorax phantom ($100 \times 50 \times 60$ mm$^3$) was generated by scaling all structure dimensions down by a factor of 2.4.

To simulate noisy projections, detector data was drawn from a Poisson distribution. The simulated source intensity (number of photons per source-detector element ray) was adjusted so that the mean number of counts exiting through the short axis of the phantom matched the mean photon counts measured on the SBDX system with a stack of acrylic in the beam. Three fluence levels were simulated, corresponding to 18.0, 27.2, and 34.6 cm thick acrylic phantoms imaged at 120 kV and full power (120 kV, 200 mA$_{peak}$).

The gFBP and gPWLS-TV reconstruction methods were applied to the projection data, since gridding-based approaches were shown to yield the most practical reconstruction times (Section 3.5). Reconstructions were performed using a $512 \times 512 \times 385$ image grid with 0.234 mm isotropic pixel pitches resulting in a 120 mm in-plane FOV and 90.2 mm axial coverage.

#### *3.6.1.2 Performance metrics*

For each reconstruction method, background noise level, edge blurring, and chamber surface accuracy were evaluated. The noise level was quantified by measuring the standard deviation of image values (HU) in three regions-of-interest defined in the thorax phantom's heart, tissue and lungs. The edge blurring of the high contrast chamber was quantified by the full width half maximum (FWHM) of a line spread function (LSF) derived from an angularly-averaged edge profile. Surface accuracy was calculated by performing intensity-based segmentation of the high contrast sphere, followed by a calculation of the distance of each segmented surface point to the closest point on the ground truth surface. The chamber surface was defined as the boundary of the voxels obtained after thresholding to half the intensity difference between the chamber and background regions. After calculating the distances to the ground truth surface, a histogram of distances was calculated and the 99th percentile distance was determined. The Sørensen-Dice coefficient was also calculated, in order to compare the segmented chamber volume with the ground truth volume.

### 3.6.1.3  *CT Detector binning, scan range and gating*

To reduce noise in projection data and decrease computation time, the 160 x 80 detector elements were binned (2 columns x 4 rows) prior to reconstruction for each of the methods examined. This resulted in a 1.32 mm in-plane pitch and 2.64 mm axial pitch (0.79 mm at isocenter). Following detector binning, the mean photon counts transmitted through the phantom were 52.2, 7.6 and 1.6 photons per ray for the three fluence levels simulated. All CT simulations employed a short-scan consisting of 180 superviews uniformly distributed over 200 degrees.

Although object motion was not simulated in this investigation, cardiac gating was simulated to explore the impact of incomplete sinogram data on image quality. To simulate gated acquisition, 71 superviews were extracted from the full set of 180 superviews resembling a 55% to 95% R-R gating window. Chapter 4 investigates cardiac gating for realistic heart rates ranging from 60 to 90 bpm.

### 3.6.2   Results

#### *3.6.2.1   Selection of regularization weight*

The β parameter in equation (3.14) controls the relative weighting between the data fidelity term and the total variation regularization function, with larger β values leading to increased noise suppression and smoothing. The segmentation of cardiac chambers depends on the tradeoff between voxel noise standard deviation and spatial resolution. Surface accuracy (99% percentile method) and Sørensen-Dice coefficient are plotted versus β parameter for the thorax phantom at the lowest simulated fluence in Figure 3.8A and Figure 3.8B, respectively. A β value of 0.1 was found to maximize surface accuracy and the Sørensen-Dice coefficient for the gPWLS-TV method. Since the focus of this work is the segmentation of large high contrast objects, this β value was assumed to be most suitable for tasks involving segmentation of large high contrast objects, and was used for the reconstructions reported below.

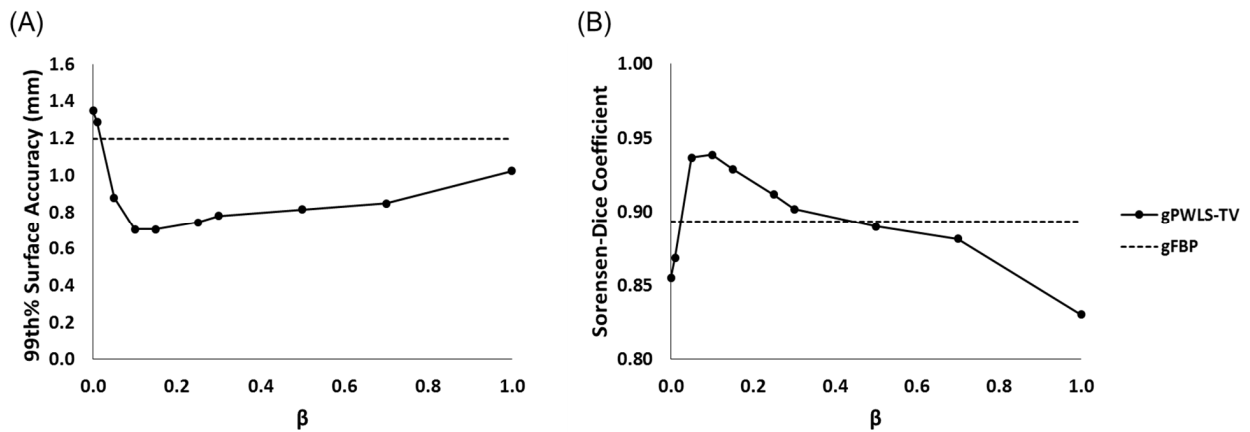

Figure 3.8: The surface accuracy (A) and Sørensen-Dice coefficient (B) are plotted versus the regularization weight, β for the miniature thorax phantom at the lowest simulated photon fluence. The dashed lines represent the surface accuracy of a gridded FBP reconstruction. A β value of 0.1 maximized surface accuracy and Sørensen-Dice coefficient.

#### *3.6.2.2   Miniature thorax phantom*

Example transverse slices of the reduced size 3D thorax phantom are shown in Figure 3.9, for three fluence levels and three reconstruction methods: i) gridded FBP, ii) gridded WLS and iii) gridded PWLS-TV (β=0.1). Figure 3.10 shows reconstructed sagittal slices under the same conditions. For the three fluence levels, the mean photon counts transmitted through the phantom were 52.2, 7.6 and 1.6 photons per ray after detector binning.

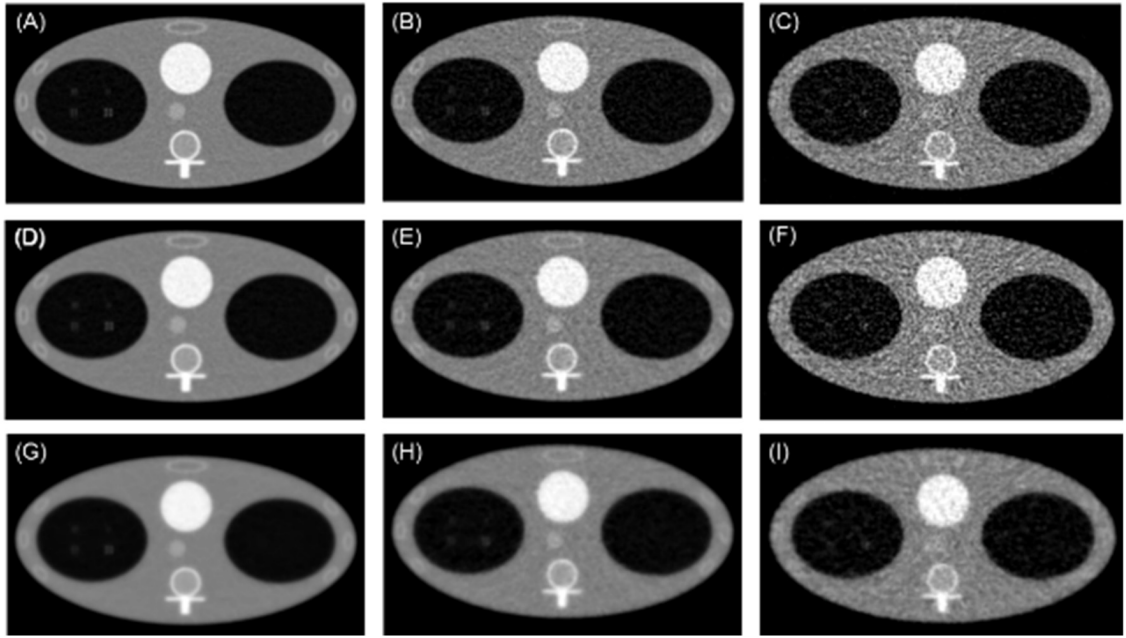

Figure 3.9: Axial slice reconstruction of the reduced-size thorax phantom using gFBP (top row), gWLS (middle row) and gPWLS-TV with β=0.1 (bottom row) reconstruction. The mean fluence transmitted through the phantoms are equivalent to imaging through 18.0, 27.2 and 34.6 cm of acrylic (left to right). Display windows is [-800, 800] HU.

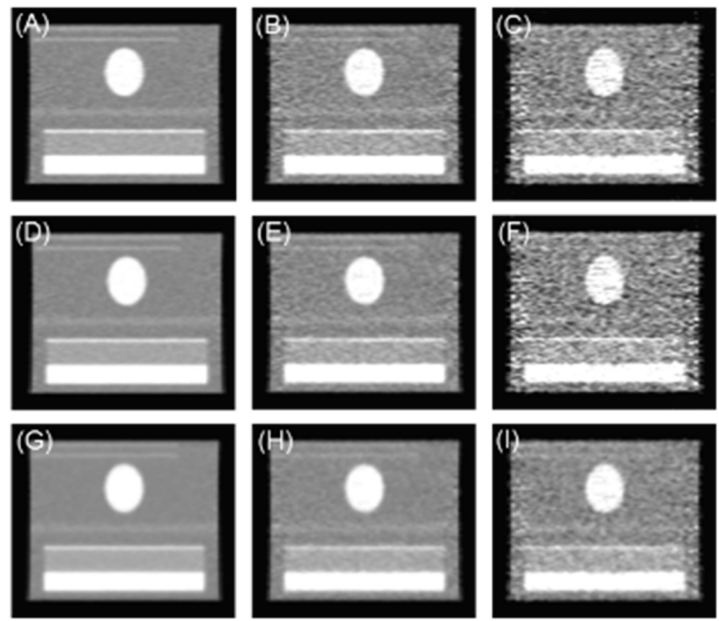

Figure 3.10: Sagittal slice reconstruction of the reduced-size thorax phantom using gFBP (top row), gWLS (middle row) and gPWLS-TV with β = 0.1 (bottom row) reconstruction. The mean fluence transmitted through the phantoms are equivalent to imaging through 18.0, 27.2 and 34.6 cm of acrylic (left to right). Display windows is [-800, 800] HU.

Table 3.3: Reconstructed surface accuracy, LSF FWHM, and noise standard deviation for three fluence levels and three reconstruction methods.

| | Surface Accuracy 99th Percentile (mm) | Sørensen-Dice Coefficient | LSF FWHM (mm) | Noise Standard Deviation (HU) | | |
| --- | --- | --- | --- | --- | --- | --- |
| | | | | Heart | Tissue | Lung |
| *18.0 cm acrylic fluence* | | | | | | |
| gFBP | 0.41 | 0.97 | 0.92 | 27.9 | 24.8 | 17.9 |
| gWLS | 0.33 | 0.96 | 1.06 | 19.7 | 19.1 | 13.5 |
| gPWLS-TV | 0.33 | 0.95 | 1.22 | 9.2 | 10.5 | 5.0 |
| *27.2 cm acrylic fluence* | | | | | | |
| gFBP | 0.53 | 0.96 | 0.94 | 64.5 | 65.7 | 47.1 |
| gWLS | 0.47 | 0.96 | 1.05 | 45.4 | 47.6 | 36.7 |
| gPWLS-TV | 0.47 | 0.95 | 1.23 | 23.4 | 24.4 | 20.0 |
| *34.6 cm acrylic fluence* | | | | | | |
| gFBP | 1.20 | 0.89 | 1.01 | 189.5 | 128.2 | 104.3 |
| gWLS | 1.35 | 0.86 | 0.98 | 217.0 | 150.3 | 123.3 |
| gPWLS-TV | 0.70 | 0.94 | 1.29 | 92.3 | 52.9 | 52.5 |

Qualitatively, the gFBP and gWLS have similar appearance, while the gPWLS-TV has reduced noise and slightly less sharp edges. The reduction in sharpness for gPWLS-TV is a consequence of the choice in β parameter, which was optimized for surface accuracy (Section 3.6.2.1). These observations are reflected in the quantitative measures of noise, spatial resolution, and surface accuracy, presented in Table 3.3. The gPWLS-TV method had the lowest voxel noise, measuring −66%, -61%, and −53% relative to gFBP for the three fluence levels. Averaging the full-width half-maximum of the LSF across fluence levels, the results for gFBP, gWLS, and gPWLS-TV were 0.96 mm, 1.03 mm, and 1.25 mm, respectively.

Under the highest noise condition (34.6 cm acrylic equivalent), the surface accuracy metrics show that the gPWLS-TV method yielded the most accurate segmentations. In this case, 99% of the segmented boundary points were within 0.70 mm of the ground truth surface, versus 1.20 mm and 1.35 mm for the gFBP and gWLS methods. In the lower noise conditions (18.0 cm, 27.2 cm acrylic equivalent) there was a slight improvement in surface accuracy for gPWLS-TV and gWLS relative to gFBP, although results were generally similar (e.g. 0.33-0.41 mm in the lowest noise case). The trends in the Sorensen-Dice coefficient followed those observed in the 99[th] percentile deviation. Overall, results suggest that, for segmentation tasks in an interventional setting, the benefits of iterative reconstruction with regularization are mainly seen in low flux imaging scenarios. For high flux imaging, the gFBP method may be more expedient.

### 3.6.2.3  *Truncated thorax phantom*

The effects of data truncation were investigated by performing 3D reconstructions of the large thorax phantom with the highest fluence level (18.0 cm acrylic equivalent). Figure 3.11 demonstrates transverse slices of the truncated thorax phantom using gPWLS-TV and gFBP.  As expected, the truncated reconstructions demonstrated bright shading artifacts toward the periphery and inaccurate CT numbers. Despite this, intensity-based segmentation of the high contrast chamber was successful. The gFBP method showed improved surface accuracy versus gPWLS-TV with 99% of the segmented surface points being within 0.58 mm of the ground truth versus 1.00 mm for gPWLS-TV. Note the β value (0.1) used for the gPWLS-TV reconstruction was optimized for the lowest fluence case, which resulted in increased edge blurring and decreased image noise versus gFBP. The LSF FWHM of gPWLS-TV measured 1.22 mm versus 0.99 mm for gFBP. The noise standard deviation within the high contrast chamber region measured 17.3 HU for gPWLS-TV versus 31.9 HU for gFBP. The Sørensen-Dice coefficient measured 0.98 for both reconstruction methods.

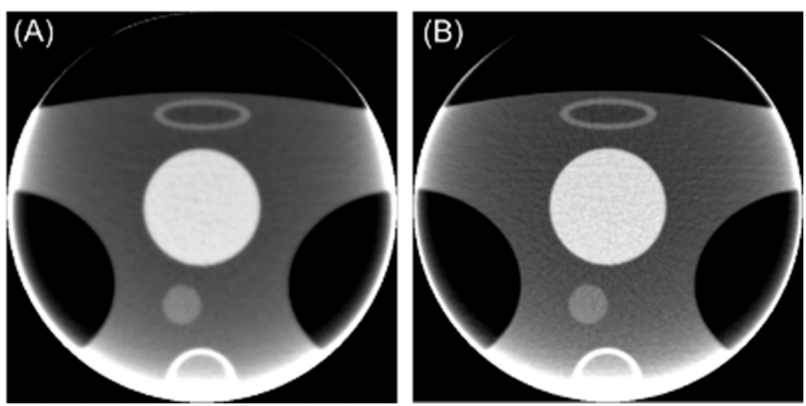

Figure 3.11: Reconstruction of the full size thorax phantom with 3D gridded PWLS-TV (A) and 3D gFBP (B). Display window is [-200, 1000] HU.

### *3.6.2.4 Truncated thorax phantom with gated projections*

Figure 3.12 demonstrates 2D gPWLS-TV and gFBP reconstruction of the full sized thorax phantom using gated projection data (18.0 cm acrylic equivalent fluence). The view angle undersampling created by gating caused streak artifacts in both images. For the gPWLS-TV case, the high contrast chamber could still be recovered by applying a gradient-based segmentation scheme which is less sensitive to gradual brightness variations. The surface segmented from the gPWLS-TV reconstruction was within 1.41 mm of the ground truth 99% of the time. The gFBP surface was not successfully segmented by simple intensity-based or gradient-based methods due to the level of artifacts present. The results demonstrate that gating and truncation together pose a major challenge when existing IGCT approaches are applied. A new solution would be desirable.

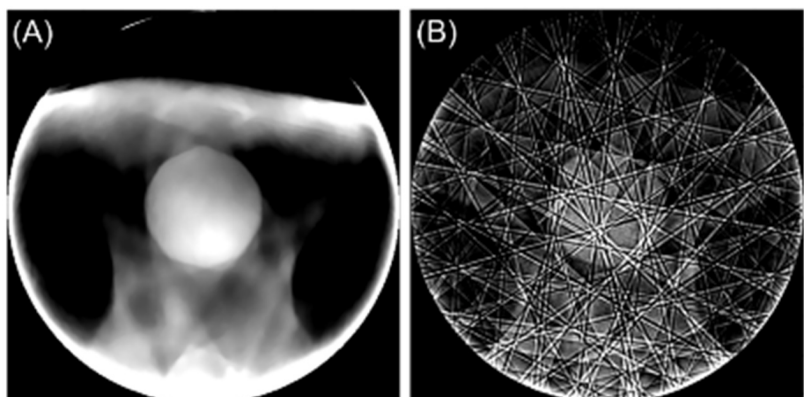

Figure 3.12: Reconstruction of gated data from the full sized thorax phantom, for 2D gridded PWLS-TV (A) and gFBP (B). The display windows were set to [-200, 1000] HU (A) and [-600, 800] HU (B) in order to emphasize the reconstruction artifacts.

## 3.7  Summary

This chapter investigated the feasibility of performing CT imaging with Scanning Beam Digital X-ray. A study of Radon space sampling during simultaneous SBDX source scanning and C-arm rotation showed that CT angular sampling requirements are satisfied by a 182.1 degree short-scan rotation. The SBDX projection data acquired at a single view angle (i.e. superview) covered a 10.1 degree by 144.2 mm region of Radon space for a single slice. The non-uniform sampling due to overlapping superviews leads to ring artifacts if not corrected for during IGCT reconstruction. Several existing IGCT reconstruction algorithms were implemented for the SBDX geometry to evaluate the suitability of these algorithms for the cardiac SBDX-CT task. A statistical iterative reconstruction method (PWLS-TV) adapted for the native SBDX geometry yielded the best reconstruction accuracy when evaluated for a numerical Shepp-Logan phantom. The rRMSE of a PWLS-TV reconstruction measured 0.58% versus 2.62% for a direct fan beam method, 2.18% for gFBP, and 1.85% for gPWLS-TV. Despite offering superior reconstruction accuracy, extending PWLS-TV to three dimensions resulted in long reconstruction times that would be unacceptable and limit workflow in an interventional setting. Reconstruction times were accelerated by a factor of 293 by first gridding the native IGCT projection data to a parallel ray dataset and then performing gPWLS-TV reconstruction.

The gridded FBP and gridded PWLS-TV reconstruction methods were investigated further for the target task of 3D cardiac chamber mapping. Both reconstruction methods were evaluated at three different noise levels, with and without data truncation and view angle undersampling. Results supported the feasibility of accurate anatomic surface mapping with the SBDX system, however several additional challenges must be addressed. The small field-of-view results in projection data truncation that leads to shading artifacts, which, if unaccounted for during image reconstruction, increased the difficulty of segmentation. Angular undersampling in a gated acquisition increases image noise and causes reconstruction artifacts that reduce surface accuracy.

Reconstruction techniques which address data truncation and gating-induced artifacts would help improve the accuracy and stability of 3D surface mapping results. Prior image constrained compressed sensing (PICCS)[134] is a reconstruction method that is capable of generating time resolved cardiac images for a slowly rotating C-arm system[29] in the presence of data truncation.[135] Chapter 4 describes the implementation and evaluation of a PICCS reconstruction algorithm adapted for inverse geometry CT. A more realistic numerical anthropomorphic phantom and cardiac motion model is used to evaluate the proposed reconstruction technique.

# 4  Method for 3D cardiac chamber reconstruction using gated and truncated SBDX projection data

*A subset of the results reported in this chapter were presented at the 2016 AAPM Annual Meeting.*[136]

## 4.1  Introduction

Chapter 3 examined the feasibility of inverse geometry CT reconstruction with Scanning Beam Digital X-ray. Long reconstruction times resulting from an increased number of back projection and forward projection operations in the scanning geometry were reduced by gridding the native SBDX data to parallel ray projections prior to reconstruction. The noise standard deviations were low enough to enable cardiac chamber segmentation under clinically relevant conditions. However, several additional reconstruction challenges were noted. First, since the C-arm rotates slowly relative to the cardiac cycle, the acquired projection data span several heart beats. A scheme is needed to mitigate motion artifacts; it was found that naive application of retrospective cardiac gating can introduce its own artifacts. Second, the SBDX CT field-of-view is small relative to the patient chest dimensions, resulting in projection truncation and cupping artifacts in reconstructed images. The combination of retrospective gating and projection truncation was shown to decrease the segmentation accuracy of high contrast objects. Third, the parallel-ray gridding parameters were selected empirically to demonstrate feasibility. Parallel ray gridding parameters should be optimized for the cardiac chamber mapping task. And fourth, the feasibility study utilized simple numerical phantoms consisting of geometric shapes such as cylinders, ellipsoids and cubes. Segmentation accuracy must be evaluated for more anatomically correct phantoms and motion models.

This chapter presents a new iterative reconstruction method for SBDX-CT which is appropriate for the SBDX geometry and which addresses the challenges associated with cardiac-gated and truncated projection data. Performance of the proposed method is evaluated in realistic numerical simulations of the patient chest with a beating heart. Specifically, the PICCS reconstruction algorithm[134] is adapted for C-arm inverse geometry CT reconstruction and evaluated using the 4D extended cardiac-torso phantom (XCAT).[137,138] PICCS has previously been investigated for cardiac C-arm CBCT and is well suited to deal with the reconstruction challenges faced in developing C-arm IGCT.[29,135] Since this work is focused on cardiac chamber imaging, chamber segmentation accuracy is evaluated as a function of the different gridding parameters and for several noise levels, using static and dynamic cardiac phantoms. The gridding parameters are optimized for the task of 3D cardiac chamber mapping. This chapter serves as the foundation for the subsequent chapters which focus on the hardware implementation and experimental validation of the proposed SBDX-CT technique.

## 4.2  Methods

### 4.2.1  Overview

The overall goal of this chapter is to develop an IGCT reconstruction method that provides sufficient image quality for cardiac chamber (e.g. left atrium) segmentation during interventional procedures such as radiofrequency catheter ablation of atrial fibrillation. Prior to reconstruction, retrospective ECG-gating is performed to reduce projection data inconsistency responsible for cardiac motion artifacts (Sec. 4.2.2). The non-gated and gated projection data are then rebinned to parallel ray datasets to deal with Scanning Beam Digital X-ray's unique inverse-geometry design and sampling of Radon space (Sec. 4.2.3). A projection extrapolation scheme is applied to mitigate truncation artifacts (Sec. 4.2.4). A region-of-interest (ROI) reconstruction technique is proposed to enable iterative reconstruction of the small SBDX-CT FOV and to minimize voxel resolution loss (Sec. 4.2.5). Cardiac motion artifacts are reduced using retrospective ECG-gating in combination with iterative reconstruction via the prior image constrained compressed sensing (PICCS) algorithm (Sec. 4.2.6). An edge-preserving low pass filter is investigated to reduce image noise and improve the robustness of the segmentation results (Sec. 4.2.7). The complete reconstruction method proposed here is referred to as multi-source prior image constrained compressed sensing or MS PICCS. The MS PICCS workflow is summarized in Figure 4.1. Specific implementation details are provided below.

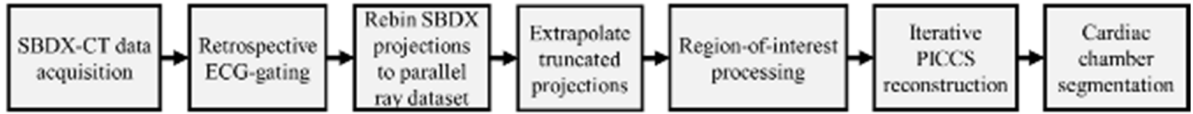

Figure 4.1: Flowchart of the proposed MS PICCS reconstruction algorithm for inverse geometry C-arm CT.

### 4.2.2 Cardiac gating

SBDX-CT projection data, $\widetilde{y}$, are retrospectively gated to select a target cardiac phase for reconstruction. The subset of SBDX superviews acquired within the gating window of a target cardiac phase (e.g. end-systole or end-diastole), defined relative to the R-R interval of an electrocardiogram (ECG) signal, are denoted as $\widetilde{y}_{gated}$. Details of the ECG-gating procedure, including the temporal width of the gating window, are outlined with a description of the numerical XCAT phantom in section 4.2.9. In clinical practice, it is straightforward to record the ECG of the patient during a rotational scan. Alternatively, one could derive the gating signal directly from the image data, similar to the approach described by *Kachelriess et al.*[139] The latter approach is explored in Sec. 7.2.2.

### 4.2.3 Parallel ray gridding procedure

Statistical iterative CT reconstruction using the native SBDX geometry was demonstrated in the previous chapter, however, the long reconstruction times may be unacceptable for interventional imaging. In order to reduce computation times, the SBDX projection data are gridded to a collection of parallel ray projections prior to reconstruction.[120] Denoting the gridding operation as function $G$, a non-gated gridded parallel-ray projection dataset is given by: $\mathbf{y} = G(\widetilde{y})$. A gated parallel ray dataset is given by: $\mathbf{y}_{gated} = G(\widetilde{y}_{gated})$. The non-gated projections, $\mathbf{y}$, are utilized to reconstruct a fully sampled prior image as described in Sec. 4.2.7 below. The gated projection data are used as input to the PICCS reconstruction procedure outlined in Sec. 4.2.6. No additional detector binning was performed prior to the gridding procedure. The gridding parameters are optimized for cardiac chamber mapping as described in section 4.2.11 below.

#### 4.2.4 Projection extrapolation

It is expected that adult subjects will have a chest that is larger than the 144.2 mm (transverse) by 130.0 mm (axial extent) SBDX CT FOV, resulting in fully truncated projection data. An image reconstructed from truncated projection data with an FDK type algorithm[140] will suffer from cupping artifacts and inaccurate CT numbers (i.e. HU). In order to suppress artifacts, this work approximates the object support using a water cylinder technique inspired by the elliptical cylinder method proposed by *Kolditz et al.*, followed by projection extrapolation.[141] A synthetic projection dataset, $\boldsymbol{y_{H2O}}$, is generated assuming a uniform elliptical water cylinder with the monoenergetic linear attenuation coefficient of water, $\mu_{H2O}$, taken from the NIST database at the mean energy of the x-ray spectrum.[142] It is common practice within the cardiac cath lab to measure the anteroposterior dimension, $d_{AP}$, of the patient prior to performing an intervention. Therefore, it's assumed that the mediolateral dimension, $d_{ML}$, can also be measured with a negligible interruption to the clinical workflow. For a projection ray at a distance $\rho$ from the water cylinder center, the projection value, p(u, v, ϕ), is given by the ray intersection length with the elliptical water cylinder, multiplied by the linear attenuation of water, $\mu_{H2O}$, as,

$$p(u,v,\phi) = \frac{2\mu_{H2O} d_{AP} d_{ML}}{d_{AP}{}^2 \cos^2\phi + d_{ML}{}^2 \sin^2\phi} \sqrt{d_{AP}{}^2 \cos^2\phi + d_{ML}{}^2 \sin^2\phi - (x_0 \cos\phi + y_0 \sin\phi - \rho)^2} \tag{4.1}$$

where, $u$ and $v$ are the parallel ray detector indices, $\phi$ represents the rotation angle (i.e. view angle) about the axis-of-rotation, and $(x_0, y_0)$ are the coordinates of the center of the ellipse.

After computing $\boldsymbol{y_{H2O}}$, the gridded parallel-ray projection data, $\boldsymbol{y}$, is extrapolated at the truncation boundary to yield a non-truncated projection dataset denoted, $\boldsymbol{y_{extrapolated}}$, using the algorithm outlined by *Lauzier et al.*[135] The extrapolation method guarantees continuity at the boundary between the gridded projection data and the extrapolated region as shown in Figure 4.2.

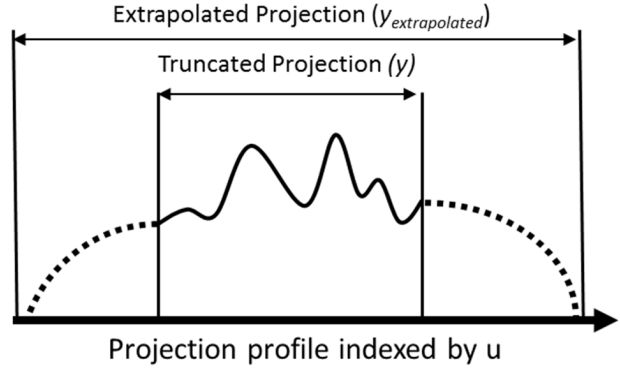

Figure 4.2: Projection extrapolation is performed to mitigate truncation artifacts that would result from FBP type reconstruction of the small SBDX CT FOV.

### 4.2.5    Region-of-interest reconstruction

Region-of-interest (ROI) reconstruction was performed to limit the voxel resolution loss which would occur from reconstructing an extended FOV, assuming the number of reconstructed image voxels remains constant due to computing resources. The method described in detail by *Ziegler et al.* was used in this work, adapted for the parallel ray geometry.[143] A brief review of the method and implementation details pertinent to this work are provided here.

Prior to iterative ROI reconstruction, the goal is to estimate the component of the projection data, $y$, which is attributed to attenuation by the object within the scanner FOV, shown as region **F** in Figure 4.3, versus attenuation due to the portion of the object extending outside of the FOV, shown as region **E**. This is accomplished by first reconstructing the EFOV using a filtered backprojection (gFBP) algorithm with the extrapolated projection data. Denoting the image over the EFOV as, $x_E$, the reconstruction procedure is expressed as, $x_E = gFBP(y_{extrapolated})$. Second, the image values within the SBDX-CT FOV plus a 2 mm transition region (region T) are set to zero. The masked image is denoted, $x_{mask}$, as shown in Figure 4.3B. Third, a forward projection is performed through the mask image, $x_{mask}$, to estimate the portion of the projection data that can be attributed to region E. Representing the forward projection process by the system matrix **A**, the projection data due to attenuation in region E, $y_E$, are estimated as, $y_E = Ax_{mask}$. Finally, the portion of projection data attributed to attenuation by the object within the SBDX CT FOV can be estimated by subtraction as, $\boldsymbol{y_F} = \boldsymbol{y} \text{-} \boldsymbol{y_E}$. The projection data $\boldsymbol{y_F}$ was then used to perform iterative reconstruction of an ROI as described in the next section.

### 4.2.6 Adapted PICCS objective function

The PICCS algorithm was implemented for SBDX-CT reconstruction as PICCS has been shown capable of reducing motion artifacts that result from the use of a slowly rotating C-arm.[29,135] The PICCS reconstruction was cast as an optimization problem,[135]

$$\hat{x} = \underset{x}{arg\,min}\, f_{PICCS}(x) \tag{4.2}$$

$$f_{PICCS}(x) = \frac{\alpha \left|\psi\big(x - x_p\big)\right|_{1,F} + (1-\alpha)|\psi(x)|_{1,T\cup F}}{\left|\psi\big(x_p\big)\right|_1} + \frac{\lambda}{2} \frac{\big(\boldsymbol{A_{T\cup F}}x - \boldsymbol{y_{F,gated}}\big)^T \boldsymbol{D}\big(\boldsymbol{A_{T\cup F}}x - \boldsymbol{y_{F,gated}}\big)}{\left\|\boldsymbol{A}x_p\right\|_2^2} \tag{4.3}$$

where $\hat{x}$ is the reconstructed image, $x_p$ is a prior image, $\boldsymbol{y_{F,gated}}$ is the component of the ECG-gated projection data in the ROI corresponding to the scanner FOV denoted by region F (Figure 4.3), $\boldsymbol{A_{T\cup F}}$ is the system matrix and $\boldsymbol{D}$ is a noise matrix. The system matrix $\boldsymbol{A_{T\cup F}}$ operates over the SBDX FOV (F) and a 2 mm transition region (T). The noise matrix $\boldsymbol{D}$ was set equal to the identity matrix. The function $\psi(x)$ is a sparsifying transform. The total variation (TV) function is used as the sparsifying transform in this work. The scalar $\alpha$ controls the relative weight between the PICCS term and the traditional compressed sensing term. The scalar $\lambda$ determines the weight of the data consistency term. The first term of the PICCS objective function is constrained to operate over only the scanner FOV (region F) as the prior image is not expected to be accurate outside of the FOV. The second term operates over the entire image volume (region T ∪ F).

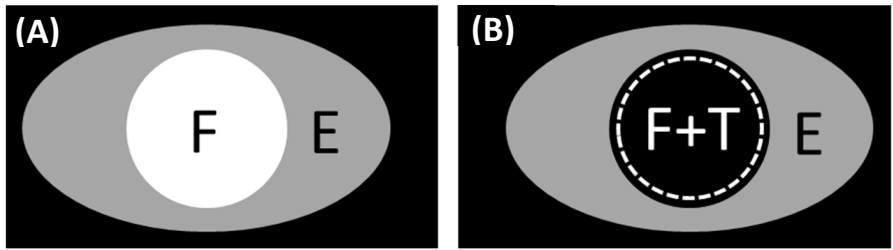

Figure 4.3: (A) SBDX CT imaging is performed with a 144.2 mm FOV shown as region F. Projection extrapolation is used to extend the scanner FOV, region E, and mitigate truncation artifacts. (B) For iterative region-of-interest reconstruction, projection contributions outside of the FOV plus a 2 mm transition region T, denoted by a dotted line, are estimated by forward projection of a mask image.

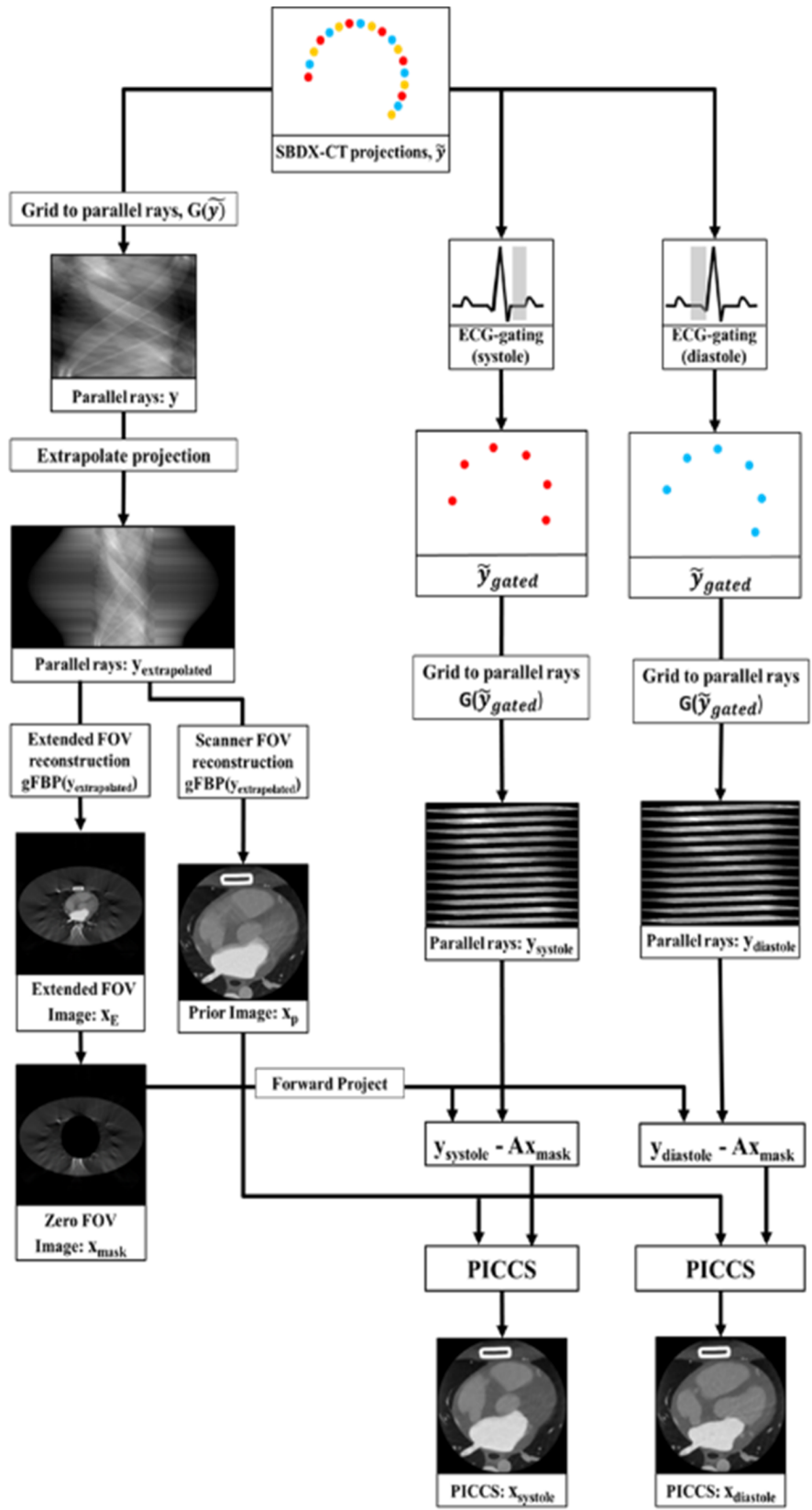

Figure 4.4: The inverse geometry ECG-gated CT reconstruction scheme is shown.

### 4.2.7 Prior image reconstruction

The prior image, $\mathbf{x_p}$, was reconstructed from the extrapolated projection data without cardiac gating as, $x_p = gFBP(y_{extrapolated})$. The reconstruction FOV was set equal to the SBDX CT FOV plus a 2 mm transition region. As the prior image is reconstructed from projection data that spans multiple heart beats, it is expected to be contaminated with motion artifacts. Following gFBP reconstruction, the prior image was filtered with an anisotropic diffusion filter to reduce noise.[144,145] The filter process was implemented using a 26-nearest neighbor finite difference approach[146] with the filtered image at iteration number $i$ given by the discrete update equation:

$$x_p^{i+1} = x_p^i + \Delta\tau \cdot g\left(|\nabla x_p|\right) \cdot \nabla x_p \tag{4.4}$$

The step size, $\Delta\tau$, was set to 3/44 as suggested by *Gerig et al.* for numerical stability.[146] The diffusion function $g\left(|\nabla x_p|\right)$ was specified as:

$$g\left(|\nabla x_p|\right) = \frac{1}{1 + \left(\frac{|\nabla x_p|}{\kappa}\right)^2} \tag{4.5}$$

Optimization of the diffusion filter parameters for the segmentation task is difficult as noise can also be reduced, with a resolution tradeoff, by changing the weight of the PICCS regularization term or by increasing the parallel ray gridding kernel widths. As a result, the diffusion filter parameters were chosen empirically but consistent across all reconstructions performed. The constant $\kappa$ was set to the maximum gradient value (i.e. $\kappa = \max(|\nabla x_p|)$) and a single diffusion iteration was performed.

### 4.2.8 PICCS implementation details

The PICCS objective function was minimized using a nonlinear conjugate gradient method with a backtracking line search.[147] The iterative reconstruction procedure was initialized with the prior image, $\mathbf{x_p}$. The reconstruction procedure was terminated following 100 iterations or a difference in the PICCS objective function between the current iteration, $k$, and two iterations prior, $k$-2, of less than 5 x 10$^{-7}$, (i.e. $f_{PICCS}(x_k) - f_{PICCS}(x_{k-2}) < 5 \times 10^{-7}$). The system matrix **A** was approximated using a ray-driven forward projector and pixel-driven backprojection with linear interpolation at the detector. For all reconstructions, the image matrix size was set to 512 x 512 x 300 voxels with dimensions of 0.28 mm x 0.28 mm x 0.43 mm. The MS PICCS reconstruction algorithm was implemented using the CUDA (NVIDIA, Santa Clara, CA) and C programming languages. All reconstructions were performed on a Windows PC with an Intel Xeon 3.40 GHz processor and an NVIDIA GeForce GTX 1080 GPU.

### 4.2.9   Numerical cardiac torso phantom

SBDX-CT projection data were simulated for the 4D extended cardiac-torso (XCAT) phantom with known ground truth to evaluate the performance of the proposed reconstruction approach.[137] The XCAT phantom was derived from segmentations of human images and presents a realistic cardiac anatomy to study SBDX-CT. Projection data were generated by forward projection through a high resolution (up-sampled 2.4 times the size of the reconstructed volume) version of the 4D XCAT phantom. The anterior-posterior and lateral dimensions of the phantom chest measured 232.2 mm and 325.2 mm respectively, resulting in fully truncated projections. A monochromatic energy spectrum was assumed to enable the investigation of reconstruction performance without confounding factors such as beam hardening. Phantom studies presented in later chapters investigate 3D cardiac chamber mapping accuracy with CT projection data acquired with the actual SBDX system, and these studies include the effects of beam hardening, scatter, and other physical phenomena not accounted for in the numerical simulations. The linear attenuation coefficient values of the XCAT phantom were selected at 52.5 keV, corresponding to the mean energy of a simulated 120 kV x-ray spectrum. The x-ray spectrum was simulated using the TASMICS based x-ray spectrum modeling toolkit SPEKTR 3.0.[148] C-arm rotation was modeled with an angular velocity of 14.5 degree/s, approximately equal to the C-arm rotation speed of the SBDX prototype system. A total of 217 projection view angles were simulated over a 210 degree short-scan rotation, consistent with the 15 frame/s SBDX imaging mode.

Physiological parameters, such as heart rate, can be specified for the XCAT phantom to evaluate reconstruction performance versus a known ground truth for different imaging scenarios. For this work,

SBDX-CT projection data were simulated for the 4D XCAT phantom at heart rates of 60, 75 and 90 beats per minute (BPM). To simulate an arterial contrast injection, the left atrium blood pool and connected pulmonary veins were modeled using the linear attenuation of water with an increased density of 2.5 g/cm³ similar to the approach by *Taubmann et al.*[149] Poisson noise was added to the simulated projection data at three realistic fluence levels corresponding to SBDX imaging at 10%, 25%, and 100% full power (24.3 kWp) at 120 kV. The simulated x-ray fluence was calibrated by matching simulated x-ray transmission through varying thicknesses (18.64 cm to 34.95 cm) of poly(methyl methacrylate) (PMMA) with experimental measurements performed on the SBDX system. Prior to reconstruction, the projection data were retrospectively gated at end-systole and end-diastole using a two-frame (i.e. 0.13 second) rectangular gating window defined within the R-R interval. The ground truth XCAT phantom is shown at end-systole and end-diastole in Figure 4.5. The left ventricle and left atrium volumes are plotted in Figure 4.5C as a function of cardiac phase for the 60 BPM case to demonstrate the variation in chamber size over the cardiac cycle. The left atrium volume varied by 1.4% (0.9 ml) at ventricular end-systole and 2.0% at end-diastole (0.7 ml) within the 0.13 s gating window. For comparison, left atrium volume varied by 87.2% (30.0 ml) across the complete cardiac cycle.

Noise-free projection data were also simulated for a static XCAT phantom at end-systole (Figure 4.5A) to compare the gated reconstructions to an imaging scenario free of cardiac motion. In order to compare SBDX-CT performance to a traditional cone beam CT (CBCT) scanner, projection data were generated for a 1240 x 960 pixel flat panel detector with 0.308 mm pixel resolution. The source-detector-distance (SDD) was 1201 mm and the source to axis-of-rotation distance (SAD) was 750 mm. Reconstruction was performed using the FDK algorithm.[140]

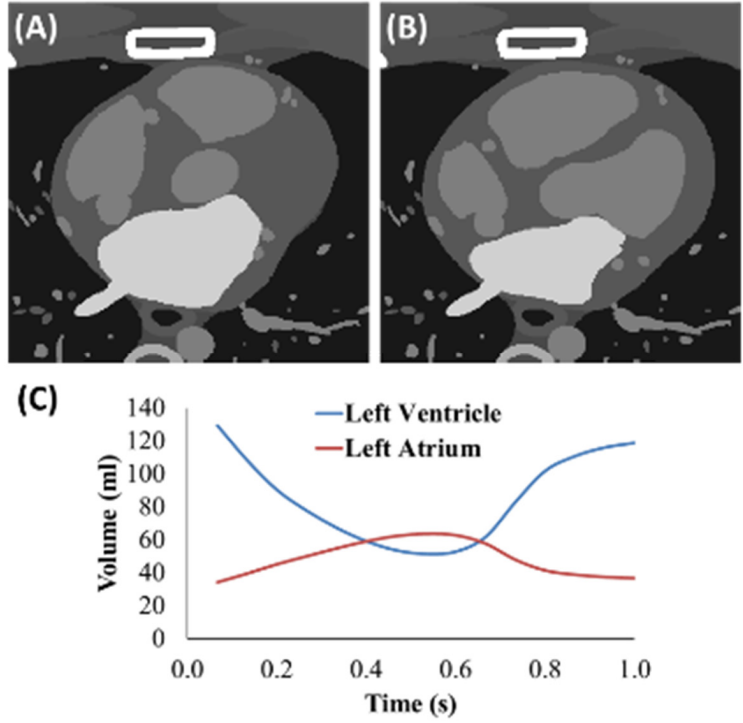

Figure 4.5: A ground truth XCAT phantom is shown at ventricular end-systole (A) and end-diastole (B). The volumes of the left ventricle and left atrium are plotted as a function of time for a single heartbeat.

## 4.2.10 Evaluation metrics

Several metrics were used to quantify image quality and left atrium segmentation accuracy.

### 4.2.10.1 Reconstruction accuracy

The relative root mean square error (rRMSE) and universal image quality index (UQI) were used to quantify the reconstruction accuracy.[150] The rRMSE of a reconstructed image, $\mathbf{x}$, was computed versus the ground truth image $\mathbf{x}^{\text{ref}}$ as,

$$rRMSE = \frac{1}{\max(\boldsymbol{x^{ref}}) - \min(\boldsymbol{x^{ref}})}\sqrt{\frac{1}{N_\Omega}\sum_{i\epsilon\Omega}(x_i - x_i^{ref})^2} \qquad (4.6)$$

where $\Omega$ designates the ROI over which the rRMSE was computed and $N_\Omega$ denotes the number of voxels within the ROI. For the numerical XCAT phantom, $\Omega$ designates the region within the SBDX CT FOV and excludes the extended FOV and transition regions.

The UQI is defined according to Ref. 135 as,

$$UQI = \frac{4\sigma_{x,ref}\mu_x\mu_{ref}}{(\sigma_x^2 + \sigma_{ref}^2)(\mu_x^2 + \mu_{ref}^2)} \tag{4.7}$$

where

$$\mu_x = \frac{1}{N_\Omega}\sum_{i\in\Omega} x_i,$$

$$\mu_{ref} = \frac{1}{N_\Omega}\sum_{i\in\Omega} x_i^{ref},$$

$$\sigma_x^2 = \frac{1}{N_\Omega - 1}\sum_{i\in\Omega}(x_i - \mu_x)^2,$$

$$\sigma_{ref}^2 = \frac{1}{N_\Omega - 1}\sum_{i\in\Omega}(x_i^{ref} - \mu_{ref})^2,$$

$$\sigma_{x,ref} = \frac{1}{N_\Omega - 1}\sum_{i\in\Omega}(x_i - \mu_x)(x_i^{ref} - \mu_{ref}).$$

The value of UQI can vary with a range of [-1, 1]. A value of 1 indicates perfect agreement between the reconstructed image and the ground truth image. The UQI was computed over the same ROI, $\Omega$, as the rRMSE.

### 4.2.10.2 Cardiac chamber segmentation accuracy

The left atrium and pulmonary veins were segmented from reconstructed images using an intensity based threshold segmentation method. The threshold value was set to half the intensity difference between an ROI centered in the left atrium and an ROI centered in the left ventricle.

The accuracy of the left atrium segmentation was quantified using the Sorensen-Dice coefficient (DC) and a histogram of segmentation errors. After segmentation of the left atrium, the Sorensen-Dice coefficient was computed versus the known ground truth atrium.

$$DC = \frac{2|A_{recon} \cap A_{ref}|}{|A_{recon}| + |A_{ref}|} \tag{4.8}$$

Here, $|A_{recon}|$ denotes the number of image voxels in the left atrium segmented from the reconstructed image, $|A_{ref}|$ denotes the number of voxels in the reference atrium, and $|A_{recon} \cap A_{ref}|$ denotes the

number of voxels between the intersection of the segmented and reference atriums. A DC value of 1.0 indicates the segmented atrium has 100% overlap with the ground truth atrium. A value of 0.0 indicates no overlap.

A second metric was developed to further quantify the segmentation error. Following the segmentation of the left atrium, the distance of each atrium surface point to the closest point on the reference atrium was computed and referred to as segmentation error. A histogram of segmentation errors was then calculated and the distance corresponding to the 99[th] percentile was determined and used as a metric for segmentation accuracy. The 99[th] percentile criteria was considered to more accurately represent image quality than maximum deviation which is sensitive to outliers. For this work, image quality was defined as sufficient for reconstructions yielding segmentation errors less than or equal to 3.3 mm, which is comparable to the diameter of a typical ablation catheter tip (1.7-3.3 mm).[99,151,152]

### 4.2.11  Selection of reconstruction parameters

The accuracy of high contrast object segmentation from SBDX-CT images was shown in chapter 3 to depend on image noise level and regularization strength. Additionally, segmentation accuracy depends on the grid dimensions and kernel sizes used during the parallel ray gridding procedure outlined in section 3.3. Optimal reconstruction parameters are difficult to derive as resolution and noise may be traded by varying several reconstruction parameters including i) parallel ray grid dimensions, ii) gridding kernel sizes, iii) PICCS reconstruction parameters ($\alpha$, $\lambda$) and iv) use of various post-processing techniques including low-pass filtering. A two-step approach was utilized here to optimize reconstruction parameters for the target task of 3D anatomic mapping. First, the gridding parameters were optimized for segmentation of the XCAT phantom's left atrium. Second, PICCS reconstruction parameters were optimized.

### *4.2.11.1 Investigation of gridding parameters for a static XCAT phantom*

The static XCAT phantom at end-systole (Figure 4.5A) was used to evaluate the impact of gridding parameters on segmentation accuracy by removing organ motion as a confounding factor. The dynamic XCAT phantom is considered later. Projection data were simulated as described in section 4.2.9 for the end-systole XCAT phantom for four different noise scenarios including noise-free, 100%, 25%, and 10% full power at 120 kV.

Reconstruction of the static projections was performed using non-gated gridded FBP for different combinations of parallel ray geometries and gridding kernel sizes. The sampling density of the gridded parallel ray datasets was varied by fixing the transverse and axial field-of-view and changing the projection sampling (pitch) along the u and v dimensions of the parallel ray dataset. The field-of-view of the estimated parallel-ray dataset (hereafter referred to as the gridded dataset) was fixed at 140 mm (transverse/u dimension) by 100 mm (axial/v dimension). As a result, the number of parallel rays in a given dataset was determined by dividing the FOV by the parallel ray sampling pitch, denoted as $\Delta u$ and $\Delta v$. For example, when the transverse sampling pitch (i.e. $\Delta u$) was set to 0.25 mm and the axial sampling pitch (i.e. $\Delta v$) was set to 0.5 mm, the grid dimension was 560 x 201 rays. The dimensions of the u and v parameters were rounded up to the nearest even and odd integer, respectively.

The $\Delta u$ parameter was varied as 0.15 mm, 0.25 mm and 0.35 mm. The $\Delta v$ parameter was varied as 0.25 mm, 0.5 mm and 0.75 mm. The gridding kernel sizes were set as a multiple of the $\Delta u$ and $\Delta v$ values. Kernel widths ($k_{\Delta u}$, $k_{\Delta v}$) were set as 2, 3, and 4 times the corresponding $\Delta u$ and $\Delta v$ values. The parallel ray dataset consisted of 480 view angles equally distributed over 180°. A 1.0° angular kernel width was used. The tilt angle, $\theta$, was set equal to 4.5° as it ensures all acquired SBDX rays are used during the gridding process. Each combination of grid dimension and kernel width was evaluated. The parameters considered are summarized in Table 4.1.

Table 4.1: Parallel ray rebinning parameters considered for the static end-systole phantom.

| | |
|---|---|
| Number of ray columns (u axis) per view | 934, 560, 400 |
| $\Delta u$ (mm) | 0.15, 0.25, 0.35 |
| $k_{\Delta u}$ (mm) | [0.3, 0.45, 0.6], [0.5, 0.75, 1.0], [0.7, 1.05, 1.4] |
| Number of ray rows (v axis) per view | 401, 201, 135 |
| $\Delta v$ (mm) | 0.25, 0.5, 0.75 |
| $k_{\Delta v}$ (mm) | [0.5, 0.75, 1.0], [1.0, 1.5, 2.0], [1.5, 2.25, 3.0] |
| Number of view angles over 180 degrees | 480 |
| Angular ($\phi$) kernel width (degrees) | 1.0 |
| $\theta$ kernel width (degrees) | 4.5 |

### *4.2.11.2 Investigation of gridding parameters for a dynamic XCAT phantom*

A dynamic XCAT phantom with realistic cardiac motion at a frequency of 75 BPM was used to investigate the impact of different grid dimensions and kernel widths on segmentation accuracy. The dynamic XCAT projection data was reconstructed using the proposed MS PICCS reconstruction technique. The grid dimensions and kernel widths studied for the XCAT thorax phantom with cardiac motion were the same as those for the static case, summarized in Table 4.1.

While the focus of this analysis was to optimize the gridding parameters for the segmentation task, reasonable iterative reconstruction parameters ($\lambda$, $\alpha$), optimized separately in the following section (4.2.11.3), must be selected. The parameter $\alpha$ was set to 0.25. The parameter $\lambda$ was fixed at 3000 for all reconstructions. For the 10% full power case, a $\lambda$ value of 300 was also considered. The $\lambda$ values were selected empirically by examining the segmentation error for a series of reconstructions (25% full power) performed with $\lambda$ spanning several orders of magnitude, $\lambda \in [3 \times 10^0, 3 \times 10^1, 3 \times 10^2, 3 \times 10^3]$.

Segmentation accuracy depends on the contrast between a targeted cardiac chamber and the surrounding structures or background regions, as well as image noise. The anisotropic diffusion filter described in section 4.2.7 was applied to the reconstructed images to determine if segmentation accuracy could be improved. A total of 1, 3, 5 and 10 iterations of the diffusion filter were considered. The application of the low-pass filter results in a tradeoff between noise reduction and edge blurring.

### *4.2.11.3 PICCS reconstruction parameters (α, λ)*

PICCS reconstructions were performed for $\alpha$ equal to 0.0, 0.25, 0.5, and $\lambda$ values ranging from 100 to 15,000. For $\alpha$ equal to 0.0, the PICCS reconstruction method is equivalent to compressed sensing with total variation regularization and referred to as TV-CS. The segmentation accuracy, quantified as the 99[th] percentile segmentation error, was computed for each combination of PICCS parameters to determine the optimal reconstruction parameters for the segmentation task. Each of the iterative reconstruction results was compared to a non-gated gFBP reconstruction.

## 4.3  Results

The results are presented in two parts. First, the optimization of parallel ray gridding parameters for cardiac chamber mapping is presented. Second, the proposed MS PICCS reconstruction approach is evaluated for three heart rates and three noise levels.

### 4.3.1  Investigation of gridding parameters for a static phantom

A total of 81 reconstructions were performed from the noise free static end-systole projection data using the different combinations of gridding parameters summarized in Table 4.1. Segmentation errors as a function of $\Delta u$ and $\Delta v$ are shown in Figure 4.6 for the different kernel widths ($k_{\Delta u}$, $k_{\Delta v}$). The segmentation error increased for large grid dimensions and large $\Delta u$ and $\Delta v$ that can be attributed to the loss of resolution resulting from use of a coarser parallel ray sampling grid. The 99th percentile segmentation error ranged from 0.56-0.77 mm. For all cases, the segmentation error was less than the typical diameter of an ablation catheter tip (1.7-3.3 mm).[99,151,152]

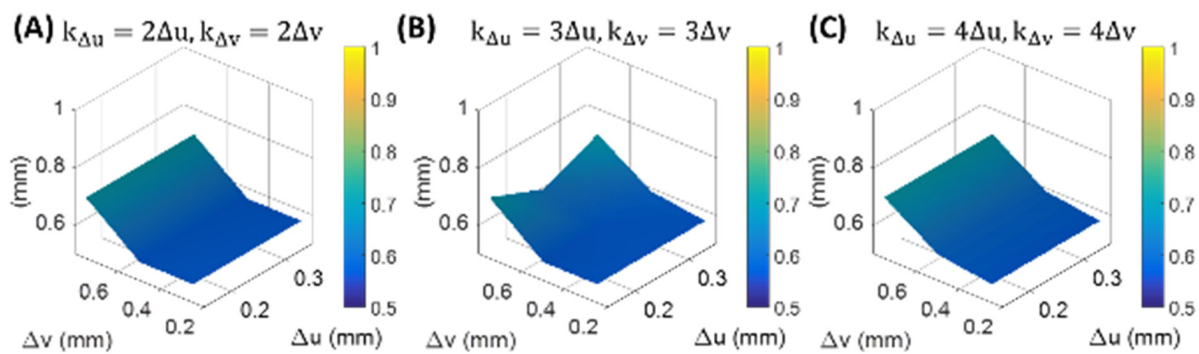

Figure 4.6: Segmentation error versus gridding parameters for noise-free, end-systole (static phantom) projection data. Gridding kernel widths increase as a function of $\Delta u$ and $\Delta v$ from left to right.

Segmentation error as a function of the gridding parameters was analyzed next for reconstructions performed from projection data with added Poisson noise equivalent to imaging at 100%, 25% and 10% full power. Figure 4.7 plots the 99[th] percentile error as a function of the gridding parameters for each of the noise levels considered.

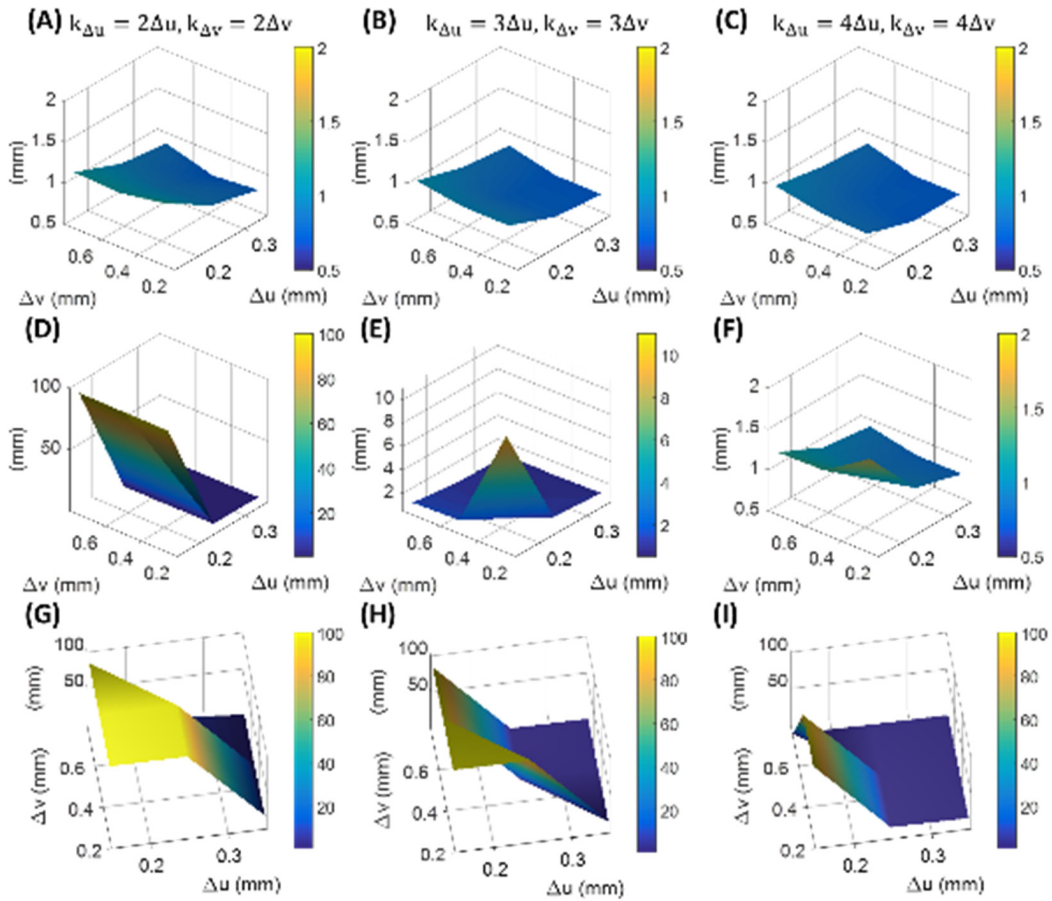

Figure 4.7: Segmentation error versus gridding parameters for end-systole (static phantom) projection data with Poisson noise corresponding to imaging at 100% (top row), 25% (middle row) and 10% full power. Gridding kernel widths increase as a function of Δu and Δv from left to right.

The left atrium was successfully segmented from each of the 81 reconstructions performed for the 100% full power case (Figure 4.7A-C) with errors ranging from 0.71-1.26 mm, depending on ray sampling density and kernel width. Segmentation errors ranged from 0.77-10.48 mm and 0.95-10.91 mm for the 25% and 10% full power cases, respectively. For the 25% full power case, 8 of the 81 (9.9%) reconstructions could not be segmented using the simple intensity based thresholding technique due to increased image noise. Segmentations with errors greater than 15 mm were categorized as failed segmentations. Left atrium segmentation failed for 33 of 81 (40.7%) reconstructions at 10% full power. Comparing the surface plots from left to right, Figure 4.7 shows that larger gridding kernel widths (i.e. panels C, F, I) were associated with improved segmentation accuracy and robustness, at each power level. Panels (D) and (G) of Figure 4.7 show large segmentation errors for the smallest Δu and Δv values considered due to an increase in image noise as the number of native SBDX projection rays used to estimate the parallel rays, and corresponding fluence, is reduced as the gridding kernel widths decrease. Example reconstructions for the central axial slice are shown in Figure 4.8 for the optimal gridding parameters at each noise level. The optimal gridding parameters are summarized in Table 4.2. The corresponding coronal slices are shown in Figure 4.9. Figure 4.10 presents example reconstructions where intensity based thresholding failed at 25% and 10% full power.

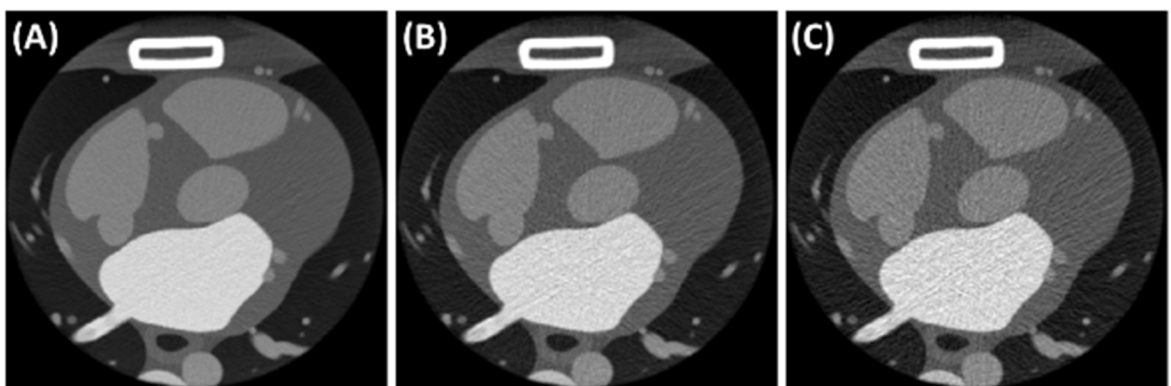

Figure 4.8: Reconstructed axial slice images for a static XCAT phantom are shown using the optimal gridding parameters for imaging at 100% (A), 25% (B), and 10% full power (C). [WL = 500, WW = 1500] HU.

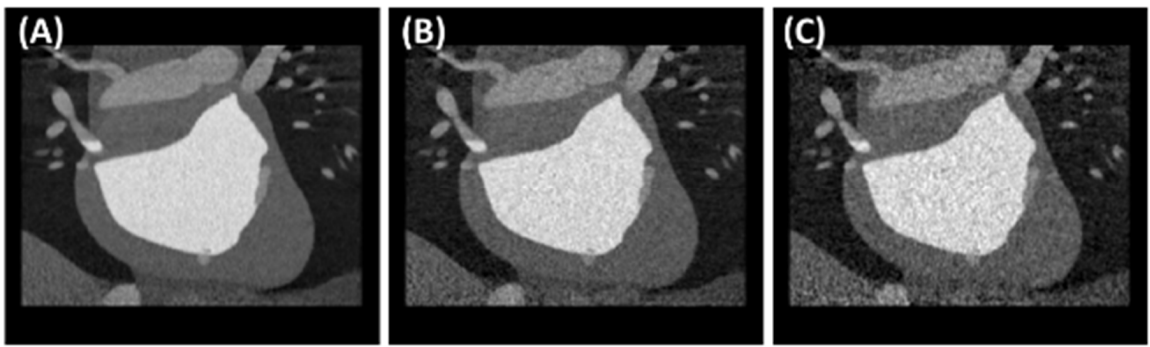

Figure 4.9: Reconstructed coronal slice images for a static XCAT phantom are shown using the optimal gridding parameters for imaging at 100% (A), 25% (B), and 10% full power (C). [WL = 500, WW = 1500] HU.

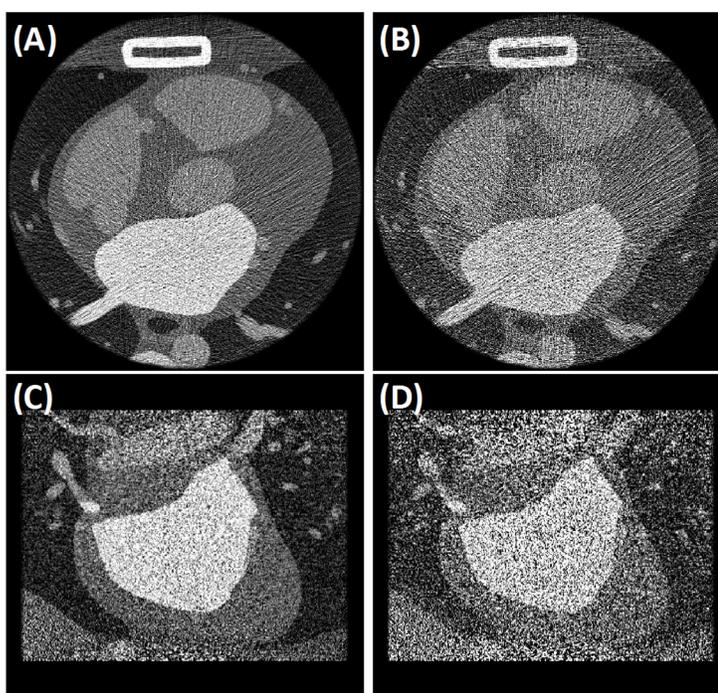

Figure 4.10: Example reconstructions where left atrium segmentation failed using an intensity-based thresholding method. Images are shown for imaging at 25% (A and C) and 10% full power (B and D).

Anisotropic diffusion filtering was investigated to improve segmentation accuracy and robustness. The number of diffusion iterations was varied as 1, 3, 5, and 10. Segmentation errors for each power level are shown in Figure 4.11 following 1 iteration of diffusion filtering which can be compared against Figure 4.7 where no filtering was performed. Table 4.2 lists the number of diffusion iterations that minimized segmentation error for each noise level considered. Segmentation errors ranged from 0.71-1.11 mm for the 100% full power case representing a modest improvement in segmentation robustness when compared to the unfiltered results which ranged from 0.71-1.26 mm. Larger reductions in segmentation errors were observed for the noisier images. For the 25% power level, errors ranged from 0.80-1.24 mm with anisotropic diffusion filtering versus 0.77-10.48 mm without. Similarly, at 10% full power, errors ranged from 0.91-2.95 mm with filtering versus 0.95-10.91 mm for the unfiltered case. For each power level, improved segmentation robustness was demonstrated by the smaller range between maximum and minimum segmentation error. Diffusion filtering reduced the maximum segmentation error at all power levels considered. The use of diffusion filtering reduced the percentage of failed segmentations from 9.9% to 0.0% at 25% full power, and from 40.7% to 1.5% at 10% full power.

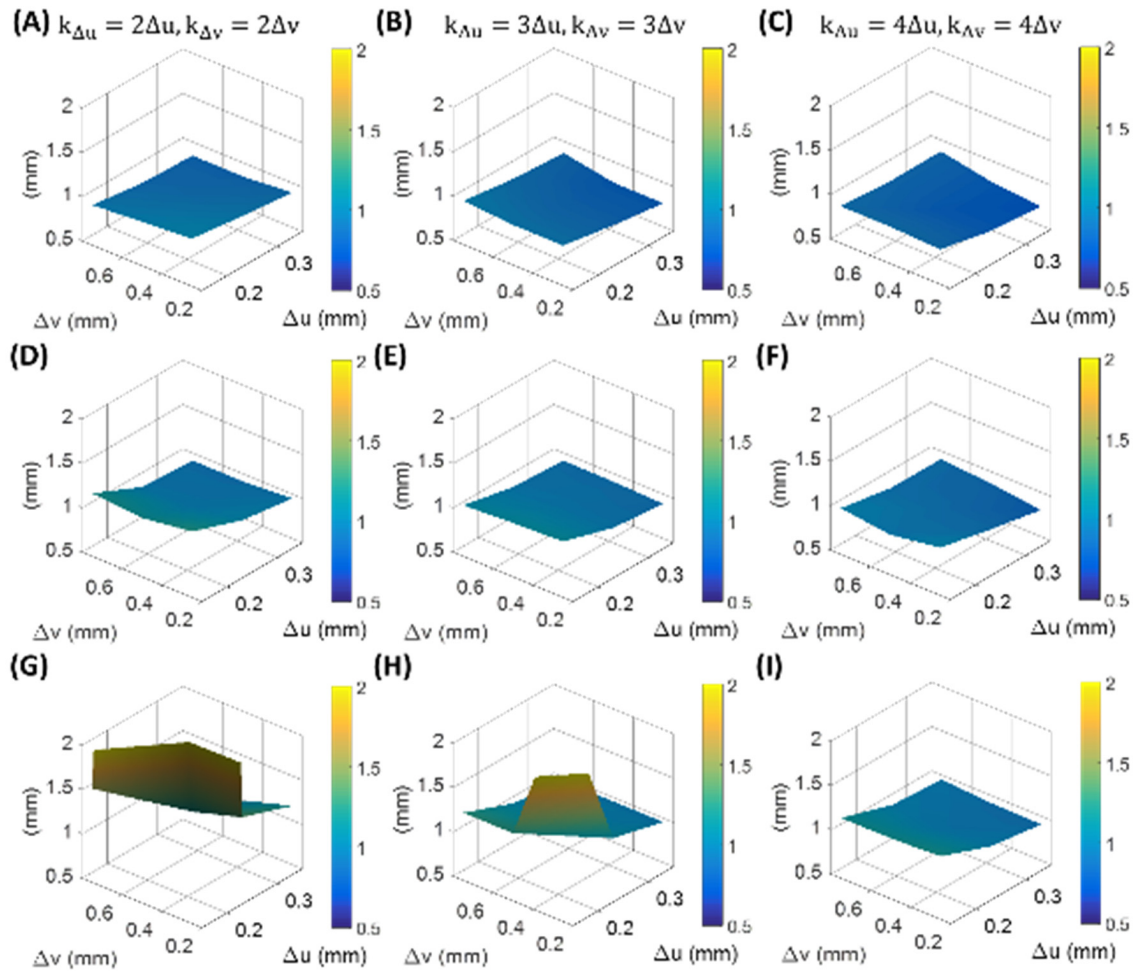

Figure 4.11: Segmentation error versus gridding parameters for end-systole (static phantom) projection data with Poisson noise corresponding to imaging at 100% (top row), 25% (middle row) and 10% full power. A single iteration of anisotropic diffusion filtering was applied to the same reconstructions used in Figure 4.7 prior to segmentation to reduce voxel noise.

### 4.3.2 Investigation of gridding parameters for a dynamic cardiac phantom

The effect of gridding parameters on segmentation accuracy for a dynamic XCAT phantom with realistic cardiac motion was investigated next. Chapter 3 presented the SBDX sampling of Radon space during rotational CT data acquisition. The SBDX Radon space sampling becomes sparser when using gated datasets as there are fewer overlapping superviews (Figure 3.3). As a result, the optimal gridding parameters for gated projection datasets may differ from those for the non-gated case. The static phantom results showed that the use of an anisotropic diffusion filter improved the segmentation accuracy and robustness for noisy images. Therefore, diffusion filtering was performed prior to segmentation for each of the dynamic phantom reconstructions. Segmentation errors for the noise-free dynamic thorax phantom are plotted as a function of the gridding parameters in Figure 4.12. Segmentation errors ranged from 1.21-1.55 mm.

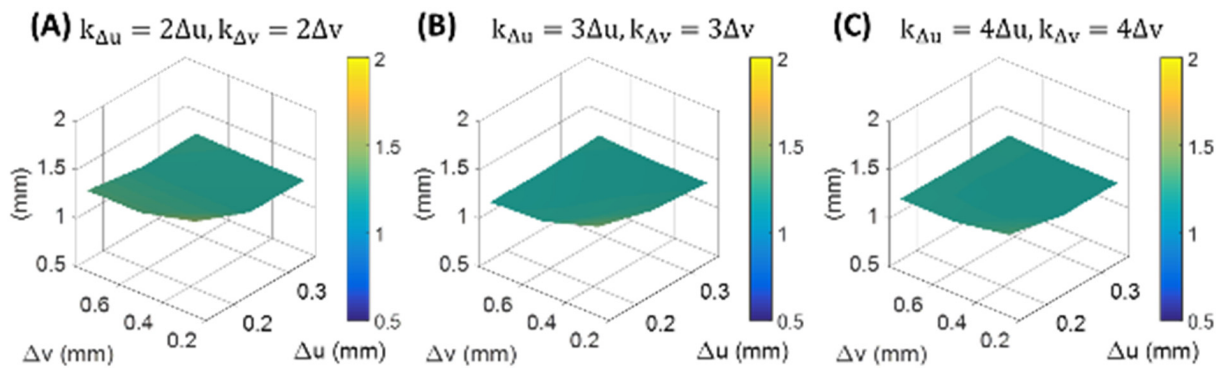

Figure 4.12: Segmentation error versus gridding parameters for reconstructions from noise-free projection data for an XCAT phantom with 75 BPM cardiac motion.

Segmentation errors from reconstructions of projection data with added Poisson noise equivalent to imaging at 100%, 25% and 10% full power, 120 kV are shown in Figure 4.13 for the dynamic thorax phantom. For the 100% full power case (Figure 4.13A-C), the 99th percentile segmentation error ranged from 1.24-3.12 mm. This was a slight increase from the static phantom results where errors measured 0.71-1.26 mm. Nevertheless, the optimal segmentation result was less than the target goal of 3.3 mm indicating image quality was sufficient for the cardiac chamber segmentation task.

For the 25% full power case (Figure 4.13D-F), segmentation errors ranged from 1.33-11.43 mm. In total, 324 images were analyzed corresponding to the different combinations of gridding parameters and the number of diffusion filter iterations. At 25% full power, segmentation failed in 12 of the 324 reconstructions (3.7%). Segmentation was successful in 100% of the reconstructions when 3 or more diffusion filter iterations were performed with errors ranging from 1.33-1.86 mm demonstrating improved segmentation robustness.

At 10% full power and $\lambda$ equal to 3000, segmentation errors ranged from 1.64-12.02 mm with segmentation failing in 108 of the 648 reconstructions (17.3%). With $\lambda$ equal to 300, all segmentations were successful with errors ranging from 1.64-4.51 mm. The results presented here demonstrate that reconstructions yielding segmentation errors less than 3.3 mm could be achieved across three noise levels by optimizing the parameters associated with the parallel ray gridding procedure and diffusion filter. The $\lambda$ reconstruction parameter was varied as $\lambda = 3000$ and $\lambda = 300$. The PICCS reconstruction parameters are optimized for the task of cardiac chamber mapping in section 4.3.5 below to determine whether further improvements in segmentation accuracy may be achieved.

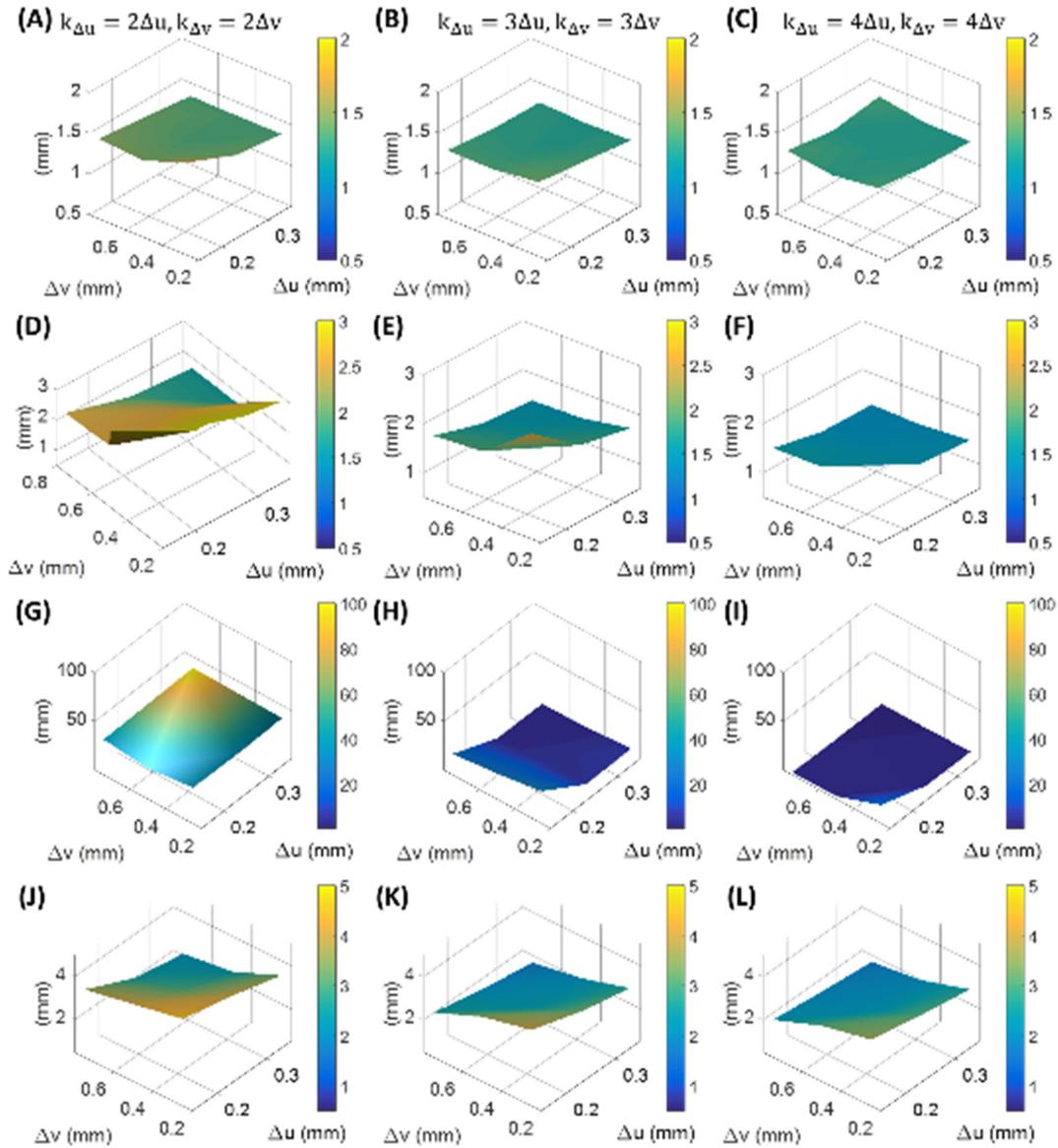

Figure 4.13: 75 BPM XCAT phantom with noisy projections. Segmentation error versus gridding parameters for reconstructions of an XCAT phantom with 75 BPM cardiac motion corresponding to imaging at 100% (first row), 25% (second row) and 10% full power (third row $\lambda$ = 3000, and fourth row $\lambda$ = 300).

### 4.3.3 Optimal gridding parameters for cardiac chamber mapping

Table 4.2 presents the optimal gridding parameters for the static and dynamic XCAT phantoms at each of the noise levels considered. A total of 2,754 images were analyzed to generate the results of Table 4.2. The optimal gridding parameters varied based on the simulated noise level and phantom motion model (i.e. static or dynamic). The dynamic phantom results are most interesting in terms of selecting gridding parameters that will perform sufficiently for a range of cardiac imaging scenarios. The optimal gridding parameters were selected as those that yielded the minimum mean segmentation error across the three noise levels studied, for the dynamic XCAT phantom. The optimal parameters were, $\Delta u = 0.35$ mm, $k_{\Delta u} = 1.4$ mm, $\Delta v = 0.50$ mm and $k_{\Delta v} = 2.0$ mm. The mean segmentation error measured 1.42 mm across the three fluence levels studied (1.24 mm at 100% power, 1.33 mm at 25% power, 1.68 mm at 10% power). These gridding parameters were used for the remainder of the work performed in this dissertation unless otherwise stated.

Table 4.2: Optimal gridding parameters are summarized for a static and 75 BPM dynamic XCAT phantom across four noise levels with and without a post-processing anisotropic diffusion filter applied. The static XCAT phantom was reconstructed using a filtered backprojection algorithm. The dynamic XCAT phantom was reconstructed from cardiac gated projection data with the PICCS algorithm. For PICCS, $\alpha = 0.25$ and $\lambda = 3000$.

| | $\Delta u$ (mm) | $k_{\Delta u}$ (mm) | $\Delta v$ (mm) | $k_{\Delta v}$ (mm) | Diffusion Iterations | Segmentation Error (mm) |
|---|---|---|---|---|---|---|
| Static, Noise free | *39 different combinations yielded 0.56 mm segmentation error* | | | | | |
| Static, 100% full power | 0.35 | 1.4 | 0.5 | 2.0 | 0 | 0.71 |
| Static, filtered, 100% full power | 0.35 | 1.4 | 0.25 | 1.0 | 1 | 0.71 |
| Static, 25% full power | 0.35 | 1.4 | 0.5 | 1.5 | 0 | 0.77 |
| Static, filtered, 25% full power | 0.35 | 1.4 | 0.25 | 1.0 | 1 | 0.80 |
| Static, 10% full power | 0.35 | 1.4 | 0.5 | 2.0 | 0 | 0.95 |
| Static, filtered, 10% full power | 0.35 | 1.4 | 0.5 | 2.0 | 3 | 0.91 |
| Dynamic, Noise free | 0.35 | 1.4 | 0.5 | 2.0 | 0 | 1.21 |
| Dynamic, 100% full power | 0.35 | 1.4 | 0.5 | 1.5 | 1 | 1.24 |
| Dynamic, 25% full power | 0.35 | 1.4 | 0.5 | 2.0 | 1 | 1.33 |
| Dynamic, 10% full power | 0.35 | 1.4 | 0.5 | 2.0 | 3 | 1.68 |

### 4.3.4   FBP reconstruction of a noise-free static phantom

SBDX-CT image quality in the presence of full FOV truncation was evaluated and compared versus CBCT with the static noise-free XCAT phantom. Filtered backprojection reconstructions are displayed in Figure 4.14 for the SBDX and CBCT geometries. Evaluation metrics are listed in Table 4.3 for both cases. CBCT reconstruction accuracy exceeded SBDX-CT demonstrated by a UQI value of 0.80 versus 0.71, respectively. Likewise, the rRMSE was 3.9% for CBCT versus 5.1% for SBDX-CT. SBDX reconstruction parameters were optimized for the segmentation task, however, and not reconstruction accuracy. Segmentation accuracy was similar as 99% of points on the segmented atrium surface were within 0.59 mm of the ground truth for SBDX-CT versus 0.52 mm for CBCT. Dice coefficients measured 0.98 and 0.99 for SBDX-CT versus CBCT, respectively. Small differences in the 99th percentile segmentation error and Dice coefficients exceeding 0.98 indicate that SBDX may be capable of providing cardiac chamber roadmaps with accuracy comparable to conventional x-ray fluoroscopy imaging systems.

Table 4.3. Evaluation metrics for SBDX-CT and CBCT reconstructions of static noise-free XCAT projections.

| C-arm system | Seg. error [mm] | Dice | rRMSE [%] | UQI |
|---|---|---|---|---|
| SBDX-CT | 0.59 | 0.98 | 5.1 | 0.71 |
| CBCT | 0.52 | 0.99 | 3.9 | 0.80 |

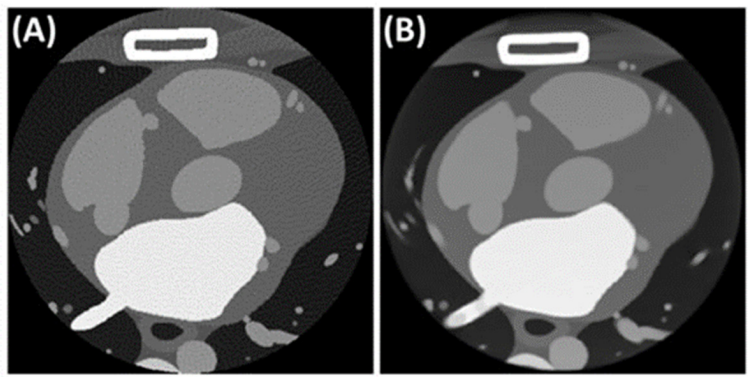

Figure 4.14: A numerical XCAT phantom without cardiac motion was reconstructed from noise free projection data for a traditional CBCT geometry (A) and the SBDX geometry (B) using gFBP. Segmentation accuracy of the left atrium was comparable indicated by Dice coefficient values of 0.99 and 0.98, respectively. [WL = 350, WW = 2700 ] HU.

### 4.3.5   Optimal iterative reconstruction parameters for cardiac chamber mapping

The iterative reconstruction parameters ($\lambda$ and $\alpha$) were optimized next to determine whether the segmentation errors could be reduced. The parameters $\lambda$ and $\alpha$ control the relative weights between the regularization terms and the data fidelity term, equation (4.3). Larger $\lambda$ values enforce greater data consistency between the measured projection data and the forward projection of the reconstructed image. Smaller $\lambda$ values increase noise suppression. Figure 4.15 plots the 99th percentile segmentation errors and UQI as a function of λ and α for the 75 BPM end-systole case. Results are presented at 100%, 25%, and 10% full power. Non-gated gFBP is also shown for comparison.

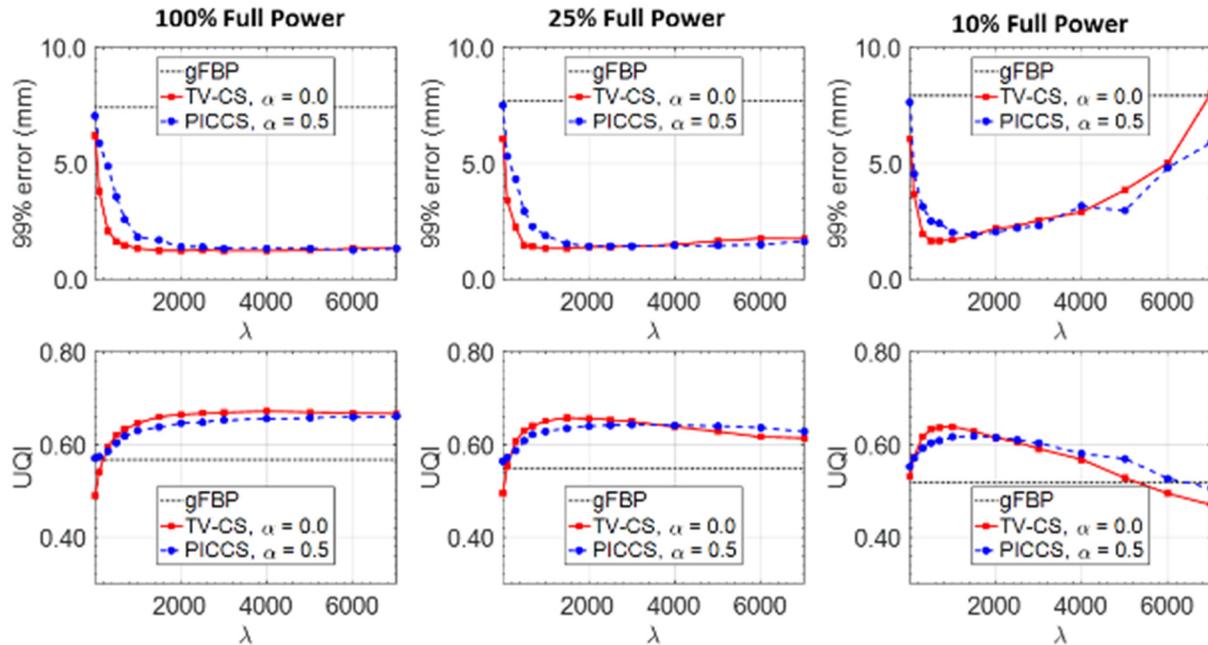

Figure 4.15: 75 BPM, end-systole results. The 99th percentile distance of segmentation errors (top row) and UQI (bottom row) are plotted as a function of λ for TV-CS (α = 0.0) and PICCS (α=0.5) reconstructions at 100%, 25% and 10% full power from left to right. Results corresponding to non-gated gFBP are shown as a dotted black line.

To demonstrate the effect of λ on image quality, example TV-CS images corresponding to small (λ = 10), optimal (λ = 500) and large (λ = 7000) data fidelity weights are shown in Figure 4.16 for the 10% full power case. Examples PICCS images are shown in Figure 4.17. The segmented left atrium is shown versus the ground truth for a single image slice. Histograms of point distances between the segmented atrium and ground truth present segmentation errors for each reconstruction. The example images demonstrate the tradeoff between data consistency and regularization imposed by the λ parameter. For instance, in the TV-CS images, the standard deviation of image values in a region-of-interest placed in the left atrium measured 188 HU for the gFBP reconstruction (Figure 4.16, left column). Iterative reconstruction with a small data fidelity weight (i.e. λ = 10) reduced the noise standard deviation to 9 HU. The 99% segmentation error was 6.1 mm. Noise reduction, due to a high regularization weight and low data fidelity weight, was achieved with a resolution tradeoff demonstrated by the blurring of the cardiac chamber borders (Figure 4.16, second column). Without a sufficient data fidelity weight, the reconstructed left atrium resembled a denoised version of the non-gated gFBP image reconstructed from projection data spanning multiple heart beats. Increasing λ from 10 to 500 decreased the segmentation error from 6.1 mm to 1.7 mm, demonstrating the importance of selecting the optimal λ value. The noise standard deviation measured 42 HU and each of the four cardiac chamber borders could be delineated (Figure 4.16, third column). Increasing the λ parameter to 7000 resulted in greater image noise (580 HU) and an 8.0 mm segmentation error (Figure 4.16, right column). At very large λ values the regularization term receives a nearly negligible weight and the reconstruction approaches the penalized weighted least squares solution. Furthermore, at large λ values the image noise increases relative to gFBP as only a fraction of the available projection data are utilized for reconstruction due to cardiac gating.

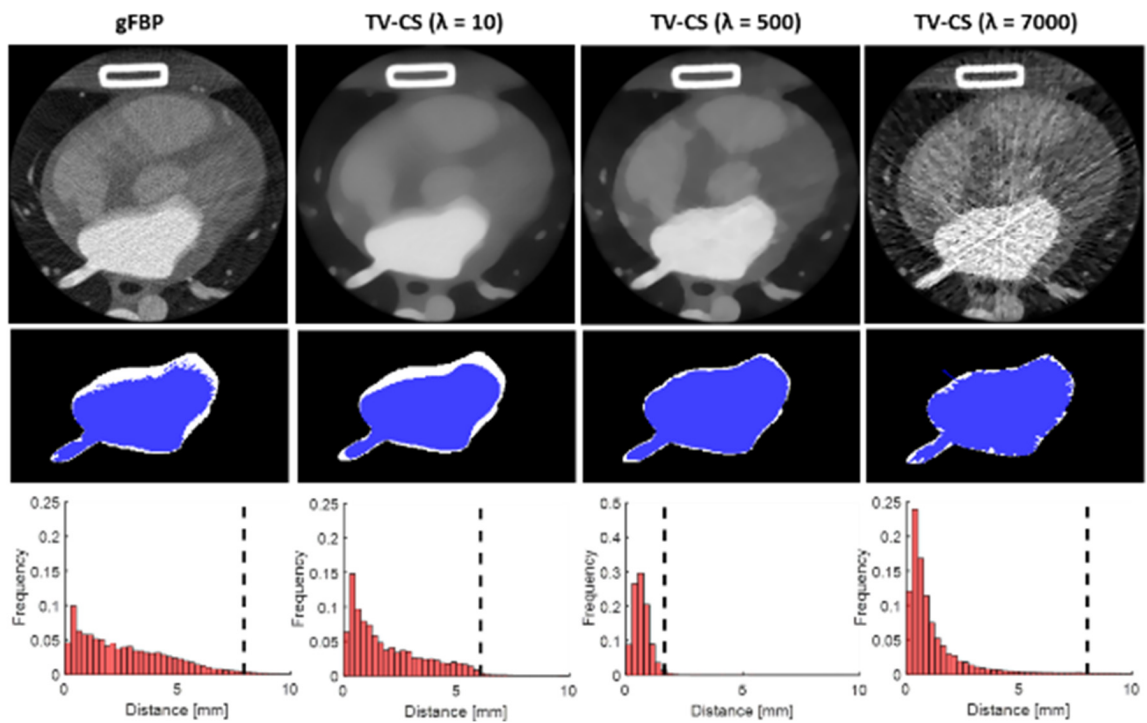

Figure 4.16: Example reconstructions from projection data gated at end-systole for a 75 BPM cardiac frequency at 10% full power are shown (top row). Reconstructions were performed using gFBP, TV-CS (λ = 10, α = 0.0), TV-CS (λ = 500, α = 0.0), and TV-CS (λ = 7000, α = 0.0) shown from left to right. The segmented left atrium is shown in blue versus the ground truth in white (middle row). Histograms show the minimum distance from each surface point of the 3D segmented atrium to the ground truth (bottom row). The dotted black line denotes the 99th percentile distance. [WL = 500, WW = 1500] HU.

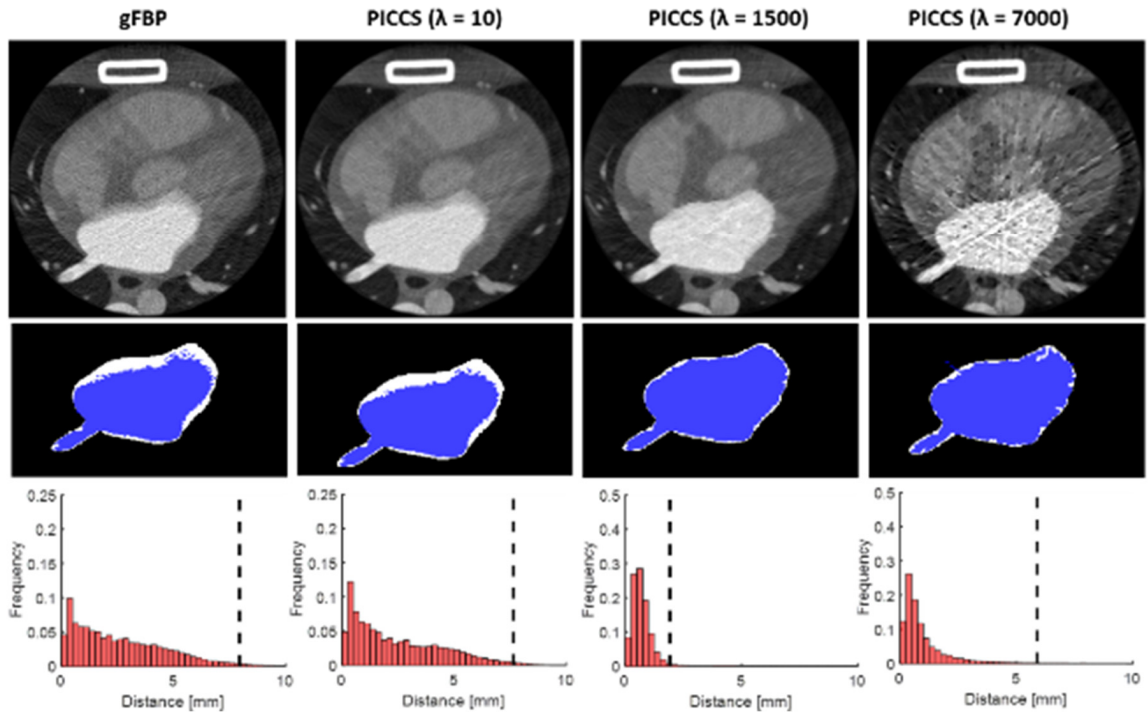

Figure 4.17: Example reconstructions from projection data gated at end-systole for a 75 BPM cardiac frequency at 10% full power are shown (top row). Reconstructions were performed using gFBP, PICCS ($\lambda = 10$, $\alpha = 0.5$), PICCS ($\lambda = 1500$, $\alpha = 0.5$), and PICCS ($\lambda = 7000$, $\alpha = 0.5$) shown from left to right. The segmented left atrium is shown in blue versus the ground truth in white (middle row). Histograms show the minimum distance from each surface point of the 3D segmented atrium to the ground truth (bottom row). The dotted black line denotes the 99th percentile distance. [WL = 500, WW = 1500] HU.

Figure 4.18 presents segmentation errors and UQI as a function of $\lambda$ and $\alpha$ for the 75 BPM end-diastole reconstructions. The trends in segmentation error and UQI versus $\lambda$ and $\alpha$ for the 60 and 90 BPM cases were similar to those presented for the 75 BPM reconstructions. The reconstruction parameters that minimized the 99th percentile segmentation error were considered optimal for the target task of cardiac chamber mapping. Thus, the reconstructions and results reported in the following section used the optimal $\lambda$ values determined here.

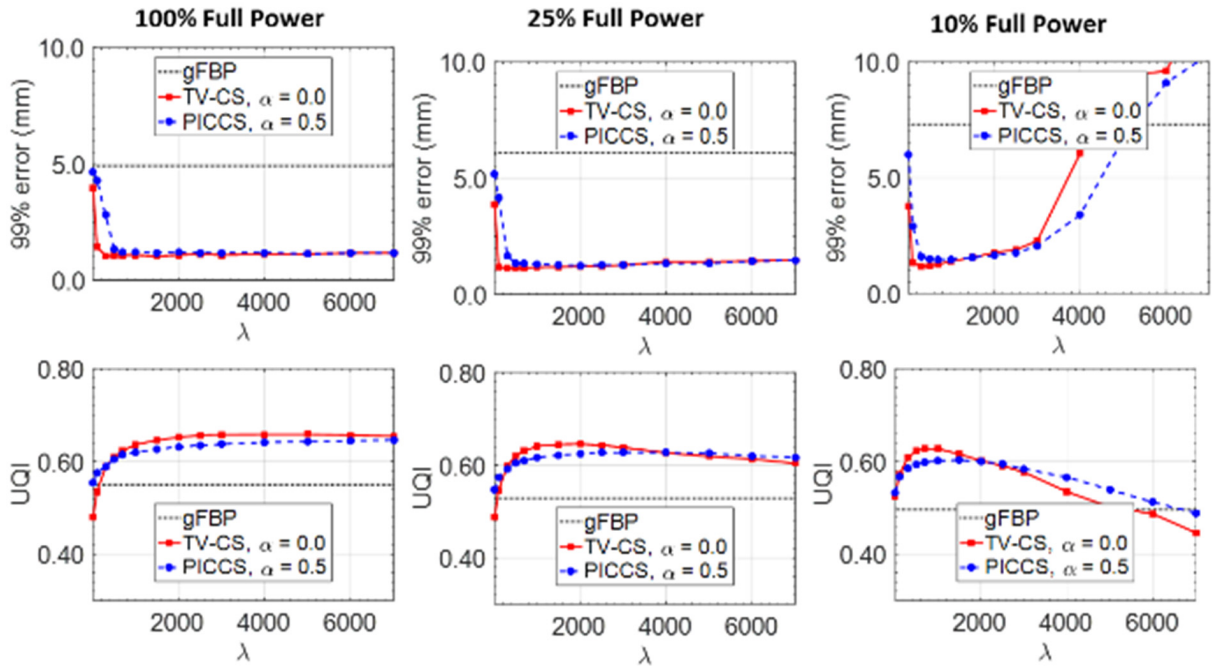

Figure 4.18: 75 BPM, end-diastole results. The 99th percentile distance of segmentation errors (top row) and UQI (bottom row) are plotted as a function of λ for TV-CS (α = 0.0) and PICCS (α=0.5) reconstructions at 100%, 25% and 10% full power from left to right. Results corresponding to non-gated gFBP are shown as a dotted black line.

### 4.3.6 PICCS reconstruction of noisy projections of a dynamic cardiac phantom

Example axial slices of the 60 BPM XCAT phantom reconstructions at ventricular end-systole are presented in Figure 4.19 for three fluence levels and three reconstruction methods: i) non-gated gFBP, ii) gated TV-CS, and iii) gated PICCS. Qualitatively, the gFBP reconstructions show blurring of cardiac chamber borders and streak artifacts due to data inconsistency introduced by cardiac motion (left column). The TV-CS (center column) and PICCS reconstructions (right column) show less streak artifacts as well as more clearly delineated cardiac chamber borders. An increase in image noise can be observed as the simulated fluence is decreased. Small vessel objects that are blurred out in the TV-CS images are preserved in the PICCS images as highlighted in Figure 4.20. Coronal and sagittal slices shown in Figure 4.21 and Figure 4.22, respectively, demonstrate successful 3D reconstructions. The reconstruction accuracy and left atrium segmentation errors are quantified in Table 4.4 for end-systole and Table 4.5 for end-diastole.

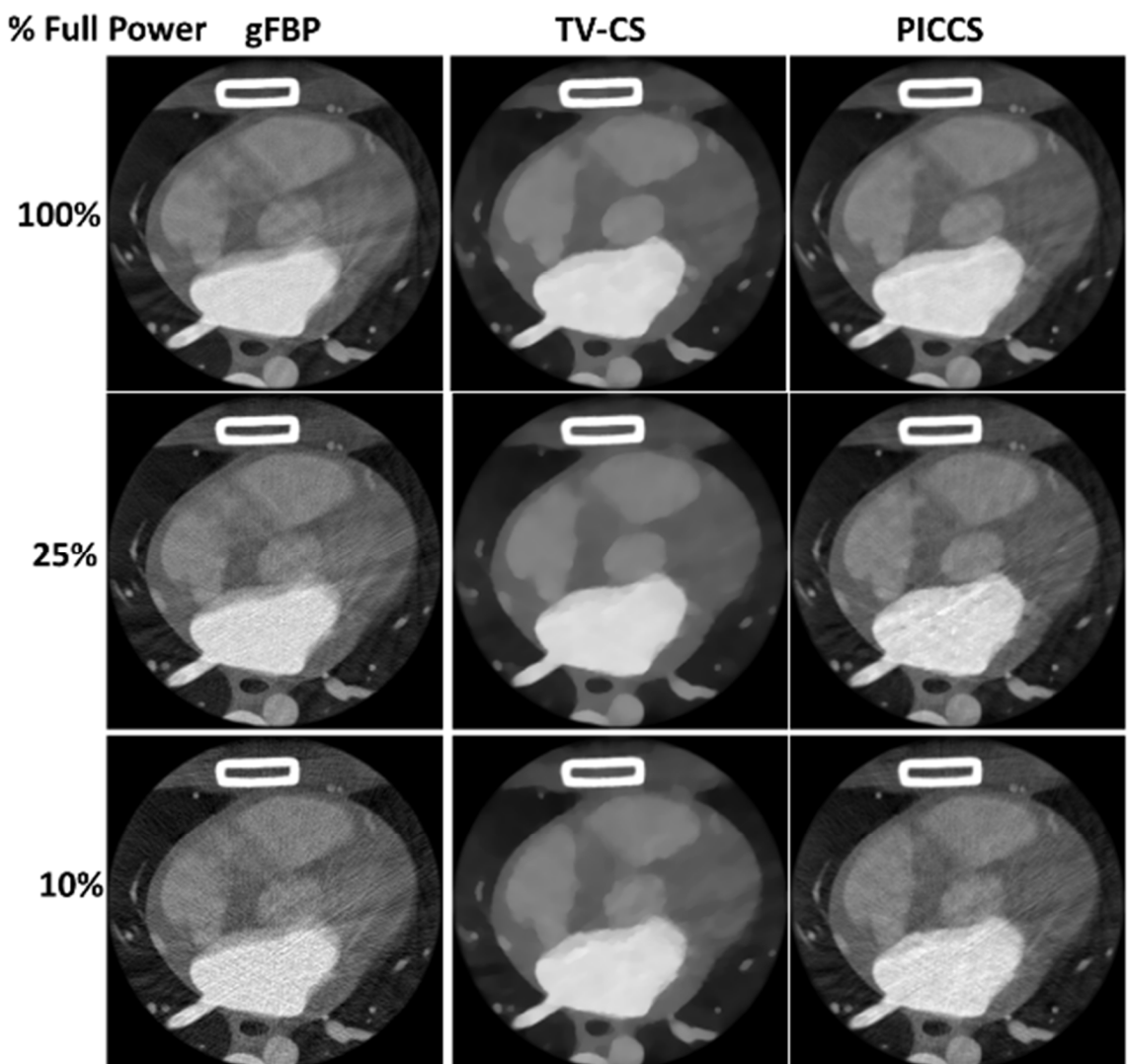

Figure 4.19: 60 BPM. End-systole. Axial slice images are shown for reconstructions performed with gFBP (left column), TV-CS (center column) and PICCS with $\alpha = 0.5$ (right column). Three noise levels are shown corresponding to imaging at 100% (top row), 25% (middle row), and 10% full power (bottom row). [WL = 500, WW = 1500] HU.

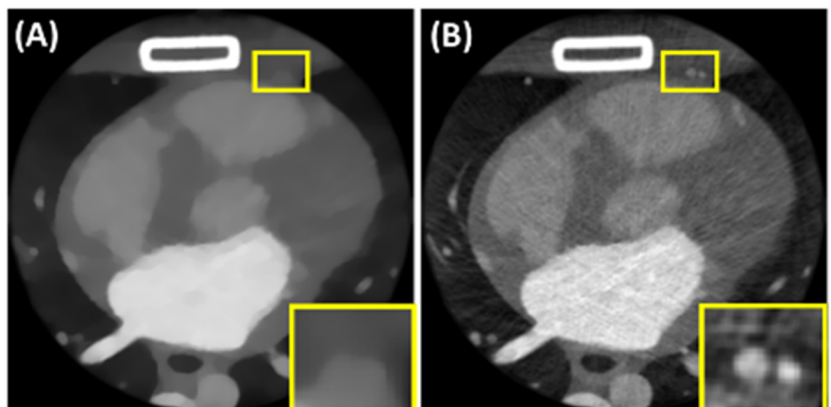

Figure 4.20: 75 BPM. End-systole. (A) Small structures may be removed in TV-CS reconstructions due to heavy regularization. (B) PICCS reconstructions preserve small structures such as the vessels shown in the zoomed region of interest.

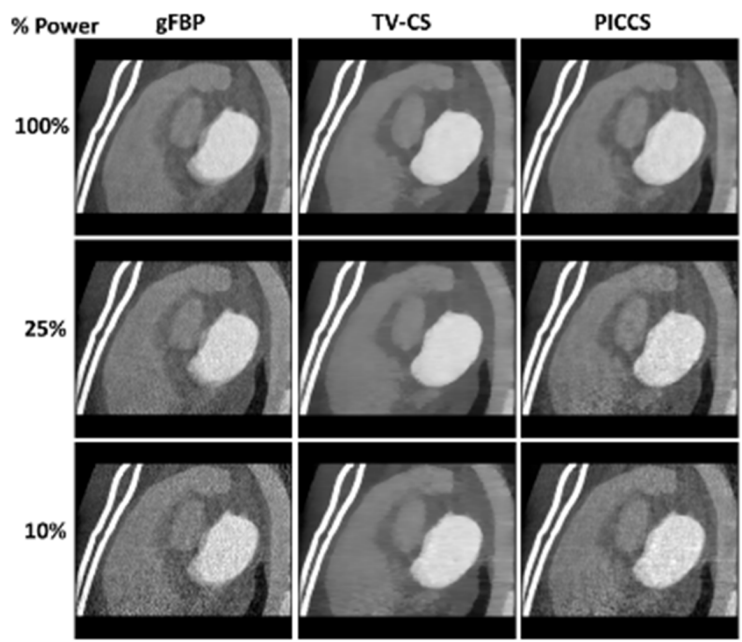

Figure 4.21: 60 BPM. End-systole Sagittal slice images are shown for reconstructions performed with gFBP (left column), TV-CS (center column) and PICCS with $\alpha = 0.5$ (right column). Three noise levels are shown corresponding to imaging at 100% (top row), 25% (middle row), and 10% full power (bottom row). [WL = 500, WW = 1500] HU.

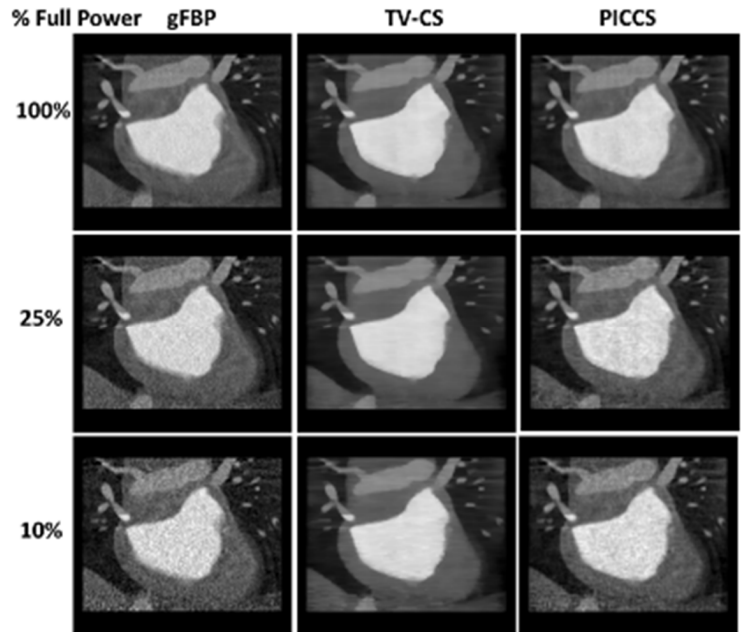

Figure 4.22: 60 BPM. End-systole. Coronal slice images are shown for reconstructions performed with gFBP (left column), TV-CS (center column) and PICCS with α = 0.5 (right column). Three noise levels are shown corresponding to imaging at 100% (top row), 25% (middle row), and 10% full power (bottom row). [WL = 500, WW = 1500] HU.

Table 4.4: End-systole results. Segmentation errors, dice coefficient and UQI for end-systole reconstructions performed with gFBP, TV-CS and PICCS at three fluence levels and three cardiac frequencies.

| Frequency (BPM) | Power Level (%) | Seg. Error 99% [mm] | | | Dice Coefficient | | | UQI | | | λ | |
|---|---|---|---|---|---|---|---|---|---|---|---|---|
| | | gFBP | TV-CS | PICCS | gFBP | TV-CS | PICCS | gFBP | TV-CS | PICCS | TV-CS | PICCS |
| 60 | 10 | 7.57 | 2.58 | 3.06 | 0.83 | 0.92 | 0.92 | 0.52 | 0.61 | 0.60 | 700 | 1500 |
| 60 | 25 | 7.46 | 2.09 | 2.41 | 0.83 | 0.94 | 0.94 | 0.55 | 0.63 | 0.62 | 1000 | 5000 |
| 60 | 100 | 6.99 | 1.98 | 2.09 | 0.83 | 0.94 | 0.93 | 0.56 | 0.64 | 0.63 | 3000 | 5000 |
| 75 | 10 | 7.93 | 1.66 | 1.92 | 0.82 | 0.94 | 0.95 | 0.52 | 0.63 | 0.62 | 500 | 1500 |
| 75 | 25 | 7.70 | 1.33 | 1.42 | 0.82 | 0.96 | 0.95 | 0.55 | 0.66 | 0.64 | 1500 | 2500 |
| 75 | 100 | 7.41 | 1.24 | 1.27 | 0.82 | 0.96 | 0.96 | 0.57 | 0.67 | 0.66 | 4000 | 6000 |
| 90 | 10 | 7.91 | 1.93 | 2.02 | 0.82 | 0.94 | 0.94 | 0.52 | 0.63 | 0.62 | 500 | 1000 |
| 90 | 25 | 7.78 | 1.80 | 1.86 | 0.82 | 0.94 | 0.94 | 0.54 | 0.64 | 0.64 | 700 | 1500 |
| 90 | 100 | 7.39 | 1.75 | 1.76 | 0.82 | 0.95 | 0.94 | 0.56 | 0.65 | 0.65 | 1000 | 2500 |

Table 4.5: End-diastole results. Segmentation errors, dice coefficient and UQI for end-diastole reconstructions performed with gFBP, TV-CS and PICCS at three fluence levels and three cardiac frequencies.

| Frequency (BPM) | Power Level (%) | Seg. Error 99% [mm] | | | Dice Coefficient | | | UQI | | | λ | |
|---|---|---|---|---|---|---|---|---|---|---|---|---|
| | | gFBP | TV-CS | PICCS | gFBP | TV-CS | PICCS | gFBP | TV-CS | PICCS | TV-CS | PICCS |
| 60 | 10 | 8.03 | 1.13 | 3.51 | 0.92 | 0.97 | 0.96 | 0.49 | 0.59 | 0.56 | 500 | 500 |
| 60 | 25 | 7.12 | 1.02 | 2.31 | 0.92 | 0.97 | 0.96 | 0.52 | 0.60 | 0.58 | 700 | 1000 |
| 60 | 100 | 5.84 | 0.95 | 1.27 | 0.92 | 0.97 | 0.97 | 0.54 | 0.61 | 0.60 | 1500 | 2500 |
| 75 | 10 | 7.30 | 1.16 | 1.47 | 0.92 | 0.96 | 0.96 | 0.50 | 0.61 | 0.60 | 300 | 1000 |
| 75 | 25 | 6.11 | 1.13 | 1.25 | 0.93 | 0.96 | 0.96 | 0.53 | 0.60 | 0.62 | 300 | 2000 |
| 75 | 100 | 4.91 | 1.04 | 1.13 | 0.93 | 0.97 | 0.96 | 0.55 | 0.59 | 0.64 | 300 | 5000 |
| 90 | 10 | 7.50 | 1.13 | 1.47 | 0.93 | 0.97 | 0.96 | 0.50 | 0.59 | 0.62 | 100 | 700 |
| 90 | 25 | 6.22 | 1.18 | 1.26 | 0.93 | 0.96 | 0.96 | 0.54 | 0.57 | 0.64 | 100 | 1500 |
| 90 | 100 | 4.89 | 1.11 | 1.16 | 0.94 | 0.96 | 0.96 | 0.56 | 0.67 | 0.67 | 2000 | 6000 |

Reconstructions for the 90 BPM XCAT phantom are shown in Figure 4.23 for end-systole and end-diastole gated datasets to demonstrate the proposed methods ability to target a specific cardiac phase. Qualitatively, the left atrium appears larger for the end-systole reconstructions compared to the end-diastole reconstructions, as expected and shown in the ground truth images (Figure 4.5). For each case, the segmented left atrium is shown in blue versus the ground truth in white. Similar to the 60 and 75 BPM results, the non-gated gFBP reconstructions show streak artifacts and poorly delineated chamber borders. For gFBP, the segmented left atrium is in poor agreement with the ground truth left atrium, shown at end-systole. Segmentation errors for gFBP ranged from 4.89-8.03 mm for the three cardiac frequencies and three noise levels considered (Table 4.4 and Table 4.5). The proposed iterative reconstruction approach reduced the range of segmentation errors to 0.95-2.58 mm for TV-CS and 1.13-3.51 mm for PICCS. The performance of each reconstruction algorithm for the task of cardiac chamber segmentation is discussed in section 4.4.

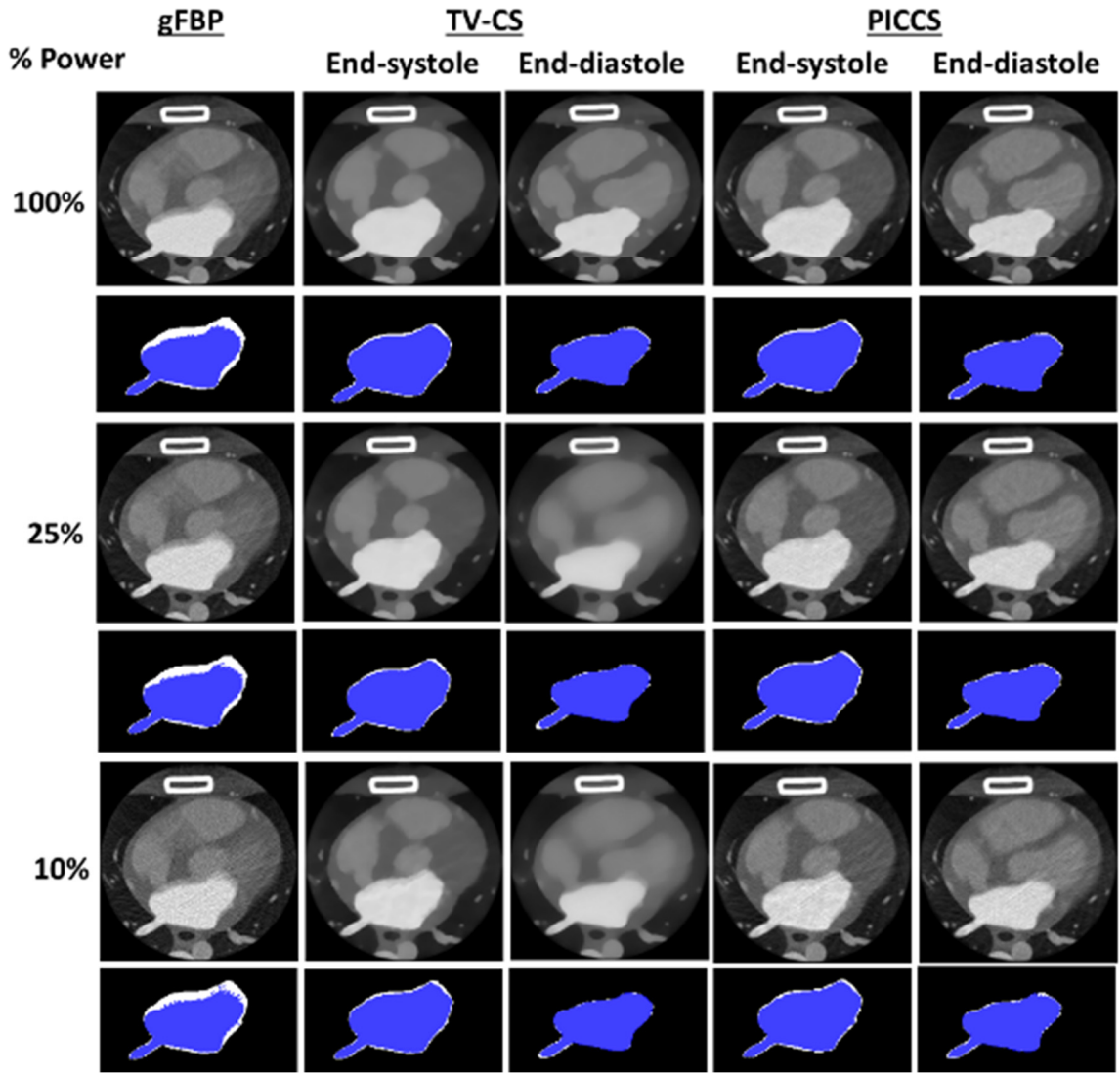

Figure 4.23: 90 BPM. Axial slice images are shown at end-systole and end-diastole for reconstructions performed with gFBP (left column versus end-systole), TV-CS (second and third column) and PICCS with α = 0.5 (fourth and fifth column). Three noise levels are shown corresponding to imaging at 100% (row 1 and 2), 25% (row 3 and 4), and 10% full power (row 5 and 6). The segmented left atrium is shown below each reconstructed image. [WL = 500, WW = 1500] HU.

## 4.4 Discussion

An iterative CT reconstruction scheme was proposed for retrospectively gated and fully truncated projection data acquired with an inverse geometry x-ray fluoroscopy system. The reconstruction technique was evaluated in the context of cardiac chamber segmentation for roadmapping during catheter-based interventions such as radiofrequency catheter ablation of atrial fibrillation. The proposed method was successfully implemented and performance was assessed for the SBDX system geometry.

### 4.4.1 Gridding parameters for cardiac chamber mapping

In the proposed scheme, SBDX projection data were rebinned to a parallel ray geometry prior to reconstruction to reduce long computation times associated with the use of 71 x 71 source focal spot positions. Parameters ($\Delta u$, $\Delta v$, $k_{\Delta u}$, $k_{\Delta v}$) related to the gridding of SBDX CT projection data to a parallel ray dataset were optimized for the task of cardiac chamber mapping using a numerical XCAT phantom without and with motion. The sampling density of the parallel ray geometry was shown to have an impact on segmentation accuracy. For noise-free projection data corresponding to a static phantom, reduced parallel ray sampling density (i.e. large $\Delta u$ and $\Delta v$) increased the segmentation error (Figure 4.6). For projection data with added Poisson noise, small kernel widths and small ray sampling pitches were associated with increased segmentation errors or complete segmentation failure (e.g. Figure 4.10). Segmentation of the left atrium failed in 40.7% of reconstructions performed for imaging at 10% full power. Increasing the kernel widths and application of a post-reconstruction anisotropic diffusion filter reduced segmentation errors and image noise. The use of a diffusion filter decreased the percentage of failed segmentations at 10% full power to 1.5%.

The use of larger kernel widths also improved the robustness of the dynamic XCAT phantom segmentations. For example, the left column of Figure 4.13 shows large segmentation errors when narrow kernel widths were used to perform the ray rebinning ($k_{\Delta u} = 2\Delta u$, $k_{\Delta v} = 2\Delta v$). The right column of Figure 4.13 shows increasing the kernel widths ($k_{\Delta u} = 4\Delta u$, $k_{\Delta v} = 4\Delta v$) reduced 99th percentile segmentation errors to less than 2 mm for all noise levels, with proper $\lambda$ selection. The smaller segmentation errors associated with larger kernel widths can be attributed to noise-reduction and increased sampling of input grid points (i.e. native SBDX rays) in sparse regions of Radon space that occur due to cardiac gating of SBDX projection data. For high spatial resolution tasks, the gridding parameters may need to be re-optimized to minimize edge blurring due to the use of large kernel widths. However, for the task of cardiac chamber mapping, the larger gridding kernel widths improved segmentation robustness which is a desirable reconstruction trait for roadmapping during interventional procedures.

### 4.4.2    Reconstruction parameters λ and α for cardiac chamber mapping

Following the selection of parallel ray gridding parameters, iterative reconstruction parameters were optimized for cardiac chamber mapping. Trends in segmentation errors as a function of λ (Figure 4.15) were similar for α=0.0 and α=0.5. Quantitatively, the segmentation error averaged over three heart rates (60, 75, 90 BPM) for imaging at 100% full power was 1.34 mm for TV-CS (α=0.0), 1.45 mm for PICCS (α=0.5), and 6.24 mm for gFBP. Segmentation errors increased as the simulated fluence decreased. For imaging at 10% full power, segmentation error averaged 1.60 mm for TV-CS (α=0.0), 2.24 mm for PICCS (α=0.5), and 7.71 mm for gFBP. The proposed iterative reconstruction scheme outperformed gFBP for all imaging scenarios considered. The average segmentation errors were less the diameter of a typical ablation catheter tip (approximately 3 mm). Segmentation errors less than 3 mm are assumed to be acceptable as the chamber model resolution would be comparable to the size of the tracked catheter tip displayed during navigation. The optimal λ value decreased as the simulated fluence decreased for both TV-CS and PICCS reconstructions in order to compensate for increased image noise. For TV-CS, the average optimal λ values were 1967, 717 and 433 for imaging at 100%, 25%, and 10% full power, respectively. For PICCS, the average optimal λ values were 4500, 2250, and 1033 for imaging at 100%, 25% and 10% full power, respectively.

TV-CS slightly outperformed PICCS in terms of segmentation error (1.41 mm versus 1.81 mm averaged across all reconstructions summarized in Table 4.4 and Table 4.5). The slight difference may be explained by noting that the left atrium of the XCAT phantom is a piecewise constant high-contrast object. Compressed sensing with total variation minimization reconstruction has been shown to perform well for piecewise constant objects.[153] Future work should investigate the segmentation accuracy of TV-CS versus PICCS in animal studies where the left atrium iodinated blood pool may be less piecewise-constant due to contrast dynamics. Additionally, all segmentations were performed using an intensity based threshold method that depends on the contrast between the left atrium and nearby background region, as well as image noise. Segmentation error was shown to increase with image noise. A study performed in Ref. 147 demonstrated that the noise standard deviation of PICCS reconstructed images decreased as the α parameter decreased.

It's worth noting that TV-CS based algorithms have been shown to exhibit patchy artifacts as the optimization procedure favors piecewise constant regions to minimize the total variation regularization term.[131] This is an important consideration in diagnostic imaging tasks as the patchy artifacts may resemble low-contrast lesions or small objects may be removed due to regularization induced blurring thereby potentially impacting diagnostic outcomes.[29,131,147] The incorporation of a prior image for PICCS reconstructions has been shown to produce reconstructions that retain small low-contrast features.[29,147] This is demonstrated in Figure 4.20 which presents example TV-CS and PICCS images for the 75 BPM, end-systole case at 10% full power. A zoomed region-of-interest outlined in yellow shows two small vessels present near the sternum. The two vessels are blurred together in the TV-CS image but can clearly be observed in the PICCS image. Nonetheless, for the cardiac chamber segmentation task considered here, noise texture and small low-contrast object visibility are less of a concern as the final display to the operator will be a virtually rendered cardiac chamber model.

Future work may investigate whether superior segmentation accuracy can be achieved using pre-processing techniques to improve the image quality of the prior image used for PICCS reconstructions. Artifacts in the prior image, such as streaks due to cardiac motion, have been shown to persist in the final PICCS image to a greater extent than TV-CS.[154] Reference 149 presents artifact reduction methods for cardiac C-arm CT that could be extended to C-arm IGCT to improve the image quality of the prior image utilized by PICCS.

### 4.4.3 Cardiac chamber segmentation results

Segmentation accuracy was evaluated as a function of heart rate (Table 4.4 and Table 4.5). For gFBP, the 99th percentile segmentation accuracy was 7.17 mm, 6.90 mm, and 6.95 mm at 60, 75, and 90

BPM averaged across noise levels at end-systole and end-diastole. Results for TV-CS were 1.62 mm, 1.26 mm, and 1.48 mm at 60, 75, and 90 BPM, respectively. Results for PICCS averaged 2.44 mm, 1.41 mm, and 1.59 mm at 60, 75, and 90 BPM, respectively.

SBDX-CT segmentation accuracy results were similar across the three heart rates evaluated. Optimal segmentation accuracy results occurred at 75 BPM. A slight increase in segmentation errors was observed at 60 BPM relative to 75 BPM and 90 BPM. Previous work with a conventional C-arm CBCT system showed superior accuracy for PICCS reconstructions performed with projection data acquired at higher heart rates versus lower heart rates.[29] In Ref. [29], the improved performance at higher heart rates was attributed to the greater number of gated view angles available as more complete heart beats were scanned within the gantry rotation period. For the SBDX-CT simulations performed here, all reconstructions utilized gated projection data from a 0.13 s gating window, corresponding to two gated superviews per simulated heartbeat. Future work could investigate the impact of varying the temporal width of the gating window on image quality. For the 14.5 s simulated scan, the 0.13 s gating window resulted in 28, 36 and 42 superviews per gated dataset at 60 BPM, 75 BPM, and 90 BPM, respectively. Despite fewer superviews at lower heart rates, SBDX-CT reconstruction performance was robust to variations in heart rate, demonstrated by a narrow range of segmentation accuracy values across cardiac frequencies (e.g. 1.26-1.62 mm for TV-CS). This may be attributed to the sampling of Radon space by the multi-source array. For a SBDX superview, projection rays from a single source-row and detector-row combination sample a 144.2 mm by 10.1° area of Radon space (Figure 3.3A). Overlapping superviews provide additional fluence that decrease image noise and fill the peripheral regions of Radon space where the sampling density is sparse (Figure 3.3B). Cardiac gating reduces the number of view angles (CBCT) and superviews (IGCT) available for reconstruction. As a result, gating reduces the total fluence available for reconstruction while also increasing the projection undersampling factor that may cause image artifacts. The use of a fast scanning-source array reduces the projection undersampling due to the associated area-sampling, which covers a larger portion of Radon space compared to cone-beam projections. Figure 4.24 presents a rebinned parallel ray sinogram for the gated end-systole projection datasets at 60 and 90 BPM to demonstrate the sampling impacts of gating. Despite using a limited subset of the total available projection data, due to cardiac gating (12.9% at 60 BPM, 19.4% at 90 BPM), sampling of the 2D Radon space is sufficient for iterative reconstructions due to the multi-source area-sampling (43% at 60 BPM, 56% at 90 BPM).

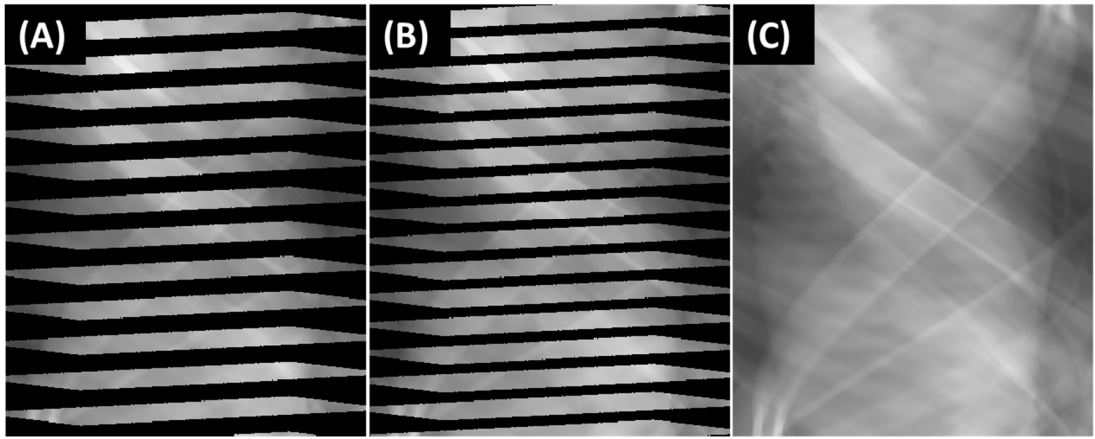

Figure 4.24: Sinograms are shown for SBDX gated projection data corresponding to imaging at a 60 BPM cardiac frequency (A) and a 90 BPM frequency (B). The 60 BPM gated dataset consisted of 28 superviews. The 90 BPM gated dataset consisted of 42 superviews. A non-gated sinogram is shown at 90 BPM (C).

## 4.5 Summary

A CT reconstruction framework for retrospectively gated and fully truncated projections was developed and successfully demonstrated for an inverse geometry x-ray fluoroscopy system. This represents the first attempt to mitigate artifacts caused by cardiac motion and projection truncation in inverse geometry CT. Image quality was evaluated and reconstruction parameters were optimized for cardiac chamber segmentation at several noise levels and cardiac frequencies.

Several limitations exist for the work performed here. First, the numerical simulations performed assumed monochromatic projection data. Furthermore, the effects of a non-ideal system (e.g. detector response and imperfect gantry rotation) were not modeled. The simulations were simplified as the purpose of the numerical simulations performed here was to evaluate a proposed reconstruction technique to accommodate cardiac gating and FOV truncation. Experimental work performed in chapters 5, 6, and 7 evaluate SBDX CT image quality under more realistic noise scenarios, with the effects of beam hardening, x-ray scatter, and the detector response included. The following chapter also presents a geometric calibration method for SBDX-CT and projection conditioning techniques to mitigate image artifacts. Development of geometric calibration and data conditioning methods will enable experimental validation of the simulations presented here. A second limitation of the simulations is that intra-frame motion was not considered. Chapter 7 validates the simulations performed here in phantom studies which include the effects of intra-frame motion.

# 5  Methods for C-arm geometric calibration and data conditioning

*A subset of the results reported in this chapter were presented at SPIE Medical Imaging 2016 and published in the Journal of Medical Imaging.*[155,156]

## 5.1  Introduction

Chapter 4 presented an IGCT reconstruction method for ECG-gated and fully truncated projection data acquired for a simulated SBDX geometry. Cardiac chamber segmentation accuracy was evaluated using numerical simulations. The reconstruction of real-world SBDX projection data requires C-arm geometric calibration and data conditioning methods to prevent image artifacts. The purpose of this chapter is to present geometric calibration and data conditioning methods to enable IGCT reconstruction of projections obtained with the SBDX prototype at the UW-Madison. The effects of geometric uncertainty (section 5.2), non-linear detector response (section 5.3), beam hardening (section 5.4), and x-ray scatter (section 5.5) are each considered. A combination of numerical simulations and rotating stage experiments are performed to isolate the various sources of artifacts considered. Correction and calibration schemes are proposed for the sources of image artifacts found to be most detrimental to the accuracy of high-contrast object segmentation from SBDX CT images.

## 5.2  Geometric calibration

In C-arm based inverse geometry CT (IGCT), uncertainties in the imaging geometry may exist due to manufacturing tolerances, non-ideal source and detector alignment, or C-arm deflection during rotation. An incorrect mapping between the assumed 3D object coordinate system and the projection acquisition system has been shown to degrade spatial resolution and can introduce image artifacts.[51]

A proposed IGCT calibration method by *Schmidt et al.* estimated four parameters describing the location and orientation of a common axis-of-rotation for the source and detector arrays of a table-top IGCT system with a rotating stage.[45] A second approach proposed by *Baek et al.* improved on the four parameter method by estimating individual source coordinates for an 8 spot x-ray source array mounted on a rotating gantry.[46] Baek's approach also determined four parameters describing the axis-of-rotation. For C-arm CT, it is desirable that geometric calibration be performed independently at each view angle. Uncertainties in the C-arm geometry may be larger due to the use of a flexion-prone C-arm compared to more stable gantry-based systems. A novel gantry calibration method for inverse geometry CT with SBDX was developed that is motivated by the projection matrix (P-matrix) technique often used in cone-beam CT (CBCT).[157] The proposed method performs a calibration for each gantry position independent of the other acquired view angles.

The ability of the proposed method to recover parameters describing SBDX system geometry is examined using numerical simulations. Experimental SBDX data acquired with a rotating stage is used to demonstrate that the proposed method reduces image artifacts caused by geometric uncertainty.

### 5.2.1   Methods

In conventional C-arm cone-beam CT, a 3D-to-2D projection matrix is established for each C-arm angle. Typically measured from the detector images of a 3D helical array of fiducials acquired during a rotational scan, each projection matrix defines a mapping between 3D patient space and 2D detector space which is needed in back projection and forward projection operations. A naïve application of this calibration method to the SBDX system would involve imaging a helix for each of the 71x71 focal spot positions, for each C-arm angle. This is impractical for a number of reasons. Beyond the major challenge of fabricating such a calibration phantom, it is unlikely that the individual raw detector images would have sufficient signal-to-noise ratio and field-of-view to enable accurate determination of a projection matrix. The key to solving this problem is the observation that the live SBDX fluoroscopy display reconstructed from the individual detector images can be interpreted as a *virtual cone beam projection* of the 3D patient space. This section begins with a description of the SBDX fluoroscopic reconstruction method and an explanation of the virtual cone beam interpretation. This is followed by application of the P-matrix method to the virtual projections, and evaluation of the geometric calibration method in SBDX-CT reconstructions.

### *5.2.1.1 SBDX fluoroscopy and the virtual cone beam projection*

SBDX is inherently a real-time tomosynthesis system due to the use of inverse geometry beam scanning. A live 15 fps fluoroscopy display is generated using a GPU-based real-time image reconstructor. Each displayed 2D image frame is generated through a two-stage reconstruction procedure. First, shift-and-add (SAA) digital tomosynthesis is performed to generate a stack of 32 single plane images spaced through the patient volume (Figure 5.1A). In this method, each detector image acquired in a frame period is backprojected and summed at the stack of reconstruction planes. The pixel width in each reconstructed plane is defined by dividing the shift distance between adjacent backprojected detector images into 10 pixels. Each detector element value is divided among the pixels it overlaps according to the area of overlap. By this convention for defining pixels, each single-plane image has a fixed number of pixels (710 x 710) and the physical pixel dimension increases as the distance from detector to reconstruction plane increases. The isocenter-plane pixel width is 0.161 mm. As shown in Figure 5.1B, the pixel centers for the stack of tomosynthesis images are defined such that a fixed pixel position (e.g. row 100, column 100) in the stack corresponds to a ray originating at the detector center.

After digital tomosynthesis, a gradient filtering procedure is applied to each of the single-plane images to identify local regions of high sharpness and contrast. The final 2D "composite" image is then formed by selecting, for each pixel position, the pixel value from the single-plane image with highest contrast and sharpness. Due to the geometry of the tomosynthesis pixel centers, this is equivalent to selecting the sharpest pixels for rays originating at the center of the detector, and the final composite image can be viewed as an inverted "virtual" cone-beam projection of the in-focus objects in the patient volume (Figure 5.1C). A virtual SBDX projection originates at the center of the detector and falls on the source plane. Noting that the lateral distance between rays drawn from the detector center to adjacent focal spot positions is always divided into 10 pixels during tomosynthesis reconstruction, the pitch of the virtual detector elements at the source plane is equal to the focal spot pitch (2.3 mm) divided by 10 pixels, 0.23 mm.

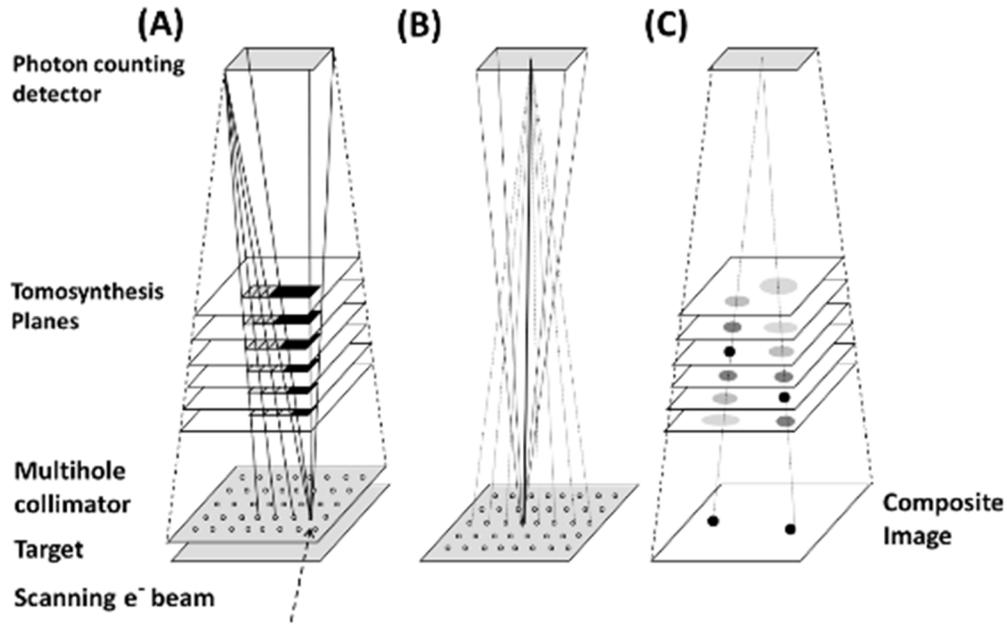

Figure 5.1: (A) SBDX performs digital tomosynthesis using shift-and-add backprojection. (B) Individual detector images are backprojected along the ray connecting each source position to the center of the detector. (C) A multi-plane composite image reconstructed from the tomosynthetic plane stack is analogous to a 2-D virtual projection image.

#### 5.2.1.2   Geometric calibration

Since the SBDX composite image can be viewed as a virtual cone-beam projection, a projection matrix approach can be used to estimate the geometric parameters of the SBDX system for calibration. The 3x4 projection matrix **P** maps a point (x, y, z) in the 3D object frame-of-reference to homogeneous coordinates ($\lambda$u, $\lambda$v, $\lambda$):

$$[\lambda u, \lambda v, \lambda]^T = \boldsymbol{P}[x, y, z, 1]^T \tag{5.1}$$

The parameter $\lambda$ is a scaling factor. The detector indices (u, v) corresponding to the projection of point (x, y, z) onto the virtual detector plane may be obtained by dividing the homogenous coordinates by $\lambda$.

The unknown projection matrix **P** is written as a product of three matrices **P = KRT**. Here **K** describes the virtual projection geometry's intrinsic parameters, **R** is a rotation matrix, and **T** is a translation matrix. **K** depends on source-detector-distance (SDD), pitch between virtual detector elements ($s_p$), and the coordinates ($u_o$, $v_o$) which define the piercing point on the virtual detector. **R** depends on three angles ($\theta_x$, $\theta_y$, $\theta_z$) describing the rotations about the three principal axes. **T** depends on the location of the virtual source point ($x_s$, $y_s$, $z_s$) in the 3D object coordinate system. Denoting $c_j = \cos(\theta_j)$ and $s_j = \sin(\theta_j)$, the projection matrix **P** is given by:

$$P = \begin{bmatrix} -u_o & SDD/s_p & 0 \\ -v_o & 0 & SDD/s_p \\ -1 & 0 & 0 \end{bmatrix} \begin{bmatrix} 1 & 0 & 0 \\ 0 & c_x & s_x \\ 0 & -s_x & c_x \end{bmatrix} \begin{bmatrix} c_y & 0 & -s_y \\ 0 & 1 & 0 \\ s_y & 0 & c_y \end{bmatrix} \begin{bmatrix} c_z & s_z & 0 \\ -s_z & c_z & 0 \\ 0 & 0 & 1 \end{bmatrix} \begin{bmatrix} 1 & 0 & 0 & -x_s \\ 0 & 1 & 0 & -y_s \\ 0 & 0 & 1 & -z_s \end{bmatrix} \tag{5.2}$$

The **P** matrix is parameterized by a vector, $\xi$, consisting of nine elements, $\xi = $ [SDD, $u_o$, $v_o$, $\theta_x$, $\theta_y$, $\theta_z$, $x_s$, $y_s$, $z_s$]. The pitch between virtual detector elements, $s_p$, equals 0.23 mm. A calibration phantom containing a known helical configuration of N high contrast point-like markers is then imaged (Figure 5.2A). The virtual detector coordinates ($u_i$, $v_i$) of the projections of the markers are then determined using a center-of-mass technique. The geometric parameters describing the IGCT system are estimated by minimizing the sum-of-squared differences between the measured positions of the markers ($u_i$, $v_i$) and the **P**-matrix-projected marker detector coordinates ($u_i(\xi)$, $v_i(\xi)$) as described by equation (5.3). The optimization was performed using a Quasi-Newton method, with initial parameters set to the nominal system geometry described in Table 2.1.

$$\hat{\xi} = \arg\min_\xi \frac{1}{N} \sum_{i=1}^{N} [(u_i(\xi) - u_i)^2 + (v_i(\xi) - v_i)^2]) \tag{5.3}$$

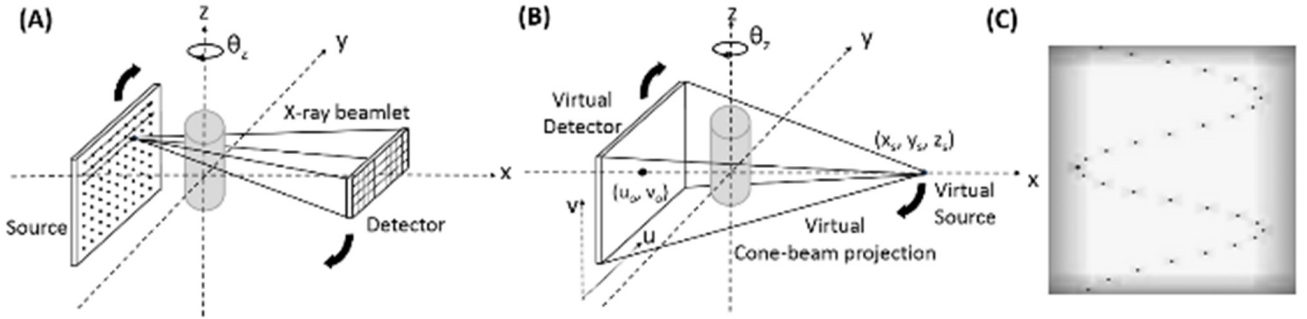

Figure 5.2: A: SBDX CT acquisition consisting of source scanning and gantry rotation about the z-axis. B: Matching virtual cone-beam projection geometry. C: Example composite image (virtual projection) of the helix phantom.

### 5.2.1.3   Simulations

The performance of the proposed geometric calibration method was evaluated through numerical simulations of SBDX CT data acquisition with simulated deviations from the nominal SBDX geometry. A noise free SBDX CT data acquisition consisting of 210 view angles uniformly distributed over 210 degrees was simulated for a known helical configuration of 30 spherical steel fiducials. The helix phantom was centered at iso-center with its long axis aligned with the axis-of-rotation (z-axis; see Figure 5.2A). The SBDX source and detector were translated 1 mm in the +y direction and the source and detector arrays were rotated 1 degree about the x-axis to mimic a hypothetical system misalignment. For each gantry orientation, a composite image (Figure 5.2C) was reconstructed and a 3 x 4 projection matrix **P** was derived.

The CT data acquisition scheme was then repeated for a numerical thorax phantom. Reconstruction was performed using a gridded filtered back projection algorithm (gFBP)[119] with and without geometric calibration to assess the proposed method's ability to reduce image artifacts. In order to quantify the accuracy of the reconstruction with and without calibration, the relative root-mean-squared-error (rRMSE) was calculated versus the known ground truth. To isolate and quantify errors related only to geometric calibration, the rRMSE was also computed versus a gFBP reconstruction of the thorax phantom accounting for the known deviations in system geometry. The rRMSE is defined as:

$$rRMSE(x) = \sqrt{\frac{1}{N} \sum_i^N (\frac{x_i - x_i^{ref}}{\max(x^{ref}) - \min(x^{ref})})^2}$$

(5.4)

### 5.2.1.4  *Sensitivity analysis*

The sensitivity of the method to uncertainties in SDD, translations and rotations was investigated through numerical simulations. The nominal SBDX system geometry parameters defined in Table 2.1 were perturbed by varying amounts for two scenarios. In the first scenario, the SDD and SAD were varied by 1 mm. The SBDX source and detector were translated 1 mm in the y-direction, 1 mm in the z-direction, and rotated 1 degree about the x-axis, 1 degree about the y-axis, and 1 degree about the z-axis. For the second scenario, the SDD and SAD were varied by 5 mm, the source and detector were translated 5 mm in the y-direction, 5 mm in the z-direction, and rotated 5 degrees about each of the x-, y-, and z-axes. For each scenario, the proposed calibration method was used to estimate the system geometry and compared versus the known perturbations. The mean error and standard deviation in the estimated geometric parameters were determined for 210 gantry view angles evenly distributed over 210 degrees.

### *5.2.1.5  Experimental validation*

The proposed calibration method was tested in three phantom studies using projection data acquired with the SBDX system. For each study, a geometric calibration procedure was followed. The SBDX gantry was rotated 90 degrees to a lateral angulation and a helix phantom was placed on a rotating stage at iso-center (Figure 5.3A). The calibration phantom was constructed out of poly(methyl methacrylate) (PMMA) and consists of a known helical configuration of 41 spherical steel fiducials.[42] The diameter of the fiducials was 1/16 inch. The angular pitch and increment between fiducials along the z-axis were 22.5 degrees and 0.15 inch respectively. The phantom contains a 3/32 inch diameter reference fiducial which is larger than the other fiducials. The larger fiducial can be used to relate the known 3D fiducial coordinates to the 2D projection coordinates using connected component analysis. Only fiducials appearing in the CT field-of-view are used for calibration. The outer diameter of the phantom measured 4 inches and the wall thickness was 3/8 inch.

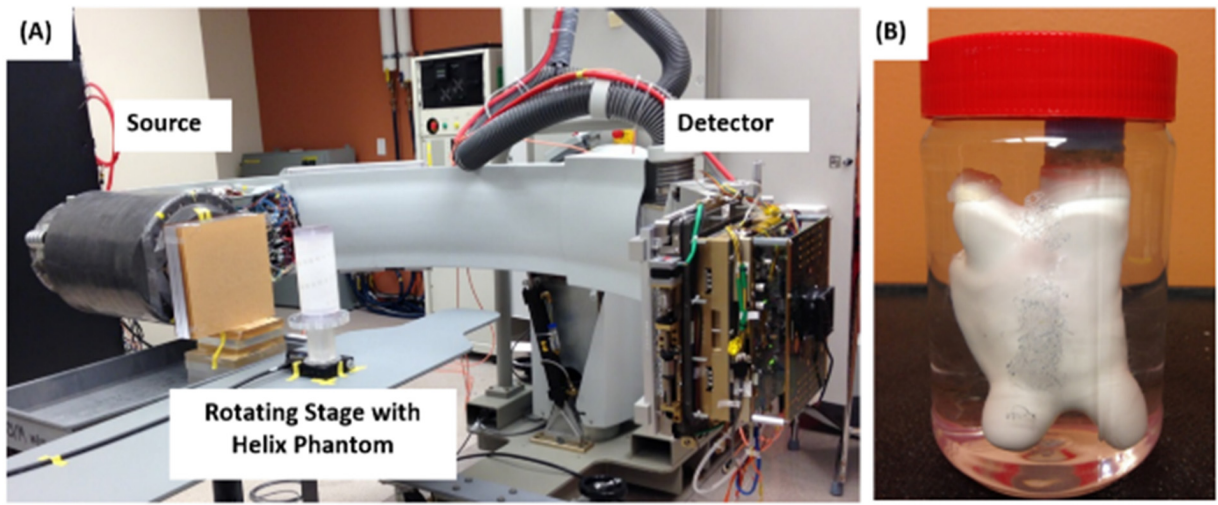

Figure 5.3: A. SBDX CT data acquisition was simulated by rotating the C-arm to a horizontal position and rotating the imaging object. The helix calibration object is shown. B. A custom-made anatomic chamber model placed in an 85 mm diameter water cylinder.

The helix phantom was imaged using a continuous stage rotation with an angular velocity of 2 degree per second over a 220 degree short-scan range. A step-and-shoot technique was used by selecting every 7[th] frame of the complete projection dataset resulting in 236 view angles with an angular increment of 0.93 degree. Imaging was performed at 100 kV tube potential, 40 mA peak tube current (19% full power), with a 71 x 71, 15 frames/sec scanning technique. Acrylic with a thickness of 7 cm was placed in the beam before the helix phantom. For each view angle, a composite image was reconstructed and a 3 x 4 projection matrix **P** was derived.

### 5.2.1.6 *Atrium phantom*

In the first phantom study, a custom-made phantom designed to resemble a left atrium was imaged. This phantom was chosen because the SBDX system is being investigated for the task of 3D imaging of a cardiac chamber during an interventional procedure. The hollow atrium was filled with 3 ml iohexol contrast agent (Omnipaque 350, GE Healthcare, Waukesha, WI, USA) diluted with 88 ml of water and placed in an 85 mm diameter cylinder of deionized water that fits entirely within the SBDX field-of-view (Figure 5.3B). Six 1/16 inch diameter aluminum fiducials were attached to the outside of the water cylinder to demonstrate the proposed method's ability to mitigate artifacts and resolve small high contrast point-like objects. SBDX projection data was then acquired using the same imaging technique used for the calibration phantom. The atrium phantom was reconstructed using gFBP with and without geometric calibration. Table 5.1 presents the rebinning algorithm parameters that were used for reconstruction, unless otherwise stated.

Table 5.1: Gridded FBP reconstruction parameters for rotating stage.

| | |
|---|---|
| Number of view angles over 180 degrees | 480 |
| Rays per 2D view [columns x rows] | 1120 x 290 |
| 2D view sampling pitch | 0.125 mm x 0.5 mm |
| Radial kernel | 1.4 mm x 2.0 mm |
| Angular kernel | 2.0 degree x 2.0 degree |
| Reconstruction grid (number of voxels) | 512 x 512 x 330 |
| Voxel dimensions | 0.27 mm x 0.27 mm x 0.27 mm |

### 5.2.1.7 *Spatial fidelity metric*

In the second study, IGCT reconstruction of the helix calibration phantom itself was performed with and without geometric calibration. The spatial fidelity of a reconstruction was validated by computing the Euclidean distance between each combination of the steel fiducials. The 30 steel fiducials within the CT image volume were segmented and a center-of-mass calculation was performed to determine the 3D Cartesian coordinates of each fiducial. The separation distance was calculated for each of the 435 pair combinations. As a reference, the separation distances were also calculated from the known helix geometry. The reference fiducial separation distances ranged from 17.0 mm to 126.8 mm.

### 5.2.1.8  Modulation transfer function

The third study measured the modulation transfer function (MTF) from an image of a wire phantom reconstructed with geometric calibration. A hollow cylindrical PMMA phantom containing a 0.154 mm diameter stainless-steel wire at the center was used to measure the LSF. The wire phantom was positioned on the rotating stage with the wire near iso-center and parallel to the rotation axis of the stage. SBDX projection data were then acquired using the same imaging technique used for the helix and atrium phantoms. The projection data were reconstructed using the rebinning parameters listed in Table 5.1 with a targeted reconstruction FOV. A 512 x 512 x 512 pixel reconstruction grid was used with isotropic pixel resolution set to half the diameter of the steel wire, 0.077 mm x 0.077 mm x 0.077 mm. The MTF was calculated following the method of Kayugawa et al.[158] A LSF was determined by integrating row by row across the image columns within a 41 x 41 pixel ROI centered on the wire. An offset correction was applied to the LSF by subtracting the mean value of the three data points at each edge of the LSF. The LSF was normalized to unit area. The normalized LSF was zero padded and the MTF was computed as the magnitude of the fast Fourier transform of the LSF.

## 5.2.2  Results

### 5.2.2.1  Simulations

The geometric parameters corresponding to the simulated SBDX CT data acquisition were determined at each view angle. The difference between the extracted value and the true value was computed for each parameter of $\xi$. Table 5.2 summarizes the mean error and standard deviation in error versus gantry angle, for each parameter. Over all angles, the maximum error in a rotation parameter ($\theta_x$, $\theta_y$, $\theta_z$) was less than 0.02 degrees. The maximum error in the virtual source point ($x_s$, $y_s$, $z_s$) was 0.4 mm, and the maximum error in SDD was -0.13 mm. Errors in the ($u_o$, $v_o$) coordinates were less than the dimension of a virtual detector element.

Table 5.2: Mean, standard deviation, and maximum errors in estimated geometric parameters for simulated 1 mm translation in $y_s$ and a 1 degree rotation in $\theta_x$.

| | $x_s$ | $y_s$ | $z_s$ | $\theta_x$ | $\theta_y$ | $\theta_z$ | SDD | $u_o$ | $v_o$ |
|---|---|---|---|---|---|---|---|---|---|
| | | (mm) | | | (degrees) | | | (mm) | |
| **Mean Error** | 0.0 | 0.0 | 0.1 | $2.4 \times 10^{-4}$ | $-6.0 \times 10^{-3}$ | $-2.0 \times 10^{-3}$ | 0.0 | 0.0 | 0.0 |
| **Std. Dev.** | 0.1 | 0.1 | 0.1 | $4.0 \times 10^{-3}$ | $6.0 \times 10^{-3}$ | $6.0 \times 10^{-3}$ | $4.1 \times 10^{-2}$ | $2.0 \times 10^{-4}$ | $2.3 \times 10^{-4}$ |
| **Max Error** | 0.3 | -0.3 | 0.4 | $-1.2 \times 10^{-2}$ | $-2.0 \times 10^{-2}$ | $-1.6 \times 10^{-2}$ | $-1.3 \times 10^{-1}$ | $-6.4 \times 10^{-4}$ | $-6.8 \times 10^{-4}$ |

A miniature thorax phantom fully enclosed within the SBDX 140 mm field-of-view was used to investigate geometry-calibration-related reconstruction artifacts without the presence of confounding truncation artifacts. The numerical thorax phantom was reconstructed without (Figure 5.4A) and with (Figure 5.4B) geometric calibration using gFBP. Figure 5.4A shows significant artifacts and distortion of structures caused by the failure to account for the translation and rotation of the system geometry. These artifacts were removed in the image reconstructed with the proposed calibration technique. The rRMSE was calculated versus the known ground truth to quantify reconstruction accuracy. The rRMSE was 1.3% using the proposed geometric calibration method and 8.0% without geometric calibration. The rRMSE was also calculated versus an image reconstructed using gFBP with exactly known geometry to isolate errors due only to geometric uncertainties. The rRMSE was 0.4% with geometric calibration, versus 7.7% without geometric calibration.

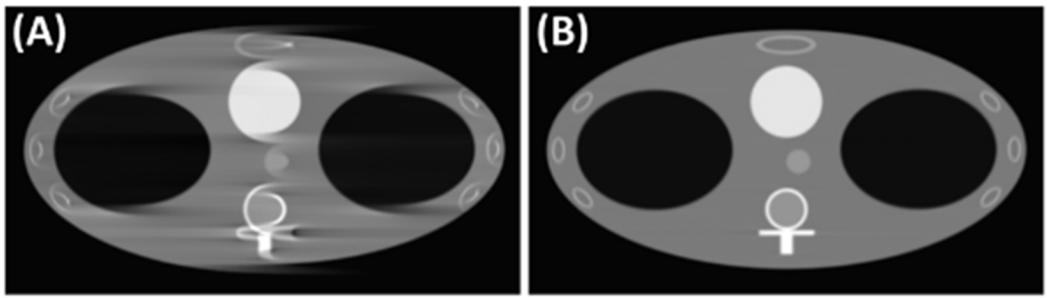

Figure 5.4: A thorax phantom was simulated with 1 mm translation in $y_s$ and a 1 degree rotation in $\theta_x$. Shown at left is a reconstruction without geometric calibration (A). At right is a reconstruction with geometric calibration (B). Display window is [-800, 1000] HU.

### 5.2.2.2  *Sensitivity analysis*

Table 5.3 and Table 5.4 summarize geometry parameter recovery for additional deviations from the nominal geometry. A total of 210 gantry view angles were evaluated for each case. Table 5.3 presents results where the SDD and SAD were varied by 1 mm, the source and detector arrays were translated +1 mm and +1 mm along the y-axis and z-axis, and rotated +1 degree about each of the x-, y-, and z-axes. Table 5.4 presents results where the SDD and SAD were varied by 5 mm, the source and detector arrays were translated +5 mm along the y-axis, +5 mm along the z-axis, and rotated +5 degree about each of the x-, y-, and z-axes.

For both scenarios, the average errors in parameters describing the virtual source point were less than or equal to 0.7 mm. The average errors observed in parameters describing rotations of the source and detector arrays were less than 0.01 degree. The piercing point coordinates ($u_o$, $v_o$) were estimated with average errors less than or equal to 0.1 mm for both cases considered. Errors in the ($u_o$, $v_o$) coordinates were less than the dimension of a virtual detector element. The maximum error observed in the SDD parameter estimation was -2.0 mm. Potential techniques to improve estimates of the SDD parameter are discussed in section 5.6.

Table 5.3: Mean and standard deviation of errors in estimated geometric parameters with simulated deviations from the nominal system geometry. The SDD and SAD were varied 1 mm from nominal values. The detector and source arrays were translated 1 mm along the y-axis and 1 mm along the z-axis. The source and detector arrays were rotated 1 degree about the y-axis, 1 degree about the x-axis, and 1 degree about the z-axis.

| $x_s$ | $y_s$ | $z_s$ | $\theta_x$ | $\theta_y$ | $\theta_z$ | SDD | $u_o$ | $v_o$ |
|---|---|---|---|---|---|---|---|---|
| | (mm) | | | (degrees) | | | (mm) | |
| $0.1 \pm 0.3$ | $-0.2 \pm 0.2$ | $0.1 \pm 0.1$ | $0.0 \pm 4.2 \times 10^{-3}$ | $0.0 \pm 5.7 \times 10^{-3}$ | $0.0 \pm 6.1 \times 10^{-3}$ | $-0.6 \pm 4.1 \times 10^{-2}$ | $0.0 \pm 2.8 \times 10^{-3}$ | $0.0 \pm 1.2 \times 10^{-2}$ |

Table 5.4: Mean and standard deviation of errors in estimated geometric parameters with simulated deviations from the nominal system geometry. The SDD and SAD were varied 5 mm from nominal values. The detector and source arrays were translated 5 mm along the y-axis and 5 mm along the z-axis. The source and detector arrays were rotated 5 degrees about the y-axis, 5 degrees about the x-axis, and 5 degrees about the z-axis.

| $x_s$ | $y_s$ | $z_s$ | $\theta_x$ | $\theta_y$ | $\theta_z$ | SDD | $u_o$ | $v_o$ |
|---|---|---|---|---|---|---|---|---|
| | (mm) | | | (degrees) | | | (mm) | |
| $0.2 \pm 1.0$ | $-0.7 \pm 0.6$ | $0.2 \pm 0.1$ | $0.0 \pm 4.9 \times 10^{-3}$ | $0.0 \pm 6.2 \times 10^{-3}$ | $0.0 \pm 1.0 \times 10^{-2}$ | $-2.0 \pm 9.8 \times 10^{-2}$ | $0.0 \pm 2.5 \times 10^{-2}$ | $-0.1 \pm 3.6 \times 10^{-2}$ |

### 5.2.2.3  Experimental geometry determined from P-matrices

The estimated geometric parameters for the experimental SBDX set-up describing SDD, piercing point ($u_o$, $v_o$) and rotation ($\theta_x$, $\theta_y$) are presented in Table 5.5, averaged across view angle. The exact geometric parameters of the SBDX system (e.g. the exact distance between the tungsten target and the surface of the CdTe detector) were not measured directly due to the invasive and destructive nature of such measurements. Instead, the experimentally derived geometry was compared to the nominal geometry of SBDX. The mean estimated SDD was $1500.0 \pm 0.1$ mm, compared to a nominal value of 1500.0 mm. The proposed calibration method estimated the source and detector rotation about the x-axis to be 0.2 degrees and rotation about the y-axis to be 0.5 degrees.

Table 5.5:  Estimated geometric parameters (SDD, $\theta_x$, $\theta_y$ $u_o$, $v_o$) for the bench-top SBDX set-up.

| Parameter | Value |
|---|---|
| SDD (mm) | $1500.0 \pm 0.1$ |
| $\theta_x$ (degrees) | $0.16 \pm 0.03$ |
| $\theta_y$ (degrees) | $0.46 \pm 0.03$ |
| $u_o$ | $355.5 \pm 0.0$ |
| $v_o$ | $355.5 \pm 0.0$ |

The estimated view angle $\theta_z$ is plotted versus the view index in Figure 5.5. A linear regression was performed to determine the relationship between view index and view angle. The slope of the linear regression line was 0.93 degree per view index and the intercept was 0.0 degree. The angular increment between view indices determined from the slope of the regression line was in good agreement with the motion controller programming of $0.93 \pm 0.05$ degree per view index.

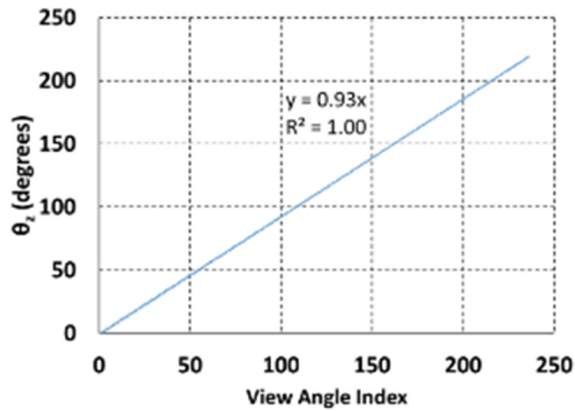

Figure 5.5: The estimated view angle ($\theta_z$) is plotted versus the view angle index.

### *5.2.2.4 Experimental phantom reconstructions and spatial fidelity*

Figure 5.6 shows reconstructions of the atrium phantom generated with the SBDX prototype and a rotating stage. The performance of the geometric calibration method can be assessed qualitatively by comparing reconstructed CT images without (Figure 5.6 A and C) and with geometric calibration (Figure 5.6 B and D). Double contour artifacts are observed near the high-contrast fiducials and object blurring is present for the image reconstructed without geometric calibration. The artifacts are reduced in the image reconstructed with geometric calibration.

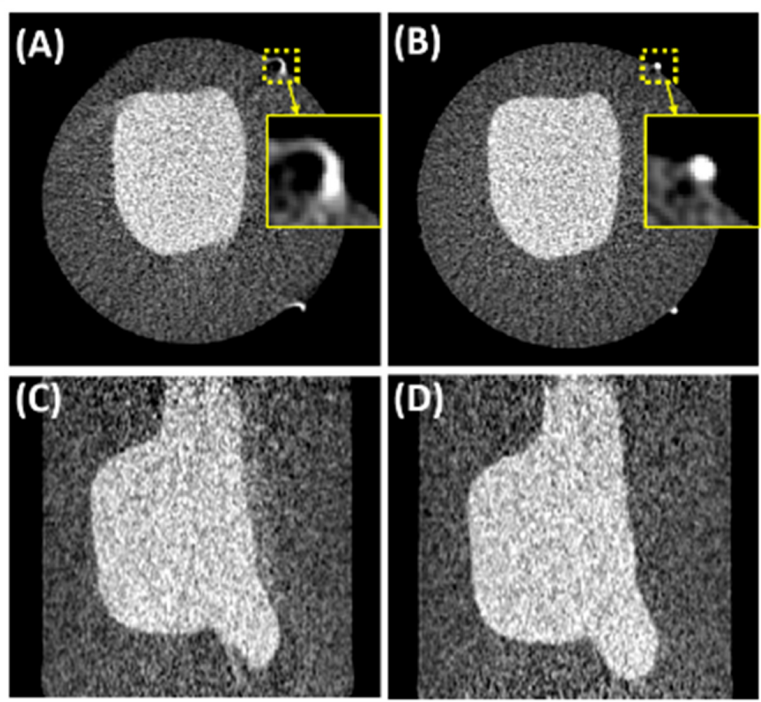

Figure 5.6: SBDX projection data reconstructed without geometric calibration (A) shows double contour artifacts and blurring of object structures. The artifacts are eliminated when geometric calibration is used during image reconstruction (B). Sagittal slices without (C) and with (D) geometric calibration are presented. Display window is [-300, 700] HU.

The reconstructed helix phantom is shown in Figure 5.7A without geometric calibration and in Figure 5.7B with geometric calibration. The PMMA cylinder supporting the helix is blurred in the image without calibration and artifacts are present near the steel fiducial. A maximum intensity projection (MIP) image shows contour artifacts for each of the steel fiducials in the image reconstructed without calibration (Figure 5.8A). Geometric calibration reduced the artifacts (Figure 5.8B).

Spatial fidelity was determined by comparing IGCT-derived fiducial separation distances to the reference distances computed from the known helix geometry. In the image reconstructed without geometric calibration, the mean error in fiducial separation distance was 0.58 mm and the standard deviation was 5.90 mm. Application of geometric calibration during reconstruction reduced the mean error to -0.04 mm and the standard deviation to 0.18 mm, indicating improved spatial fidelity.

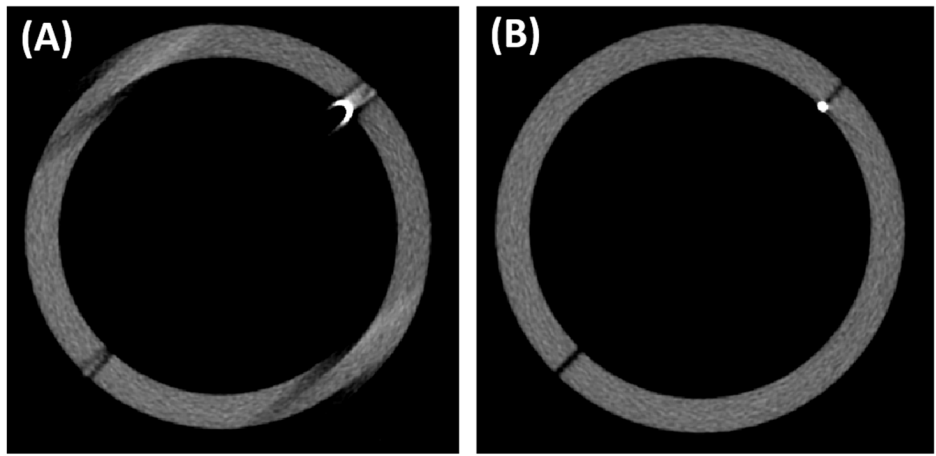

Figure 5.7: IGCT reconstruction without geometric calibration (A) shows artifacts. Contour artifacts around an aluminum fiducial are reduced in an image reconstructed with calibration (B). Display window is [-600, 1000] HU.

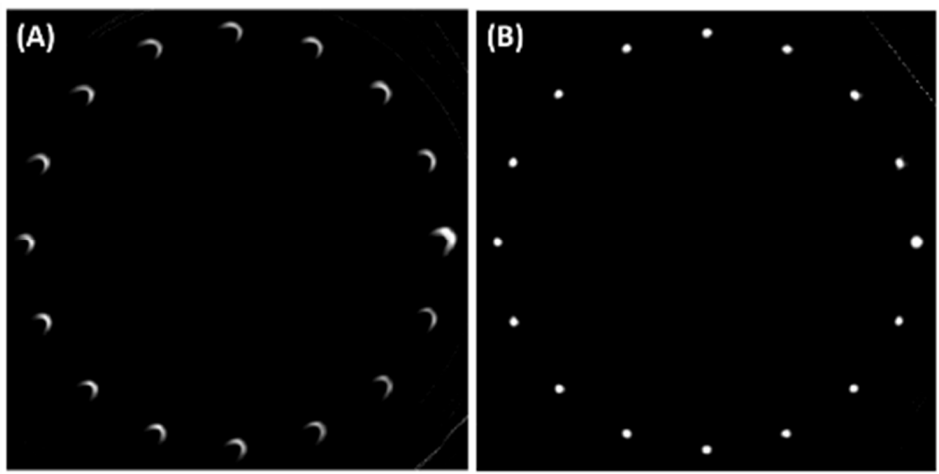

Figure 5.8: A maximum intensity projection image demonstrates contour artifacts around high-contrast aluminum fiducials for an image reconstructed without geometric calibration (A). Application of the proposed calibration technique reduced the artifacts (B).

### *5.2.2.5 Modulation transfer function*

Figure 5.9 shows the LSF derived from an SBDX CT image of a wire phantom using the proposed geometric calibration method. The full width at half maximum (FWHM) of the LSF measured 0.81 mm. For context, we note that SBDX CT is being investigated for its potential to provide 3D anatomic maps during catheter-based interventions. The measured FWHM value (0.81 mm) of the LSF is less than the diameter (2-3 mm) of a typical ablation catheter tip that would be displayed with the 3D anatomic map.[152]

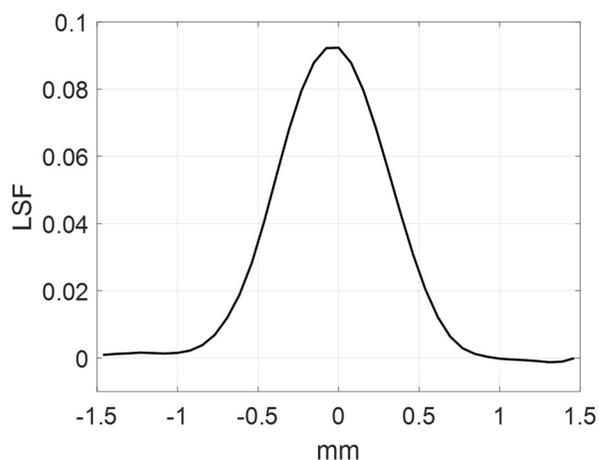

Figure 5.9: A line spread function derived from imaging of a 0.154 mm diameter steel wire. The full width at half maximum value measures 0.81 mm.

Figure 5.10 presents the MTF curve. The MTF value was 50% at a spatial frequency of 5.8 cm$^{-1}$ and 10% at 10.3 cm$^{-1}$. The LSF and MTF are presented as example resolution metrics in the presence of geometric calibration. Reconstruction parameters were not optimized to maximize spatial resolution. Chapter 6 presents results that demonstrate spatial resolution may be improved by adjusting the gridding parameters associated with the gFBP technique.

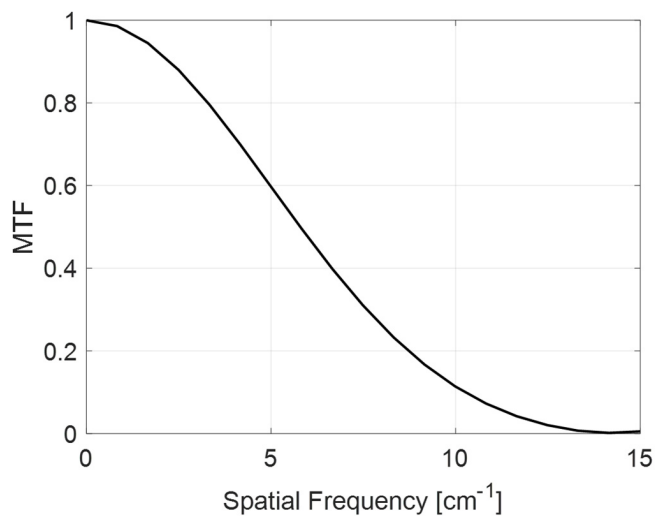

Figure 5.10: The modulation transfer function using IGCT geometric calibration is shown. The 10% MTF value corresponds to a spatial frequency of 10.3 cm$^{-1}$.

## 5.3 Detector non-linearity

As discussed in section 2.4.3, the SBDX photon counting detector exhibits a non-linear response at high fluences. For CT imaging, a non-linear detector response can lead to artifacts and inaccurate CT numbers. Therefore, the SBDX detector response was measured for a range of incident detector fluences and a simple calibration procedure was implemented to correct for detector non-linearity.

### 5.3.1 Methods

The detector nonlinearity calibration procedure can be summarized in three parts. First, SBDX source output and incident detector fluence versus tube current linearity was verified experimentally. Second, the SBDX detector response was measured for a range of incident fluences and compared versus measurements performed with a dose meter. Third, polynomial functions were fit to the linear and non-linear regions of the measured detector response data. A mapping function was determined from the polynomial fits and used to linearize the SBDX detector response. Each step of the detector calibration procedure is detailed further below.

### 5.3.1.1  *Source output versus tube current*

The incident fluence on the detector is expected to scale linearly with peak tube current ($mA_{peak}$). The fluence/tube current linearity assumption was validated experimentally using the set-up shown in Figure 5.11. A dose-area-product (DAP) meter (Radcal DAPcheck PDC, Radcal Corporation, Monrovia, CA, USA) was placed directly on top of the source collimator to measure the total x-ray output of the tube. A solid state digital dose meter (Solidose Model 300, R100 probe, RTI Electronics AB, Mölndal, Sweden) was secured immediately adjacent to the detector entrance window to measure the exposure incident on the detector. The probe's active area was centered on the detector and aligned parallel with the long axis of the rectangular detector. Varying thicknesses of PMMA (6.99 cm, 9.32 cm, 11.65 cm, 13.98 cm, and 18.64 cm) were placed on the table, one stack at a time, with the center of each stack at iso-center (i.e. 40 cm from the collimator exit). The thicknesses of PMMA were chosen to generate a range of incident detector fluences. SBDX imaging was then performed at 100 kV with peak tube currents of 10, 20, 30, 40, 60, 80, 100, 120, 140, 160, 180, 200, and 210 $mA_{peak}$ for each thickness of PMMA. A total of five measurements were performed for each operating point with acquisition times of 15 seconds per measurement. The DAP was recorded in units of $\mu Gy \cdot m^2/min$. The exposure rate was recorded in units of mR/s.

To evaluate the linearity between tube output and tube current, the SBDX peak tube current ($mA_{peak}$) was plotted versus exposure rate and DAP for each of the five PMMA thicknesses. A linear regression was performed for each dataset and the coefficient of determination, $R^2$, was computed to assess linearity. Linear was defined as regressions (detector response versus tube output) yielding $R^2$ values greater than or equal to 0.99. An additional regression was performed to determine the relationship between the DAP measured at the collimator exit and the exposure rate measured at the detector entrance.

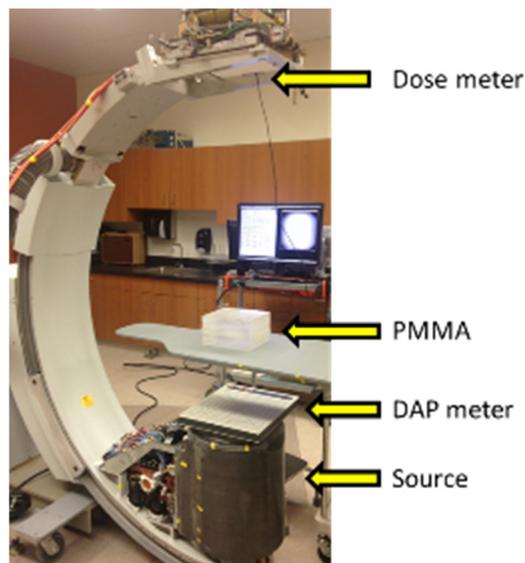

Figure 5.11: A DAP meter placed on the SBDX collimator exit and a dose meter positioned at the detector entrance were used to assess the linearity of the SBDX source output versus peak tube current.

### 5.3.1.2 *Detector response measurement*

The experimental set-up shown in Figure 5.11 was modified by removing the dose meter from the detector prior to evaluating the detector response. The DAP meter was left in place on the source array to monitor tube output during imaging. SBDX imaging was then performed at 100 kV with peak tube currents of 10, 20, 30, 40, 60, 80, 100, 120, 140, 160, 180, 200, and 210 mA$_{peak}$ for each of the five thicknesses of PMMA (6.99-18.64 cm). The PMMA thicknesses were chosen to yield a range of incident fluences spanning both the linear and nonlinear response regions of the detector. Finally, a 95 frame no power acquisition was performed for dark field offset correction.

For each peak tube current and PMMA thickness combination, the number of fluoroscopic frames required to yield 5% or less relative uncertainty in the mean photon counts per detector element per frame with 95% confidence was determined based on Poisson statistics. The native detector images were then averaged across the total number of imaging frames and a frame averaged dark-field was subtracted. Next, the mean counts per detector element were averaged over a subset of the 71 x 71 native detector images (one per focal spot) to reduce noise. The median number of photons detected per frame was computed for all 5,041 native detector images. Detector images with averaged photon counts per frame within +/- 3% of the median value were averaged yielding a single 160 x 80 element detector image. Finally, for each operating point and combination of PMMA thickness, the mean detected counts per element was computed in an ROI centered on the averaged detector image approximately equal in size to the active area of the solid state dose meter. The ROI was chosen to select a relatively uniform incident fluence region and to exclude the detector edges where the fluence falls off due to beam collimation.

The mean number of detected photons per detector element per frame was then plotted versus the incident detector exposure rate and tube current for each operating point and PMMA thickness to evaluate the detector response. Next, the DAP values recorded during image acquisition were converted to an incident exposure rate using the regression derived during the source output linearity validation. The purpose of this was twofold. First, the DAP measurement accounts for potential variations in tube output that may occur with long imaging times. Second, converting the DAP to an incident exposure rate at the detector accounts for the varying transmission of the different PMMA thicknesses. The mean detected photons per detector element per frame were then plotted versus the measured exposure rate incident on the detector. A linear regression was performed to validate the assumption that the detector response is non-linear at high incident fluences.

### 5.3.1.3  *Detector response calibration*

The goal of the SBDX detector response calibration is to determine a mapping function that converts the number of photon counts recorded by the detector to the true number of photons incident on the detector. The mapping function was determined from the measured photons per detector element per frame ($\overline{D}$) versus exposure rate ($\dot{X}$) data. The incident exposure rates on the detector ranged from 0.2 mR/s to 63.7 mR/s. A third-order polynomial regression was performed to determine coefficients ($a_0, a_1, a_2, a_3$) that yielded the detector response, $\overline{D}$, as a function of incident exposure, $\dot{X}$, as described by equation (5.5).

$$\overline{D}(\dot{X}) = a_0 + a_1\dot{X} + a_2\dot{X}^2 + a_3\dot{X}^3 \tag{5.5}$$

At low enough exposure rates, the detector exhibits linear response. In this scenario, a linear regression was performed to determine coefficients ($b_0$, $b_1$) that define the expected linear detector response, $\overline{D_L}$, as shown in equation (5.6).

$$\overline{D_L}(\dot{X}) = b_0 + b_1\dot{X} \tag{5.6}$$

The linear regression was constrained to use data within the detector's linear response region defined as exposure rates less than or equal to a threshold value $\tau_L$. The optimal threshold, $\tau_L$, was determined by performing linear regressions with $\tau_L$ varied from 0.5 mR/s to 10.0 mR/s in increments of 0.1 mR/s. The $\tau_L$ that yielded the maximum $R^2$ value of the linear fits was selected to define the upper boundary of the linear response region.

After determining coefficients ($a_0, a_1, a_2, a_3$) and ($b_0, b_1$), a detector non-linearity correction was performed by mapping the raw recorded detector counts, $D$, to an exposure rate $\dot{X}$ on the measured nonlinear response curve $\overline{D}(\dot{X})$, and then mapping from the exposure rate $\dot{X}$ to the expected linear-response counts using the curve $\overline{D_L}(\dot{X})$. The correction technique was validated by performing a linear regression on the calibrated detector data. The static end-systole XCAT phantom (Figure 4.5A) presented in section 4.2.9 was used to investigate the impact of the non-linear detector response on segmentation accuracy. Projection data were simulated at 100% full power. The simulated detector counts were then transformed according to equations (5.5) and (5.6), after determining the coefficients from measured data, to model the SBDX detector response. SBDX CT reconstruction was performed using the gFBP algorithm with the linear and nonlinear projection datasets. Following reconstruction, the left atrium was segmented from both sets of projection data and the 99th percentile segmentation error was computed to quantify the impact of detector response on segmentation accuracy.

### 5.3.2 Results

#### 5.3.2.1 *Source output versus tube current*

The SBDX source output measured in terms of DAP at the collimator exit is plotted as a function of peak tube current in Figure 5.12. The $R^2$ of the linear regression was 1.000. Figure 5.13 presents the incident detector exposure rate with 6.99 cm of PMMA (left) and 18.64 cm of PMMA (right) present as a function of peak tube current. The $R^2$ was 0.999 for both thicknesses of PMMA considered. Source output versus peak tube current can be assumed linear for the SBDX scanning source array demonstrated by linear regressions with coefficients of determination greater than or equal to 0.99. Figure 5.14 plots the measured DAP at the collimator exit versus the measured exposure rate at the detector. A linear relationship between source output at the collimator exit and detector entrance is demonstrated by $R^2$ values of 0.999.

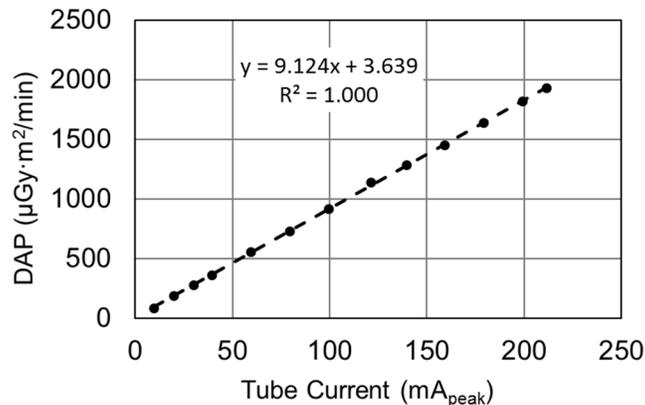

Figure 5.12: DAP measured at the collimator exit versus peak tube current for imaging at 100 kV.

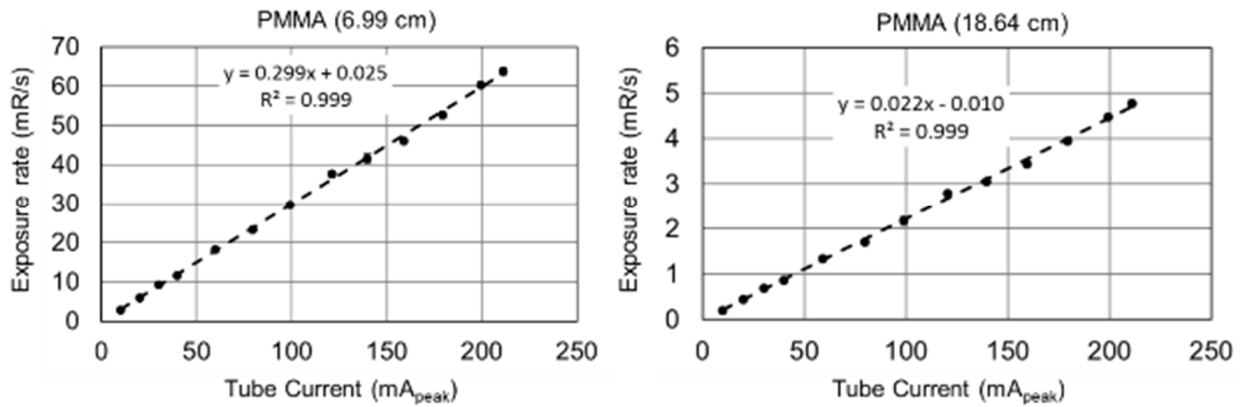

Figure 5.13: Incident detector exposure rate is plotted versus peak tube current for imaging at 100 kV. PMMA (6.99 cm at left, 18.64 cm at right) was added for x-ray attenuation.

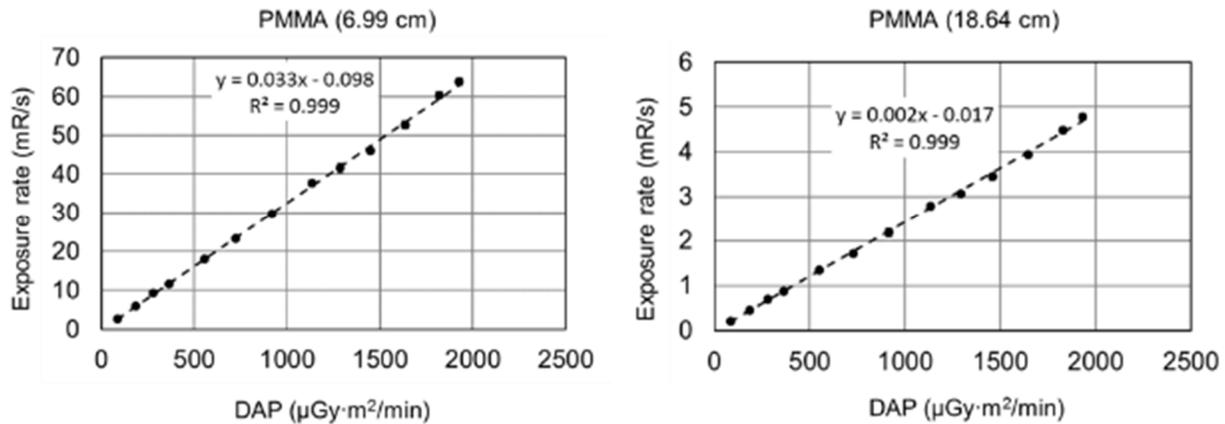

Figure 5.14: The incident detector exposure rate is plotted versus DAP at the collimator exit for imaging at 100 kV. PMMA (6.99 cm at left, 18.64 cm at right) was added for x-ray attenuation.

#### 5.3.2.2 Detector response measurement

The mean number of photons recorded per detector element per frame as a function of DAP at the source collimator exit is plotted in Figure 5.15 for varying amounts of PMMA attenuation (6.99-18.64 cm). To normalize the detected fluence for transmission, DAP was converted to an exposure rate at the detector according to the linear relationships determined in the previous section. The mean number of photons detected per detector element per frame is plotted versus the incident detector exposure rate in Figure 5.16. The $R^2$ value of the linear regression equaled 0.699 indicating the SBDX detector response is non-linear at high incident fluence rates.

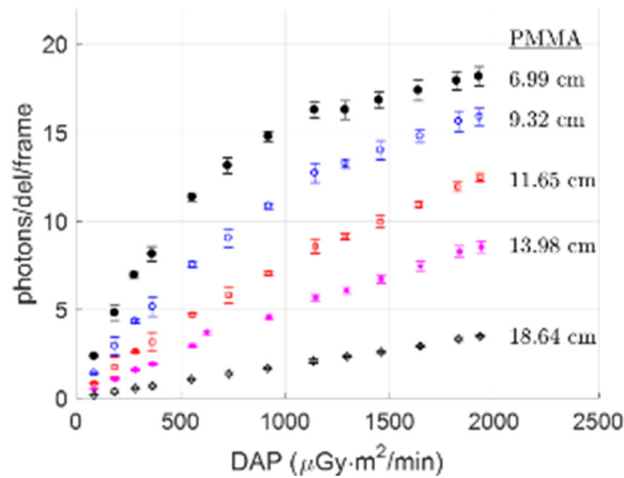

Figure 5.15: The mean number of detected photons per detector element per imaging frame is plotted versus DAP for imaging at 100 kV and varying thicknesses of PMMA ranging from 6.99 cm to 18.64 cm.

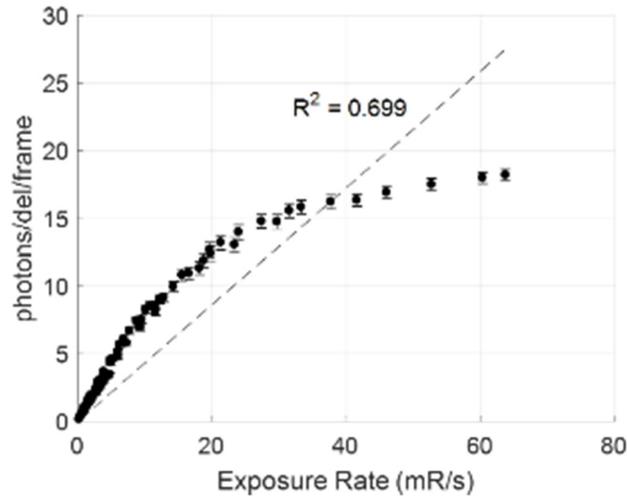

Figure 5.16: The mean number of detected photons per detector element per imaging frame as a function of the incident detector exposure rate is shown for imaging at 100 kV. A non-linear detector response is indicated by the low $R^2$ value of 0.699 for a linear regression.

### 5.3.2.3  *Detector response calibration*

The third-order polynomial coefficients derived to fit the measured detector data are shown in equation (5.7). The expected linear detector response as a function of incident detector exposure rate is defined by equation (5.8). The linear regression was performed from the data shown in Figure 5.16 for exposure rates less than or equal to 8.8 mR/s ($\tau_L$). The $R^2$ values were 0.997 for both the linear and third-order polynomial regressions. Both the linear and third-order functions are plotted with the raw calibration data in Figure 5.17. Detector non-linearity correction was performed by mapping raw detector counts to the expected linear values using equations (5.7) and (5.8). The calibrated counts data are also shown in Figure 5.17. An $R^2$ value of 0.997 demonstrates that the proposed method effectively linearized the measured detector data.

$$\overline{D}(\dot{X}) = 0.0 + 0.94\dot{X} - 0.018\dot{X}^2 + (1.25 \times 10^{-4})\dot{X}^3 \qquad (5.7)$$

$$\overline{D_L}(\dot{X}) = 0.0 + 0.85\dot{X} \qquad (5.8)$$

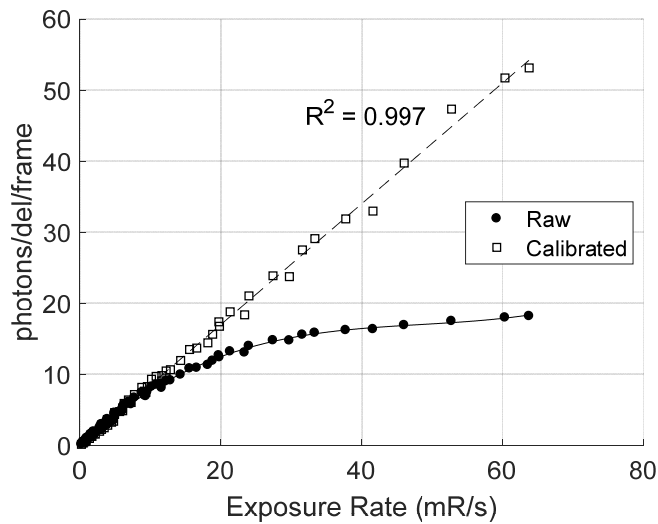

Figure 5.17: The mean number of detected photons per detector element per imaging frame as a function of the incident detector exposure rate is shown for imaging at 100 kV. A linear regression constrained to use data corresponding to an exposure rate less than or equal 8.8 mR/s is shown. A third-order polynomial regression utilizing all of the data is also shown.

The XCAT phantom reconstructed from projection data assuming a linear detector response is shown in Figure 5.18 at left. A reconstruction performed from the simulated non-linear detector response is shown at right. Contrast degradation at the myocardium and lung interface attributed to detector non-linearity is indicated by the yellow arrows. The 99th percentile segmentation errors were 0.71 mm and 0.77 mm for the linear and non-linear detector responses, respectively. The 0.06 mm difference in segmentation accuracy is negligable compared to the diameter of a typical ablation catheter tip (1.7-3.3 mm).

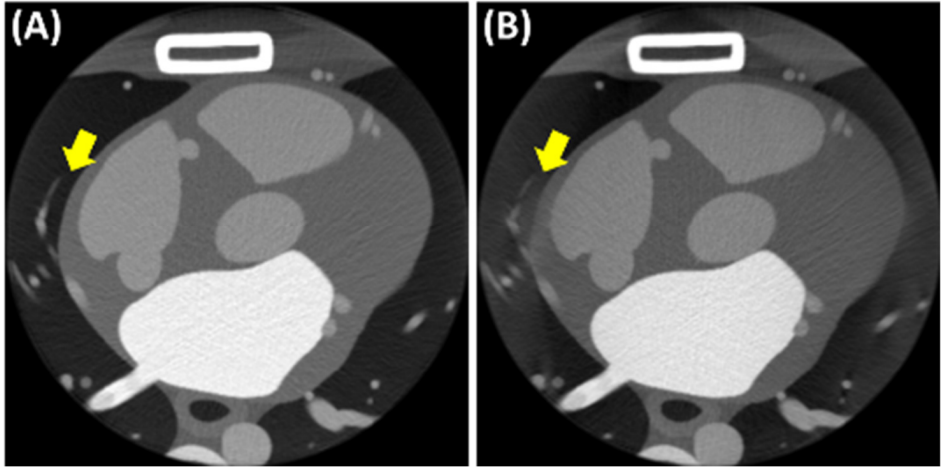

Figure 5.18: An XCAT phantom reconstructed from projections assuming a linear detector response (A) is shown versus a reconstruction performed assuming a non-linear detector response (B). The non-linear detector response results in contrast loss at the tissue air interface indicated by the yellow arrow [WL = 350 HU, WW = 2700 HU].

## 5.4 Beam hardening

The third source of image artifacts considered was beam hardening. The CT reconstruction methods presented in chapters 3 and 4 assumed x-ray attenuation follows the monoenergetic Beer-Lambert law as shown in equation (5.9).

$$\mathrm{I} = I_0 e^{-\int_L dl\mu(\vec{r})} \tag{5.9}$$

Here, $I$ represents the number of incident photons on the detector, $I_0$ represents the number of photons that would be incident on the detector if the object were removed, and $\mu(\vec{r})$ is the spatially variant linear attenuation coefficient describing the object to reconstruct. The integral is performed over the photon path L. Equation (5.9) assumes a monochromatic x-ray spectrum and attenuation coefficients at a specific energy. X-ray beams used in medical imaging are polychromatic however. Defining the x-ray beam spectrum as $\Omega(E)$, and noting that the linear attenuation coefficient is a function of energy, x-ray attenuation is more properly described by equation (5.10).

$$\text{I} = I_0 \int\limits_0^{E_{max}} dE \Omega(E) e^{-\int_L \ dl \mu(\vec{r}, E)} \tag{5.10}$$

The incorrect assumption by equation (5.9), that $\mu(\vec{r})$ is independent of energy, introduces a spectral inconsistency in the log transformed CT projection data that is the source of beam hardening artifacts. Beam hardening manifests as cupping or streak artifacts in the reconstructed image.[159] The impact of beam hardening on left atrium segmentation accuracy is investigated below.

### 5.4.1 Methods

The static end-systole XCAT phantom was used in this study. SBDX CT projections were generated from the 120 kV polychromatic x-ray spectrum shown in Figure 5.19. The spectrum was modeled from an EGSnrc simulation of a transmission-style SBDX target consisting of 12 µm tungsten-rhenium followed by 10 µm Nb (provided by the author of Ref. 160). Additional inherent filtration of 5 mm Be, 1.2 mm $H_2O$, and 0.812 mm Al was included using the SPEKTR 3.0 toolkit.[148] Noise-free projections were simulated to remove noise as a confounding source of segmentation error. SBDX CT reconstruction was performed using the gFBP algorithm for the polychromatic projections and monochromatic projections at the mean spectrum energy. Following reconstruction, the left atrium was segmented for both sets of projection data and the 99th percentile segmentation error was computed to quantify the impact of beam hardening on segmentation accuracy.

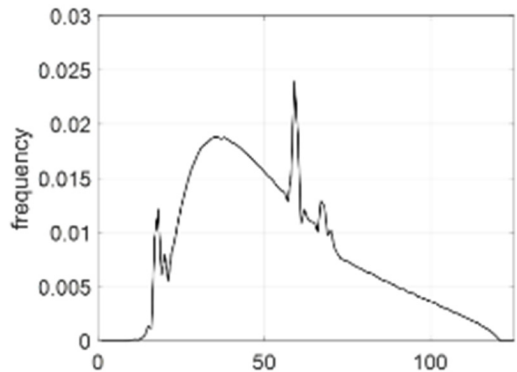
Figure 5.19: A 120 kV SBDX source spectrum.

### 5.4.2 Results

The XCAT phantom reconstructed from projections assuming a monoenergetic x-ray spectrum is shown in Figure 5.20 at left. The same phantom reconstructed from projections assuming a polyenergetic spectrum is shown in Figure 5.20 at right. The 99[th] percentile segmentation error was 0.59 mm for the reconstruction performed from monoenergetic projection data versus 0.56 mm for the reconstruction performed from polyenergetic projection data. The slight difference in segmentation accuracy between the monoenergetic and polyenergetic cases is negligible relative to the 0.30 mm dimension of the reconstructed image voxels or ~3.3 mm diameter of a typical ablation catheter tip. While large iodine blood pools may potentially cause beam hardening artifacts, the results presented here indicate the impact on segmentation accuracy would likely be small relative to other sources of artifacts such as streaks from data inconsistency caused by cardiac motion.

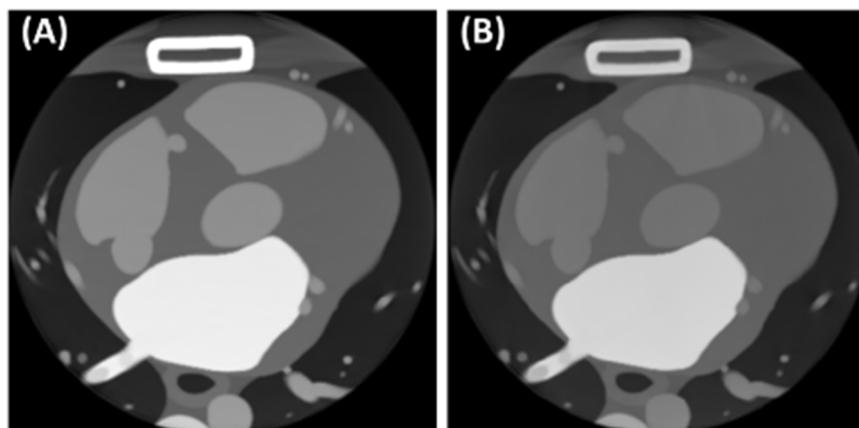

Figure 5.20: An XCAT phantom reconstructed from projections assuming a monoenergetic x-ray spectrum (A) is shown versus a reconstruction performed assuming a polyenergetic spectrum (B). [WL = 350 HU, WW = 2700 HU].

## 5.5 X-ray scatter

Scattered radiation is another potential source of CT image artifacts. Scattered photons reaching the detector contaminate the primary signal. X-ray scatter is associated with shading and streak artifacts and may cause inaccurate CT numbers and loss of contrast.[161] Scatter-to-primary ratios exceeding 120% (scatter fraction ~55%) have been reported for CBCT.[162] Scatter fraction increases with increasing patient thickness and increasing cone angles. In contrast, SBDX is a low scatter system with measured scatter fraction ranging from 2.8-7.8%, under a range of realistic imaging conditions, depending mainly on phantom thickness.[36,117] Therefore scatter is not expected to be a major source of artifacts in SBDX-CT. Nonetheless, to be complete, the potential impact of scatter on segmentation accuracy was considered for a simple numerical phantom.

### 5.5.1 Methods

A 3D numerical phantom consisting of a 4 cm diameter sphere of iodine material (comparable in size to a reference male atrium) centered within a 36 cm diameter water cylinder was used to investigate the impact of scatter on segmentation accuracy. The simple phantom was chosen for its symmetry to reduce potentially long Monte Carlo computation times that are further increased due to the large number of projection rays in SBDX CT.[163] Primary photons were computed using an analytical ray tracer developed for the SBDX geometry. A scatter signal was estimated from a 36 cm water cylinder using a Monte Carlo code (MC-GPU) adapted by *Dunkerley et al.* for the SBDX geometry and superimposed with the primary signal.[117] Scatter fractions computed with the Monte Carlo code were shown by *Dunkerley et al.* to be in good agreement with experimental values measured on the SBDX system. The iodine cylinder phantom was then reconstructed from projections with and without the scatter component using gFBP. The iodine cylinder was segmented and the segmentation accuracy was quantified versus a reference phantom using the 99$^{th}$ percentile metric outlined in section 4.2.10.2.

### 5.5.2 Results

The iodine cylinder reconstructed from projections with (left) and without (right) scatter modeled is shown in Figure 5.21. As expected, no major scatter artifacts were observed due to the low scatter property of SBDX. Segmentation accuracy was equivalent for both phantoms with 99% of the segmented iodine sphere surface points being with 0.30 mm of the ground truth.

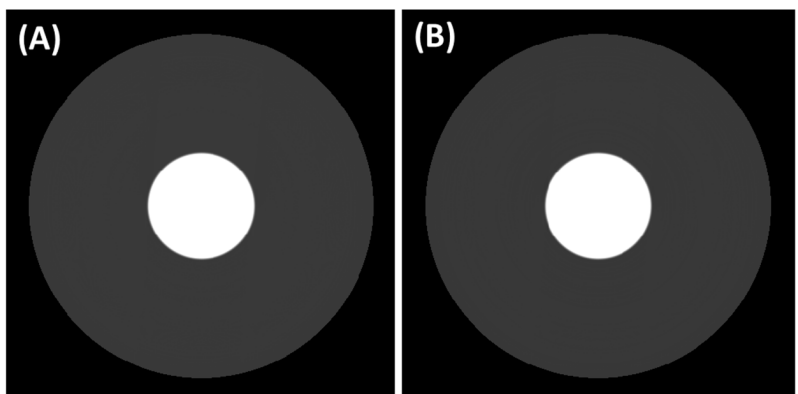

Figure 5.21: A water cylinder containing an iodine sphere was reconstructed from primary photons only (A) and primary with scatter photons included (B).

## 5.6  Summary

This chapter investigated the impact of image artifacts on IGCT image quality. Four sources of image artifacts were considered: geometric uncertainty, non-linear detector response, beam hardening, and x-ray scatter. Calibration techniques were developed for geometric uncertainty and detector non-linearity. X-ray scatter and beam hardening were shown to have a negligible effect on segmentation accuracy and not investigated further in this work.

SBDX CT reconstructions performed without geometric calibration showed severe double contour artifacts. A novel single-view geometric calibration method for inverse geometry C-arm CT was developed to mitigate artifacts due to geometric uncertainty. Single-view calibration is an important step towards enabling IGCT on flexion-prone C-arm systems. The calibration method presented in this chapter was inspired by the **P**-matrix approach used in conventional CT.[157] The SBDX composite image is analogous to a virtual cone-beam projection originating at the center of the detector array. This was exploited to estimate geometric parameters by parameterizing the SBDX system properties used to construct the projection matrix. A limitation of this approach is that the 2D detector array is reduced to a single point referred to as the virtual source point. As a result, the proposed calibration technique assumes that rotations or translations of the detector and source arrays occur in unison. Nonetheless, experimental data demonstrated the proposed method's ability to reduce artifacts caused by geometric uncertainties. Future work could investigate extending the **P**-matrix approach presented here to use a stereoscopic imaging method to estimate rotations and translations of the source and detector arrays independent of one another.[42]

The geometric calibration method was evaluated through numerical simulations and experimentally acquired projection data. For both scenarios, the proposed method reduced the image artifacts that were observed in reconstructions performed without calibration. The rRMSE was reduced from 7.7% to 0.4% for the reconstruction of a numerical thorax phantom. The sensitivity of the method to uncertainties in SDD, translations, and rotations were examined through numerical simulations. The sensitivity analysis showed that uncertainties in parameters describing translation or rotation could be estimated with average errors on the order of 0.7 mm and 0.01 degrees. The most challenging parameter to estimate accurately was the SDD. The proposed method considered only information contained in the single multi-plane composite image during calibration. Future work could investigate using the 3D information contained in the tomosynthetic plane stack[38] to reduce errors observed in the SDD parameter estimation.

A potential alternative to the technique examined here is to determine a P-matrix for each individual source and detector pair using raw IGCT projection data. Although this would provide a more complete characterization of the imaging geometry, accurate localization of fiducials in the many individual low-flux detector images is an expected challenge. An advantage of the approach pursued here is that fiducials can be easily detected and localized in the "virtual projection" composite image which is formed following shift-and-add tomosynthesis reconstruction.

The incident detector exposure was shown to vary linearly with tube output as expected. Measurements also showed a non-linear detector response at high incident fluence rates. A mapping function was derived from third-order and first-order polynomial fits to effectively linearize the SBDX detector response. Left atrium segmentation errors were within 0.06 mm for XCAT reconstructions performed from projections simulated with ideal and non-linear detector responses. The small difference in segmentation errors indicate that detector non-linearity correction may not be critical for the left atrium segmentation task. Detector non-linearity correction may however be necessary for SBDX-CT applications requiring accurate CT numbers. A limitation of the method presented in this chapter is that a single mapping function was assumed sufficient for each of the individual SBDX detector elements. An area for future work could be to extend the proposed technique to compute an independent mapping function for each detector element.

The third source of image artifacts considered was beam hardening. SBDX CT projection data were simulated for a numerical XCAT phantom assuming a monoenergetic spectrum and a 120 kV polychromatic spectrum. The XCAT phantom was reconstructed and the left atrium was segmented for both simulations to investigate the impact of beam hardening on cardiac chamber segmentation accuracy. The simulation showed only a 0.03 mm difference in segmentation accuracy between the two reconstructions. For the cardiac chamber segmentation task, beam hardening does not appear to degrade segmentation accuracy. A limitation of the beam hardening study performed in this chapter is that a single x-ray spectrum was considered at 120 kV. The severity of beam hardening artifacts is known to increase at lower energies. As a result, future work may consider further investigation of the sensitivity of segmentation accuracy to beam hardening with varying x-ray spectrums and patient sizes.

X-ray scatter was the final source of image artifacts considered in this chapter. A Monte Carlo simulation of an iodine sphere in water was performed to investigate the impacts of scatter on segmentation accuracy. Equivalent segmentation accuracy was observed for CT reconstructions performed from projections with scatter superimposed compared to scatter-free projections. The scatter study results support the assumption that scatter should not be a major detracting factor from cardiac chamber segmentation accuracy as SBDX is a low scatter system. Nonetheless, future work could perform additional scatter studies as the work here was limited to a simple two material, symmetrical, phantom. Additionally, the simulated scatter component was estimated from a single SBDX superview of a water cylinder, due to computational limitations. Future work should compute the scatter component independently at each superview using a more realistic phantom such as the XCAT.

This chapter established calibration and data conditioning techniques that are essential for implementation of CT reconstruction on an experimental SBDX system. The initial experimental work performed here used a rotating phantom stage to mimic C-arm rotation with precisely known rotation increments. The next chapter applies the proposed methods to actual SBDX C-arm rotational scans of phantoms.

# 6 SBDX CT with a rotating C-arm: Initial results

*A subset of the results reported in this chapter were presented at the 2017 AAPM Annual Meeting.*[164]

## 6.1 Introduction

Chapter 5 presented geometric calibration and data conditioning methods to enable IGCT with a rotating C-arm and Scanning Beam Digital X-ray system. Previous IGCT work performed by others, as well as the results presented in chapters 4 and 5, focused on numerical simulations and bench-top experiments with rotary stages.[45,128] Recently, the successful implementation of IGCT with a diagnostic CT gantry-based system with 8 and 32 focal spot positions was reported.[46,47] Implementation of IGCT on an interventional C-arm presents unique challenges not encountered with more stable gantry-based systems, including a flexion prone gantry, the use of 71 x 71 (5041) focal spot positions, and projection truncation from objects extending outside of the CT field-of-view.

The purpose of this chapter is to present initial inverse geometry CT results obtained by true rotational C-arm scans with the SBDX system. A method of encoding C-arm angle is developed and the single-view geometric calibration method presented in the previous chapter is validated. SBDX-CT performance is characterized in terms of gantry rotation reproducibility, computed tomography dose index, artifacts, image uniformity, and spatial resolution quantified by a modulation transfer function (MTF). Chapter 7 will present phantom studies that characterize SBDX-CT performance for the task of cardiac chamber mapping.

## 6.2 Methods

### 6.2.1 SBDX CT acquisition procedure

SBDX CT projections were acquired by performing simultaneous source scanning (fluoroscopic imaging at 15 frames per second) and C-arm rotation over a 190 degree short-scan arc. The prototype SBDX C-arm can be rotated with a nearly constant angular velocity of 14.2 degree per second, resulting in a 13.4 sec. scan time. A slight acceleration is observed near the end-points of the rotation trajectory, partially due to safeguards that limit C-arm rotation to a maximum 190 degree angular range. A complete SBDX-CT dataset consists of 201 superviews. Table 6.1 summarizes SBDX-CT acquisition parameters.

Table 6.1: SBDX CT acquistion scan parameters.

| Parameter | Value |
|-----------|-------|
| Number of superviews | 201 |
| Total rotation angle | 190$^\circ$ |
| Mean rotation velocity | 14.2$^\circ$/s |
| Total scan time | 13.4 s |

Figure 6.1 demonstrates C-arm rotation for SBDX-CT. The C-arm is mounted to a mobile platform on wheels. Four feet on the platform were lowered to the ground to stabilize the system during rotational acquisitions. The gantry was rotated to position the source array at an initial position of 100$^\circ$ left anterior oblique (LAO) prior to data acquisition using the joystick shown in Figure 6.2. The gantry position is queried from the Controller Area Network (CAN) bus and displayed. The displayed CAN Bus precision is limited to 1$^\circ$ intervals. Before each CT acquisition, the gantry starting position was manually adjusted using the joystick to align two orthogonal lasers centered on two 90 degree cross hairs of the gantry to improve reproducibility (Figure 6.3). Continuous gantry rotation then proceeds until reaching a final position of 90$^\circ$ right anterior oblique (RAO) as shown in Figure 6.1D.

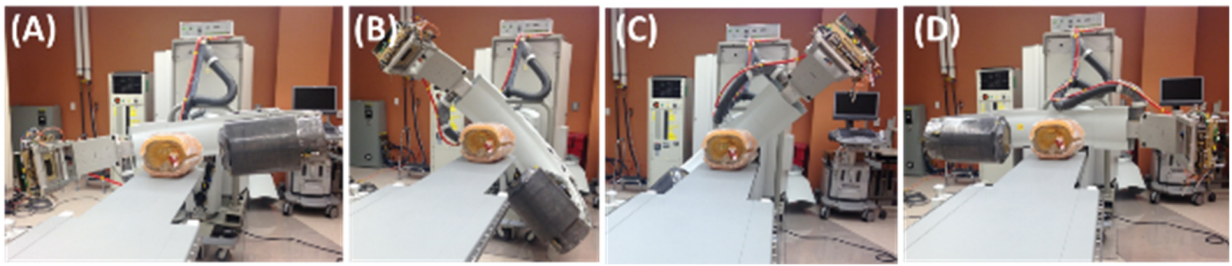

Figure 6.1: An example SBDX-CT data acquisition by C-arm rotation is portrayed. A: Prior to C-arm rotation, the SBDX source is rotated to 100$^\circ$ LAO. B-C: Gantry rotation proceeds clockwise with an average angular velocity of 14.2 degree per second. D: CT data acquisition is complete when the source array reaches the 90$^\circ$ RAO position.

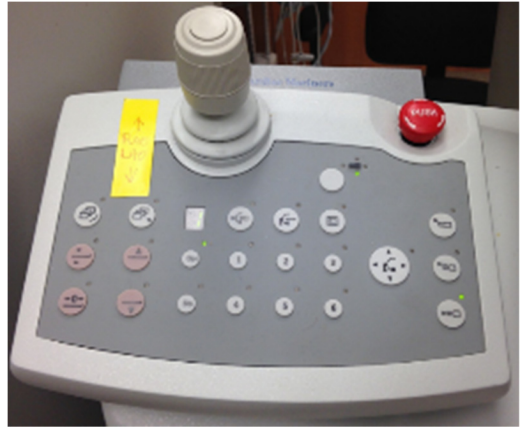

Figure 6.2: The SBDX C-arm control panel is shown. C-arm position is controlled via a joystick or buttons that constrain C-arm motion to a rotation plane.

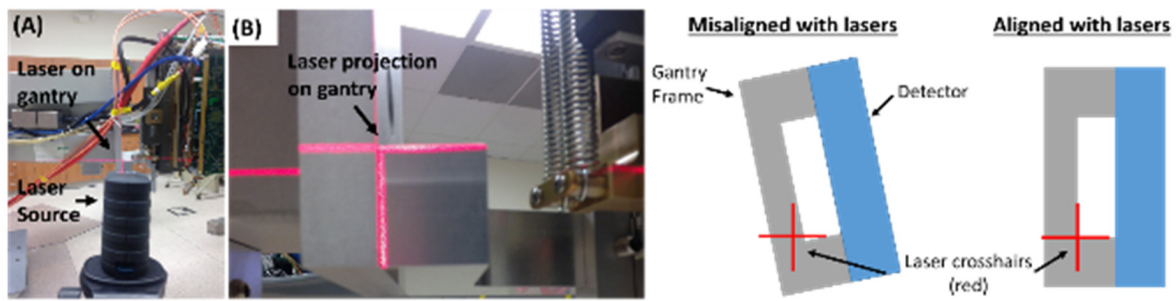

Figure 6.3: A. Two orthogonal laser beams are projected onto the steel frame of the SBDX gantry. B. The initial position of the gantry, prior to rotation, was adjusted to align the gantry crosshairs with the two laser beams to improve initial position reproducibility. C. A schematic portrays a misaligned and aligned gantry relative to the fixed laser position.

The gantry position that each IGCT superview (i.e. frame of projection data) was acquired at is required input for CT reconstruction. The SBDX C-arm was not originally developed with CT data acquisition in mind. Therefore, a method to sync the x-ray on signal with the CAN bus output of gantry angle was developed with technical support provided by Triple Ring Technologies, Inc. (Newark, CA). SBDX x-ray imaging is triggered by stepping on a foot switch. When the foot switch is pressed down for x-ray imaging, a voltage signal is triggered and read out over a USB multifunction data acquisition device (U6 series, Labjack Corporation, Lakewood, CO). The CAN bus signal containing the gantry position is simultaneously queried from an RS-232 serial connection. The x-ray on and gantry position were both queried to a single PC and recorded for input to the SBDX-CT reconstruction algorithm.

## 6.2.2 Geometric calibration

The geometric calibration technique developed in chapter 5 was employed for C-arm IGCT. The helix calibration phantom (Figure 6.4A) was placed near system iso-center. The helix phantom was inserted in a cylinder of soft tissue material (200 mm outer diameter, 104 mm inner diameter) to prevent detector saturation (Figure 6.8A). Continuous fluoroscopic imaging was performed during a 190 degree gantry rotation at 100 kV, 20 mA peak tube current. A projection matrix was derived from each of the 2D fluoroscopic images (Figure 6.4C).

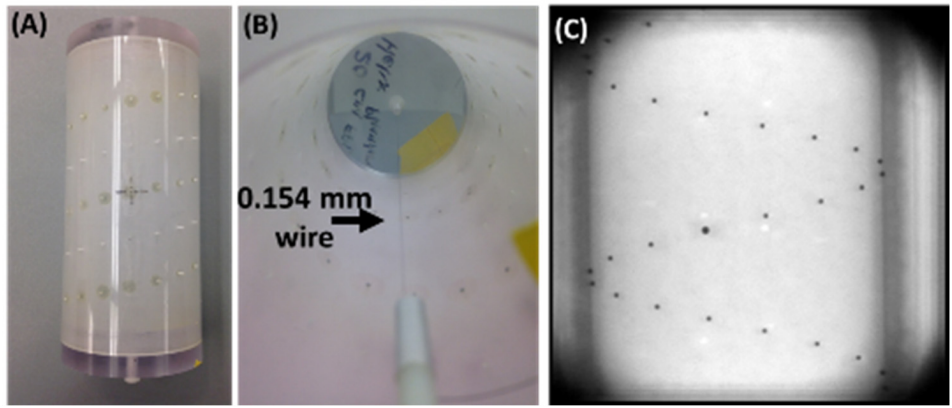

Figure 6.4: A: A helical configuration of steel fiducials embedded in a 101.6 mm diameter PMMA cylinder is used for SBDX-CT calibration. B: A 0.154 mm diameter steel wire was placed at the center of helix to measure the modulation transfer function. C: An example SBDX fluoroscopic image of the helix phantom is shown. A projection matrix is derived during the geometric calibration procedure that maps the known 3D fiducial coordinates to the 2D fluoroscopic image coordinates.

## 6.2.3 Image reconstruction

Image reconstruction was performed using gridded FBP with a Hanning windowed ramp filter as described in section 3.3. The gridded FBP algorithm was adapted to the SBDX geometry and extended for reconstruction of fully truncated projection data. A reconstruction grid of 512 x 512 x 261 pixels with pixel dimensions of 0.254 mm x 0.254 mm x 0.5 mm was used unless otherwise specified.

### 6.2.4 Gantry rotation reproducibility

Geometric calibration is performed offline prior to imaging objects or subjects. Offline calibration assumes that any geometric deviations, for example due to flex of the C-arm during gantry rotation, are reproducible from one rotation to the next. Gantry rotation reproducibility was evaluated from five SBDX CT scans of the helix phantom. The phantom was not moved between acquisitions. Single-view calibration was performed to estimate geometric parameters from each projection dataset. At each view angle, the standard deviation of each geometric parameter was computed across the five calibrations.[165] The mean value of the standard deviations was then computed over the 201 superviews to quantify reproducibility.

The sensitivity of image quality to errors in geometric parameters is difficult to evaluate from absolute deviations in system geometry parameters. For example, small errors in the source-detector-distance may have only a minor effect on image quality. Meanwhile, source and detector misalignments, such as simulated in Figure 5.4, may be catastrophic. A spatial fidelity metric was computed to further evaluate reproducibility. Each of the five sets of helix projection data were reconstructed using each of the five calibration files. In total, 25 reconstructions of the helix phantom were performed. The separation distance between the center-of-mass of each of the segmented helix fiducials was computed. Separation distances were then compared to reference values derived from the known helix geometry. The mean difference in separation distances versus the reference values was computed for each reconstruction. Mean separation differences should be similar for each combination of projection data and calibration file if gantry rotation is reproducible.

### 6.2.5 Computed Tomography Dose Index

The radiation dose associated with SBDX CT was assessed using the weighted computed tomography dose index (CTDI$_w$).[166] CTDI$_w$ is defined in equation (6.1) as the weighted sum of average dose measurements performed during CT data acquisition at the center ($D_0$) and periphery ($\overline{D_p}$) of a cylindrical PMMA (polymethyl methacrylate) phantom. The dose at the periphery, $\overline{D_p}$, is the average dose measured at four positions centered 1 cm from the cylinder edge, labeled A-D in Figure 6.5.

$$CTDI_w = \frac{1}{3}D_0 + \frac{2}{3}\overline{D_p} \qquad (6.1)$$

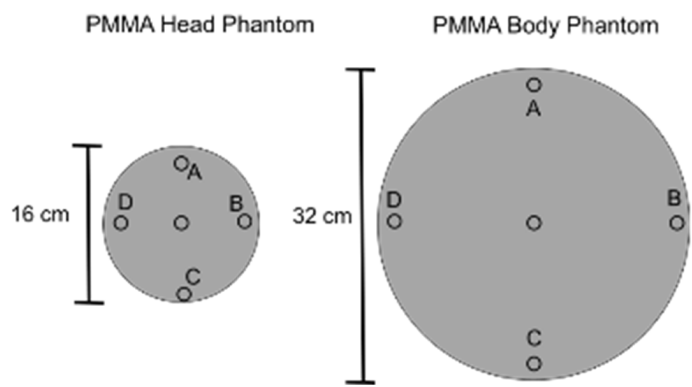

Figure 6.5: CTDI$_w$ is measured using cylindrical PMMA phantoms. A 16 cm head phantom (left) and 32 cm body phantom (right) are shown. Dose is measured at five hole positions, one located at the center of the phantom, and four located 1 cm from the cylinder edge positioned north, south, west and east (labeled A-D).

To assess the SBDX CT dose, a PMMA CTDI body phantom (32 cm diameter, 14.5 cm long) was placed on a thin piece of foam on the table at system iso-center (Figure 6.6). A pencil-type ionization chamber (Radcal Model 10X9-3CT, Monrovia, CA, USA) was placed inside each of the five insert locations one at a time (Figure 6.5). The remaining four holes were filled with PMMA plugs. SBDX-CT data acquisition was then performed as described in section 6.2.1 at 100 kV, 100 mA$_{peak}$, and the integrated exposure was recorded. The measurement was repeated three times for each of the five hole locations and the mean exposure was computed.[167] The mean exposure for each hole position was converted to dose in units of cGy with a conversion factor of 0.876 cGy/R. CTDI$_w$ was then computed according to equation (6.1). The measurement procedure was repeated for a CTDI head phantom (16 cm diameter, 15.0 cm long). The CTDI$_w$ is reported at three operating points corresponding to 20, 60, and 100 mA$_{peak}$ by scaling the measured CTDI$_w$ at 100 mA$_{peak}$ by the ratio of the operating tube current and 100 mA$_{peak}$.

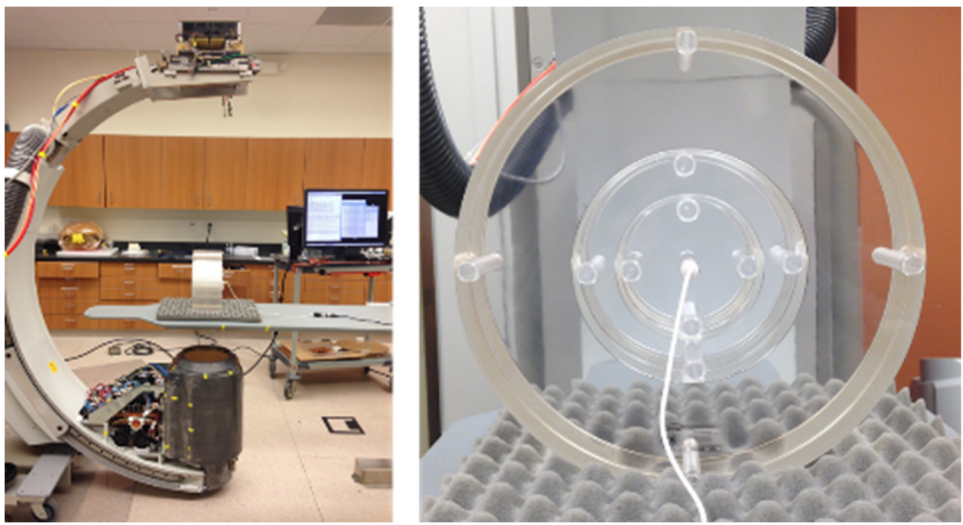

Figure 6.6: The CTDI body phantom (32 cm diameter) is shown positioned on the SBDX table near system iso-center. A close up view of the body phantom at right shows the pencil ionization chamber placed in the center hole location.

### 6.2.6 Anthropomorphic thorax phantom

An anthropomorphic thorax phantom designed to resemble a 73 kg male (ATOM adult male phantom, Model 701, CIRS Inc., Norfolk, VA) was imaged to demonstrate C-arm IGCT with SBDX (Figure 6.7). The thorax dimensions measured 23 cm (anterior/posterior) x 32 cm (medial/lateral). A 3.81 cm diameter QA insert containing cylindrical (10 HU) and spherical objects (20 HU) in lung background was placed in section #18 of the thorax phantom (Lung insert, Model 700-QA, CIRS Inc., Norfolk, VA). The diameters of the cylindrical objects measured 7, 5, 3.5, 2.5, 1.8 and 1.2 mm (three of each size). The diameter of the spherical objects measured 10, 8, 6.5, 5, 4, 3.2, 2.5 and 2 mm. The thorax phantom was placed on the SBDX table and positioned with the QA insert located near system iso-center. Imaging was performed at 100 kV and three power levels (40, 80, 120 $mA_{peak}$). Image reconstruction was performed for a 100 mm CT FOV to show the small QA objects. The results are presented qualitatively here to demonstrate IGCT with a rotating C-arm.

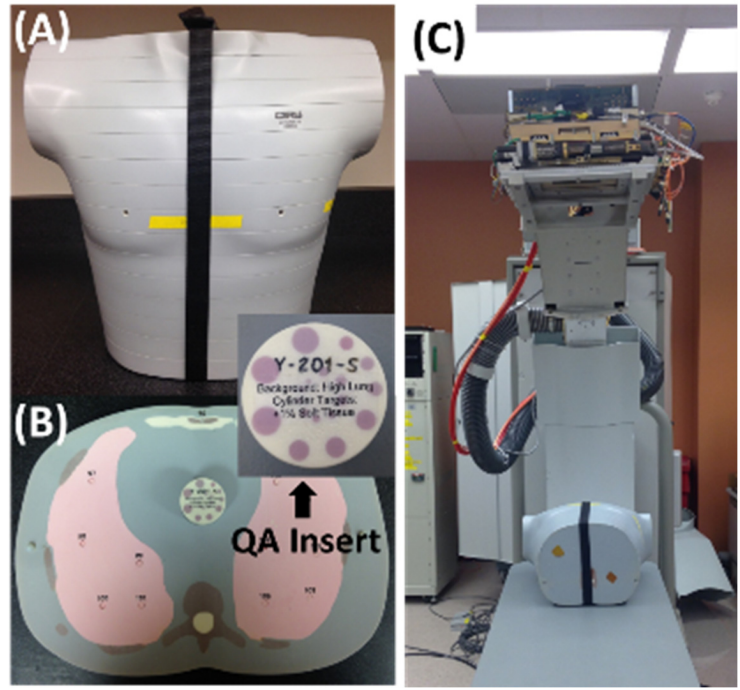

Figure 6.7: A. An anthropomorphic thorax phantom was imaged with SBDX. B. The phantom contained a QA insert with small diameter cylinders (1.2-7 mm diameter, 10 HU in lung background). C. The QA cylinder was placed near the system isocenter prior to CT imaging.

### 6.2.7   Image uniformity

A 90 mm diameter water cylinder was scanned at 100 kV, 20 mA$_{peak}$ and 40 mA$_{peak}$ to assess the uniformity of response. Two scans were completed at each operating point. The water cylinder was surrounded by a hollow soft tissue equivalent cylinder phantom (200 mm outer diameter, 104 mm inner diameter) to prevent detector saturation (Figure 6.8A). A three-dimensional reconstruction was performed to assess response uniformity. Response uniformity, $U$, was quantified as the absolute value of the difference between the mean in an ROI located at the center of the reconstructed image and the mean value of four ROIs located near the periphery.

$$U = \left| x_{center} - x_{periphery} \right|$$

(6.2)

The diameter of each ROI was 8 mm. The ROIs were positioned one diameter (8 mm) from the edge of the water cylinder phantom. The locations of the ROIs are shown in Figure 6.8B.

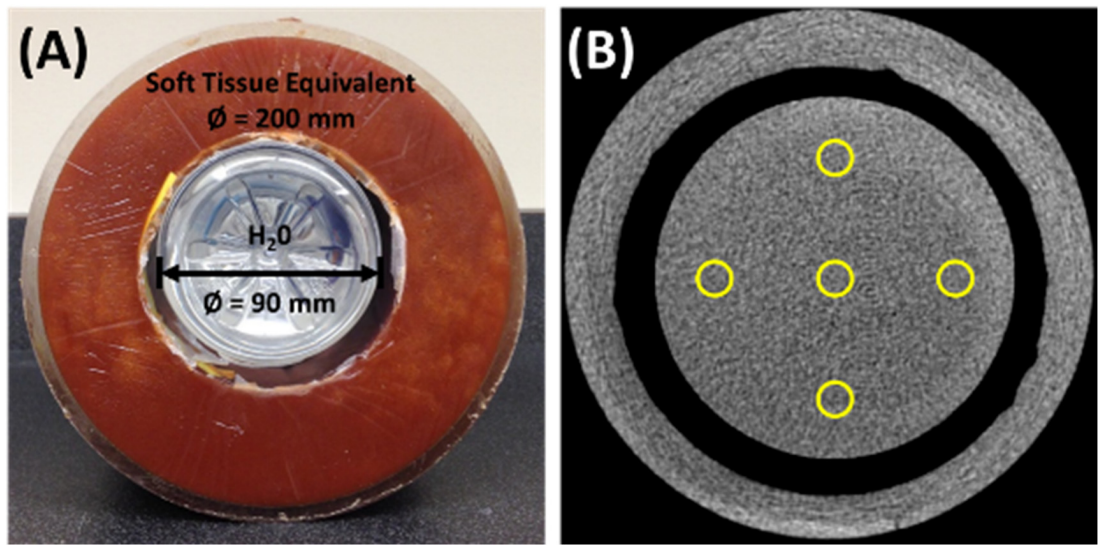

Figure 6.8: A. A 90 mm diameter water cylinder was imaged. The water cylinder was placed inside of a 200 mm diameter cylinder of soft tissue equivalent material to prevent detector saturation. B. Regions-of-interest were placed at the center and periphery of the water phantom as shown to evaluate image uniformity.

### 6.2.8 Modulation transfer function (MTF)

The helix calibration phantom contains a 0.154 mm diameter steel wire that was used to measure the system MTF (Figure 6.4B). The phantom was positioned on the patient table with the wire placed near iso-center and parallel to the axis-of-rotation of the C-arm. SBDX CT imaging of the wire was performed at 100 kV, 20 mA$_{peak}$. Five wire images were reconstructed from a single set of projections using gridded FBP with each of the calibration files to further validate gantry rotation reproducibility. A targeted reconstruction was performed with a 39.4 mm FOV. A 512 x 512 x 512 pixel reconstruction grid was used with isotropic pixel resolution set to half the diameter of the steel wire, 0.077 mm x 0.077 mm x 0.077 mm. The MTF was calculated for each wire image using the method outlined by Kayugawa et al.[158] Image slices were averaged over a 0.5 mm extent about the central image slice to reduce image noise. A Hann window was applied to reduce high frequency artifacts and noise.[168] A line-spread-function (LSF) was then determined by integrating row by row across the image columns within a 41 x 41 pixel ROI centered on the wire. An offset correction was applied to the LSF by subtracting the mean value of the three data points at each edge of the LSF. The LSF was normalized to unit area. The normalized LSF was zero padded and the MTF was computed as the magnitude of the fast Fourier transform of the LSF. The full-width-half-maximum (FWHM) of the LSF was also computed.

The gridding kernels determined in chapter 4 were optimized for the task of cardiac chamber mapping. Since these gridding kernels were not optimized to maximize spatial resolution, the MTF was also computed for a second set of gridding kernels, referred to as "sharp kernels" and summarized in Table 6.2. These kernels are used to demonstrate that spatial resolution can be improved for high-resolution imaging tasks by adjusting the gridding kernels. Image slices were averaged over a 1.0 mm extent prior to computing the MTF due to increased noise resulting from the use of narrower kernels. The LSF was computed from a 25 x 25 pixel ROI centered on the wire. Note that no attempt was made here to optimize the gridding kernels to maximize spatial resolution.

Table 6.2: Reconstruction kernels used for gridded FBP.

|  | Smooth Kernels | Sharp Kernels |
| --- | --- | --- |
| Number of ray columns (u axis) per view | 378 | 662 |
| $\Delta u$ (mm) | 0.35 | 0.2 |
| $k_{\Delta u}$ (mm) | 1.4 | 0.6 |
| Number of ray rows (v axis) per view | 301 | 601 |
| $\Delta v$ (mm) | 0.5 | 0.25 |
| $k_{\Delta v}$ (mm) | 2.0 | 1.0 |

## 6.3  Results

### 6.3.1  Gantry rotation reproducibility

The standard deviations of estimated geometric parameters across five calibration scans at each view angle, averaged over 201 view angles, are presented in Table 6.3. The standard deviation of the source-detector distance was less than 1.0 mm. Errors in the piercing point coordinates ($u_o$, $v_o$) were less than a single detector index. The standard deviation in coordinates describing the virtual source point ($x_s$, $y_s$, $z_s$) ranged from 0.23-4.9 mm. For context, variations in the virtual source point coordinates were on the order of 0.02-0.46% of the nominal virtual source point to iso-center distance (1050 mm). The small variations in geometric parameters across independent calibration files suggest gantry reproducibility may be suitable for CT data acquisition.

Table 6.3: Mean standard deviation of geometric parameters derived from five independent SBDX-CT calibrations.

| Parameter | Value | % of nominal value |
|---|---|---|
| SDD | 0.68 mm | 0.05% |
| $u_o$ | 0.01 | 0.003% |
| $v_o$ | 0.01 | 0.003% |
| $x_s$ | 4.88 mm | 0.46% |
| $y_s$ | 4.68 mm | 0.45% |
| $z_s$ | 0.23 mm | 0.02% |
| $\theta_x$ | 0.01° | -- |
| $\theta_y$ | 0.01° | -- |
| $\theta_z$ | 0.41° | -- |

A MIP image of a SBDX-CT reconstruction of the helix phantom is shown in Figure 6.9. The SBDX-CT system geometry was estimated from five calibration phantom (i.e. helix) scans. Each of the five acquisitions was then reconstructed from each calibration procedure resulting in a total of 25 reconstructed images. Table 6.4 displays the mean differences in separation distance between the segmented fiducial coordinates and reference coordinates for each combination of calibration file and projections. Differences computed using calibration files derived from the same set of projections used for reconstruction are shown in gray. The mean difference in separation distance was -0.29 ± 0.35 mm for reconstructions that utilized calibration files derived from the same set of projection data. The average difference was similar to the dimension of a single image voxel (0.254 mm). The mean difference in separation distances was also -0.29 ± 0.36 mm for reconstructions performed with independent calibration files (i.e. calibration files determined from different acquisitions than the projections) suggesting the reproducibility of gantry rotation is acceptable for CT reconstruction.

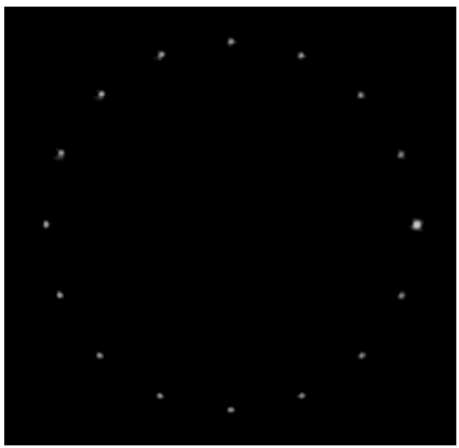

Figure 6.9: A maximum-intensity-projection image of the reconstructed helix phantom containing steel fiducials.

Table 6.4: The mean difference in separation distance between segmented fiducials and reference coordinates are shown for 25 reconstructions performed using each combination of calibration file and projection data from five SBDX-CT scans. Mean differences were similar for each combination.

| | | Acquisition Number | | | | |
|---|---|---|---|---|---|---|
| | | 1 | 2 | 3 | 4 | 5 |
| **Calibration Number** | **1** | -0.29 ± 0.35 | -0.29 ± 0.36 | -0.29 ± 0.35 | -0.29 ± 0.36 | -0.29 ± 0.35 |
| | **2** | -0.28 ± 0.36 | -0.29 ± 0.35 | -0.28 ± 0.35 | -0.28 ± 0.35 | -0.28 ± 0.35 |
| | **3** | -0.29 ± 0.36 | -0.29 ± 0.35 | -0.28 ± 0.35 | -0.29 ± 0.35 | -0.29 ± 0.36 |
| | **4** | -0.29 ± 0.36 | -0.30 ± 0.36 | -0.28 ± 0.35 | -0.29 ± 0.35 | -0.29 ± 0.36 |
| | **5** | -0.28 ± 0.36 | -0.28 ± 0.36 | -0.28 ± 0.35 | -0.28 ± 0.35 | -0.29 ± 0.36 |

### 6.3.2   Computed Tomography Dose Index

Table 6.5 presents $CTDI_w$ results for the body phantom. Table 6.6 presents results for the head phantom. For SBDX CT imaging at 100 kV, and tube currents ranging from 20-100 $mA_{peak}$, the measured $CTDI_w$ ranged from 0.40 mGy to 2.01 mGy for the body phantom. For the head phantom, $CTDI_w$ ranged from 1.18 mGy to 5.90 mGy.

Table 6.5: CTDI$_w$ for a 32 cm PMMA body phantom and SBDX CT imaging at 100 kV.

| Tube Current (mA$_{peak}$) | CTDI$_w$ (mGy) |
|---|---|
| 20 | 0.40 |
| 60 | 1.21 |
| 100 | 2.00 |

Table 6.6: CTDI$_w$ for a 16 cm PMMA head phantom and SBDX CT imaging at 100 kV.

| Tube Current (mA$_{peak}$) | CTDI$_w$ (mGy) |
|---|---|
| 20 | 1.18 |
| 60 | 3.54 |
| 100 | 5.90 |

### 6.3.3 Anthropomorphic thorax phantom

The anthropomorphic thorax phantom was successfully reconstructed at 40, 80 and 120 mA peak tube current as shown in Figure 6.10 (panels B-D. Figure 6.10A shows a reconstruction performed without geometric calibration. Severe artifacts are present and the small cylindrical objects within the QA insert cannot be resolved. Application of the developed single-view calibration technique eliminated the double contour artifacts (Figure 6.10B-D). Each of the 1.2-7 mm diameter cylindrical objects was resolved at each of the three noise levels considered.

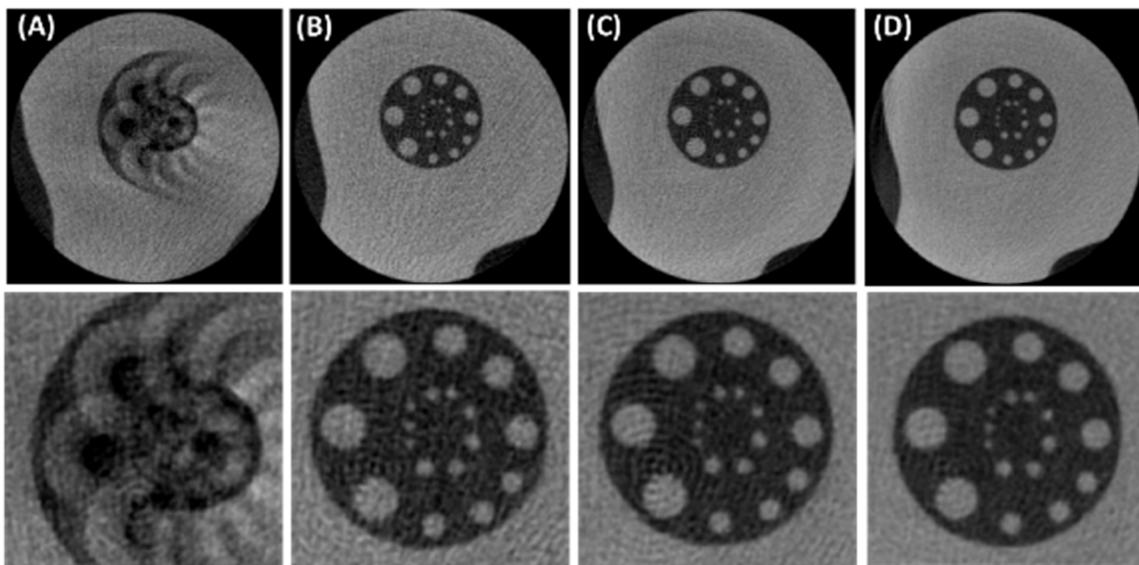

Figure 6.10: (A). Inverse geometry C-arm CT without gantry calibration shows severe image artifacts. Single-view geometric calibration reduced artifacts for imaging at (B) 40 mA$_{peak}$, (C) 80 mA$_{peak}$, and (D) 120 mA$_{peak}$. The corresponding zoomed images (bottom row) show cylindrical objects in groups of three with diameters of 7 mm, 5 mm, 3.5 mm, 2.5 mm, 1.8 mm, and 1.2 mm.[WL = 200, WW = 1500]

### 6.3.4   Image uniformity

Figure 6.11 presents reconstructions of two scans of the water cylinder at 20 mA$_{peak}$. Figure 6.12 presents reconstructions at 40 mA$_{peak}$ (the water cylinder is centrally located). Horizontal profiles through the center of phantom are shown for each of the reconstructions. Image uniformity measured from regions-of-interest at the periphery and center of the image averaged 3 HU for the 20 mA$_{peak}$ reconstructions. Image uniformity averaged 6 HU for the 40 mA$_{peak}$ reconstructions.

No severe artifacts were observed in the water images. A slight bright spot was observed at the periphery (upper right quadrant) of the water image, relative to the center region. The artifact is currently under investigation but may be due to field-of-view truncation caused by the patient couch.[169]

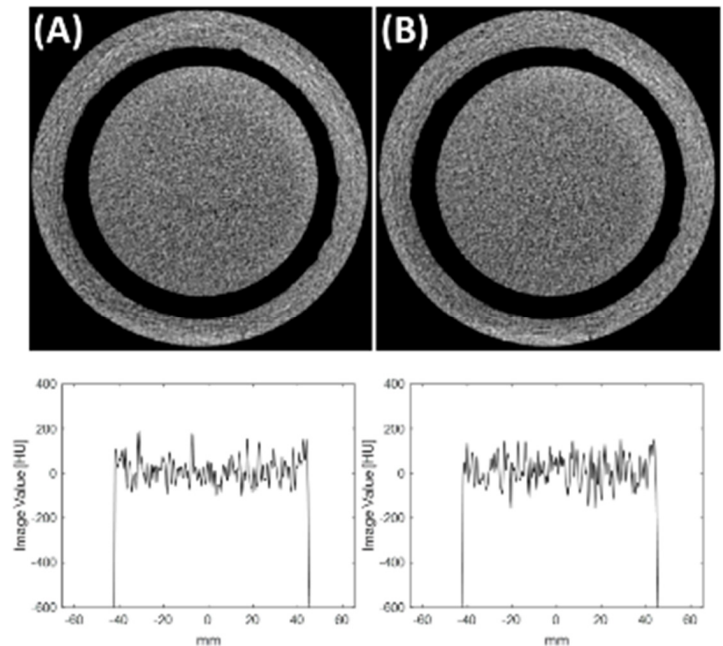

Figure 6.11: Two scans and reconstructions of a 90 mm water cylinder are shown for imaging at 100 kV, 20 mApeak. Horizontal profiles are shown through the center of the image. [WL = 40, WW = 600]

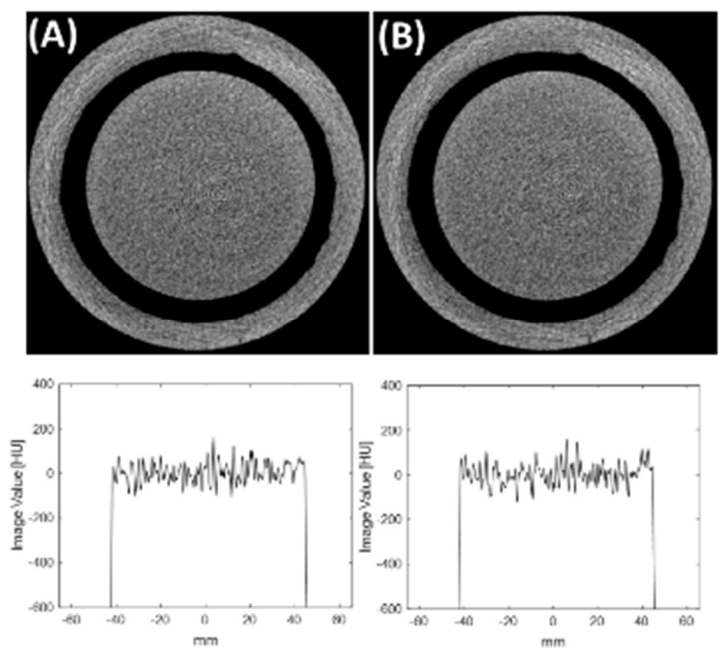

Figure 6.12: Two scans and reconstructions of a 90 mm water cylinder are shown for imaging at 100 kV, 40 mA$_{peak}$. Horizontal profiles are shown through the center of the image. [WL = 40, WW = 600]

### 6.3.5   Modulation transfer function

The modulation transfer function is plotted in Figure 6.13. Error bars represent +/- one standard deviation computed across five reconstructions. Limiting resolution at 10% was 11.1 cm$^{-1}$ for the smooth kernels averaged over five reconstructions. Limiting resolution at 10% improved to 17.0 cm$^{-1}$ for the sharp kernels averaged over five reconstructions. The MTF measured 6.6 cm$^{-1}$ and 9.7 cm$^{-1}$ at 50% for the smooth and sharp kernels, respectively. The average FWHM of the derived LSF was 0.73 mm for smooth kernel reconstructions and 0.48 mm for sharp kernel reconstructions.

Table 6.7 presents 50% and 10% MTF values for each of the ten reconstructions. Resolution at 10% MTF ranged from 11.0-11.2 cm$^{-1}$ for the smooth kernels. The narrow range in MTF values, for reconstructions performed from five independent calibrations, suggests gantry reproducibility is acceptable for C-arm IGCT imaging. Slight differences in MTF values may be due to image noise. Limiting resolution at 10% ranged from 16.8-17.5 cm$^{-1}$ for the sharp kernel reconstructions. The increased range compared to the smooth kernels may be due to increased image noise resulting from the use of narrower gridding kernels

(Table 6.2).

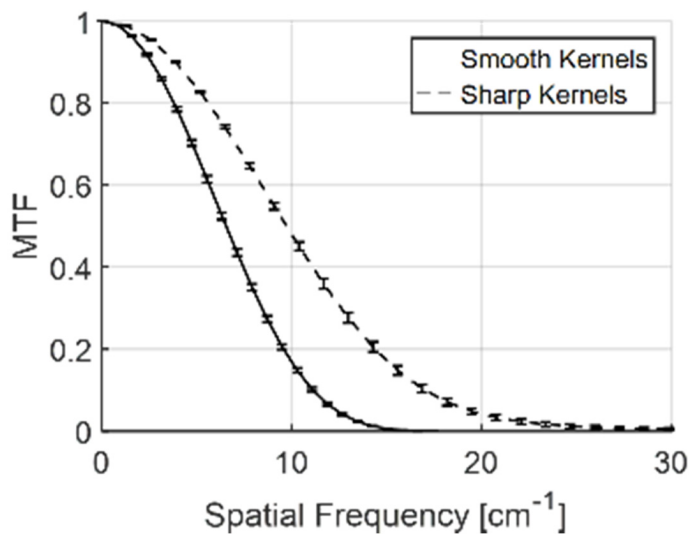

Figure 6.13: The MTF measured from a 0.154 mm diameter steel wire is shown for reconstruction parameters optimized for a segmentation task (11.1 cm$^{-1}$ at 10%, smooth kernels) and adjusted for higher resolution imaging tasks (17.0 cm$^{-1}$ at 10%, sharp kernels).

Table 6.7: MTF and FWHM values are summarized for five SBDX-CT reconstructions of a wire phantom performed with two sets of reconstruction parameters.

| | **Calibration Number** | | | | | | |
| | 1 | 2 | 3 | 4 | 5 | Mean Value | Std. Dev. |
|---|---|---|---|---|---|---|---|
| **Smooth Kernels** | | | | | | | |
| LSF FWHM (mm) | 0.74 | 0.73 | 0.73 | 0.73 | 0.74 | 0.73 | 0.01 |
| 50% MTF (cm$^{-1}$) | 6.5 | 6.6 | 6.5 | 6.7 | 6.5 | 6.6 | 0.08 |
| 10% MTF (cm$^{-1}$) | 11.1 | 11.2 | 11.1 | 11.2 | 11.0 | 11.1 | 0.09 |
| **Sharp Kernels** | | | | | | | |
| LSF FWHM (mm) | 0.47 | 0.49 | 0.49 | 0.47 | 0.49 | 0.48 | 0.01 |
| 50% MTF (cm$^{-1}$) | 9.7 | 9.8 | 9.7 | 9.9 | 9.6 | 9.7 | 0.13 |
| 10% MTF (cm$^{-1}$) | 17.3 | 16.8 | 16.8 | 17.5 | 16.8 | 17.0 | 0.35 |

## 6.4 Summary

Inverse geometry CT imaging on a rotating C-arm gantry (SBDX-CT) was successfully demonstrated. The single-view calibration method that was initially evaluated with rotary stage data was successfully deployed to calibrate the flexion prone SBDX gantry. The reproducibility of SBDX gantry rotation was shown to be sufficient for CT data acquisition and reconstruction. The SBDX gantry position is output and displayed from a CAN bus with $1°$ precision. To improve the reproducibility of the gantry starting position, orthogonal lasers were utilized to manually adjust the gantry to the same starting location prior to C-arm rotation for CT data acquisition. Future work could investigate automated gantry control and feedback mechanisms to improve precision and eliminate manual gantry position adjustments. Additionally, a clinical SBDX system would be mounted to the floor and likely be more stable than the mobile platform utilized for the prototype system here.

Radiation dose is an ongoing concern in medical imaging and should be kept "As Low As Reasonably Achievable" while still providing sufficient image quality for task completion. Weighted CTDI was measured using body and head phantoms to quantify the dose associated with SBDX-CT. For the body phantom, $CTDI_w$ ranged from 0.40 mGy to 2.01 mGy for imaging at 100 kV and tube currents ranging from 20-100 $mA_{peak}$. It's important to acknowledge the limitations of using $CTDI_w$ to quantify SBDX-CT dose. First, $CTDI_w$ is not a measurement of effective dose. The risk of biological effects from radiation exposure varies by organ and tissue type.[170] Future work could consider estimating the effective dose from SBDX CT using a MC-GPU Monte Carlo code that has been adapted for the SBDX geometry.[117] A second limitation of $CTDI_w$ is that it was originally proposed to measure the dose for single (narrow) axial slice imaging. Cone-beam CT has evolved to image large volumetric regions that cover full organs. As a result, CTDI has been criticized for use in CBCT as the 100 mm long ionization chamber used may underestimate the dose outside of the 100-mm long active region.[171] A third limitation is that the weighting coefficients 1/3 and 2/3 used to compute $CTDI_w$ were not validated for IGCT. Nevertheless, $CTDI_w$ was chosen as the metric used to quantify SBDX-CT dose as $CTDI_w$ measurements can be easily repeated by others with equipment and phantoms that are readily available.

Spatial resolution was quantified by the modulation transfer function. Spatial resolution achieved with the rotating C-arm was comparable to that attained with a rotary stage. The MTF value at 10% was 11.1 $cm^{-1}$ for the rotating gantry versus 10.3 $cm^{-1}$ for the rotary stage, with reconstruction parameters optimized for a high-contrast segmentation task. The MTF value at 10% indicates that SBDX spatial resolution is likely sufficient for the task of cardiac chamber mapping. The measured 1.1 cycle per mm MTF at 10% is similar to resolving 1.1 line pair per mm which suggests a 0.55 mm object, smaller than the diameter of a typical ablation catheter tip, could be resolved. Improved resolution is available if necessary and demonstrated by a MTF value at 10% of 17.0 cm$^{-1}$ using the sharp gridding kernels.

The results presented in this chapter represent the first demonstration of inverse geometry CT with a rotating C-arm. SBDX-CT using the gFBP reconstruction algorithm and truncation correction was investigated for static phantoms. Imaging performance was characterized in terms of gantry rotation reproducibility, radiation dose, artifacts due to geometric uncertainty, image uniformity, and spatial resolution. The following chapter evaluates SBDX-CT performance for the target task of cardiac chamber mapping during interventional procedures.

# 7 Evaluation of cardiac chamber roadmap accuracy from Scanning Beam Digital X-ray CT

## 7.1 Introduction

The overall goal of this work is to develop a CT capability for the Scanning-Beam Digital X-ray system that enables 3D imaging of cardiac chambers in an interventional procedure. Segmented cardiac chamber roadmaps could be displayed with real-time SBDX 3D catheter tracking to guide complex structural heart interventions or electrophysiology procedures. The early chapters of this dissertation presented methods for SBDX CT image reconstruction. Techniques were developed to accommodate a C-arm short scan acquisition, field-of-view truncation, and data inconsistency introduced by cardiac motion, and the non-ideal geometry of an actual rotating C-arm. The purpose of this chapter was to evaluate SBDX CT performance for the target task of cardiac chamber mapping.

Phantom studies were performed to investigate the impacts of several factors on cardiac chamber map accuracy. The effects of noise, motion frequency, and field-of-view truncation were each examined. Finally, the results of a porcine animal study are presented to demonstrate SBDX CT imaging for a large subject with realistic anatomy and cardiac motion. Previous work in IGCT has focused on small animal studies, including rats and rabbits, due to limited source output, without consideration of cardiac motion.[47] The demonstration of *in vivo* IGCT in a large animal, with realistic cardiac motion, is an important milestone towards the clinical implementation of IGCT and SBDX CT.

## 7.2 Methods

### 7.2.1 Validation studies

A total of four phantom studies and one animal study were performed to evaluate SBDX CT cardiac chamber roadmap accuracy. Table 7.1 summarizes the experimental conditions and data acquisition method for each of the studies. Each experiment is described in detail in the following sections.

Table 7.1: Summary of SBDX-CT validation studies.

| Study Name | Motion | FOV Truncation | Data Acquisition |
|---|---|---|---|
| Static non-truncated atrium | Static (0 BPM) | None | Rotary stage |
| Dynamic non-truncated atrium | Dynamic (60-90 BPM) | None | Rotary stage |
| Static truncated atrium in thorax | Static (0 BPM) | Full | C-arm rotation |
| Dynamic truncated atrium in thorax | Dynamic (60-88.2 BPM) | Full | C-arm rotation |
| Animal | Dynamic (72–75 BPM) | Full | C-arm rotation |

### 7.2.1.1  *Static non-truncated atrium phantom*

The first experiment performed focused on a stationary custom-made atrium phantom placed on a rotating stage and designed to fit entirely within the SBDX-CT FOV. The purpose of the static non-truncated atrium phantom study was to assess SBDX CT cardiac chamber segmentation accuracy without object motion, FOV truncation, or geometric uncertainty induced by C-arm deflection present. In this regard, the first phantom study was designed to provide reference SBDX CT images obtained under ideal imaging conditions that more complicated experiments with programmed motion and FOV truncation could be compared against.

The atrium phantom shown in Figure 7.1A was filled with 57 ml of iodine contrast agent (Omnipaque 350, GE Healthcare, Waukesha, WI, USA) diluted with deionized water to 5% by volume. The atrium was then placed in a 73 mm diameter cylinder container (125 mm height) of deionized water. Eight 1/16" diameter spherical aluminum fiducials were attached to the outside of the water cylinder in a helical configuration. The aluminum fiducials provide point-like high contrast objects that may be used to assess reconstructed CT images for artifacts. Furthermore, the aluminum fiducials were employed to register the SBDX CT images of the atrium with a reference CT scan, discussed further in section 7.2.4 below.

As the purpose of the first experiment was to establish reference SBDX CT images without errors due to confounding factors such as geometric uncertainty, the SBDX gantry was rotated 90 degrees to a lateral angulation and a rotary stage was used for CT data acquisition. The atrium phantom was placed at system iso-center on a motorized linear stage (Model MOX-02-50, Optics Focus Instruments Co., Beijing,

China) controlled by a motion controller (MOC-01-4-110, Optics Focus Instruments Co., Beijing, China) as shown in Figure 7.1B. The linear stage was then placed on top of a motorized rotary stage (Model M-495CC, Newport Corp., Evry, France) controlled by a second motion controller (Model ESP 300, Newport Corp., Irvine, CA, USA). Both motion stages were controlled simultaneously via serial commands sent over RS-232 from a Windows PC using custom-made software implemented in the C++ programming language. To simulate typical beam hardening encountered during patient imaging, acrylic with a thickness of 6.99 cm and copper with a thickness of 0.5 mm was placed in the beam immediately adjacent to the source array.

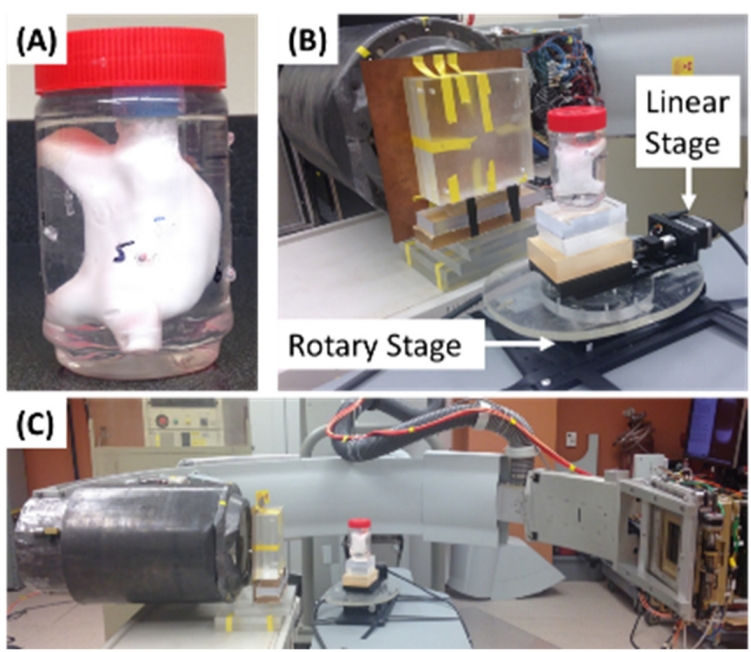

Figure 7.1: A. The custom-made atrium phantom containing iodine contrast agent is shown. B. The atrium phantom was placed on a linear stage with programmed in-plane motion trajectories. C. The SBDX gantry was rotated to a horizontal position and the linear stage was placed on a rotary stage for SBDX CT data acquisition.

For CT data acquisition, the atrium phantom was imaged using a continuous stage rotation with an angular velocity of 2.0 degree per second over a 188 degree short-scan range. A step-and-shoot technique was used by selecting every 7[th] frame of the complete projection dataset resulting in 201 superviews with an angular increment of 0.93 degree. The rotating stage acquisition parameters were selected to approximately match the SBDX CT rotating gantry acquisition protocol outlined in Table 6.1. The motorized linear stage remained stationary during data acquisition for the static non-truncated atrium phantom experiment.

SBDX CT imaging was performed at 100 kV tube potential and three tube currents, 10 mA$_{peak}$, 20 mA$_{peak}$, and 40 mA$_{peak}$. Imaging was performed using the 71 x 71, 15 frame/sec scanning technique for all experiments performed in this chapter. The single-view calibration procedure described in chapter 5 was performed for rotary stage calibration prior to CT data acquisition.

### 7.2.1.2 Dynamic non-truncated atrium phantom

The second phantom study performed investigated the impact of object motion on cardiac chamber segmentation accuracy. The dynamic non-truncated atrium phantom experiment utilized the same set-up as the static non-truncated atrium phantom study (Figure 7.1). As a result, the effects of projection undersampling due to cardiac-gating and high-frequency motion on segmentation accuracy could be investigated independently of FOV truncation and geometric uncertainty.

The motorized linear stage was programmed to move the atrium phantom over a cyclical 8 mm peak-to-peak (16 mm total translation distance per cycle) linear trajectory with a static phase corresponding to 20% of the period length as shown in Figure 7.2. The motion trajectory was programmed in the same plane as the direction of stage rotation, in a pattern similar to the methods used in Ref. [29]. The period of linear motion was determined by computing the number of full heart beats that would occur during a 13.4 second SBDX CT rotational acquisition assuming cardiac frequencies of 60, 75, and 90 beats per minute. The assumed cardiac frequencies correspond to periods of 1 s, 0.8 s, and 0.667 seconds resulting in 13.4, 16.8, and 20.1 heartbeats per scan, respectively. The programmed linear motion period was then set to provide the same number of complete cycles (i.e. heart beats) per rotary stage CT acquisition as would be expected for a C-arm rotational scan at frequencies of 60, 75, and 90 beats per minute.

SBDX CT data acquisition of the atrium phantom was performed with simultaneous linear stage motion and rotation following the same methodology used for the static non-truncated atrium experiment. Imaging was performed at a tube potential of 100 kV and peak tube currents of 10 mA$_{peak}$, 20 mA$_{peak}$, and 40 mA$_{peak}$ for each of the three frequencies. Nine rotary stage CT acquisitions were performed in total for the dynamic non-truncated atrium phantom.

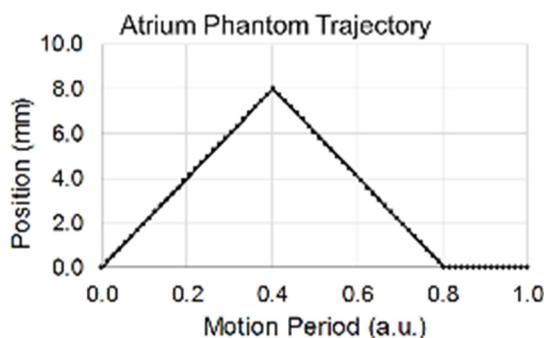

Figure 7.2: The programmed atrium motion trajectory is shown for an arbitrary period length. The atrium was translated 8 mm from its' starting position with a 20% stationary phase. The period length is defined in seconds for frequencies ranging from 60-90 cycles per minute.

### 7.2.1.3   *Static truncated atrium in thorax phantom*

The purpose of the third phantom study was to evaluate SBDX CT segmentation accuracy for fully truncated projection data obtained with C-arm rotational scans. The left atrium phantom was secured to a piece of acrylic and attached to the Washington University 4D Motion Phantom.[172]  The left atrium was then inserted inside of an anthropomorphic thorax phantom with chest plates (Chest Plates 41337-010, Kyoto Kagaku, Kyoto, Japan) attached to represent a larger patient as shown in Figure 7.3. The anthropomorphic thorax phantom measured 30 cm in the anteroposterior (AP) dimension and 38 cm in the mediolateral (ML) dimension, including the chest plates, at the plane of the atrium phantom, resulting in full FOV truncation. The total tissue thickness measured 15.6 cm in the AP dimension and 11.8 cm in the ML dimension. The table position was adjusted to place the atrium phantom at system isocenter. A solidstate dose meter was used to measure x-ray transmission at 100 kV tube potential through the iso-centered phantom containing the atrium model. Transmission was 1.08% in the AP dimension and 0.50% in the ML dimension. A transmission to mass conversion equation provided by *Ogden et al.* indicates the phantom resembles a typical 233 lb (105.7 kg) male patient.[173]

SBDX CT imaging of the static truncated atrium in thorax phantom was performed following the acquisition procedure outlined in section 6.2.1. SBDX CT projections were acquired by C-arm rotation over a 190 degree arc resulting in 201 superviews per CT dataset. Imaging was performed at a tube potential of

100 kV and three fluence levels corresponding to 20 mA$_{peak}$, 60 mA$_{peak}$, and 100 mA$_{peak}$.

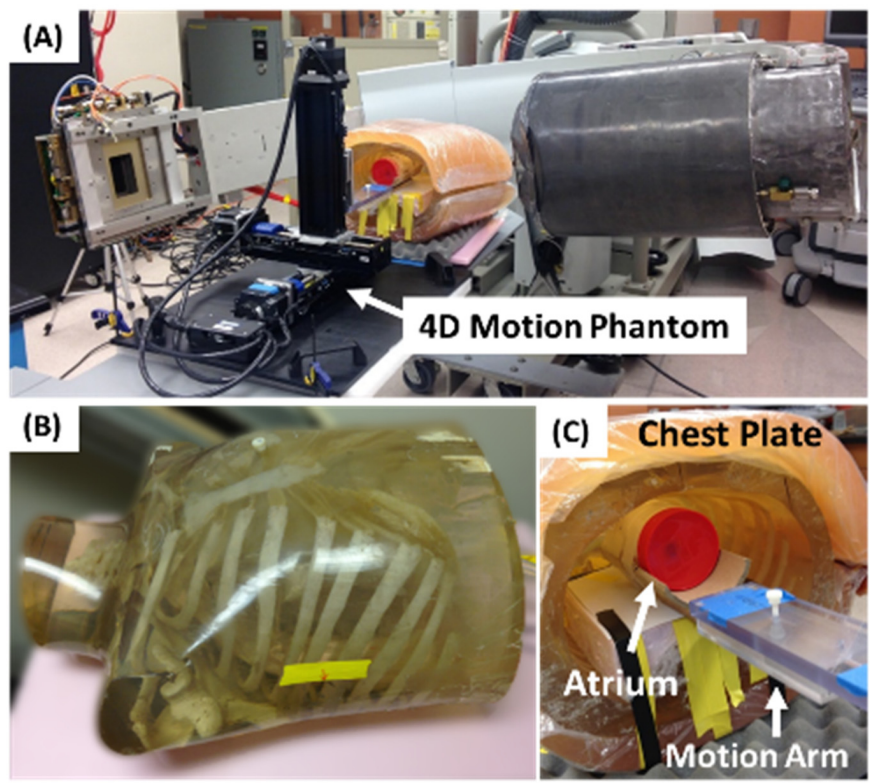

Figure 7.3: A. The atrium phantom attached to a motion phantom and inserted within a thorax phantom is shown for the rotating C-arm experiments. B. The anthropomorphic thorax phantom is shown without the chest plates to demonstrate the detailed bony anatomy. C. A close-up view shows the chest plates placed over the thorax containing the atrium phantom. The atrium phantom is attached to the motion phantom with an acrylic arm. Foam placed below the atrium provides additional support that limits irreproducible motion due to vibrations.

#### 7.2.1.4 *Dynamic truncated atrium in thorax phantom*

The fourth and final phantom study performed evaluated C-arm rotational scans of the atrium phantom undergoing two-dimensional motion within the anthropomorphic thorax phantom. The purpose of the fourth phantom study was to evaluate the accuracy of atrium segmentations from SBDX-CT images reconstructed from projections that were i) fully truncated, ii) undersampled due to retrospective gating, iii) contaminated with data inconsistency resulting from high-frequency object motion, and iv) acquired by SBDX C-arm rotational acquisition. Additionally, intra-frame motion not accounted for in the slowly rotating rotary stage experiments is considered in C-arm rotational acquisitions.

The set-up shown in Figure 7.3 used for the static phantom experiment was also used for the dynamic truncated atrium in thorax phantom experiments. Atrium phantom motion trajectories were programmed in two-dimensions using the Washington University 4D Motion Phantom.[172] Similar to the dynamic atrium experiments performed with the rotary stage, peak-to-peak motion of 7.3 mm to 7.5 mm (i.e. 14.5 mm – 15.0 mm total distance per cycle) was programmed with a 20% stationary phase within the motion period as shown in Figure 7.4. The same motion trajectory was programmed simultaneously within the coronal and axial planes resulting in a 2D elliptical motion pattern. The motion pattern was implemented at frequencies of 60, 75, and 88.2 cycles per minute. The slight differences in the atrium translation (14.5-15.0 mm) and frequency (88.2 cycles per minute instead of 90 cycles per minute) were due to hardware constraints imposed by the motion phantom's rotary stepper motors.

SBDX CT data acquisition was performed using the same procedure outlined for the static truncated atrium phantom. Imaging was performed at a tube potential of 100 kV and three fluence levels corresponding to 20 mA$_{peak}$, 60 mA$_{peak}$, and 100 mA$_{peak}$, for each of the three programmed motion frequencies. A total of nine SBDX CT acquisitions were performed to investigate the effects of varying noise levels and motion frequency on segmentation accuracy.

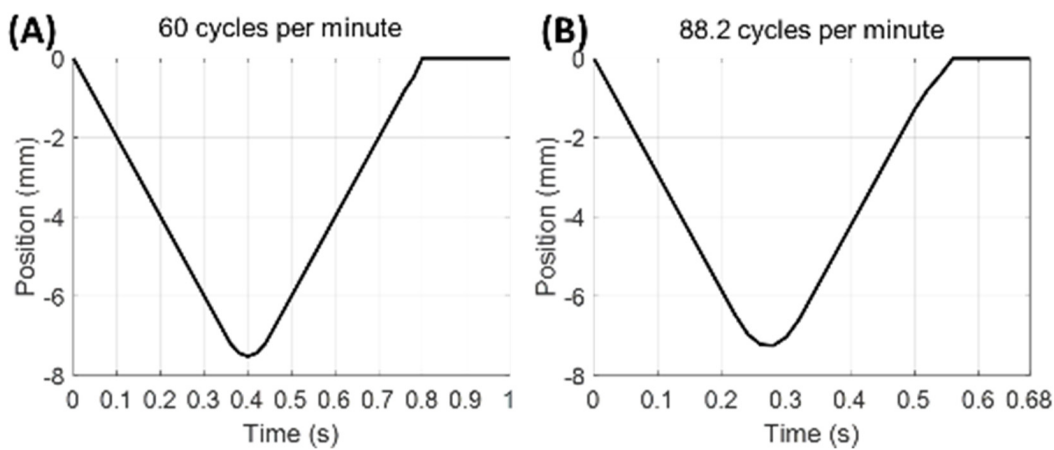

Figure 7.4: Example motion periods are shown for the atrium in thorax phantom at 60 cycles per minute (A) and 88.2 cycles per minute (B). The motion trajectory was programmed in two orthogonal planes resulting in an elliptical motion profile at realistic cardiac frequencies.

### 7.2.1.5  Animal study

The purpose of the four phantom studies described was to evaluate SBDX CT image quality and segmentation accuracy for a range of motion frequencies and noise levels in a well-controlled manner. Ultimately, clinical implementation will require extensive animal studies to demonstrate SBDX CT performance for a more realistic anatomy and cardiac motion. Therefore, an *in vivo* porcine study was performed to demonstrate SBDX CT in a large animal model.

A 55 kg swine measuring 255 mm (AP) by 300 mm (ML) was placed on the SBDX table with the animal's heart located approximately at the SBDX system isocenter. The animal's heart rate was recorded at 75 beats per minute immediately prior to SBDX CT data acquisition. The heart rate measured 72 beats per minute at the end of SBDX CT data acquisition. SBDX CT projections were acquired at 100 kV, 80 $mA_{peak}$, over a 13.4 s C-arm rotation period with simultaneous contrast injection (370 mgI/ml). To simulate a breath hold, mechanical ventilation was suspended during CT data acquisition. Contrast agent (80 ml) was administered manually (i.e. without use of a power injector) through an ear vein over a 40 second period. A delay prior to C-arm rotation allowed the contrast agent to reach the animal's heart before initiation of CT data acquisition. Because a power injector was not available, the manual contrast injection protocol required switching to a second syringe mid-data acquisition. This injection discontinuity presents a potential limitation that may have impacted the contrast dynamics and mixing within the cardiac chambers.

SBDX CT image quality was assessed qualitatively versus a reference CT image obtained on a commercial CT scanner (Discovery CT750 HD, GE Healthcare, Waukesha, WI, USA) immediately prior to the SBDX CT imaging study. The animal was scanned with a venous contrast injection at 100 kV and 182.4 mAs. Image reconstruction was performed using the scanner's filtered backprojection algorithm from projection data spanning a 0.228 s temporal window.

### 7.2.2 Intrinsic projection gating

In the interventional setting, phase correlated CT reconstruction has been proposed to reduce artifacts that may occur from cardiac motion induced data inconsistency, to improve procedure planning, or for functional evaluation including cardiac chamber volume measurements. Retrospective projection gating was performed as input to the SBDX CT reconstruction algorithm for the dynamic phantom experiments and animal study.

For the rotating stage experiments, the programmed linear stage motion was synced with the x-ray on signal using a USB multifunction data acquisition device (U6 series, Labjack Corporation, Lakewood, CO). When the hand switch was pressed down to initiate x-ray imaging, linear stage motion was automatically triggered. The acquired SBDX CT projections were then retrospectively gated by selecting a single SBDX superview acquired during the static portion of the programmed motion profile from each motion cycle for the gated dataset. The total number of superviews in each gated projection dataset were 13, 16, and 20 for the 60, 75, and 90 cycles per minute frequencies respectively. The gated datasets corresponding to higher motion frequencies consisted of more superviews as the number of motion cycles completed within the constant CT data acquisition timeframe is larger.

The x-ray on signal was not synced with the 2D motion programmed for the rotating gantry experiments as the phantom was controlled with proprietary software. As a result, an intrinsic projection gating technique, motivated by the kymogram detection method,[139] was developed and used for the rotating gantry phantom and animal studies. The kymogram method provides a cardiac motion signal that serves as a motion surrogate when an ECG signal is not available or potentially corrupted.

Here, the motion signal was derived from the mean image value computed in an ROI located in the SBDX composite image (Figure 7.5) at each gantry view angle using a series of signal processing steps. Plotting the mean composite image value in the ROI as a function of gantry position showed a bias in the corresponding motion signal (Figure 7.6A). The bias is introduced by the edge of the x-ray table which slowly translates across the SBDX composite image as the x-ray source is rotated under the table. An unbiased signal is obtained by convolving the raw, biased, motion signal with a 1.0 s wide rectangular filter and subtracting the result from the biased signal (Figure 7.6B). Finally, a bandpass filter is applied to remove extremely low (i.e. less than 40 cycles per minute) or high (i.e. greater than 150 cycles per minute) frequencies. An estimate of the motion frequency derived from the peak-to-peak spacing was in exact agreement with the programmed motion frequency of 88.2 cycles per minute.

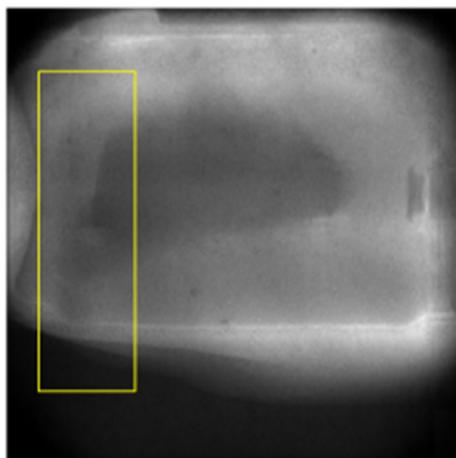

Figure 7.5: A motion signal for projection gating was derived from the mean composite image value computed within the ROI shown in yellow.

SBDX CT projection data were retrospectively gated by selecting superviews with a constant offset from the intrinsic gating signal peak positions corresponding to the static phase of the programmed motion profile (Figure 7.4). The total number of superviews in each gated projection dataset were 14, 16, and 20 for the 60, 75, and 88.2 cycles per minute frequencies, respectively. Lower motion frequencies are associated with greater undersampling.

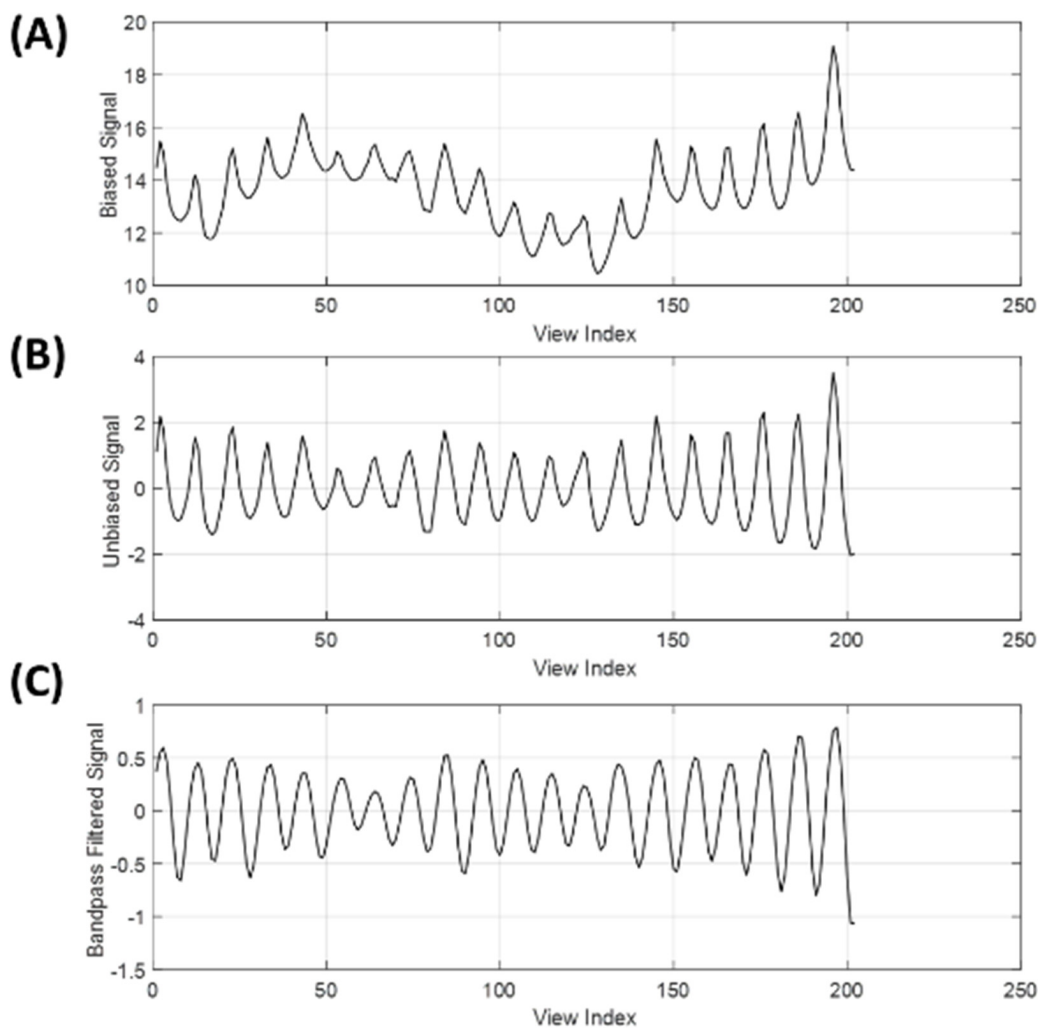

Figure 7.6: (A) An example motion signal derived from the 88.2 cycles per minute phantom study is shown. (B) An unbiased signal is computed by convolving a 1.0 s rectangular function with the biased signal and subtracting the result from the biased signal. (C) A bandpass filter is applied to remove potential signals at extremely low (40 cycles per minute) or high (150 cycles per minute) frequencies.

### 7.2.3   Image reconstruction

For each of the validation studies performed, the SBDX CT projection data were reconstructed using gridded FBP as described in section 3.3.1 with the complete set (i.e. non-gated) of projection data. The projection extrapolation method described in section 4.2.4 was applied prior to gFBP reconstruction of the truncated atrium in thorax phantom and animal study.  Three additional reconstruction techniques, constrained to use the retrospectively gated projection datasets, were applied for each of the dynamic phantom experiments and animal study. Projection data were gated as described in the previous section to mimic the selection of a target cardiac phase for reconstruction, such as end-diastole, that could be selected during an interventional procedure. The gated projection data were then reconstructed using gFBP, MS PICCS with α = 0.5, and TV-CS. The reconstructed image volume was 300 x 300 x 280 voxels with isotropic pixel dimensions $(0.5 \text{ mm})^3$ for each of the reconstruction methods considered. The iterative reconstructions were performed for λ values spanning several orders of magnitude (10-10,000). The λ value that minimized the 99[th] percentile segmentation error was considered optimal. The noise matrix, **D**, in equation (4.3) was set equal to the identity matrix. The incorporation of a noise model in the PICCS framework has been investigated previously for CT projections acquired on a diagnostic scanner.[145,174] Incorporation of a noise model in the PICCS framework may lead to *increased* image noise for circular objects, *decreased* image noise for non-circular objects,[174] or anisotropic spatial resolution that could impact left atrium segmentation accuracy.[145] A subset of the dynamic atrium in thorax phantom projections (88.2 bpm, 20 $mA_{peak}$) were reconstructed by setting the diagonal elements of the noise matrix, **D**, equal to the number of photons rebinned for a given parallel ray to investigate the impact of incorporating a statistical noise model in the reconstruction procedure.

### 7.2.4   Reference atrium CT image

The left atrium phantom (Figure 7.1A) was scanned on a commercial helical CT scanner (Discovery CT750 HD, GE Healthcare, Waukesha, WI, USA) to establish a reference surface model. The atrium was imaged at 100 kV tube potential and 250 mAs in an axial scan mode. A 512 x 512 x 224 (0.29 mm x 0.29 mm x 0.625 mm) voxel image volume was reconstructed using the scanner's filtered backprojection algorithm (standard kernel). The reconstructed image was interpolated to generate a 300 x 300 x 280 voxel resolution image with pixel dimensions that matched the SBDX CT reconstructions. Example left atrium reference images are shown in Figure 7.7 for axial, coronal, and sagittal slices.

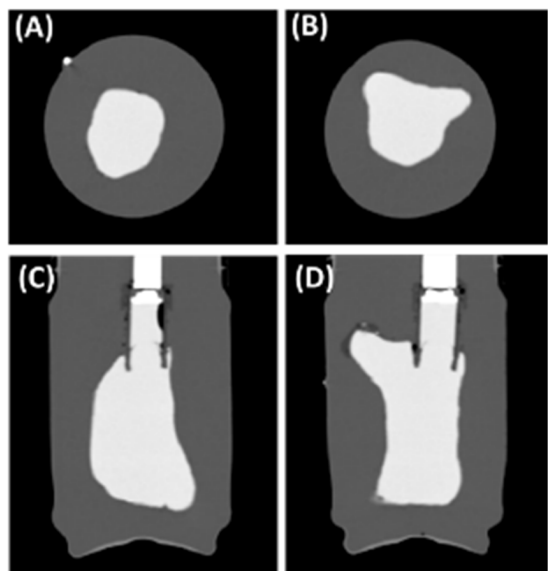

Figure 7.7: Reference atrium phantom images obtained with a commercial helical CT scanner are shown for axial slices (A-B), a coronal slice (C), and sagittal slice (D).

### 7.2.5   Evaluation metrics

SBDX CT image quality was evaluated in terms of noise, contrast-to-noise, image sharpness, and segmentation accuracy.

#### *7.2.5.1   Noise*

Image noise level was shown in chapter 3 to impact segmentation accuracy. The standard deviation of image values within uniform regions-of-interest located in atrium and background areas was computed to quantify the local noise level near the atrium surface boundary (Figure 7.8).

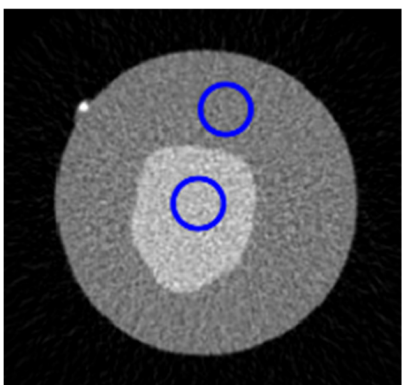

Figure 7.8: The locations of regions-of-interest used to compute noise standard deviation and signal difference are shown for atrium and background regions.

### *7.2.5.2  Contrast-to-noise ratio*

The signal difference between the left atrium iodine contrast agent and background water region, as well as noise level, influence cardiac chamber segmentation accuracy. The contrast-to-noise ratio (CNR) was computed as the difference of the mean image value in an ROI centered in the atrium and an ROI in a nearby background region divided by the standard deviation of image values in the background ROI, according to equation (7.1).

$$CNR = \frac{\bar{x}_{atrium} - \bar{x}_{bkgd}}{\sigma_{bkgd}} \tag{7.1}$$

Here, $\bar{x}_{atrium}$ represents the mean image value in an ROI centered in the atrium chamber and $\bar{x}_{bkgd}$ and $\sigma_{bkgd}$ denote the mean and standard deviation of image values for an ROI located in a nearby background region. Figure 7.8 presents the location of the atrium and background regions-of-interest.

### *7.2.5.3  Image sharpness*

The spatial resolution of SBDX CT images should be characterized for the cardiac chamber segmentation task to assess the impact on accuracy. Edge blurring or image sharpness changes as the reconstruction parameters (e.g. α, λ) are adjusted. The methodology described by *Li et al.* was adapted to quantify edge blurring present in SBDX CT reconstructions.[175]

Measurement of the image sharpness metric can be described in nine steps. First, a reduced noise image was determined by averaging axial slices +/- 2.0 mm about a reference fiducial. Second, an ROI was placed within the reduced-noise image at an air-water edge interface as shown in Figure 7.9A. Third, the magnitude of the gradient image of the ROI was computed to identify edge boundaries (Figure 7.9B). Fourth, the edge center line was extracted by selecting the maximum gradient value for each image column. Next, a $5^{th}$ order polynomial, $f(x)$, was fit to the edge center line. Sixth, line profiles normal to the edge center line were determined by differentiating $f(x)$ with respect to $x$ and computing the normal slopes as, $\frac{-1}{df(x)/dx}$. Seventh, an edge spread function (ESF) was determined by averaging each of the line profiles (Figure 7.9C). Eighth, the ESF was differentiated to produce a line-spread-function. Finally, image sharpness was specified as the full width half maximum value (FWHM) of a Gaussian function fit to the measured LSF by a least squares procedure.

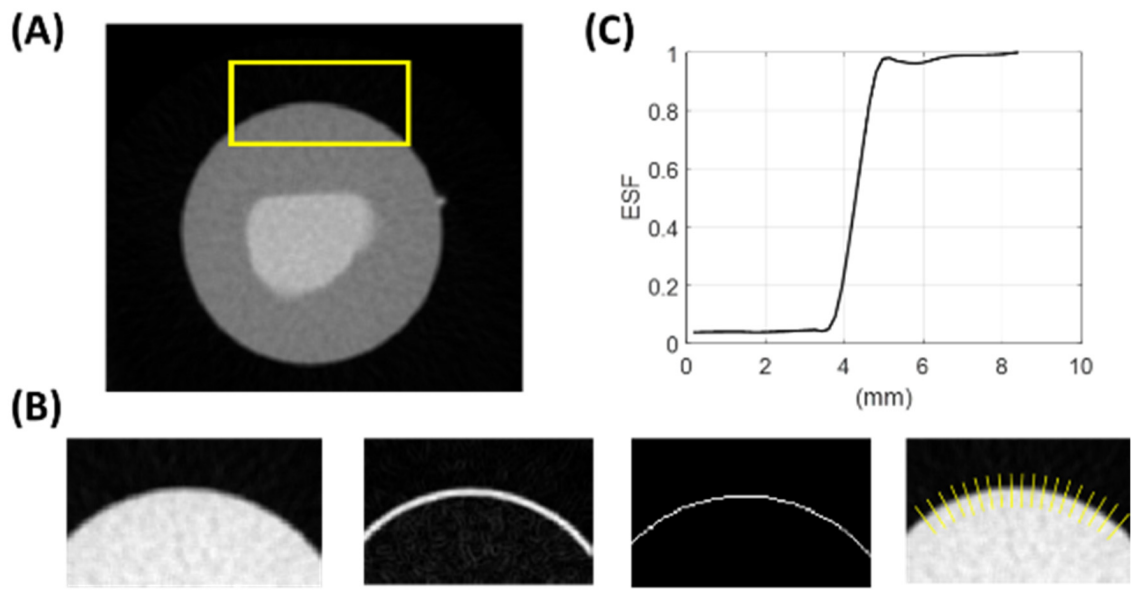

Figure 7.9: (A) Image sharpness was specified by the full-width-at-half-maximum of a point spread function derived from an estimated edge spread function. An ROI was placed at an air-water interface to estimate the ESF. (B) The workflow used to compute the image sharpness metric is shown. (C) An example ESF.

### 7.2.5.4   *Segmentation accuracy*

For cardiac chamber roadmapping during interventional procedures, the left atrium would be segmented from an SBDX CT image and volume rendered within the SBDX catheter tracking and/or fluoroscopic imaging display. The accuracy of the left atrium phantom segmentation was quantified using the Sorensen-Dice coefficient (DC) and a histogram of segmentation errors computed versus an atrium model derived from the reference CT scan (section 7.2.4).

A three-dimensional atrium model was segmented from each SBDX CT image using an intensity based thresholding method. The atrium surface was defined as the boundary of voxels obtained after thresholding to half the intensity difference between the atrium and background regions shown in Figure 7.8. A connected component analysis was performed following intensity based segmentation to select the largest object (i.e. the atrium) surviving the threshold operation and to discard small objects that remained due to image noise.

Following segmentation, the SBDX CT derived atrium model was registered with an atrium model segmented from the reference CT scan (section 7.2.4). The registration was performed using a rigid transformation that minimized the fiducial registration error (FRE) of the eight aluminum fiducials located on the water cylinder between the SBDX CT image and the reference CT image. Three-dimensional fiducial coordinates were determined from the CT volumes using a semi-automated segmentation method. First, the initial center-of-mass coordinates were estimated for each fiducial by manually placing a point at the fiducial center in the image volume. Next, a local region-of-interest (31 x 31 x 15 image voxels) was specified about the estimated center-of-mass. The high-contrast fiducials were then segmented using a thresholding method where image values within the ROI less than the fourth Otsu threshold level were set to 0 and image voxels greater than the threshold were set to 1. A connected component analysis was performed on the binary image to select the largest object present corresponding to the aluminum fiducial. The center-of-mass of each segmented fiducial was then computed. Fiducial segmentation defaulted to the manually designated coordinates for cases where the automated approach failed (e.g. due to high noise at large $\lambda$ values). Denoting the SBDX derived fiducial coordinates by a vector, $p_i$, reference fiducial coordinates by, $q_i$, rotation and translation matrices by $R$ and $T$, respectively, the FRE was computed according to equation (7.2).

$$FRE = \sqrt{\frac{1}{8}\sum_{i=1}^{8}|(Rp_i + T) - q_i|^2} \tag{7.2}$$

The rotation and translation matrices that minimized the FRE were determined using an iterative closest point algorithm.[176] An FRE value of 0.0 indicates perfect overlap between the SBDX CT fiducial coordinates and the reference fiducial coordinates. However, in practice, non-zero FRE values are observed due to fiducial localization errors. Fiducial localization errors may be attributed to image noise, programmed phantom motion, excessive edge-blurring due to a large regularization weight, or from discretization errors that result from representing continuous objects by finite-sized image voxels.

After registering the reference atrium model with an SBDX derived model, the atrium segmentation accuracy was quantified using the Sorensen-Dice coefficient (DC) and a histogram of segmentation errors. The Sorensen-Dice coefficient was computed versus the reference atrium model according to equation (7.3). A Dice coefficient of 1.0 indicates the segmented atrium has 100% overlap with the reference atrium (Figure 7.10). A value of 0.0 indicates no overlap.

$$DC = \frac{2\left|x_{recon} \cap x_{reference}\right|}{\left|x_{recon}\right| + \left|x_{reference}\right|} \tag{7.3}$$

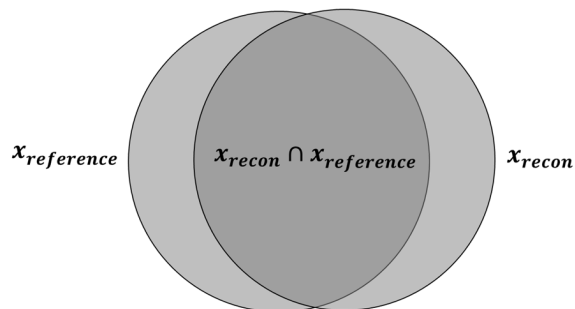

Figure 7.10: The Sorensen-Dice coefficient was used to compare an SBDX derived atrium surface with a reference surface.

To further quantify the segmentation error, the distance of each segmented atrium surface point to the closest point on the reference atrium model was computed. A histogram of segmentation errors was then calculated and the distance corresponding to the 99th percentile was determined and used as a metric for segmentation accuracy. The mean and standard deviation of the segmentation errors were also reported to more completely describe the distribution.

## 7.3 Results

### 7.3.1 Static non-truncated atrium phantom

Reconstructions of the static non-truncated atrium phantom are presented in Figure 7.11 for two different axial positions within the 3D image volume at three different imaging fluences. For each fluence level, the left atrium phantom was segmented from the surrounding water container. Figure 7.11 shows the segmented atrium in blue relative to the reference segmentation in white. Figure 7.12 presents an example sagittal slice of the reconstructed atrium phantom. An increase in image noise can be observed in the superior and inferior portions of the field-of-view. The increased noise is expected as the number of source elements contributing fluence to a point in the image volume decreases towards the superior and inferior portions of the SBDX CT FOV.

Image quality is summarized in Table 7.2 for the static non-truncated atrium phantom reconstructions. Segmentation errors at the 99[th] percentile increased slightly from 1.24 mm to 1.53 mm as the peak tube current decreased from 40 mA$_{peak}$ to 10 mA$_{peak}$, respectively, due to an increase in image noise from 78 HU to 123 HU. Mean segmentation errors were less than 0.7 mm for each of the fluence levels considered. Figure 7.13 shows histograms to characterize the distribution of the magnitude of segmentation errors at the highest and lowest fluence levels. Dice coefficients exceeded 0.95 for all reconstructions indicating good agreement between the SBDX CT derived atrium models and the reference atrium model. The fiducial registration error averaged 0.83 mm across fluence levels. The results presented in this section represent a lower bound in terms of the segmentation error that may be expected at the fluence levels considered in this chapter, since confounding factors such as FOV truncation and atrium motion are removed.

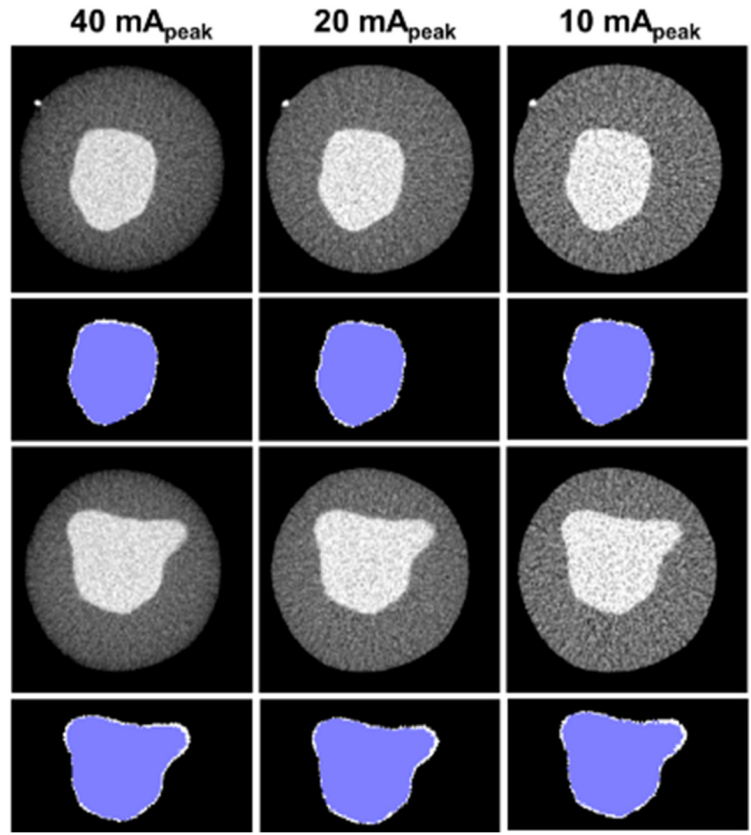

Figure 7.11: Non-gated gFBP reconstructions of the static non-truncated atrium phantom are shown for SBDX imaging at 100 kV and peak tube currents of 40 mA$_{peak}$, 20 mA$_{peak}$, and 10 mA$_{peak}$ from left to right. The segmented left atrium is shown in blue versus a reference segmentation shown in white. [WL = 0, WW = 1000] HU.

Table 7.2: Non-gated gFBP with static non-truncated atrium phantom.

| Tube Current (mA$_{peak}$) | Segmentation Error (mm) | | | Dice | Noise Std. Dev. (HU) | | PSF FWHM (mm) | CNR | FRE (mm) |
|---|---|---|---|---|---|---|---|---|---|
| | Mean | σ | 99% | | σ$_{LA}$ | σ$_{Bkgd}$ | | | |
| 40 | 0.58 | 0.34 | 1.24 | 0.96 | 78 | 61 | 0.87 | 8.9 | 0.83 |
| 20 | 0.59 | 0.35 | 1.44 | 0.96 | 93 | 75 | 0.88 | 6.8 | 0.83 |
| 10 | 0.64 | 0.38 | 1.53 | 0.95 | 123 | 104 | 0.87 | 4.7 | 0.84 |

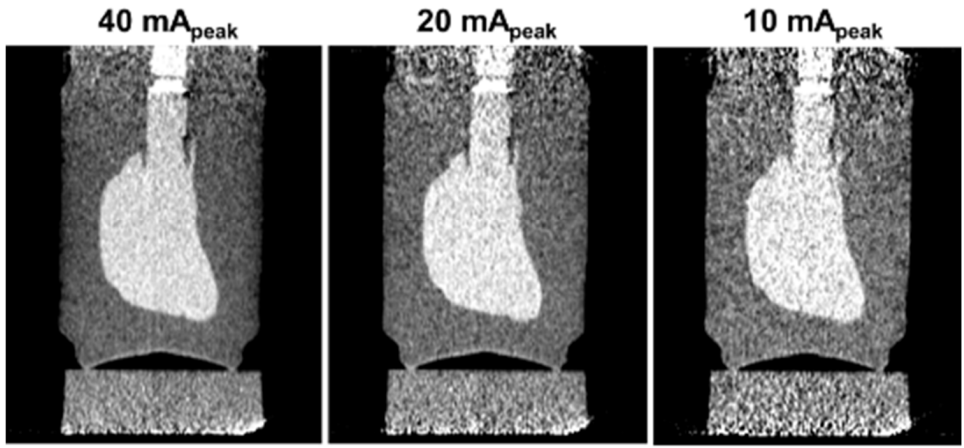

Figure 7.12: Sagittal slice reconstructions of the static non-truncated atrium phantom are shown for SBDX imaging at 100 kV and peak tube currents of 40 mA$_{peak}$, 20 mA$_{peak}$, and 10 mA$_{peak}$ from left to right. [WL = 0, WW = 1000] HU.

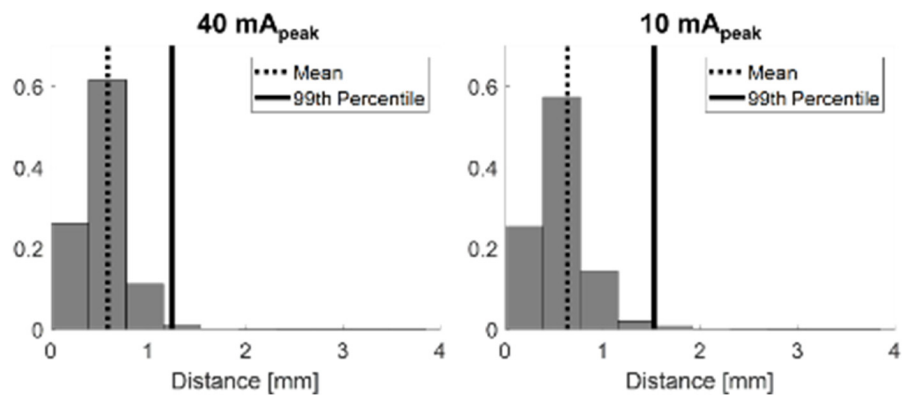

Figure 7.13: Histograms present the minimum distance from each surface point of the 3D atrium segmented from SBDX CT images versus a reference segmentation for the static non-truncated atrium phantom. The dotted black line denotes the mean point-to-point distance. The solid black line denotes the 99th percentile distance.

## 7.3.2   Dynamic non-truncated atrium phantom

### 7.3.2.1   Non-gated gFBP

The second phantom study performed investigated the impact of high frequency motion and projection undersampling resulting from retrospective gating on segmentation accuracy. Figure 7.14 presents non-gated gFBP reconstructions of the atrium phantom at a programmed motion frequency of 90 cycles per minute. As expected for a non-gated technique, blurring of an aluminum fiducial and the atrium boundary can be observed. Table 7.3 presents image quality metrics for the non-gated gFBP reconstructions at each of the three motion frequencies and three imaging fluences considered. Atrium segmentation errors increased compared to the static phantom results presented in the previous section. The 99[th] percentile segmentation error measured 3.2 mm, averaged across fluence levels and motion frequencies, versus 1.4 mm for the static non-truncated atrium phantom. Similarly, the mean FRE increased from 0.83 mm for the static phantom to 1.42 mm for the dynamic phantom. The increase in FRE can be attributed to fiducial localization errors resulting from the programmed motion pattern. The motion trajectory of an aluminum fiducial may be observed in Figure 7.14.

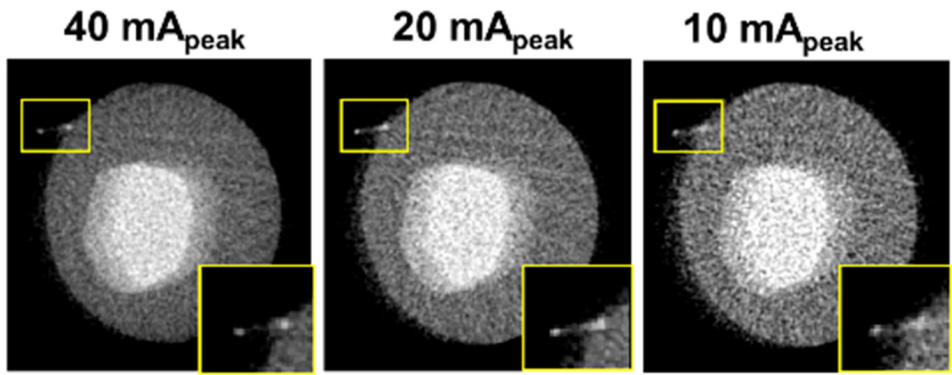

Figure 7.14: Non-gated gFBP reconstructions of the dynamic non-truncated atrium phantom at a frequency of 90 cycles per minute are shown for SBDX imaging at 100 kV and peak tube currents of 40 $mA_{peak}$, 20 $mA_{peak}$, and 10 $mA_{peak}$ from left to right. A zoomed view of the aluminum fiducial shows object blurring due to the programmed atrium motion. [WL = 0, WW = 1000] HU.

Table 7.3: Non-gated gFBP with dynamic non-truncated atrium phantom.

| Tube Current ($mA_{peak}$) | Frequency ($min^{-1}$) | Segmentation Error (mm) | | | Dice | Noise Std. Dev. (HU) | | PSF FWHM (mm) | CNR | FRE (mm) |
|---|---|---|---|---|---|---|---|---|---|---|
| | | Mean | $\sigma$ | 99% | | $\sigma_{LA}$ | $\sigma_{Bkgd}$ | | | |
| | 90 | 1.05 | 0.65 | 2.62 | 0.93 | 73 | 68 | 1.28 | 8.0 | 1.53 |
| 40 | 75 | 1.00 | 0.62 | 2.62 | 0.93 | 78 | 74 | 1.32 | 7.5 | 1.47 |
| | 60 | 1.03 | 0.67 | 2.77 | 0.94 | 75 | 69 | 1.36 | 8.0 | 1.47 |
| | 90 | 1.04 | 0.64 | 2.67 | 0.93 | 94 | 80 | 1.24 | 6.6 | 1.48 |
| 20 | 75 | 1.09 | 0.68 | 3.00 | 0.93 | 93 | 82 | 1.17 | 6.2 | 1.48 |
| | 60 | 0.92 | 0.58 | 2.36 | 0.94 | 89 | 88 | 1.37 | 6.0 | 1.32 |
| | 90 | 1.73 | 1.12 | 4.66 | 0.90 | 128 | 116 | 1.18 | 4.2 | 1.22 |
| 10 | 75 | 1.46 | 0.95 | 4.01 | 0.91 | 135 | 110 | 1.35 | 4.4 | 1.44 |
| | 60 | 1.44 | 0.93 | 3.92 | 0.91 | 136 | 113 | 1.39 | 4.4 | 1.38 |

### 7.3.2.2   Gated gFBP

Simple application of gating to the gFBP method generally gives poor results, as was demonstrated in the simulations in Chapter 3 (see Fig 3.12B). To demonstrate this experimentally, Figure 7.15A presents an example gated gFBP reconstruction of the dynamic atrium phantom. The gated gFBP reconstruction shows severe artifacts due to projection undersampling. The atrium phantom could not be segmented using the intensity based thresholding method due to the high level of image artifacts present (Figure 7.15B). Table 7.4 summarizes the evaluation metrics for the gated gFBP reconstructions performed at 40 mA$_{peak}$. The 99$^{th}$ percentile segmentation error ranged from 20.7 to 97.2 mm. The mean FRE decreased slightly from 1.49 mm for non-gated gFBP (at 40 mA$_{peak}$) to 0.96 mm for gated gFBP which indicates retrospective gating may decrease object (i.e. aluminum fiducial) blurring attributed to the programmed motion. Nonetheless, improvements in FRE were negated by the strong degree of image artifacts present.

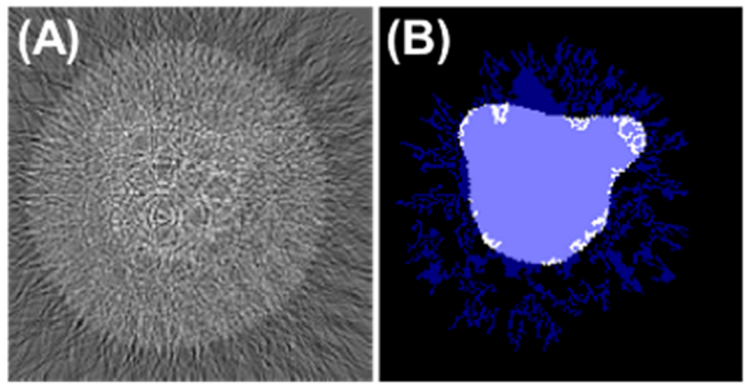

Figure 7.15: (A) Gated gFBP reconstruction of the dynamic non-truncated atrium at 90 cycles per minute shows severe artifacts due to projection undersampling. (B) The intensity based segmentation method failed due to the reconstruction artifacts.

Table 7.4: Dynamic non-truncated atrium phantom image quality metrics. Reconstruction via gated gFBP.

| Tube Current (mA$_{peak}$) | Frequency (min$^{-1}$) | Segmentation Error (mm) | | | Dice | Noise Std. Dev. (HU) | | CNR | FRE (mm) |
|---|---|---|---|---|---|---|---|---|---|
| | | Mean | σ | 99% | | σ$_{LA}$ | σ$_{Bkgd}$ | | |
| 40 | 90 | 9.37 | 5.52 | 20.81 | 0.37 | 1162 | 948 | 0.5 | 0.91 |
| 40 | 75 | 9.24 | 5.54 | 20.73 | 0.38 | 1317 | 1011 | 0.5 | 1.10 |
| 40 | 60 | 68.33 | 12.94 | 97.24 | 0.04 | 1580 | 1147 | 0.5 | 0.89 |

### 7.3.2.3 Gated TV-CS and PICCS

The gated projection data were reconstructed using the TV-CS and PICCS algorithms for each of the motion frequencies and noise levels considered. The parameter $\lambda$ controls the relative weighting between the regularization and data consistency terms as shown in equation (4.3). Figure 7.16 plots the noise standard deviation in the atrium phantom for TV-CS reconstructions (90 cycles per minute, 40 mA$_{peak}$) versus the FWHM of the LSF, for several $\lambda$ values to demonstrate the tradeoff between noise suppression and edge blurring. Larger $\lambda$ values were associated with greater image noise and superior spatial resolution indicated by smaller FWHM values.

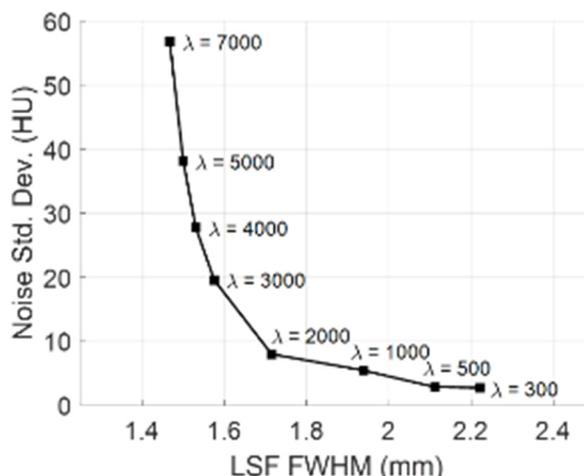

Figure 7.16: The standard deviation of image values for a region-of-interest centered in the atrium phantom is plotted versus the full-width-half-maximum value of a line spread function for several values of the $\lambda$ reconstruction parameter. The results presented correspond to imaging at 100 kV and 40 mA$_{peak}$ with 90 cycles per minute programmed motion.

The 99th percentile segmentation error is plotted versus $\lambda$ in Figure 7.17. As the goal of this work was to evaluate SBDX CT for cardiac chamber mapping, the $\lambda$ value that minimized the 99th percentile segmentation error was considered optimal and selected for the results presented below. The optimal $\lambda$ value for each programmed motion frequency and imaging fluence combination is listed in Table 7.5 for the TV-CS reconstructions and Table 7.6 for the PICCS reconstructions. The non-gated gFBP reconstruction (e.g. Figure 7.14) was utilized as the prior image for the PICCS reconstructions. In general, Figure 7.17 shows large segmentation errors at small $\lambda$ values due to increased edge blurring and also at large λ values due to increased image noise. Optimal segmentation results were obtained for λ values that balanced edge blurring and data consistency with the gated projection data versus noise suppression. Results were similar for each of the motion frequencies, consistent with the numerical simulations performed in chapter 4.

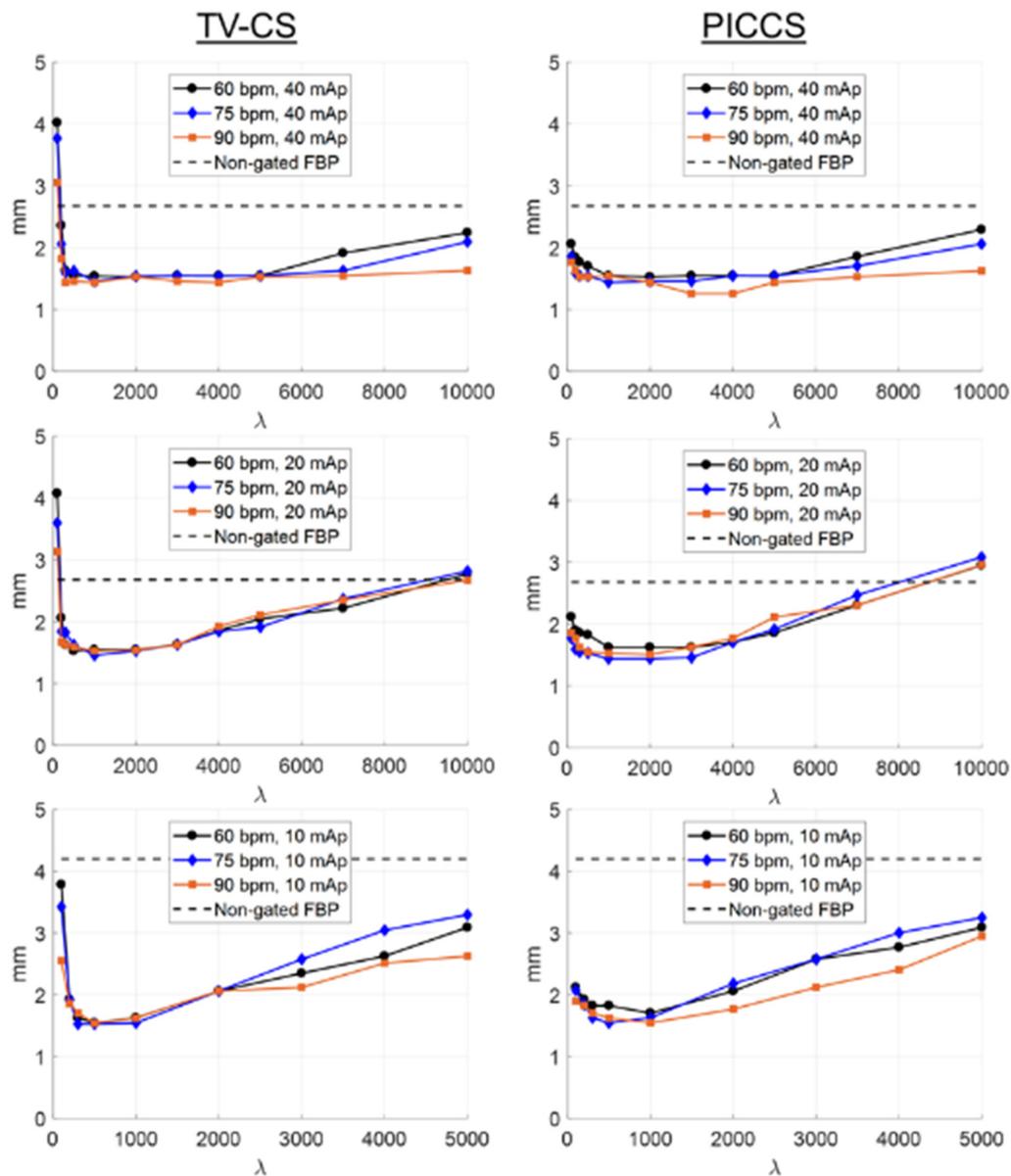

Figure 7.17: Segmentation error (99th percentile distance) is plotted versus λ for TV-CS (left) and PICCS reconstructions (right) of the dynamic non-truncated atrium phantom. The reconstructions were performed from SBDX projection data obtained at tube currents of 40 mA$_{peak}$ (top row), 20 mA$_{peak}$ (middle row), and 10 mA$_{peak}$ (bottom row).

Figure 7.18 presents TV-CS (top row) and PICCS (bottom row) reconstructions, with λ optimized for the segmentation task, for the 90 cycles per minute motion program at each of the three imaging fluences. For comparison, the non-gated gFBP results are shown in Figure 7.14. The aluminum fiducial appears blurred out in the optimal TV-CS reconstructions as the segmentation process favored, smooth, low-noise images. Qualitatively, the aluminum fiducials shown in Figure 7.18 appear sharp for the PICCS images compared to the non-gated gFBP results presented in Figure 7.14. Improved fiducial localization is supported by average FRE values of 0.92 mm and 0.89 mm for the TV-CS and PICCS reconstructions (Table 7.5 and Table 7.6), respectively, versus a 1.42 mm average FRE for the non-gated gFBP reconstructions.

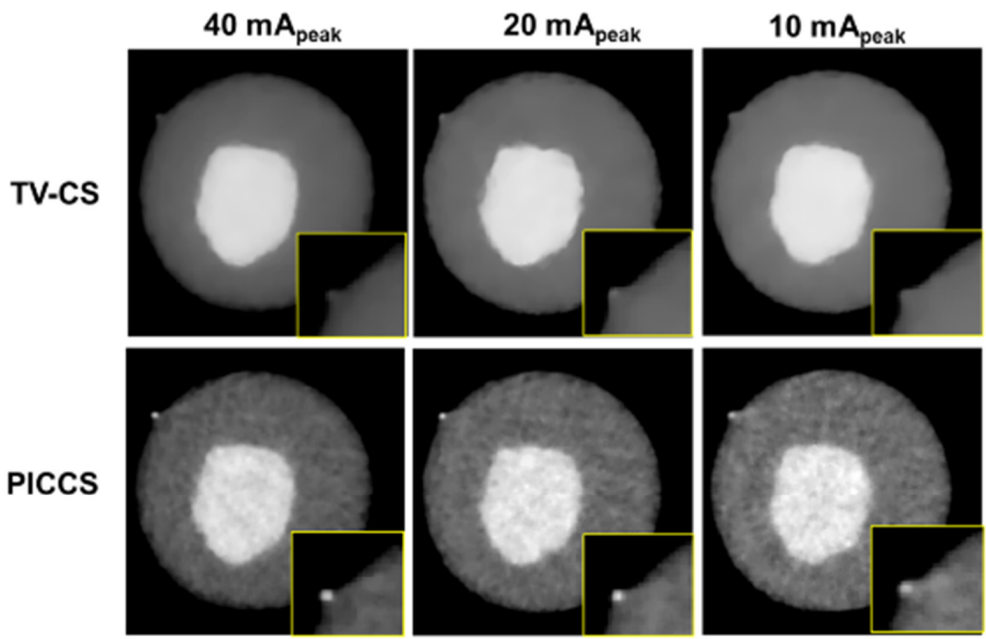

Figure 7.18: Dynamic atrium phantom at 90 cycles per minute. TV-CS (top row) and PICCS (bottom row) reconstructions are shown with the λ reconstruction parameter optimized for the atrium segmentation task (top row). [WL = 0, WW = 1000] HU.

Figure 7.19 presents example reconstructions and segmentations at each of the programmed motion frequencies and imaging fluences for the TV-CS algorithm. The corresponding PICCS results are shown in Figure 7.20. Segmentation error is quantified in Table 7.5 and Table 7.6. The mean segmentation errors, averaged across fluence levels and motion frequency, measured 1.51 mm for both the TV-CS and PICCS reconstructions. The iterative reconstruction method improved segmentation accuracy compared to the non-gated gFBP results where segmentation errors averaged 3.18 mm. Segmentation errors were similar to the 1.40 mm average result measured for the reference static non-truncated atrium phantom indicating the iterative reconstruction technique effectively reduced errors attributed to high frequency object motion.

The TV-CS technique reduced the image noise standard deviation in the atrium to an average value of 6 HU compared to 100 HU for the non-gated gFBP reconstructions. The decreased image noise increased the average CNR from 6.2 for gFBP to 46.2 for TV-CS.

Segmentation errors were similar for each of the motion frequencies considered. Averaging across fluence levels, the TV-CS 99[th] percentile segmentation error decreased slightly with increasing motion frequency from 1.53 mm to 1.50 mm for the 60 and 90 cycles per minute cases, respectively. The PICCS 99[th] percentile segmentation error decreased slightly with increasing motion frequency from 1.62 mm to 1.44 mm for the 60 and 90 cycles per minute cases, respectively. The slight decrease in segmentation errors is likely due to the increased number of superviews within each gated dataset at higher motion frequencies.

As mentioned previously, PICCS performance was equivalent to TV-CS with mean 99[th] percentile segmentation errors of 1.51 mm. The non-gated gFBP image was used as a prior image in this study. Due to the phantom study design, the entire left atrium phantom underwent periodic motion, leading to a prior image with artifacts. Future work could potentially improve the PICCS performance by pre-processing the prior image to reduce image artifacts. While the PICCS image texture closely resembles the prior image, the TV-CS images exhibited a patchy texture.

Table 7.5: Dynamic non-truncated atrium phantom image quality metrics. Reconstruction via TV-CS.

| Tube Current (mA_peak) | Frequency (min⁻¹) | Segmentation Error (mm) | | | Dice | Noise Std. Dev. (HU) | | PSF FWHM (mm) | CNR | FRE (mm) | λ |
|---|---|---|---|---|---|---|---|---|---|---|---|
| | | Mean | σ | 99% | | σ_LA | σ_Bkgd | | | | |
| 40 | 90 | 0.59 | 0.35 | 1.44 | 0.95 | 5 | 15 | 1.94 | 35.8 | 0.98 | 1000 |
| | 75 | 0.65 | 0.40 | 1.46 | 0.95 | 4 | 14 | 1.97 | 37.8 | 0.90 | 1000 |
| | 60 | 0.62 | 0.39 | 1.53 | 0.95 | 11 | 10 | 1.90 | 52.3 | 0.86 | 2000 |
| 20 | 90 | 0.63 | 0.39 | 1.53 | 0.95 | 7 | 9 | 1.80 | 58.6 | 0.96 | 1000 |
| | 75 | 0.62 | 0.37 | 1.46 | 0.95 | 8 | 11 | 1.86 | 48.5 | 0.89 | 1000 |
| | 60 | 0.61 | 0.38 | 1.53 | 0.95 | 5 | 16 | 2.05 | 30.7 | 0.89 | 500 |
| 10 | 90 | 0.64 | 0.39 | 1.54 | 0.95 | 5 | 7 | 1.85 | 70.5 | 0.97 | 500 |
| | 75 | 0.67 | 0.38 | 1.53 | 0.95 | 3 | 13 | 2.00 | 37.6 | 0.94 | 300 |
| | 60 | 0.68 | 0.43 | 1.55 | 0.95 | 4 | 11 | 1.94 | 43.7 | 0.88 | 500 |

Table 7.6: Dynamic non-truncated atrium phantom image quality metrics. Reconstruction via PICCS.

| Tube Current (mA_peak) | Frequency (min⁻¹) | Segmentation Error (mm) | | | Dice | Noise Std. Dev. (HU) | | PSF FWHM (mm) | CNR | FRE (mm) | λ |
|---|---|---|---|---|---|---|---|---|---|---|---|
| | | Mean | σ | 99% | | σ_LA | σ_Bkgd | | | | |
| 40 | 90 | 0.56 | 0.34 | 1.26 | 0.96 | 32 | 28 | 1.71 | 19.1 | 0.86 | 3000 |
| | 75 | 0.58 | 0.36 | 1.44 | 0.96 | 21 | 24 | 2.12 | 23.1 | 0.87 | 1000 |
| | 60 | 0.58 | 0.38 | 1.53 | 0.96 | 28 | 19 | 2.04 | 29.1 | 0.83 | 2000 |
| 20 | 90 | 0.60 | 0.37 | 1.51 | 0.96 | 36 | 31 | 1.93 | 16.6 | 0.85 | 2000 |
| | 75 | 0.58 | 0.36 | 1.44 | 0.96 | 47 | 32 | 2.18 | 16.1 | 0.83 | 2000 |
| | 60 | 0.63 | 0.42 | 1.62 | 0.96 | 32 | 37 | 2.05 | 14.7 | 0.94 | 2000 |
| 10 | 90 | 0.62 | 0.41 | 1.55 | 0.95 | 38 | 39 | 1.98 | 12.7 | 0.92 | 1000 |
| | 75 | 0.70 | 0.42 | 1.55 | 0.95 | 39 | 37 | 1.96 | 13.5 | 0.99 | 500 |
| | 60 | 0.69 | 0.43 | 1.70 | 0.95 | 41 | 42 | 1.94 | 11.8 | 0.91 | 1000 |

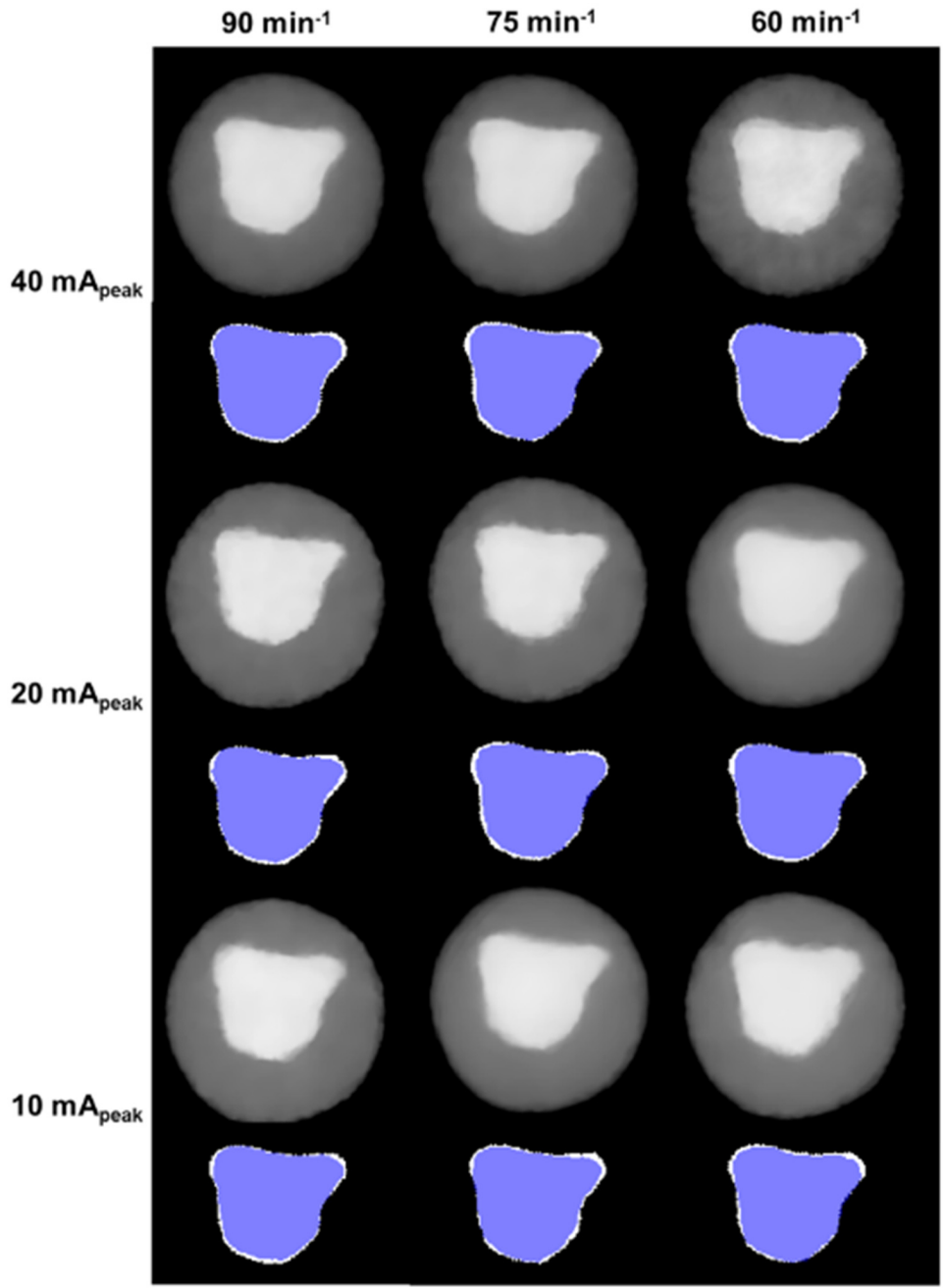

Figure 7.19: Dynamic atrium phantom. TV-CS reconstructions and example segmented atrium surfaces are shown in blue versus the reference surface in white for programmed motion frequencies of 60, 75, and 90 cycles per minute at three dose levels. The $\lambda$ reconstruction parameter was selected that minimized the 99th percentile segmentation error as listed in Table 7.5. [WL = 0, WW = 1000] HU.

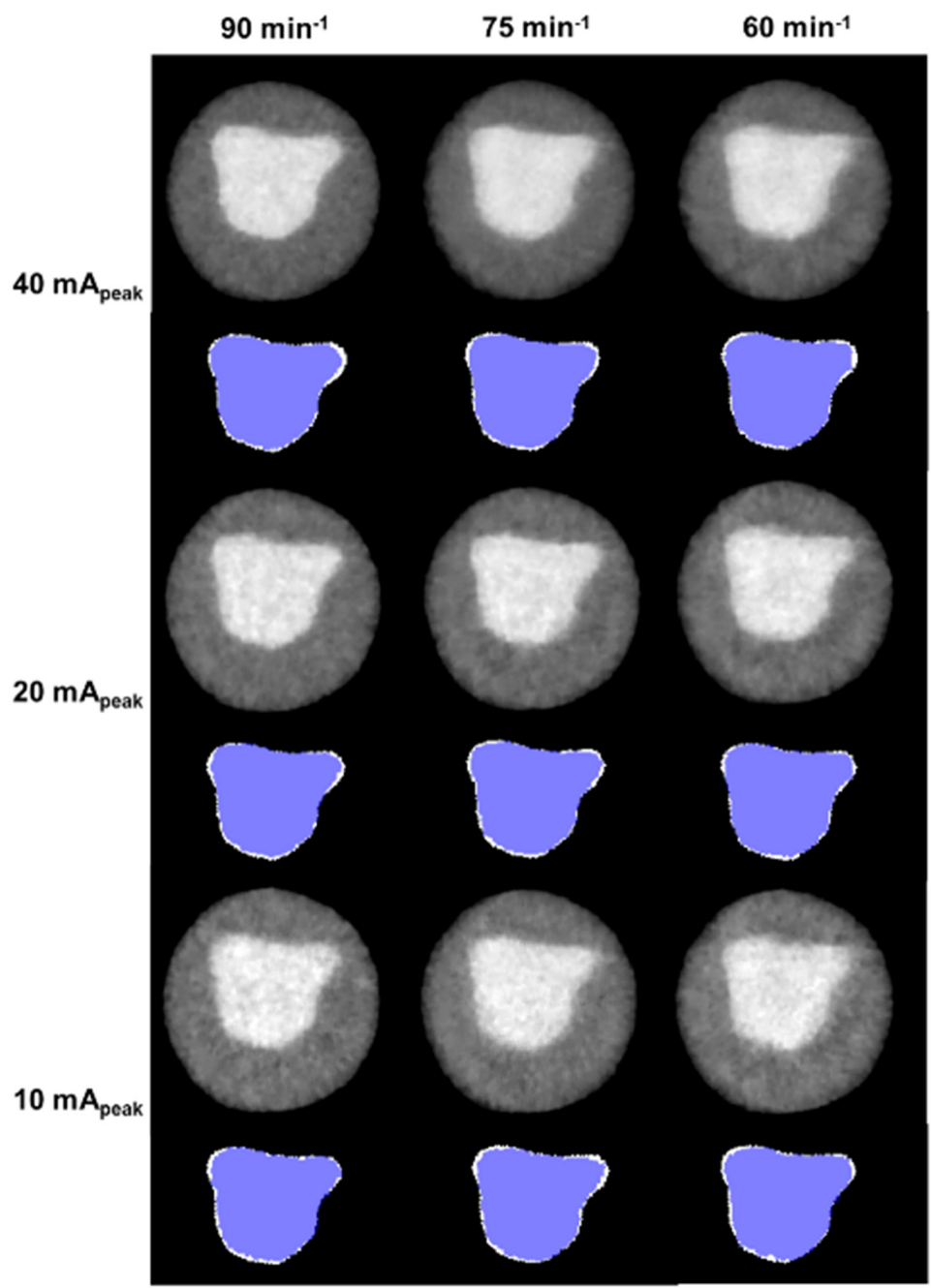

Figure 7.20: Dynamic atrium phantom. PICCS reconstructions and example segmented atrium surfaces are shown in blue versus the reference surface in white for programmed motion frequencies of 60, 75, and 90 cycles per minute at three dose levels. The $\lambda$ reconstruction parameter was selected that minimized the 99th percentile segmentation error as listed in Table 7.6. [WL = 0, WW = 1000] HU.

### 7.3.3   Static truncated atrium in thorax phantom

Non-gated gFBP reconstructions and example segmentations of the static truncated atrium in thorax phantom are shown in Figure 7.21 for two axial slice locations. Figure 7.22 presents example sagittal and coronal images. Table 7.7 summarizes the image quality metrics and segmentation errors. Segmentation errors at the 99[th] percentile ranged from 2.22 mm to 2.77 mm. The distribution of segmentation errors are presented in histograms shown in Figure 7.23 for the highest and lowest dose levels considered. Segmentation error at the 99[th] percentile averaged 2.46 mm across the three imaging fluences which corresponds to a 1.06 mm increase from the 1.40 mm average segmentation error measured for the static non-truncated atrium phantom. The noise standard deviation measured in the atrium phantom was similar ranging from 78 HU to 123 HU for the static non-truncated atrium phantom versus 76 HU to 128 HU for the static truncated atrium in thorax phantom. Similarly, the CNR ranged from 4.7 to 8.9 versus 4.4 to 8.8 for the non-truncated versus truncated phantoms, respectively.  As a result, the slight increase in segmentation errors can likely be attributed to a combination of field-of-view truncation, geometric uncertainty introduced by C-arm deformation, and heterogeneities introduced by the anthropomorphic thorax phantom that were not present for the static non-truncated phantom experiment performed with the rotating stage. Despite the slight increase in segmentation error compared to the non-truncated atrium phantom, segmentation errors at the 99[th] percentile were less than the typical diameter of an ablation catheter tip indicating sufficient performance for the task of cardiac chamber mapping.

Table 7.7: Static truncated atrium in thorax phantom image quality metrics.

| Tube Current ($mA_{peak}$) | Segmentation Error (mm) | | | Dice | Noise Std. Dev. (HU) | | PSF FWHM (mm) | CNR | FRE (mm) |
|---|---|---|---|---|---|---|---|---|---|
| | Mean | $\sigma$ | 99% | | $\sigma_{LA}$ | $\sigma_{Bkgd}$ | | | |
| 100 | 0.93 | 0.51 | 2.39 | 0.93 | 76 | 61 | 0.96 | 8.8 | 0.65 |
| 60 | 0.90 | 0.46 | 2.22 | 0.93 | 86 | 68 | 0.96 | 7.9 | 0.67 |
| 20 | 1.02 | 0.56 | 2.77 | 0.92 | 128 | 124 | 0.95 | 4.4 | 0.66 |

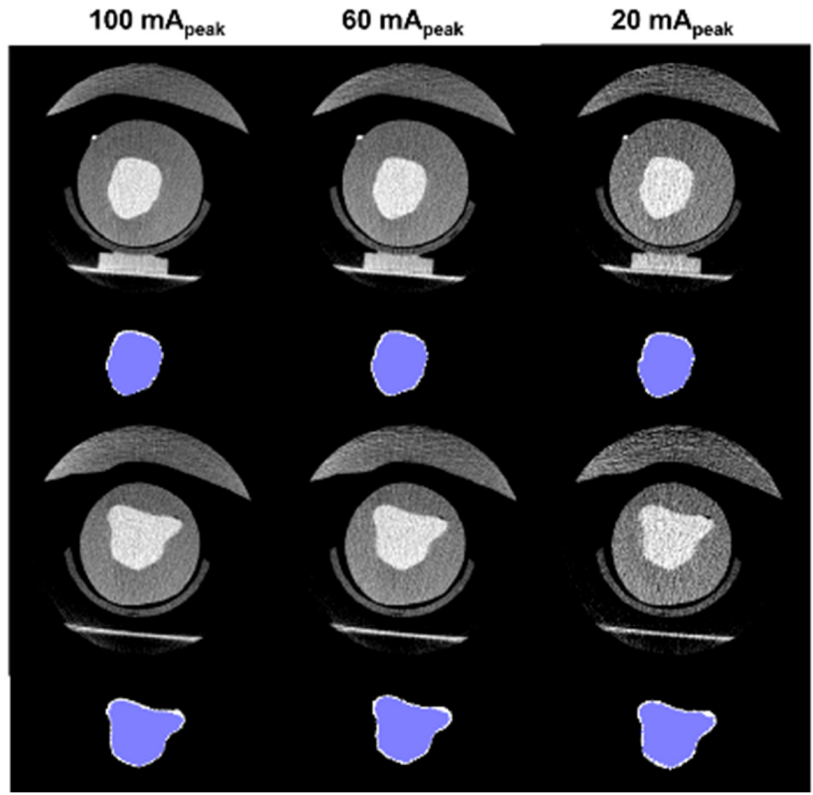

Figure 7.21: Static truncated atrium phantom imaged with rotating C-arm. Non-gated gFBP reconstructions are shown for SBDX operated at 100 kV and peak tube currents of 100 mA$_{peak}$, 60 mA$_{peak}$, and 20 mA$_{peak}$ from left to right. The segmented left atrium is shown in blue versus a reference segmentation shown in white. [WL = 0, WW = 1000] HU.

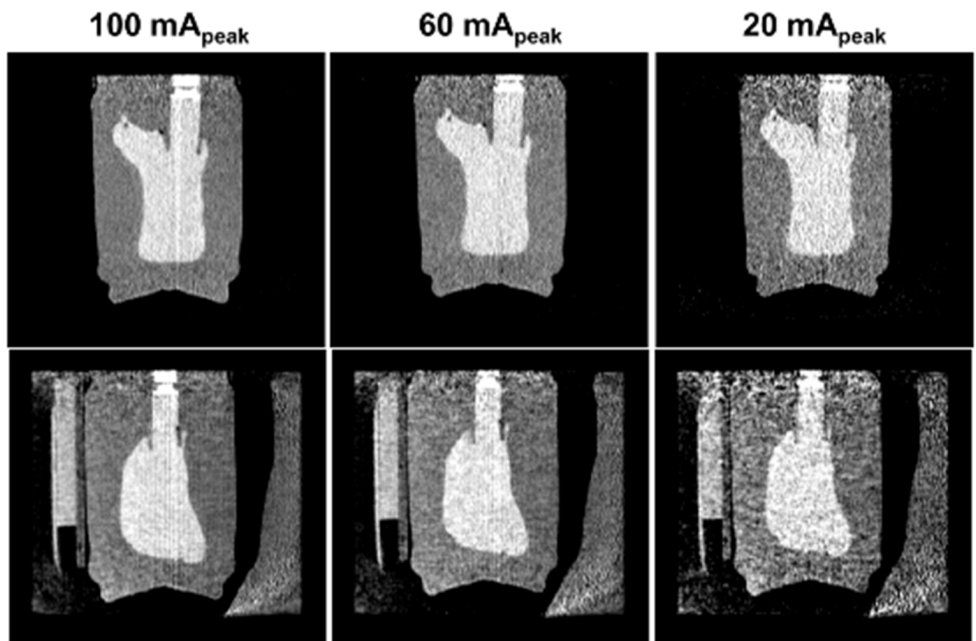

Figure 7.22: Coronal slice (top row) and sagittal slice (bottom row) reconstructions of the static truncated atrium in thorax phantom are shown for SBDX imaging at 100 kV and peak tube currents of 100 mA$_{peak}$, 60 mA$_{peak}$, and 20 mA$_{peak}$ from left to right. [WL = 0, WW = 1000] HU.

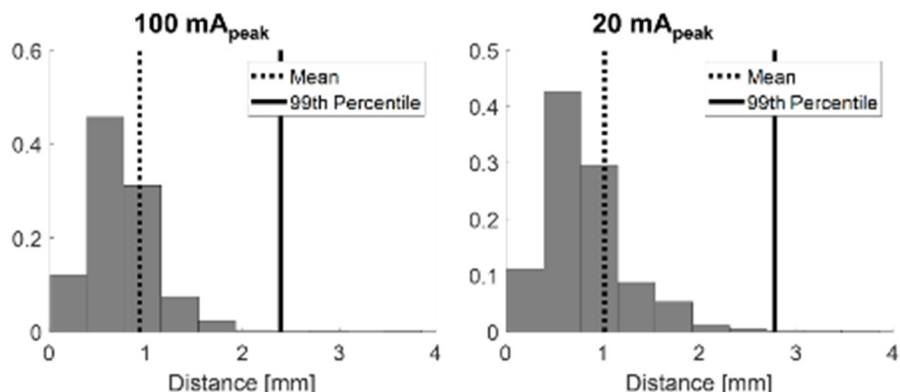

Figure 7.23: Histograms present the minimum distance from each surface point of the 3D atrium segmented from SBDX CT images of the static truncated atrium in thorax phantom versus a reference segmentation. The dotted black line denotes the mean point-to-point distance. The solid black line denotes the 99th percentile distance.

### 7.3.4 Dynamic truncated atrium in thorax phantom

The final phantom study performed to evaluate SBDX CT imaging for cardiac chamber mapping consisted of C-arm rotational acquisitions of the atrium undergoing two-dimensional high-frequency motion within the thorax phantom.

#### 7.3.4.1 Non-gated gFBP

Figure 7.24 presents non-gated gFBP reconstructions for axial and coronal image slices of the truncated atrium in thorax phantom undergoing 88.2 cycles per minute motion at each of the three imaging fluences considered. Image artifacts and blurring of the aluminum fiducial resulting from the programmed motion may be observed. Table 7.8 summarizes the image quality metrics for the non-gated gFBP reconstructions. The 99th percentile segmentation error averaged 5.76 mm across the three imaging fluences and three motion frequencies. The introduction of object motion increased the average segmentation error by 3.30 mm from 2.46 mm to 5.76 mm compared to the static truncated atrium in thorax phantom experiments (section 7.3.3).

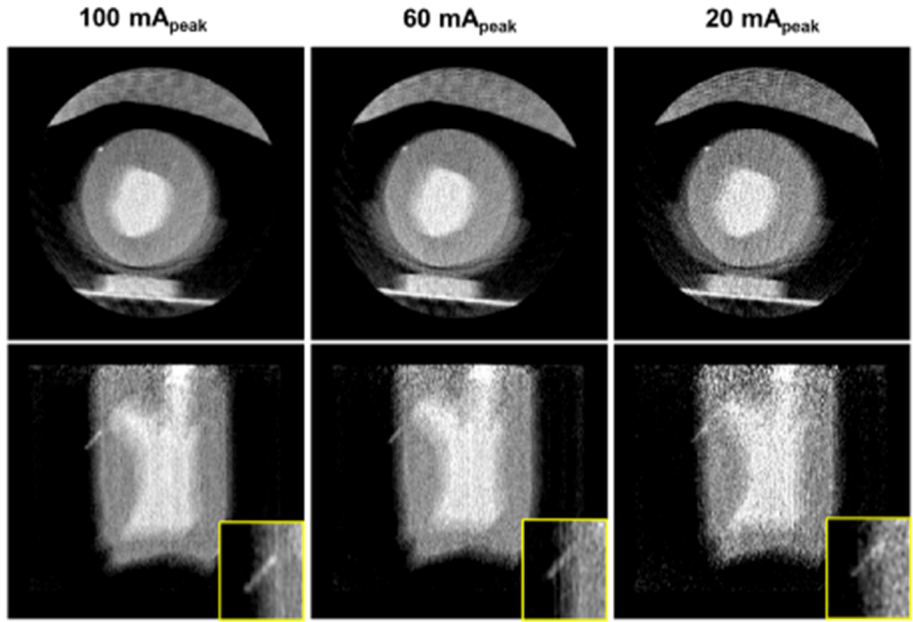

Figure 7.24: Dynamic truncated atrium phantom imaged with rotating C-arm and reconstructed with non-gated gFBP. Results for 88.2 cycles per minute are shown for axial (top row) and coronal (bottom row) slices, for SBDX operated at 100 kV and peak tube currents of 100 $mA_{peak}$, 60 $mA_{peak}$, and 20 $mA_{peak}$ from left to right. A zoomed view of the aluminum fiducial shows object blurring due to the programmed atrium motion. [WL = 100, WW = 1400] HU.

Table 7.8: Dynamic truncated atrium in thorax phantom image quality metrics. Reconstruction via non-gated gFBP.

| Tube Current ($mA_{peak}$) | Frequency ($min^{-1}$) | Segmentation Error (mm) | | | Dice | Noise Std. Dev. (HU) | | PSF FWHM (mm) | CNR | FRE (mm) |
|---|---|---|---|---|---|---|---|---|---|---|
| | | Mean | $\sigma$ | 99% | | $\sigma_{LA}$ | $\sigma_{Bkgd}$ | | | |
| | 88.2 | 2.30 | 1.38 | 5.92 | 0.82 | 63 | 73 | 2.34 | 5.1 | 0.70 |
| 100 | 75 | 2.29 | 1.27 | 5.54 | 0.83 | 66 | 78 | 2.17 | 5.5 | 1.07 |
| | 60 | 2.23 | 1.29 | 5.76 | 0.83 | 66 | 60 | 2.09 | 7.1 | 0.65 |
| | 88.2 | 2.21 | 1.37 | 5.75 | 0.83 | 74 | 78 | 1.94 | 5.0 | 0.88 |
| 60 | 75 | 2.35 | 1.32 | 5.64 | 0.83 | 79 | 80 | 2.06 | 5.2 | 1.09 |
| | 60 | 2.28 | 1.31 | 5.72 | 0.83 | 75 | 87 | 2.05 | 4.8 | 0.87 |
| | 88.2 | 2.19 | 1.32 | 5.54 | 0.84 | 118 | 91 | 2.29 | 4.3 | 0.81 |
| 20 | 75 | 2.50 | 1.44 | 6.03 | 0.82 | 126 | 109 | 2.09 | 4.1 | 0.78 |
| | 60 | 2.51 | 1.42 | 5.92 | 0.82 | 131 | 111 | 2.21 | 4.0 | 0.71 |

### 7.3.4.2   Gated gFBP

An example gated gFBP reconstruction is shown in Figure 7.25 for the dynamic truncated atrium in thorax phantom. The atrium could not be segmented using the intensity based thresholding method due to severe image artifacts.

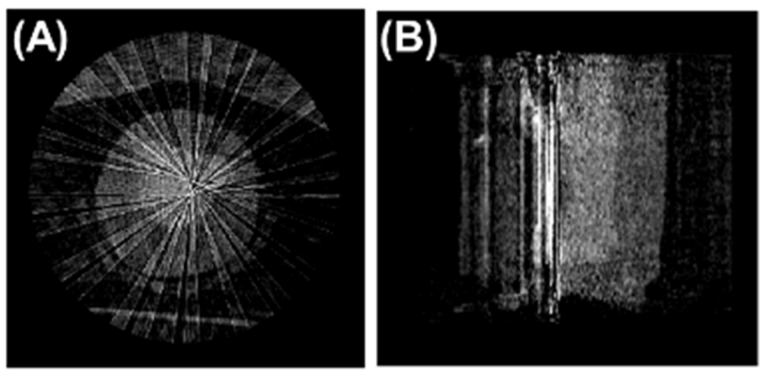

Figure 7.25: Gated gFBP reconstruction of the dynamic truncated phantom at 88.2 cycles per minute is shown for an axial slice (A) and sagittal slice (B). Severe artifacts are present due to projection undersampling.

### 7.3.4.3   Gated TV-CS and PICCS

Gated TV-CS and PICCS reconstructions were performed for each of the three fluence levels and three programmed motion frequencies. The 99th percentile segmentation error is plotted as a function of $\lambda$ in Figure 7.26. For the TV-CS reconstructions, the optimal $\lambda$ value, averaged at each motion frequency, was 667, 367, and 200 at peak tube currents of 100 $mA_{peak}$, 60 $mA_{peak}$, and 20 $mA_{peak}$, respectively. The optimal $\lambda$ value decreased as the peak tube current was reduced to compensate for increased image noise. A similar trend in segmentation error versus $\lambda$ was observed for the PICCS reconstructions. The $\lambda$ value that minimized the segmentation error was assumed optimal for the task of cardiac chamber mapping. For all cases, a $\lambda$ value existed that yielded superior segmentation accuracy relative to the non-gated gFBP reconstructions. Table 7.9 summarizes the image quality metrics for the TV-CS reconstructions at the optimal $\lambda$ value for each imaging scenario. Table 7.10 presents the corresponding results for the PICCS reconstructions.

Table 7.9: Dynamic truncated atrium in thorax phantom image quality metrics. Reconstruction via TV-CS.

| Tube Current (mA$_{peak}$) | Frequency (min$^{-1}$) | Segmentation Error (mm) | | | Dice | Noise Std. Dev. (HU) | | PSF FWHM (mm) | CNR | FRE (mm) | λ |
|---|---|---|---|---|---|---|---|---|---|---|---|
| | | Mean | σ | 99% | | σ$_{LA}$ | σ$_{Bkgd}$ | | | | |
| 100 | 88.2 | 1.17 | 0.69 | 3.17 | 0.92 | 29 | 43 | 1.96 | 7.7 | 0.73 | 1000 |
| | 75 | 1.24 | 0.69 | 3.04 | 0.90 | 8 | 32 | 1.66 | 10.3 | 0.71 | 300 |
| | 60 | 1.29 | 0.74 | 3.64 | 0.90 | 19 | 21 | 1.78 | 16.3 | 0.66 | 700 |
| 60 | 88.2 | 1.27 | 0.75 | 3.26 | 0.90 | 17 | 21 | 2.06 | 16.0 | 0.83 | 300 |
| | 75 | 1.28 | 0.73 | 3.36 | 0.90 | 5 | 30 | 1.79 | 12.2 | 0.74 | 300 |
| | 60 | 1.41 | 0.86 | 4.17 | 0.90 | 25 | 18 | 1.96 | 17.1 | 0.73 | 500 |
| 20 | 88.2 | 1.68 | 1.07 | 5.09 | 0.88 | 24 | 15 | 1.87 | 27.6 | 1.20 | 200 |
| | 75 | 1.58 | 0.96 | 4.67 | 0.88 | 57 | 35 | 1.89 | 12.3 | 0.80 | 200 |
| | 60 | 1.54 | 0.89 | 4.25 | 0.88 | 45 | 27 | 2.10 | 19.0 | 2.31 | 200 |

Table 7.10: Dynamic truncated atrium in thorax phantom image quality metrics. Reconstruction via PICCS.

| Tube Current (mA$_{peak}$) | Frequency (min$^{-1}$) | Segmentation Error (mm) | | | Dice | Noise Std. Dev. (HU) | | PSF FWHM (mm) | CNR | FRE (mm) | λ |
|---|---|---|---|---|---|---|---|---|---|---|---|
| | | Mean | σ | 99% | | σ$_{LA}$ | σ$_{Bkgd}$ | | | | |
| 100 | 88.2 | 1.24 | 0.66 | 3.00 | 0.91 | 47 | 40 | 2.01 | 8.6 | 0.81 | 400 |
| | 75 | 1.39 | 0.74 | 3.48 | 0.89 | 48 | 43 | 1.97 | 8.8 | 0.71 | 200 |
| | 60 | 1.54 | 0.88 | 4.24 | 0.89 | 49 | 35 | 2.06 | 9.5 | 0.78 | 200 |
| 60 | 88.2 | 1.38 | 0.76 | 3.48 | 0.90 | 59 | 43 | 2.09 | 8.1 | 0.86 | 200 |
| | 75 | 1.45 | 0.79 | 3.83 | 0.89 | 59 | 46 | 2.03 | 8.4 | 0.84 | 300 |
| | 60 | 1.83 | 1.06 | 4.85 | 0.88 | 63 | 51 | 2.03 | 6.3 | 0.78 | 200 |
| 20 | 88.2 | 1.76 | 1.05 | 5.09 | 0.88 | 95 | 62 | 2.03 | 5.5 | 0.77 | 100 |
| | 75 | 1.91 | 1.10 | 5.11 | 0.86 | 97 | 76 | 2.09 | 4.7 | 0.76 | 50 |
| | 60 | 1.90 | 1.16 | 5.28 | 0.87 | 104 | 84 | 2.10 | 4.7 | 0.77 | 100 |

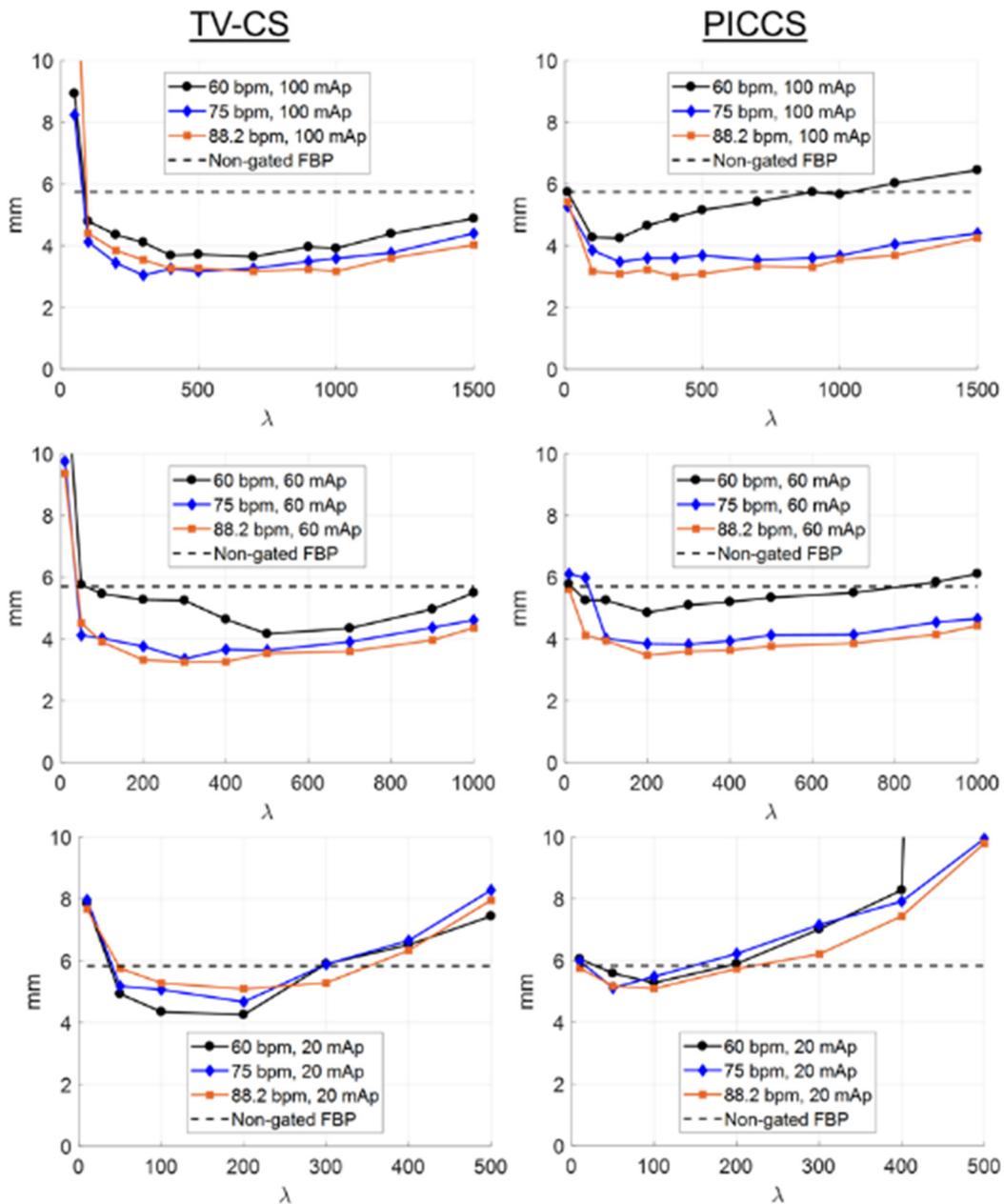

Figure 7.26: Segmentation error (99th percentile distance) is plotted versus λ for TV-CS (left) and PICCS reconstructions (right) of the dynamic truncated atrium in thorax phantom. The reconstructions were performed from SBDX projection data obtained at tube currents of 100 mA$_{peak}$ (top row), 60 mA$_{peak}$ (middle row), and 20 mA$_{peak}$ (bottom row).

The TV-CS reconstructions and example atrium segmentations are shown in Figure 7.27. PICCS reconstructions and example segmentations are shown in Figure 7.28. Qualitatively, the TV-CS reconstructions demonstrate the patchy texture commonly associated with total variation regularization. The PICCS reconstructions more closely resemble the prior image (i.e. non-gated gFBP). It's worth noting that the prior image showed considerable motion artifacts as nearly the entire object within the SBDX CT FOV underwent programmed motion (the portion of the thorax at the superior region of the FOV was the only stationary object).

Mean segmentation errors measured 1.39 mm for TV-CS and 1.60 mm for PICCS reconstructions averaged across each of the three dose levels and three motion frequencies (Table 7.9, Table 7.10). Segmentation errors increased, as expected, at lower tube currents. For example, the $99^{th}$ percentile segmentation errors were 3.04 mm, 3.36 mm, and 4.67 mm for the TV-CS reconstructions at 100 $mA_{peak}$, 60 $mA_{peak}$, and 20 $mA_{peak}$, respectively (75 cycle per minute frequency). PICCS segmentation errors were 3.48 mm, 3.83 mm, and 5.11 mm at 100 $mA_{peak}$, 60 $mA_{peak}$, and 20 $mA_{peak}$, respectively. Segmentation accuracy improved slightly for the 88.2 cycles per minute motion program compared to the 60 cycles per minute motion program. At 60 cycles per minute and 100 $mA_{peak}$, the $99^{th}$ percentile segmentation error was 3.64 mm (TV-CS) and 4.24 mm (PICCS) versus 3.17 mm (TV-CS) and 3.00 mm (PICCS) at 88.2 cycles per minute. The 60 cycles per minute reconstructions were performed from 14 gated superviews versus 20 gated superviews for the 88.2 cycles per minute case. Incorporation of a statistical noise model for the 20 $mA_{peak}$ reconstructions did not improve segmentation results. The 99th percentile segmentation errors were 5.62 mm (TV-CS) and 5.57 mm (PICCS) with the noise model versus 5.09 mm (TV-CS and PICCS) without for the 88.2 cycle per minute case.

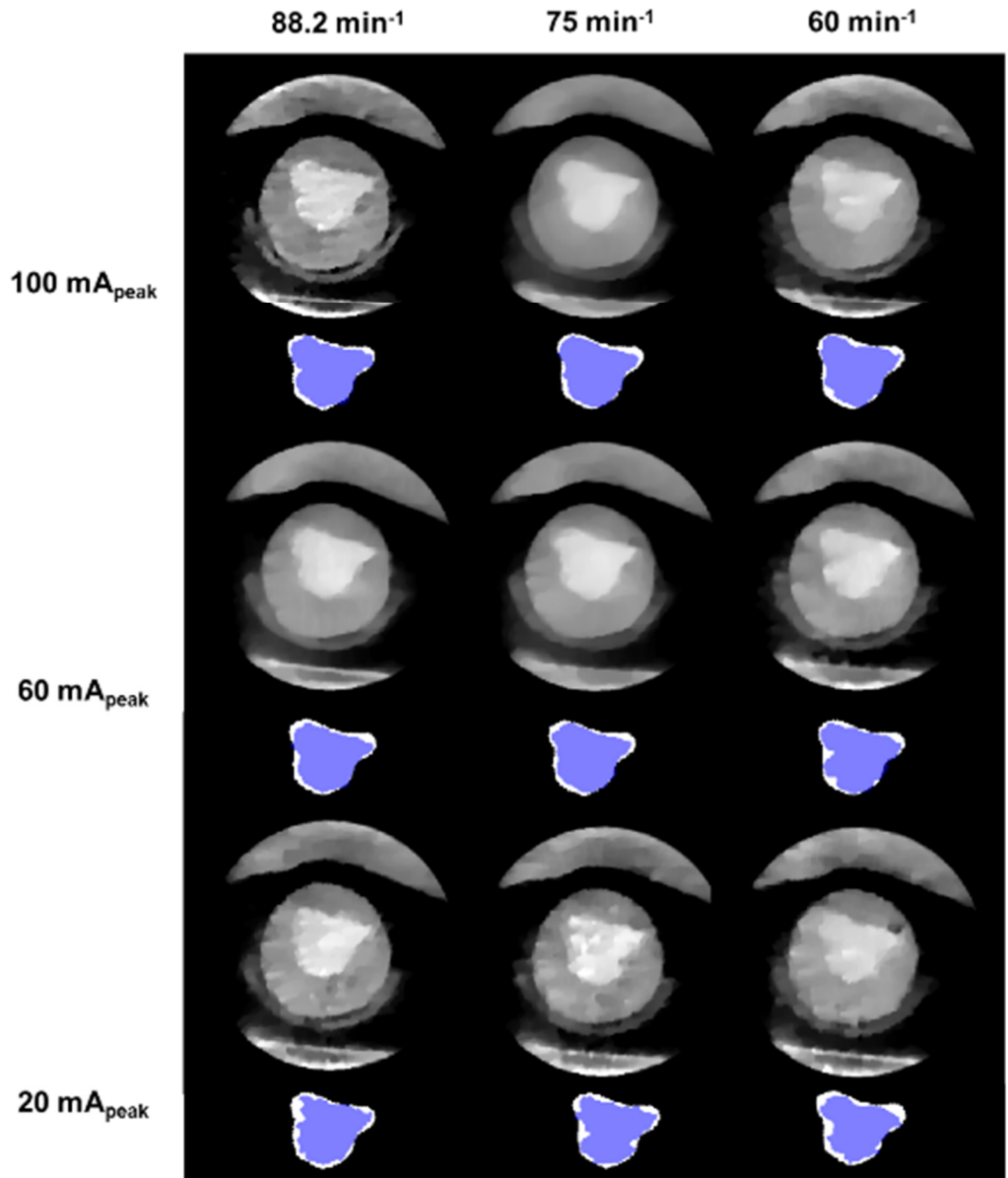

Figure 7.27: Dynamic truncated phantom imaged with rotating C-arm. TV-CS reconstructions and example segmented atrium surfaces are shown in blue versus the reference surface in white for programmed motion frequencies of 60, 75, and 88.2 cycles per minute at three dose levels. The $\lambda$ parameter was selected that minimized the $99^{th}$ percentile segmentation error as listed in Table 7.9. [WL = 100, WW = 1400] HU.

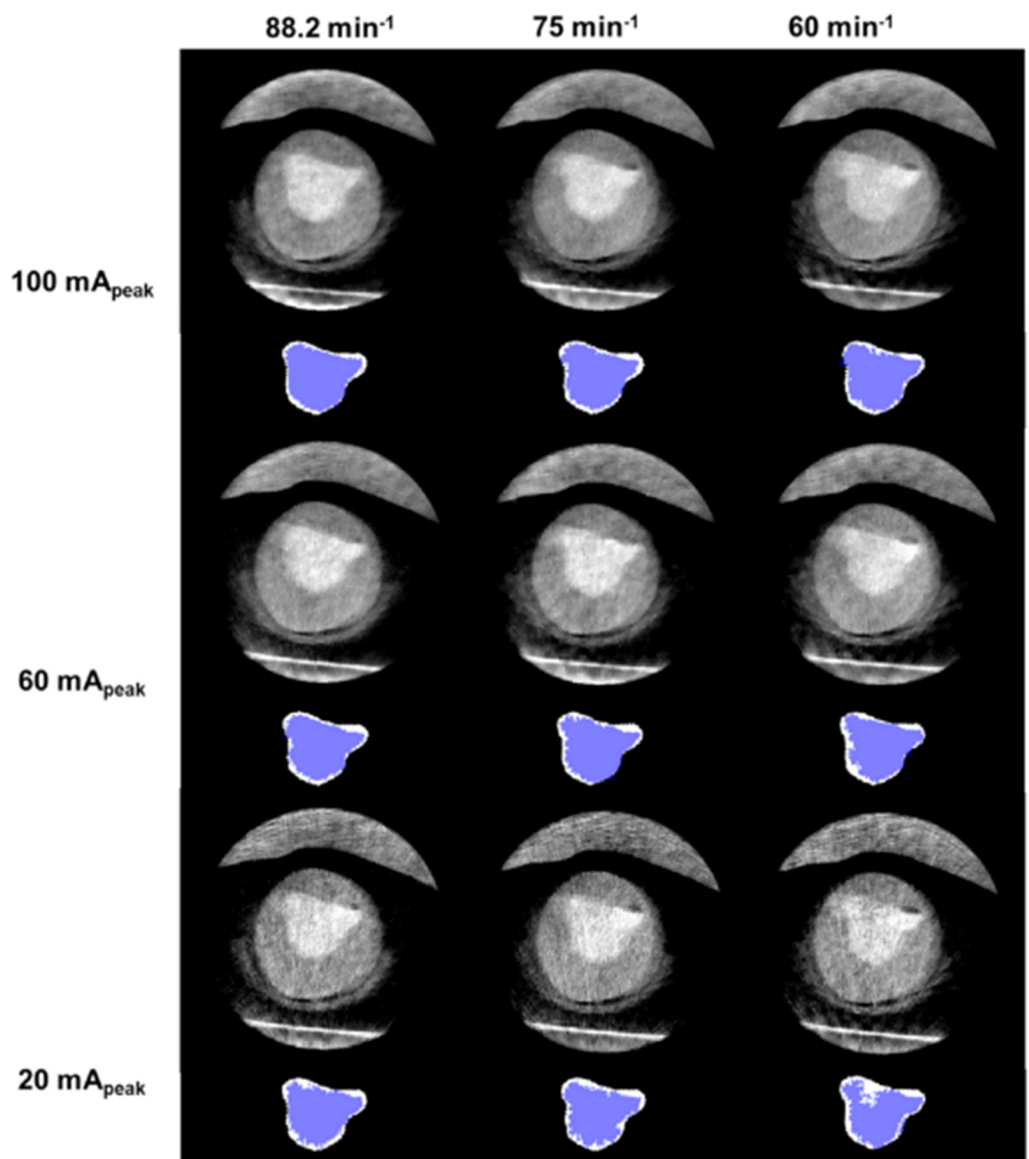

Figure 7.28: Dynamic truncated phantom imaged with rotating C-arm. PICCS reconstructions and example segmented atrium surfaces are shown in blue versus the reference surface in white for programmed motion frequencies of 60, 75, and 88.2 cycles per minute at three dose levels. The $\lambda$ parameter was selected that minimized the 99[th] percentile segmentation error as listed in Table 7.10. [WL = 100, WW = 1400] HU.

### 7.3.5   Demonstration in an animal model

The non-gated porcine projection data were reconstructed using the gFBP algorithm as shown for two orthogonal image slices (axial and coronal) in Figure 7.29. As expected, the non-gated images show blurring around the chambers due to cardiac motion occurring during the 13.4 second CT data acquisition period. Corresponding reference images obtained on a commercial CT scanner are shown at end-systole (40% RR) and end-diastole (80% RR) in Figure 7.30 for comparison. The reference images were reconstructed from a much shorter, 0.228 second, acquisition window. Additionally, differences between the SBDX CT and reference images may exist due to physiological changes (e.g. respiratory or cardiac phase), anatomic changes due to organ motion or animal positioning between the time of the pre-procedure and intra-procedural imaging, or from inaccuracies of the imaging systems.

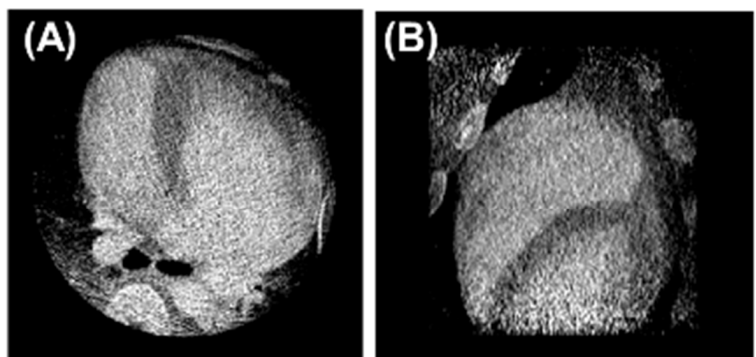

Figure 7.29: SBDX CT images of a 55 kg porcine are shown for axial (A) and coronal slices (B). Reconstruction was performed using non-gated gFBP.

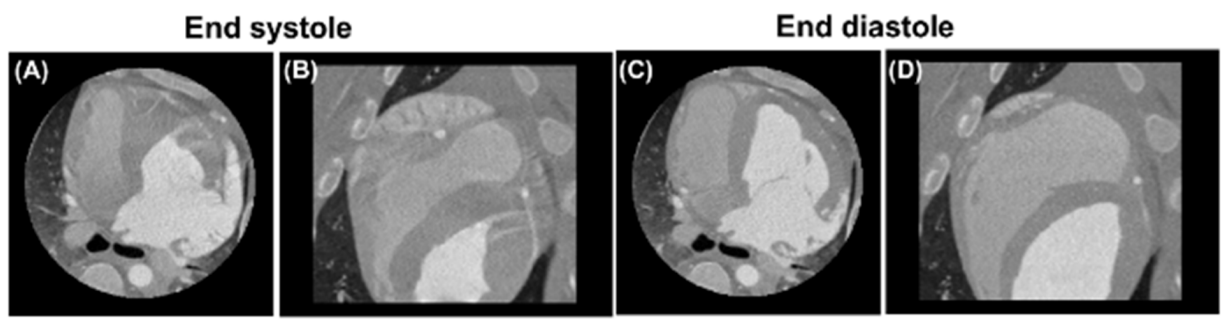

Figure 7.30: Reference CT images of a 55 kg porcine obtained on a commercial scanner are shown for: (A) End systole, axial slice. (B) End systole, coronal slice. (C) End diastole, axial slice. (D) End diastole, coronal slice.

Figure 7.31A-D presents three different gated SBDX-CT approaches. The top row shows simple gated gFBP, which, similar to the phantom studies, displays a high level of streaking artifacts resulsting from angular undersampling. The middle row shows the gated PICCS reconstructions (Figure 7.31E-H) and the bottom row shows gated TV-CS reconstructions (Figure 7.31I-L). The PICCS and TV-CS results show a reduced level of streak artifacts relative to gated gFBP. Note the PICCS reconstructions utilized the non-gated FBP image shown in Figure 7.29 as a prior image. Qualitatively, the PICCS and TV-CS reconstructions show reduced image blur due to cardiac motion. The ability to reconstruct a particular cardiac phase can be appreciated by comparing the cardiac chambers at end systole (left columns) and end diastole (right columns). The TV-CS images show the characteristic patchy artifact and the individual cardiac chambers appear nearly piecewise constant. The PICCS images show a more natural noise texture resembling the gFBP image.

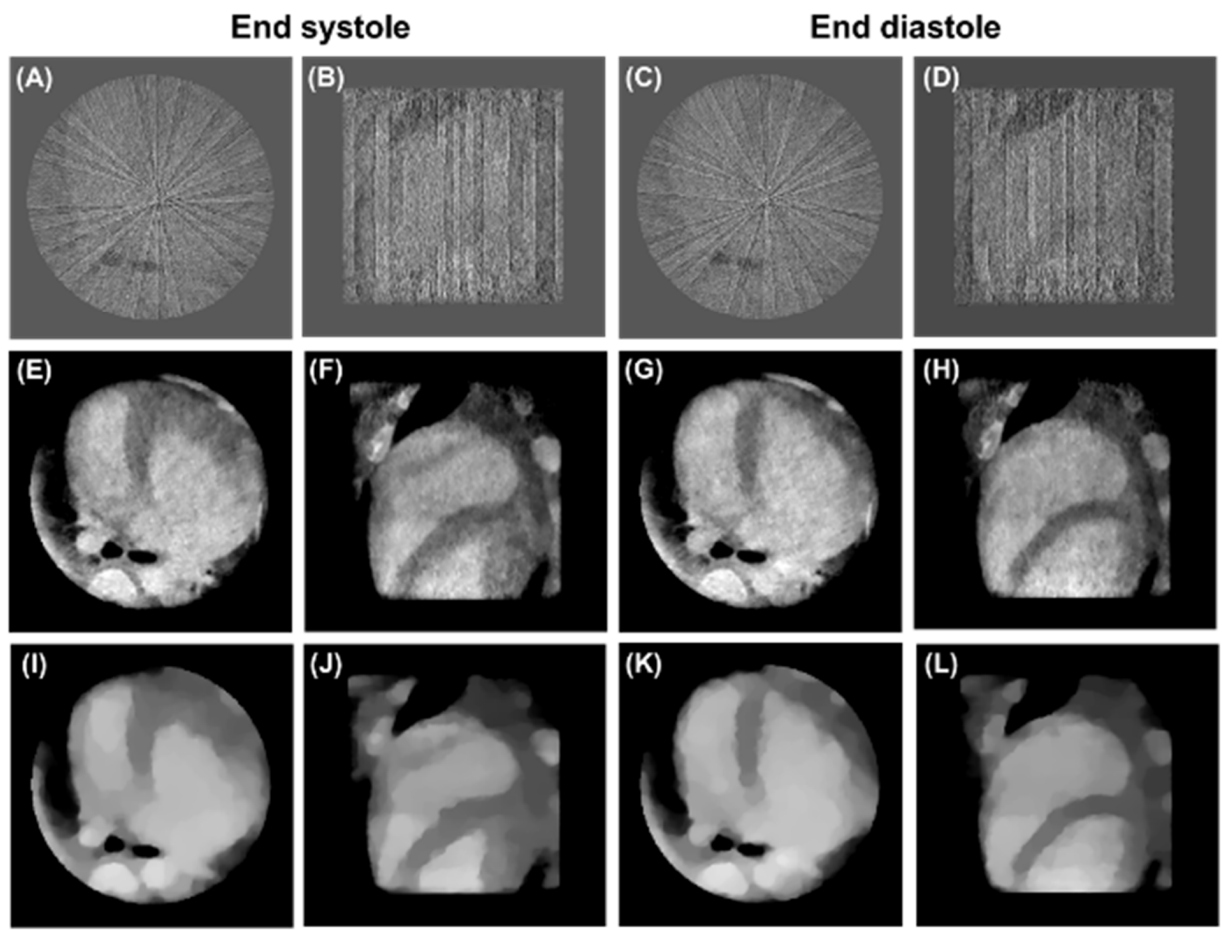

Figure 7.31: SBDX CT images of a 55 kg swine are shown for axial and coronal slices at end systole and end diastole for three reconstruction techniques. (A) Gated gFBP, axial, end systole. (B) Gated gFBP, coronal, end systole. (C) Gated gFBP, axial, end diastole. (D) Gated gFBP, coronal, end diastole. (E) PICCS, axial, end systole. (F) PICCS, coronal, end systole. (G) PICCS, axial, end diastole. (H) PICCS, coronal, end diastole. (I) TV-CS, axial, end systole. (J) TV-CS, coronal, end systole. (K) TV-CS, axial, end diastole. (L) TV-CS, coronal, end diastole.

## 7.4  Discussion

This chapter evaluated the accuracy of segmenting a high-contrast atrium phantom from SBDX CT scans under a range of imaging conditions. The effects of field-of-view truncation, C-arm rotational data acquisition, object motion frequency, and image noise were each considered in isolation and in combination to identify and quantify the sources of segmentation errors for four different reconstruction methods (gated gFBP, non-gated gFBP, TV-CS, PICCS).

SBDX CT segmentation errors were computed versus a reference atrium model segmented from an image obtained on a state-of-the-art helical CT scanner. A fraction of the SBDX CT segmentation error can be attributed to the registration procedure between SBDX CT derived surfaces and the reference surface. A bench-top SBDX CT experiment with a rotating stage was performed to establish baseline segmentation accuracy numbers to compare the more realistic phantom studies against. The atrium phantom was imaged without programmed motion or field-of-view truncation present. The 99[th] percentile segmentation error averaged 1.40 mm for reconstructions performed using gFBP. The 1.40 mm error may be attributed to the imperfect segmentation procedure, imperfect registration with the reference surface, image noise, and limited SBDX system and image voxel resolution. The FRE, for reference, was 0.84 mm.

The second experiment evaluated segmentation accuracy for SBDX CT images reconstructed with projection data acquired during atrium motion. CT data acquisition was performed with the rotary stage set-up to remove field-of-view truncation and geometric uncertainty resulting from C-arm deflection as confounding factors. The introduction of high-frequency object motion increased the 99[th] percentile segmentation error from 1.40 mm for the static atrium to 3.18 mm for *non-gated* gFBP reconstructions, averaged across all noise levels and motion frequencies. However, the application of retrospectively gated TV-CS and PICCS reconstruction methods to SBDX CT reduced the average 99[th] percentile segmentation error to 1.51 mm. Compared to the static atrium phantom, high-frequency object motion had a nearly negligible 0.11 mm (3.3% of the diameter of an ablation catheter tip) impact on segmentation accuracy, using the proposed iterative reconstruction scheme. The frequency of the object motion profile was varied as 60, 75, and 90 cycles per minute to evaluate the impact of different motion frequencies on segmentation accuracy. Varying the object frequency changes the number of superviews that may be gated at a particular cardiac phase for a constant CT data acquisition time. For example, 13 superviews were acquired at the gated phase for the 60 cycles per minute frequency versus 20 superviews for the 90 cycles per minute case. The TV-CS 99[th] percentile segmentation errors measured 1.53 mm, 1.46 mm, and 1.44 mm for imaging at 40 mA$_{peak}$ and motion frequencies of 60, 75, and 90 cycles per minute. A nearly negligible difference in segmentation error across motion frequencies was also observed in the numerical simulations presented in chapter 4 and discussed in section 4.4.3. Segmentation error increased slightly as the peak tube current decreased. The 99[th] percentile segmentation error averaged 1.47 mm, 1.50 mm, and 1.54 mm for TV-CS reconstructions of projection data acquired at peak tube currents of 40 mA$_{peak}$, 20 mA$_{peak}$, and 10 mA$_{peak}$. Segmentation error was maintained near 1.5 mm for each fluence by adjusting the λ reconstruction parameter. The optimal λ value, averaged across motion frequencies, decreased as 1333, 833, and 433 for the TV-CS reconstructions of projection data acquired at 40 mA$_{peak}$, 20 mA$_{peak}$, and 10 mA$_{peak}$, respectively.

For the third phantom experiment, C-arm rotational data acquisition was performed with the atrium phantom placed inside of an anthropomorphic thorax phantom. The purpose of the third phantom experiment was to evaluate the impact of field-of-view truncation and geometric uncertainty from C-arm deflection during rotation on segmentation accuracy, without object motion as a confounding factor. Segmentation errors averaged 2.46 mm for non-gated gFBP reconstructions across the three imaging fluences considered. Thus, segmentation errors increased by 1.06 mm on average compared to the static non-truncated atrium phantom results obtained with a rotary stage. The noise standard deviation measured in the atrium phantom was comparable ranging from 78-123 HU for the non-truncated phantom and 76-128 HU for the truncated phantom. The increase in segmentation error may be attributed to the combined effects of field-of-truncation and geometric uncertainty associated with C-arm rotation. The 1.06 mm increase in segmentation error exceeded the 0.11 mm increase attributed to high-frequency motion for the second phantom experiment. Future work may attempt to improve the geometric calibration method developed in chapter 5 to reduce the segmentation errors associated with C-arm rotational acquisitions. A brief discussion of methods to improve the calibration procedure is provided in section 5.6.

The final phantom study investigated the combined effects of field-of-truncation, C-arm rotational acquisition, and object motion on segmentation accuracy. For the non-gated gFBP reconstructions, 99% of the segmented surface points were within 5.74 mm of the reference surface for the 100 mA$_{peak}$ operating point. High-frequency motion increased the segmentation error by 3.35 mm on average compared to the 2.39 mm segmentation error observed for the static truncated atrium phantom at 100 mA$_{peak}$. Segmentations from TV-CS and PICCS reconstructions showed improved accuracy. Segmentation error averaged 3.28 mm for TV-CS and 3.57 mm for PICCS, at 100 mA$_{peak}$. In general, segmentation errors were comparable to the target goal of 3.3 mm. The PICCS performance, surprisingly, slightly underperformed TV-CS here.

The slight underperformance for PICCS relative to TV-CS is thought to be explained by two factors. First, reconstruction parameters were optimized for the high-contrast object segmentation task. The simple intensity based thresholding method favored low noise and smooth images. TV-CS images showed significant noise reduction and blurring of low contrast structures that would be unacceptable for most diagnostic imaging tasks but are advantageous for the high contrast segmentation task. Second, the prior image utilized by PICCS was corrupted by severe image artifacts. The atrium phantom and motion arm undergoing high-frequency motion nearly filled the entire SBDX CT FOV. As a result, artifacts present in the prior image persisted in the PICCS image. To improve PICCS performance, future work may consider pre-processing the prior image to reduce the artifacts associated with cardiac motion.[149]

Following the phantom studies, a porcine animal study was performed to demonstrate the feasibility of SBDX CT imaging for a realistic anatomy and cardiac motion. The acquired SBDX CT projections were successfully reconstructed at end-systole and end-diastole using the proposed reconstruction technique. Previous animal studies performed in IGCT have been limited by source output and thus focused on small animals such as rats or rabbits. The successful demonstration of IGCT for a large animal represents an important milestone towards clinical implementation of IGCT. Nonetheless, several limitations of the porcine study must be acknowledged. First, contrast injection was performed with a syringe by hand, leading to a lower than desired injection rate. Future work should use a power injector to improve chamber opacification and ensure constant flow. As a starting point, the injection protocols currently used in conventional C-arm CT with a power injector for left atrium imaging could be applied for SBDX C-arm IGCT. Second, no ECG signal was recorded for gating purposes. While the intrinsic gating scheme developed appears to have been successful, future work should investigate whether ECG-gating can provide superior results. Third, the impacts of tube potential, dose level, and C-arm rotation time on image quality should be investigated further. A single tube potential (100 kV) was considered for the results presented in this chapter. Segmentation errors and/or radiation dose could potentially be reduced by performing SBDX-CT imaging at a different tube potential. C-arm CT data acquisition required a 13.4 s rotational acquisition period for the results presented here. The C-arm rotation time could be decreased to lower the required breath-hold interval or to limit residual respiratory motion. A tradeoff exists, however, as decreasing the C-arm rotation time would decrease the number of complete heart beats occurring during CT data acquisition thereby increasing the angular undersampling within the gated dataset. Finally, in vivo SBDX-CT performance should be investigated over a range of animal sizes and heart rates and segmentation errors should be quantified versus reference scans performed on a diagnostic CT scanner. The simple intensity based thresholding segmentation algorithm employed for the phantom studies may not yield sufficient accuracy for *in vivo* images. Future work may consider alternative segmentation methods as a means to improve performance. For example, a recent comparison of left atrium segmentation accuracy for several algorithms concluded that performance could be improved using a combination of statistical shape models and region growing methods.[177]

## 7.5 Conclusion

This chapter evaluated SBDX CT performance in phantom studies for the task of 3D anatomic chamber mapping. The feasibility of SBDX CT for a realistic anatomy and cardiac motion was demonstrated in a porcine animal model. Ultimately, the motivation of performing SBDX C-arm CT is to derive a 3D cardiac chamber model that may be utilized in the SBDX catheter tracking display. Future animal studies are planned that will acquire an SBDX CT image in a porcine animal model, segment and display a cardiac chamber model in the SBDX catheter tracking display, and assess the system's utility in guiding a catheter ablation procedure.

# 8  Summary and future work

This work reports the development and successful demonstration of inverse geometry CT with a C-arm mounted Scanning Beam Digital X-ray system. Scanning Beam Digital X-ray is an inverse geometry fluoroscopy system being developed to improve image guidance and reduce radiation dose in cardiac interventions. Increasingly complex structural heart interventions and electrophysiology procedures require that catheter devices be navigated to precise anatomic targets located in relatively large cardiac chambers. The navigation task in these procedures is complicated compared to coronary based procedures as the devices are no longer constrained by narrow vasculature. Recently, a tomosynthesis based three dimensional catheter tracking algorithm was implemented in real-time on the SBDX system. Phantom studies showed real-time catheter tracking and display improved the target accuracy of navigation to points in a 3D space.[40] Previously acquired 3D models may be integrated with the SBDX catheter tracking display to supplement the poor soft tissue contrast associated with fluoroscopic imaging. However, potential changes in anatomy between the time of a pre-procedure imaging study and the interventional procedure, along with registration errors, motivate the development of *intra-procedural* CT imaging on the SBDX platform. The work completed in this dissertation will allow for three-dimensional models or roadmaps to be segmented from SBDX CT images and imported within the SBDX catheter tracking display. An alternative display option exists on SBDX to overlay the projection of an imported anatomic model with 2D fluoroscopy.

The major contributions of this work can be summarized as follows. First, reconstruction methods were developed to accommodate the unique SBDX inverse geometry design. The proposed reconstruction scheme mitigated image artifacts caused by field-of-truncation and projection undersampling resulting from cardiac gating. Second, a novel single-view geometric calibration method was implemented and shown to reduce artifacts associated with geometric uncertainty arising from C-arm deflection during rotational acquisitions. The single-view geometric calibration method enabled implementation of IGCT on the C-arm mounted SBDX system. Third, initial inverse geometry C-arm CT results were presented and evaluated in terms of gantry rotation reproducibility, dose index, image uniformity, and spatial resolution. The demonstration of C-arm based IGCT represents an important milestone as previous results were reported for bench-top systems or more stable closed-gantry based solutions. Fourth, SBDX CT performance was characterized for a high-contrast segmentation task to mimic the cardiac chamber roadmapping task. A series of phantom studies measured the segmentation accuracy of an atrium phantom for projection data obtained under varying imaging conditions including field-of-view truncation and programmed high-frequency motion. In general, segmentation accuracy was comparable to the diameter of a typical ablation catheter tip. The phantom studies also revealed areas that future work could concentrate on to improve results. And fifth, the feasibility of SBDX CT imaging for a realistic anatomy and cardiac motion was demonstrated in an animal model. The proposed reconstruction scheme was applied to successfully reconstruct a cardiac chamber at end systole as well as end diastole. The porcine study, to the best of the author's knowledge, represents the first *in-vivo* demonstration of IGCT in a large animal model.

Several areas for future work presented themselves over the course of this project. SBDX CT reconstructions were performed on an offline workstation. Projection data transfer times, and potentially long reconstruction times, could be reduced by performing SBDX CT reconstruction on the system's GPU based real-time image reconstructor. The real-time image reconstructor contains a total of 11 GPU devices that could deployed for CT reconstruction. The single-view geometric calibration method presented in chapter 5 should be extended as discussed to determine the detector orientation independent of the source array. Additional work should be performed to optimize the SBDX CT data acquisition protocol. The work presented in this dissertation was performed at a tube potential of 100 kV. Other tube potentials may offer improved segmentation accuracy or dose reduction and should be explored. Related to protocol optimization, one of the major motivating factors for diagnostic IGCT is the idea of a virtual or electronic bowtie to improve dose efficiency.[178] The regional adaptive exposure control method designed for SBDX fluoroscopic imaging could be investigated for CT imaging.[109] And finally, additional animal studies must be performed to validate SBDX CT performance. The use of SBDX CT derived roadmaps in combination with real-time catheter tracking should be explored in animal models.